\documentclass[acmsmall,screen,format=manuscript,nonacm]{acmart}
\usepackage{tabu}
\usepackage{graphicx}
\usepackage{float}
\usepackage{subfigure}
\usepackage{algpseudocode}
\usepackage[ruled, lined, linesnumbered, commentsnumbered, longend]{algorithm2e}
\usepackage{multirow}
\usepackage{caption}
\usepackage{comment}
\usepackage{soul}
\usepackage{bm} 
\usepackage{todonotes}
\usepackage{tcolorbox}
\usepackage{enumitem}
\usepackage{tikz}
\usepackage{etoolbox}
\usepackage{nicefrac}
\tcbuselibrary{skins, breakable, theorems}
\usepackage{multicol}
\usepackage{tablefootnote}
\usepackage{threeparttable}

\newcommand*\circled[1]{\tikz[baseline=(char.base)]{
            \node[shape=circle,draw,inner sep=0.8pt] (char) {#1};}}

\AtBeginDocument{%
  }

\setcopyright{acmcopyright}
\copyrightyear{2023}
\acmYear{2023}
\acmDOI{XXXXXXX.XXXXXXX}

\acmJournal{TOSEM}
\acmVolume{0}
\acmNumber{1}
\acmArticle{1}
\acmMonth{1}

\begin{document}
\title{Context-Aware Fuzzing for Robustness Enhancement of Deep Learning Models}

\thanks{The official version of this paper is to appear in ACM Transactions on Software Engineering and Methodology (accepted in July 2024).}

\author{HAIPENG WANG}
\email{haipewang5-c@my.cityu.edu.hk}
\orcid{0000-0002-7410-393X}
\affiliation{%
  \institution{City University of Hong Kong}
  \city{Kowloon Tong}
  \state{Hong Kong}
  \country{China}
}

\author{ZHENGYUAN WEI}
\email{zywei4-c@my.cityu.edu.hk}
\orcid{0000-0001-5966-1338}
\affiliation{%
  \institution{City University of Hong Kong}
  \city{Kowloon Tong}
  \state{Hong Kong}
  \country{China}
}

\author{QILIN ZHOU}
\email{qilin.zhou@my.cityu.edu.hk}
\orcid{0000-0003-2289-9849}
\affiliation{%
  \institution{City University of Hong Kong}
  \city{Kowloon Tong}
  \state{Hong Kong}
  \country{China}
}

\author{WING\,KWONG CHAN}
\authornote{Corresponding Author}
\email{wkchan@cityu.edu.hk}
\orcid{0000-0001-7726-6235}
\affiliation{%
  \institution{City University of Hong Kong}
  \city{Kowloon Tong}
  \state{Hong Kong}
  \country{China}
}

\renewcommand{\shortauthors}{Wang et al.}

\begin{abstract}
In the testing-retraining pipeline for enhancing the robustness property of deep learning (DL) models, many state-of-the-art robustness-oriented fuzzing techniques are metric-oriented. 
The pipeline generates adversarial examples as test cases via such a DL testing technique and retrains the DL model under test with test suites that contain these test cases. 
On the one hand, the strategies of these fuzzing techniques tightly integrate the key characteristics of their testing metrics.
On the other hand, they are often unaware of whether their generated test cases are different from the samples surrounding these test cases and whether there are relevant test cases of other seeds when generating the current one. 
We propose a novel testing metric called \textit{Contextual Confidence} (CC).
CC measures a test case through the surrounding samples of a test case in terms of their mean probability predicted to the prediction label of the test case.
Based on this metric, we further propose a novel fuzzing technique \textsc{Clover} as a DL testing technique for the pipeline.
In each fuzzing round, \textsc{Clover} first finds a set of seeds whose labels are the same as the label of the seed under fuzzing. 
At the same time, it locates the corresponding test case that achieves the highest CC values among the existing test cases of each seed in this set of seeds and shares the same prediction label as the existing test case of the seed under fuzzing that achieves the highest CC value. 
\textsc{Clover} computes the piece of difference between each such pair of a seed and a test case. 
It incrementally applies these pieces of differences to perturb the current test case of the seed under fuzzing that achieves the highest CC value and to perturb the resulting samples along the gradient to generate new test cases for the seed under fuzzing.
\textsc{Clover} finally selects test cases among the generated test cases of all seeds as even as possible and with a preference to select test cases with higher CC values for improving model robustness. 
The experiments show that \textsc{Clover} outperforms the state-of-the-art coverage-based technique \textsc{Adapt} and loss-based fuzzing technique \textsc{RobOT} by 67\%--129\% and 48\%--100\% in terms of robustness improvement ratio, respectively, delivered through the same testing-retraining pipeline.
For test case generation, in terms of numbers of unique adversarial labels and unique categories for the constructed test suites, \textsc{Clover} outperforms \textsc{Adapt} by $2.0\times$ and $3.5\times$ and \textsc{RobOT} by $1.6\times$ and $1.7\times$ on fuzzing clean models, and 
also outperforms \textsc{Adapt} by $3.4\times$ and $4.5\times$ and \textsc{RobOT} by $9.8\times$ and $11.0\times$ on fuzzing adversarially trained models, respectively.
\end{abstract}

\begin{CCSXML}
<ccs2012>
   <concept>
       <concept_id>10011007.10011074.10011099.10011102.10011103</concept_id>
       <concept_desc>Software and its engineering~Software testing and debugging</concept_desc>
       <concept_significance>500</concept_significance>
       </concept>
   <concept>
       <concept_id>10003033.10003083.10003095</concept_id>
       <concept_desc>Networks~Network reliability</concept_desc>
       <concept_significance>500</concept_significance>
       </concept>
 </ccs2012>
\end{CCSXML}

\ccsdesc[500]{Software and its engineering~Software testing and debugging}
\ccsdesc[500]{Networks~Network reliability}
\keywords{context-awareness, fuzzing algorithm, robustness, assessment, metric}

% \received{}
% \received[revised]{}
% \received[accepted]{}
\maketitle

\section{Introduction}
\label{sec: introduction}
Deep learning (DL) has been widely applied in many application domains, such as autonomous driving systems \cite{deeproad, deep_billboard}, protein structure prediction \cite{protein_prediction, improved_protein_prediction}, and healthcare \cite{dl_medical_image, dl_covid19, dl_healthcare}.
However, it is well-known that even slight perturbations on clean samples may trigger DL models to misbehave. 
If an application adopts a DL software component with inadequate ability to protect against adversarial examples, it can lead to severe consequences (e.g., accidents with causality \cite{google_accident, telsta_accident} in the domain of autonomous driving).

A popular way to rectify model misbehavior with respect to the robustness (i.e., the concerned quality attribute of the model under test) is to retrain the concerned DL model with a mixture of the original training data and those failing test cases that trigger the misbehavior of the DL model.
The pipeline of generating failing test cases for the DL model under test followed by retraining the model with the original training data and a test suite that contains these test cases \cite{robot}, which we refer to as the \textit{testing-retraining pipeline} or the pipeline for short, presents a scenario different from how test cases in traditional testing on traditional programs (e.g., C++/Java programs) assist in program testing \cite{coverage_metric_book}, repair \cite{context_patch_repair}, and retesting \cite{regression_test}.
In the traditional program testing domain, typically, (failing) test cases that expose failures are also used to retest the repaired programs to validate that the repaired programs can pass them.
Thus, developers can use test cases for at least two purposes: failure exposure of the program under test and validation of the behavior of the repaired program.
In particular, the retest task on the repaired program with these test cases meaningfully serves the purpose of validating the behavior of the repaired program, albeit the validation also uses the inputs that the program under test should pass.
Furthermore, developers often use these test cases for the program repair task between the test and retest tasks.

In the DL model testing domain, however, if a DL model has been retrained with the test cases produced by a DL testing technique (e.g., \textsc{RobOT} \cite{robot}) so that the expected prediction for the test cases should be learned, the resulting retrained DL model has been taught to fit its behavior to these trained (and originally failing) test cases well due to the memorization effect \cite{memorization_pmlr, memorization_neurips} of the DL model through a retraining process.
Thus, retesting the retrained model with these test cases becomes not very meaningful (and is likely misleading to indicate the extent of robustness improvement achieved by the retrained model).
To address this issue, apart from using the testing-retraining pipeline, developers validate the DL model under test and the retrained model with a standalone set of samples (which we refer to as the \emph{robust validation dataset}) that can be provided by users or generated by some standalone techniques probably independent of the DL testing technique in the pipeline \cite{robot}%
\footnote{
We note that a testing-retraining pipeline (e.g., the ones presented in experiments such as \cite{robot, deeprepair, deepgini, ATS2022}) targets to smooth those ``bugs'' (model misclassification behavior exposed by adversarial examples generated by the DL testing technique in the pipeline) by a retraining process. 
There are other types of ``bugs'', such as code implementation bugs, platform bugs, and buggy (non-optimized) neural network architecture.
These ``bugs'' are exposed and improved by other pipelines/workflows with other types of SE techniques: E.g., AutoTrainer \cite{autotrainer} for training issues that prevent a training process from normal execution or updating the weights, Duo \cite{duo_differential_fuzzing} for testing the presence of bugs in DL libraries, DeepDiagnosis \cite{DeepDiagnosis} for performance bug detection, and DeepPatch \cite{deeppatch} and DeepRepair \cite{deeprepair} for patching the network architecture or smoothing the parameters to produce the ones with higher performance for accuracy-robustness tradeoff, to name a few. Fuzzing techniques, including our proposed one, complement these works to make the testing of DL models for the improvement of respective quality attributes via the respective pipelines more comprehensive.
}.

In other words, a DL testing technique in the pipeline has at least two main purposes: failure exposure from the DL model under test and supplying test cases for model retraining for the purpose of improving the robustness property of the DL model under test.
Since the role of validation has been taken up by a robust validation dataset in the overall workflow \cite{robot}, like the related works \cite{robot, sensei, seed_select_dl_test, deepgini, quote_robot_extend, surprise_adequacy, deeptest, test_selection_tosem} in the software engineering (SE) literature, one of the main values of a DL testing technique in the pipeline is to generate test cases for model retraining to mitigate the threat of inadequate robustness generalized from these test cases exposed by the DL testing technique.
To know the relative impact of the test cases on the robustness property of the DL model under test, the robust accuracies before and after the retraining task can be compared to ensure that a gain in robust accuracy can be observed after model retraining, which serves as a validation of whether the test cases producible by the DL testing technique can lead to positive effects on the robustness property of the DL model under test delivered through the pipeline.
For instance, the \textsc{QuoTe} framework \cite{quote_robot_extend} presents a workflow with the testing-retraining pipeline, where the workflow applies the pipeline to generate test cases via a DL fuzzing technique (e.g., \textsc{RobOT} \cite{robot}) guided by a testing metric followed by retraining the DL model using a set of these test cases and the original training dataset. After that, the workflow measures the robust accuracy improvement achieved by the retrained model on a standalone robust validation dataset (which is a dataset generated by attacker techniques \cite{fgsm, pgd} or fuzzing techniques \cite{adaptfuzz, dlfuzz}).

Indeed, there has been great interest in the SE community to formulate SE techniques aiming at quality assurance and improvement of DL models via testing \cite{quote_robot_extend}.
Among those quality attributes for DL models, robustness is widely studied by the community to address threats like adversarial examples by generating test cases and improving the robustness of DL models with these test cases delivered through testing-retraining pipelines \cite{deepconclic, deepcorrect, deeppatch, deepgauge, deepfault, deepgini, deephunter, deepmutation++, robot, adaptfuzz, sensei, deeprepair, deeproad, ATS2022, test_selection_tosem}. 

To help developers assess the quality of DL models with the aim of enhancing the robustness property of DL models, apart from DL testing techniques, DL verification techniques \cite{formal_verification, deeppac, scalable_verification} are also under active research. DL verification techniques, such as formal verification \cite{formal_verification}, approximation verification \cite{deeppac}, statistical verification \cite{scalable_verification}, and certification \cite{patchcensor24, crosscert, zhou2023majority}, check whether a DL model satisfies or violates a given sample-level property and provide a strong guarantee for a verified sample. But, verifying an arbitrary DL model for all possible inputs remains an open problem. 

Recently, DL testing techniques \cite{deepxplore, dlfuzz, deepgini, robot, deeptest, surprise_adequacy, quote_robot_extend, sensei, test_selection_tosem} with the purpose of robustness improvement delivered through the testing-retraining pipeline are emerging to give insights into how to judge samples more likely to carry informative clues for the misbehavior or robustness of DL models under test.

In particular, many fuzzing techniques (e.g., \cite{robot, quote_robot_extend, deepxplore, drfuzz, sensei}), as a kind of DL testing technique, in such pipelines exhibit at least two key characteristics \cite{quote_robot_extend}. 
First, they include a testing metric to assess a given testing property of a test case or a test suite. 
Existing works have formulated both coverage-based testing metrics, such as measuring the number of neurons activated in the forward pass for a test case or a test suite (aka measuring the neuron coverage) ~\cite{deepxplore, adaptfuzz} and non-coverage-based ones, such as measuring the loss value of a test case (but the loss values of different test cases are independent with one another) \cite{robot, sensei}.
They select or generate samples by increasing the selective coverage over internal neural network states (e.g., neurons in ~\cite{deepxplore, adaptfuzz} and outliners in ~\cite{surprise_adequacy}), selecting a subset among many variants of the same seed (e.g., variants incurring largest absolute \cite{sensei} and smallest relative \cite{robot, quote_robot_extend} losses), or distinguishing more or diverse model mutants~\cite{deepmutation, deepmutation++, prima, effimap}. 

The second characteristic is to use the testing metrics to guide their procedures for various testing steps, such as seed selection \cite{deephunter}, sample mutation \cite{deepmutation++}, or the number of evolution attempts \cite{robot}.
However, applying coverage-based testing metrics to guide DL testing techniques may not be consistently effective. 
Existing empirical findings show that their adopted coverage criteria could be either too easy to satisfy or require demanding tuning to reach high coverage rates \cite{is_nc_useful, deepgini}. Still, the correlations between the coverage achieved by a test suite constructed by these techniques and the failure proneness of the test suite demonstrated via their testing-retraining pipelines are weak, and the robustness improvement is small \cite{structual_coverage, coverage_model_quality}. 
Loss-based fuzzing techniques \cite{robot, quote_robot_extend, sensei} gradually evolve individual seeds toward the side of larger values of an adopted loss function.
They deem the generated samples with either the largest absolute loss \cite{sensei} or the smallest relative loss ~\cite{robot, quote_robot_extend} to be test cases of higher quality or pick test cases from different parts of an ordered set of test cases (e.g., via the \textsc{km-st} strategy in ~\cite{robot}).
For instance, in our experiment (Section \ref{sec: result_in_configuration_A}), we have compared \textsc{km-st} using its original loss-based metric (known as the first-order loss) \cite{robot} with random selection over the same pool of adversarial examples (aka a pool of test cases) using the same testing-retraining pipeline for robustness improvement.
We find that their effects on robust accuracy improvement are similar to each other, indicating that the guiding effect of using the loss-based testing metric may not be observable compared to the random selection, which is consistent with the previous finding that ``the Random selection strategy performs surprisingly well in some cases, which is close to KM-ST'' expressed by Chen et al. on discussing their experimental results \cite{quote_robot_extend}.
Mutation-based fuzzing techniques \cite{duo_differential_fuzzing, effimap} are computationally expensive.

Following the fuzzing works \cite{deepxplore, dlfuzz, robot, deeptest, sensei} for DL testing techniques in the literature, in this paper, we adopt the testing-retraining pipeline with the purpose of improving the robustness property of DL models via testing to be the scope of our proposed technique. We focus on the DL testing technique in the pipeline.
We refer to the robustness improvement achieved by the retrained model with the test suite constructed by a DL testing technique and the original training dataset as the robustness improvement delivered through the pipeline.

We propose \textsc{Clover}, a novel fuzzing technique for DL testing.
The technique is built on two main ideas.

First, we design a novel testing metric called \textit{Contextual Confidence}, CC for short (see Eq. (\ref{eq: cc_short})) --- CC measures a test case through the consensus achieved by the surrounding samples of a test case, which is their mean probability predicted to the prediction label of the test case. 
Like the testing metric in DeepGini \cite{deepgini}, CC is a black-box non-coverage-based metric. 
The general background of CC is that it is pretty easy for an iterative gradient-based strategy (e.g., PGD \cite{pgd}) to generate many test cases with high prediction confidence (e.g., very close to 1) from the given samples if time allows. 
Our insight is that adding small uniform perturbations \cite{deepmutation, deepmutation++, sensei} to a test case of a seed can easily and efficiently produce perturbed samples with different degrees of consensus with respect to the test case that produces them, and some test cases generated from a seed may obtain a stronger consensus (in the sense of higher CC value) than some other test cases of the same seed. 
Suppose a DL model is confident in predicting the labels of some test cases. In that case, those test cases with higher CC values likely indicate higher chances that the DL model generalizes its predictions from these test cases to their surrounding samples.
(Our experimental result in Section \ref{sec: result_in_configuration_A} also shows that the test suite construction strategy that prefers test cases with higher CC values significantly outperforms random selection in robustness improvement delivered through the pipeline.)

\begin{figure} [tb]
\includegraphics[width=0.75\textwidth]{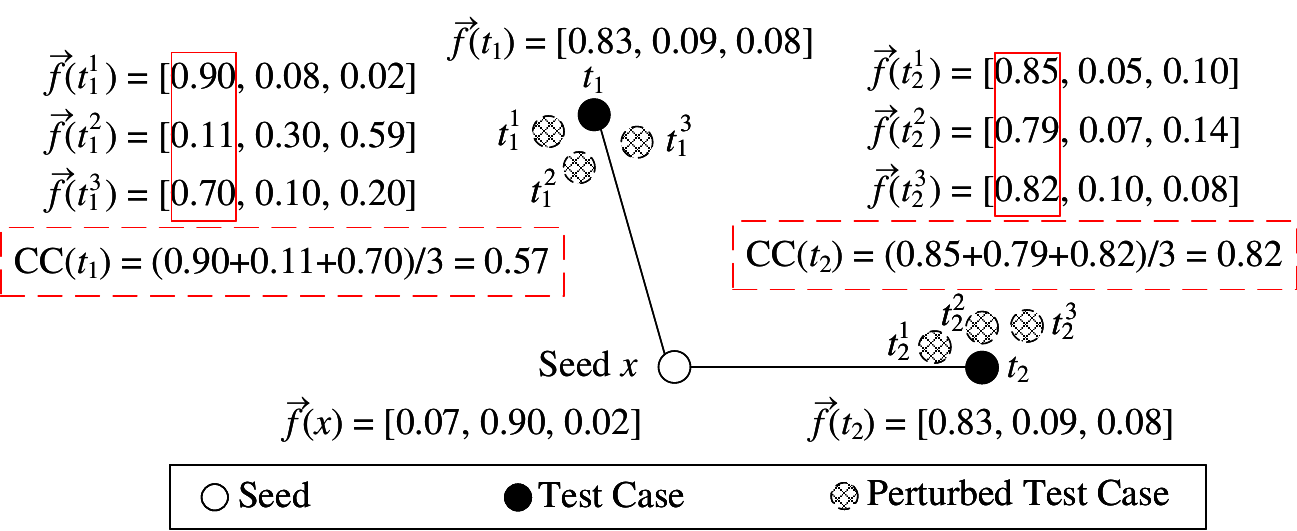}
\caption{Illustration for Calculating CC Values for Different Test Cases Generated from the Same Seed.}
\label{fig: cc_values}
\end{figure}

Fig. \ref{fig: cc_values} illustrates how to calculate the CC values for different test cases depicted by filled circles (i.e., $t_1$ and $t_2$) generated from the same fuzzing seed $x$ depicted by an empty circle.
The vector, in the style of $\vec{f}(s)$, depicts the prediction vector of a DL model $f$ on predicting the label of a sample $s$.
It has three values standing for the probabilities of three classes (1, 2, and 3). 
The prediction labels of $x$, $t_1$, and $t_2$ are 2, 1, and 1, respectively. 
We refer to a test case's prediction label as the test case's \textit{adversarial label}.
The samples (perturbed test cases) surrounding $t_1$ and $t_2$ are depicted by dashed filled circles: $t_1^1$, $t_1^2$, and $t_1^3$ for $t_1$; and $t_2^1$, $t_2^2$, and $t_2^3$ for $t_2$. 
The prediction probabilities of the three samples surrounding $t_1$ for the adversarial label of $t_1$ (i.e., class 1) are 0.90, 0.11, and 0.70. As illustrated in Fig. \ref{fig: cc_values}, the CC value for $t_1$ is computed as 0.57. Similarly, the CC value for $t_2$ is computed as 0.82.
The example illustrates that the two test cases have the same prediction vector, but their CC values could be quite different.

Second, we aim to evolve a test case of a seed (under fuzzing in the current fuzzing round) with a lower CC value into a test case (of the same seed) with a higher CC value incrementally.
As such, we can choose the latter test case instead of the former one in constructing a test suite for improving the robustness property of the DL models.
Specifically, to generate test cases in the current fuzzing round on a seed (denoted by $x$), we first retrieve the test case that yields the highest CC value among all already generated test cases of the seed $x$. 
We further retrieve the seeds that are ``similar'' to $x$: their labels are the same as the label of $x$, and their test cases with the highest CC values are all predicted to the same prediction label of the above test case of $x$. We refer to these seeds as seed-equivalent to one another at the current fuzzing round.
We then find the piece of difference for each such seed-equivalent seed of $x$ (denoted by $y$) by subtracting the seed $y$ from the test case that yields the highest CC value among all already generated test cases of the seed $y$ (denoted by $\overline{y}$), which determines the seed being seed-equivalent to $x$, i.e., the difference is $\overline{y} - y$.
We then retrieve the test case of the current fuzzing seed $x$ that achieves the highest CC value and perturb the test case with one of such pieces of differences, followed by further perturbing the resultant sample to generate a new test case.
We repeat the above test case generation procedure with the remaining pieces of differences one after another until the budget for the current fuzzing round exhausts.
During this iterative process, we update $x$'s test case with the highest CC value if found. 
Our insight is that, as the fuzzing campaign continues, the seeds that are seed-equivalent to one another at the respective fuzzing rounds gradually capture the fuzzing technique's most successful relevant experiences to evolve them toward the sharp end (in the sense of generating test cases with high CC values).
Moreover, through seed equivalence, we can also explore different ``successful'' directions when perturbing a seed.
As such, we use their corresponding pieces of differences to guide the generation of the test cases of the current fuzzing seed.
The above-mentioned test case generation produces test cases in ascending order of CC values for the same seed. 
Thus, to obtain a test suite containing test cases with higher CC values, developers can assign a longer time budget to allow the generation procedure to generate a larger test pool.

We implement the two main ideas in our fuzzing technique \textsc{Clover} and configure the same testing-retraining pipeline adopted by \textsc{robot} \cite{robot} to compare \textsc{Clover} with peer techniques.
The experimental results show the high effectiveness of \textsc{Clover}.
In terms of the number of generated test cases as well as the number of unique adversarial labels \cite{adaptfuzz, ATS2022, drfuzz} and the number of unique categories \cite{adaptfuzz, deephunter, ATS2022, drfuzz} of test suites with the same sizes, \textsc{Clover} outperforms the peer techniques significantly --- 
(1) on clean models, by $2.5\times$, $2.0\times$, and $3.5\times$ compared to \textsc{Adapt} and $3.6\times$, $1.6\times$, and $1.7\times$ compared to \textsc{RobOT}, respectively, and
(2) on adversarially trained models, by $2.1\times$, $3.4\times$, and $4.5\times$ compared to \textsc{Adapt} and $9.2\times$, $9.8\times$, $11.0\times$ compared to \textsc{RobOT}, respectively. 
Moreover, configuring \textsc{Clover} with CC is generally more effective than its two variants configuring with the testing metric (i.e., the Gini index) proposed in \textsc{DeepGini} \cite{deepgini} and the first-order loss (FOL) testing metric proposed in \textsc{RobOT} \cite{robot, quote_robot_extend} in terms of the above three measurements.
\textsc{Clover} also shows a greater potential in improving the robustness property of clean models via the testing-retraining pipeline\,%
\footnote{
We note that a vast majority of existing works \cite{adaptfuzz, robot, tensorfuzz, deepxplore, dlfuzz, sensei, tensorfuzz, deephunter, deeptest, ATS2022, drfuzz} only measure the performance of the fuzzing techniques on clean models.} than the peer techniques by large extents ---
(1) our testing metric CC boosts the robustness improvements by 11\%--18\% atop \textsc{Random}, 40\%--45\% atop \textsc{DeepGini} \cite{deepgini}, 17\%--39\% atop \textsc{be-st} \cite{robot}, and 12\%--18\% atop \textsc{km-st} \cite{robot},  
(2) \textsc{Clover} outperforms \textsc{Adapt} \cite{adaptfuzz} by 67\%--129\%  and \textsc{RobOT} \cite{robot} by 48\%--100\% in terms of robustness improvement ratio,
(3) configuring \textsc{Clover} with CC is more effective than its two variants that are configured with the two testing metrics in \textsc{DeepGini} and \textsc{RobOT}, respectively, and
(4) configuring \textsc{Clover} with the same loss-based metric of \textsc{RobOT} also outperforms the original \textsc{RobOT}.
We have also included a case study that compares \textsc{Clover} with peer techniques on fuzzing adversarially trained models in terms of the ability to generate test cases using the three measurement metrics. 
The results further consolidate the above finding on clean models that \textsc{Clover} generates more diverse robustness-oriented test suites than \textsc{Adapt} and \textsc{RobOT} in addition to generating significantly more test cases.

The main contribution of this paper is threefold: 
(1) It proposes the novel testing metric \textit{Contextual Confidence}.
(2) It proposes the novel fuzzing technique \textsc{Clover} to improve the robustness property of DL models. \textsc{Clover} can be used together with a retraining task to constitute the testing-retraining pipeline for the robustness improvement of DL models via testing.
(3) It shows the feasibility and high effectiveness of \textsc{Clover} through a comprehensive experiment.

We organize the rest of this paper as follows.
Section 2 revisits the preliminaries. 
Section 3 presents \textsc{Clover}.
Sections 4 to 8 report an experiment, its results, and data analysis to evaluate \textsc{Clover}.
We review the closely related work in Section 9 and conclude this work in Section 10.

\section{Preliminaries}
\label{sec: preliminaries}
\subsection{Deep Neural Networks}
A deep learning (DL) model $f$ is a function that takes a sample $x$ from an input domain as input and outputs a \textbf{prediction vector} $\vec{f}(x)$, in which the probability of each component $c$ in the vector is denoted by $f_{c}(x)$ and $c$ is an element in a label set $\mathbb{C}$.
The \textbf{prediction label} $f(x)$ is the component with the highest probability, defined as $f(x) = argmax_c \ f_c(x)$.
We denote the ground truth of $x$ by $g_x$. A sample $x$ is called \emph{correct} if $f(x) = g_x$, otherwise \textbf{failing}. 

An adversarial example $t$ \cite{hard_ae, fgsm, pgd, evaluate_robustness_nn, robot} is an input sample $x$ added with a minor (such as human-imperceptible) difference, yet the model $f$ predicts it with a different label (i.e., $f(x) \neq f(t)$).
We refer to the prediction label of $t$, denoted as $\boldsymbol{v}$, as an \textbf{adversarial label}, i.e., $v = f(t) = argmax_c f_{c}(t)$.

The \textbf{accuracy} of $f$ on a dataset $D$ is the proportion of $D$ that $f$ correctly predicts, i.e., $\textsc{acc}(f) = \nicefrac{| \{ x \in D | f(x) = g_x \}|}{|D|}$.
A robustness-oriented dataset is a dataset containing adversarial examples for a DL model \cite{robot}.
The training, test, or robust accuracy is the accuracy of $f$ on a training, test, or robustness-oriented dataset, respectively.
Suppose a model $f'$ is produced by retraining $f$ with adversarial examples. 
The robust accuracy difference, i.e., $\textsc{acc}(f') {-} \textsc{acc}(f)$, on a robustness-oriented dataset is called \textbf{robustness improvement}.

The \textbf{gradient} $\nicefrac{\partial \textit{obj}}{\partial x}$ of a model for an optimization objective \textit{obj} on a sample $x$ indicates the change direction of $x$ that increases the objective value most quickly \cite{fgsm, pgd}.

A DL fuzzing technique accepts a seed list $X$ as input and outputs a test suite $A$, and a test case selection technique accepts a test pool $P$ and outputs a subset $A \subseteq P$. 
A \textbf{testing-retraining pipeline} (pipeline for short) configured with a fuzzing technique or a selection technique as the DL testing technique is to apply the DL technique to produce a test suite $A$ followed by retraining $f$ with $A$ and the original training dataset.

A \textbf{robust validation dataset} is a set of perturbed samples aiming at validating the extent of a DL model to be resilient to predicting its samples to adversarial labels \cite{quote_robot_extend}.
We note that the set of adversarial examples for retraining, the robustness-oriented dataset for measuring the robustness improvement, and the robust validation dataset mentioned in Section \ref{sec: introduction} refer to three different concepts:
The first one is for model retraining; 
the second one is a test dataset to measure the robustness improvements in the experiments; 
the third one is a dataset outside the testing-retraining pipeline used to check against the user-specified requirement in the overall workflow \cite{robot}.

\subsection{DL Testing with the Target of Robustness Improvement for Deep Learning Models}
\subsubsection{Why Selecting Test Cases for Improving the Robustness Property of DL Models?}
Adversarial examples \cite{hard_ae, fgsm, pgd, evaluate_robustness_nn, robot} widely exist in the input domains of DL models. Any DL model $f$ should be quality assured and improved, if needed, against them prior to deployment. The SE community has shown a strong interest in applying SE techniques to enhance the quality of DL models through quality assurance and improvement. As we have presented in Section~\ref{sec: introduction}, many DL testing techniques focus on assisting DL models in mitigating the threat of adversarial examples through improving the robustness property of the DL models via a testing-retraining pipeline \cite{quote_robot_extend}.

Suppose that the robustness of $f$ has been found to be unsatisfactory when measured through a robust validation dataset.
One approach to robustness improvement of $f$ is first to apply a set of seed samples for a DL fuzzing technique to produce a test suite $A$ that contains their perturbed versions.
A typical strategy is to strengthen the empirical defense of $f$ against $A$, such as by retraining $f$ with $A$ and the original training dataset \cite{robot, quote_robot_extend, ae_train_free, fast_ae_train, adversarial_training_1, adversarial_training_2, deeptest, test_selection_tosem}, developing a better defender/input validator to guard $f$ against samples demonstrated by $A$ \cite{nic, dissector, ood_detect, ae_detect}, or patching $f$ through a maintenance technique \cite{deeppatch, deepcorrect, arachne, PatchNAS}. 
Among them, model retraining is the normal and most widely practiced option to mitigate the threats of the exposed adversarial examples on the robustness property of $f$.
Retraining $f$ with more adversarial examples will reduce the test accuracy, further imposing a natural limit on the number of adversarial examples contained in $A$, albeit popular to make a tradeoff to improve the robust accuracy \cite{robot, quote_robot_extend, ae_train_free, fast_ae_train}. 
For instance, in the SE literature, the test accuracies before and after retraining are compared to validate that the robustness improvement is observed and the accuracy-robustness tradeoff is not severe (e.g., controlled within 1\% difference in test accuracy \cite{robot, quote_robot_extend, sensei} or through retraining the model with a test suite whose size is at most 10\% of a training dataset \cite{deepgini, robot, quote_robot_extend, seed_select_dl_test}).
The nature of software testing for DL testing in the testing-retraining pipeline further complicates these constraints on $A$.

A testing-retraining pipeline configured with a fuzzing technique as the DL testing technique can easily generate many test cases from many different seeds.
A larger fuzzing budget will enable the DL testing technique to generate more test cases.
At the same time, the size of the subset $A$ is constrained to avoid producing a severe tradeoff between clean accuracy and robust accuracy, and the time budget for the pipeline is often limited.
These factors inherently limit such a DL testing technique from exploring the effects of many different test case compositions of $A$ on the tradeoff by trial and error.
This is vital to determine and include more representative test cases when constructing the subset $A$, even for the same seed.

The fundamental research question is:
\begin{center}
\emph{
How do we select more representative test cases in constructing a test suite $A$ for robustness improvement through a testing-retraining pipeline? 
}
\end{center}

\subsubsection{Related Methods for Selecting Test Cases and their Limitations}
\label{sec: related_methods_limitations}
A wide range of testing techniques and testing metrics have been developed in the literature for the purpose of answering the research question posed in the last subsection. In this section, we revisit the representative ones closely related to our proposed technique as our experiment compares our technique with them. We note that a detailed description of these techniques can be found in the appendix of this paper.

Following the terminology used in the DL fuzzing domain, we refer to the input sample to a technique as a \textbf{seed} and a perturbed sample output by the technique as a \textbf{test case}.

FGSM \cite{fgsm} and PGD \cite{pgd} are popular attacker techniques. They have frequently been used as the baselines to validate other techniques \cite{robot, quote_robot_extend, deepgini}.
They can easily produce many hard adversarial examples \cite{fgsm, pgd, hard_ae}, which are samples predicted to the adversarial labels with high confidence (e.g., 99.99\%).
For instance, for the testing-retraining pipeline presented in \cite{quote_robot_extend, robot}, these attacker techniques are used to generate a combined dataset to compare the robust accuracies of the original models and the respective retrained versions in their experiments.

However, these attacker techniques do not \emph{select} \emph{more representative} test cases in their generation processes. 
FGSM merely adds perturbations along the gradients to individual seeds and does \emph{not} discriminate the effects of the resultant (failing) test cases. 
PGD is a compositional gradient-based variant of FGSM. It incrementally adds a series of perturbations (along the gradients) to a seed without any selection or discrimination of intermediate perturbed sample versions during the construction process.

Unlike the above two techniques, \textbf{\textsc{DeepGini}} \cite{deepgini} is a representative metric-based technique that does not generate any test cases ---
It purely selects test cases among the given ones. 
It computes the Gini index \cite{theoretical_comparison} of the prediction vector for each test case.
\textsc{DeepGini} hypothesizes a test case with a higher Gini score is more likely to be misclassified, thereby ranking test cases in descending order of the Gini score. The top-ranked test cases are selected to construct a test suite.
Their experiment \cite{deepgini} shows that such test suites in the testing-retraining pipeline produce higher robust accuracy than those produced by coverage-based techniques \cite{deepxplore, deepgauge, surprise_adequacy}, showing its test case discrimination effect with respect to the robustness property of DL models.
Nonetheless, if the given test cases are already failing, the above hypothesis becomes inapplicable, making the selection of the representative ones among the given failing test cases unclear in concept.
Our experimental results in Section \ref{sec: result_in_configuration_A} also could not find the evidence to support test suites containing failing test cases with higher Gini scores producing larger robustness improvement than randomly constructed test suites from the same test pool through a testing-retraining pipeline.
It indicates that the technique may not effectively discriminate failing test cases yet.

A popular approach to introducing the discrimination effect in test case generation is incrementally adding a series of \emph{selective} perturbations to a seed. 
To our knowledge, many fuzzing techniques for fuzzing DL models fall into this category.
For instance, a coverage-based fuzzing technique \cite{adaptfuzz, dlfuzz} often adds one or more perturbations to each intermediate version of a test case to produce one or more candidates for the next version and selects the candidate covering more not-yet-covered coverage items (e.g., more uncovered neurons) \cite{deepxplore, deepgauge, surprise_adequacy} as the next intermediate version of the test case.
In essence, these techniques consider test cases of the same seed covering more coverage items of a coverage criterion more representative.
Nonetheless, as presented in the literature \cite{is_nc_useful, structual_coverage, coverage_model_quality}, the test suites thus produced by these coverage-based techniques are not obviously correlated with the exposure of failures or the robustness of DL models.
The results in the literature indicate that the metrics to guide the generation of test cases and the metrics for test case selection should be carefully designed to be effective.

\textbf{\textsc{RobOT}} \cite{robot, quote_robot_extend} is a state-of-the-art loss-based fuzzing technique. 
It incrementally generates intermediate test cases per seed with an increasingly larger loss distance.
Such a loss distance is measured by the first-order loss (FOL) \cite{robot}, computed as the Euclidean norm \cite{euclidean_norm} of the difference between an intermediate test case and its corresponding seed, where a smaller FOL value represents a smaller distance.
All intermediate test cases that are failing are output as test cases for selection. 
Like PGD, \textsc{RobOT} adds a perturbation along the corresponding gradient of a sample (seed or not) to that sample to produce such an intermediate test case.
The \textsc{RobOT} algorithm further includes an optimization to terminate the fuzzing round on each seed earlier: 
It limits the number of trials to generate test cases per seed to a small value (i.e., either at most three trials or the change in FOL value between the latest two intermediate test cases smaller than a small value $10^{-18}$ in their experiment) to cut off the long tail of the potential intermediate test cases for each seed.
This poses two limitations: 
First, suppose that the series of intermediate test cases converge their FOL values quickly and yet the prediction labels for these intermediate test cases may remain the same as that of the original seed. In this case, the technique can hardly generate adversarial examples from the seed.
Second, suppose that the series of intermediate test cases converge their FOL values slowly. Plenty of intermediate test cases will be added to the seed list with the priority for further fuzzing before exploiting the next original seed, leading to the low-efficiency problem in processing the seeds in the original seed list.
In our experiment, we observe that if a model under fuzzing is a model after adversarial training, \textsc{RobOT} significantly losses its ability to generate test cases.

Besides discriminating the intermediate test cases, like \textsc{DeepGini}, \textsc{RobOT} has its own test case selection strategies: \textsc{be-st} and \textsc{km-st}. \textsc{RobOT} sorts the generated test cases in descending order of first-order loss into a sorted test pool $P$.
To construct a test suite $A$ containing $n$ test cases, its \textbf{\textsc{be-st}} strategy selects the top-$\nicefrac{n}{2}$ and bottom-$\nicefrac{n}{2}$ test cases from $P$.
On the other hand, its \textbf{\textsc{km-st}} strategy equally divides $P$ into $k$ sublists and randomly picks $\nicefrac{n}{k}$ samples from every sublist to construct the test suite $A$. \textsc{km-st} is generally better than \textsc{be-st} in our experiment.
 
Although \textsc{RobOT} generates test cases with larger losses (measured in FOL) per seed, to construct the final test suite $A$, it selects test cases with diverse loss distances (e.g., the largest and smallest few if using \textsc{be-st} or the random few from different sublists if using \textsc{km-st}). 
\textsc{RobOT} has no clear (loss-based) direction between the test cases in $A$ and the test cases of the same seed it generates.
Thus, test cases in $A$ may be difficult for developers to interpret when compared with those in $P$.

\textsc{RobOT} is later generalized into the \textsc{QuoTe} framework \cite{quote_robot_extend} to make the testing-retraining pipeline clear for the purpose of robustness or fairness improvement. In their experiment, they apply the FOL metric to select test cases generated by existing techniques to show that FOL has a better effect on selecting more valuable test cases to improve the model's robustness. 
The experiment also shows that FOL-guided fuzzing has a better guidance effect in test case generation than existing techniques with respect to improving the model's robustness.

\textbf{\textsc{Adapt}} \cite{adaptfuzz} is a state-of-the-art coverage-based technique. It uses a genetic algorithm approach to explore a neuron-based coverage space.
A chromosome is a sequence of real numbers, one for a measurement metric.
\textsc{Adapt} designs a long list of such measurement metrics to measure a neuron against a chromosome (e.g., two metrics are whether the activation value of a neuron under measure is in top 10\%--20\% and top 40\%--50\% in the same forward pass of $f$, respectively).
It first populates a set $R$ of random chromosomes per seed.
Let $x^{(i)}$ be the working sample in the $i^{th}$ iteration to generate an intermediate test case for a given seed $x$ where $x^{(0)} = x$.
For each chromosome $p \in R$, \textsc{Adapt} computes a score as the dot product of $p$ and the feature vector of each neuron when predicting the label of $x^{(i)}$.
It then selects the top-$m$ (where $m$ = 10 in their experiment) neurons with the highest score and perturbs $x^{(i)}$ into $x^{(i+1)}$ against the loss of these $m$ neurons.
If $x^{(i+1)}$ covers not-yet-covered elements (e.g., the set of activated neurons is not covered by previous test cases with respect to the adopted neuron coverage metric) and is within the fuzzing boundary, it will be kept in the fuzzing list as a candidate seed for further fuzzing process.
\textsc{Adapt} measures the coverage, such as the neuron coverage \cite{deepxplore}, achieved by $x^{(i+1)}$.
After having processed all the chromosomes for the working sample, it reduces the chromosome set $R$ to a minimal subset $S$ that retains the same coverage as $R$ followed by expanding $S$ by adding these chromosomes in $R$ that cover most coverage items until a threshold on the size of $S$ is met.
\textsc{Adapt} then iteratively crossovers two randomly-picked chromosomes in $S$ followed by adding Gaussian noise to construct a new chromosome and place this new chromosome into $S$ until $|S| = |R|$. 
The resultant $S$ is assigned to $R$, and the sample $x^{(i+1)}$ that covers new coverage items becomes the working sample.
The iteration to process $R$ repeats until the candidate seed list is empty or the fuzzing budget is exhausted. 

It is unclear how to determine which test cases of the same seed are more representative in \textsc{Adapt}.
This is because the retained test cases across different iterations are not measured against the same baseline (e.g., the set of $m$ neurons used in the optimization objective varies across iterations). The list of measurement metrics is also specifically designed to make no test case able to obtain non-zero scores from the same kind of metrics simultaneously (e.g., the activation value of a neuron in the same pass cannot be located in two non-overlapping ranges of activation values simultaneously).

\subsection{Model Retraining in the Testing-Retraining Pipeline}
In the traditional program testing domain, after the testing of a program has exposed bugs in the program, the correctness of the program is improved by repairing the exposed bugs, which produces a repaired version of the program. 
The test cases that expose the bugs and those that the original program can pass are used to retest the repaired program. 
Developers usually expect that the repaired program should not only pass all these test cases but also be revised so that other inputs do not trigger the same fixed bugs.

In the DL model domain, as presented in Section~\ref{sec: introduction}, after obtaining the test cases that expose the vulnerability of a DL model, the procedure to obtain and validate an improved version can be different.
Suppose a purpose is to improve a DL model's robustness by correcting the DL model's wrong predictions exposed by a test suite $A$ generated by a DL testing technique.
In that case, model retraining \cite{deepxplore, dlfuzz, deepgini, robot, surprise_adequacy, quote_robot_extend, sensei, deeptest, test_selection_tosem, ATS2022, drfuzz} is a typical means to achieve it. 
In a typical retraining procedure for this purpose, these test cases in $A$ are added to the original training dataset to retrain the DL model while keeping all other training settings, such as hyperparameters, unchanged \cite{quote_robot_extend}. 

Nonetheless, unlike repairing a traditional program, retraining a DL model makes the retrained model memorize the training samples and their labels (so-called the memorization effect \cite{memorization_neurips, memorization_pmlr}). 
Thus, although the retrained model may ``recall from the memory'' to pass all these test cases of $A$ in the retesting procedure, it only presents a false sense of robustness to developers. 

Therefore, the retrained model should be retested with some other samples that the model under test fails to predict correctly and the retraining task in the pipeline does not use them for retraining the model simultaneously.
Since the productions of the model under test and the retrained model mainly differ in whether the test cases in the test suite $A$ are used as additional training samples in the training/retraining task, the difference in robust accuracy before and after the retraining task shows the impact of the test suite on mitigating the threat of robustness on the DL model under test, and, consequently, implies the robustness-oriented value of the DL testing technique that constructs the test suite.

\subsection{Testing-Retraining Pipeline for Testing DL Models for Robustness Improvement}
\label{sec: AM methodology}
\textsc{RobOT} \cite{robot} presents an \textbf{assessment methodology} (\textbf{AM}) to support the testing-retraining pipeline for improving the robustness property of a DL model under test $f$.
The key distinction of this pipeline from the classical testing pipelines in the DL testing literature is that it incorporates the retraining task \cite{robot, quote_robot_extend}.
Different from the pipeline of pure adversarial retraining \cite{adversarial_training_1, adversarial_training_2}, the testing-retraining pipeline in AM includes a testing step to generate test suites guided by a testing metric.
Similar methodologies are also used in the experiments of other related works \cite{deepgini, mode, quote_robot_extend, seed_select_dl_test}, albeit proposed independently.

The workflow of AM is as follows.
Users first provide a robust validation dataset $D_{ass}$ and a user-specified requirement $r$ on the model robustness (e.g., 80\% of the samples (adversarial examples) in the robust validation dataset can be predicted correctly \cite{quote_robot_extend}). 
For instance, in \cite{robot}, $D_{ass}$ is composed of the adversarial examples generated by two attacker techniques, i.e., FGSM \cite{fgsm} and PGD \cite{pgd}, on $f$.
The workflow then invokes a testing-retraining pipeline.
The pipeline first calls a metric-oriented DL testing technique to generate a test suite (where the generation process is guided by the testing metric of the technique) or a test case selection technique guided by a testing metric to select test cases from a given test pool provided by the user to construct a test suite.
In either case, it then calls a retraining task to retrain the current DL model under test with the constructed test suite and the original training dataset.
The workflow then validates whether the accuracy of the resulting model on $D_{ass}$ satisfies $r$.
If this is the case, the workflow outputs the resulting model and terminates the whole process. 

AM also implicitly requires that the clean accuracy is not severely compromised. For instance, in their experiment \cite{robot, quote_robot_extend}, the difference in the test accuracies before and after the testing-retraining pipeline is controlled to be within 1\%, or the fuzzing time budget is limited to a short period, such as 5--20 minutes, so that a fuzzing technique can only generate a relatively limited number of test cases (e.g., 4023 test cases on ResNet20 trained on CIFAR10 \cite{quote_robot_extend}) compared to the training dataset (50000 samples for CIFAR10) so that the technique can apply all generated test cases for the retraining task to improve the model robustness.

In essence, the same testing-retraining pipeline configured with different DL testing techniques can be compared in two aspects.
The first aspect is to select test cases from the same given test pool $P$ using the testing metric and its associated test case prioritization technique of such a DL testing technique to prioritize $P$ before the selection. 
The second aspect is to apply a DL fuzzing technique to generate a test pool $P$ followed by selecting a subset of $P$ to construct a test suite based on the test case prioritization/selection technique of such a DL testing technique.
We refer to these two aspects as Configurations $A$ and $B$ of the pipeline, respectively.
In either configuration, \textsc{Clover} prioritizes $P$ via the ContextSelect strategy (see Algorithm \ref{alg: context_selection}), and the peer techniques prioritize $P$ according to the respective selection strategy (e.g., \textsc{be-st} and \textsc{km-st} of \textsc{RobOT}) or prioritization technique (e.g., \textsc{DeepGini}). 

It should be worth noting that replacing the testing-retraining pipeline with pure adversarial retraining can also improve the robustness of $f$ against adversarial examples. With this alternative pipeline (which consists of a single task), a fuzzing technique in the resulting workflow can be used as a test case generation technique to assure the robustness quality of the adversarially trained model.

\subsection{Auxiliary functions}
\label{sec: helper function}
To aid our presentation, we define the following three auxiliary functions for a fuzzing technique that takes a set of seeds $X$ as input and fuzzes on a sequence $Z$, the occurrences of these seeds.

\begin{itemize}
\item
The function \mbox{\textbf{\textsc{semantic}}($s$)} returns the ground truth label of a sample $s$ if applicable. Otherwise, it returns the prediction label $f(s)$ of $s$. We refer to the returned label as the \textbf{seed label} of $s$.

\item
The function \mbox{\textbf{\textsc{isAdversarial}}($s, t$)} is a Boolean function accepting a seed $s$ and a test case $t$ generated from $s$. It returns true if the seed label of $s$ is different from the prediction label of $t$ (i.e., $\textsc{semantic}(s) \neq f(t)$); otherwise, it returns false.

\item
The function \mbox{\textbf{\textsc{source}}($z$)} returns the seed $x \in X$ for its occurrence $z$ in $Z$, i.e., \mbox{\textsc{source}}($z$) = $x$ satisfying the condition $x \in X \land z = x$. 
It is the mapping between the elements in $Z$ and $X$.
\end{itemize}

Since $Z$ contains the occurrences of the seeds in $X$, we assume that each element in $Z$ and its corresponding element in $X$ have the same seed label (i.e., $\forall z \in Z, \exists x \in X, \textsc{source}(z) = x$ $\land \textsc{semantic}(z) = \textsc{semantic}(x)$).

\section{Clover}
\label{sec: Clover}
In this section, we present \textsc{Clover}. As we have presented in Section~\ref{sec: introduction}, the scope of \textsc{Clover} is to be applied as a DL testing technique in the testing-retraining pipeline. 
For brevity, we simply refer to the robustness improvement achieved by a retrained model output by the testing-retraining pipeline configured with \textsc{Clover} to be the DL testing technique as the robustness improvement achieved by \textsc{Clover} delivered through the pipeline.

\subsection{Intuition}
Let $X$ be a set of seeds and $P$ be a test suite generated from $X$ to test a model $f$.
Each such test case in $P$ is within the $\epsilon$-ball of its corresponding seed measured in a $p$-norm distance.
Let $t$ and $t'$ in $P$ be two test cases perturbed from the same seed $x$ (denoted by $t \approx t'$). 
Suppose further $t'$ is in $\delta$-ball of $t$ (i.e., $\parallel t - t'\parallel_p < \delta$) where $\delta \ll \epsilon$.

Suppose we find a subset $A \subset P$ such that $A$ simulates $P$ in the sense that each sample $t'$ in the set $P - A$ is in the $\delta$-ball of the same sample $t$ in $A$, and the prediction vectors $\vec{f}(t)$ and $\vec{f}(t')$ are similar. 
Intuitively, thanks to the linearity of the activation functions in $f$, (re)training $f$ with $t$ has a better chance to generalize $f$ around $t$ to make the (re)trained model correctly infer $t'$ rather than some other test cases with their prediction outputs differing much. 

Thus, given the same fuzzing budget $|P|$, by taking $t$ but not $t'$ into $A$, a fuzzing technique can allocate the remaining budget quota $|P| - |A|$ to generate and select other test cases that are not in $P$. 
In practice, we cannot analytically and efficiently determine the simulation relation between $P$ and $A$ due to the statistical nature of deep learning models and the large size of $|P|$. 
We thus look for an efficient and downscaled approximation of the relation. 

Our basic idea is to generate $t \in A$ with the samples that surround it resembling $t$ in high chance and includes $t$ (without its surrounding samples) in the constructing test suite $A$. 
In other words, we aim to study the representativeness of a test case through the perturbation sets closely surrounding the former test cases and study how to generate more representative test cases.

\subsection{Measuring the Representativeness of Test Cases}
\label{sec: contextual_confidence}
In fuzzing, adding a perturbation, denoted by $\Omega$, to a test case $t$ generates a new sample (denoted by $t + \Omega$) within the $p$-norm bound $\delta$ (i.e., ${\parallel}t - (t+\Omega){\parallel}_p<{\delta}$).
Moreover, as fuzzing is computationally expensive, we aim to measure the representativeness of test cases efficiently. 
\textsc{Clover} chooses the uniform noise as the type of perturbation because we are inspired by the $\textit{sign}(.)$ function used in many techniques (e.g., \cite{fgsm, make_nn_robust}) to perturb samples in a black box manner (e.g., without computing the gradients) efficiently. Such noises are fair to all test cases so that a comparison between test cases can be more objective.
Suppose we have a test case $t$ and a sample $y \in$ $\delta$-ball of $t$ where $\Omega = y - t$. We refer to $\Omega$ as a contextual perturbation of $t$.
Since enumerating all contextual perturbations of $t$ is intractable, \textsc{Clover} samples a set of $k$ contextual perturbations, denoted by $\textit{ctx}^k(t)$.

We propose the metric \textit{\textbf{contextual confidence}}, \textit{\textbf{CC}} for short, to measure the representativeness of a test case.
CC measures the degree of consensus among the surrounding samples of a test case.
It has four parameters $(f, t, v, k)$: a model $f$, a test case $t$, the adversarial label $v$ of $t$, and the cardinality of $\textit{ctx}^k(t)$. We note that CC is not a test adequacy criterion as it does not formulate the requirement of a coverage item.
It computes the mean of the set $\{Pr_v(\vec{f}(t + \Omega)) \mid \Omega \in \textit{ctx}^k(t)\}$, where $Pr_v(\vec{f}(t + \Omega))$ is the probability value of the label $v$ in the prediction vector $\vec{f}(t + \Omega)$.
We use the shorthand notation $CC(t)$ instead of the full notation $CC(f, t, v, k)$, and Eq. (\ref{eq: cc_short}) presents its formula.
\begin{equation}
CC(t) = \frac{\sum_{\Omega \in \textit{ctx}^k(t)}Pr_v(\vec{f}(t + \Omega))}{|\textit{ctx}^k(t)|}
\label{eq: cc_short}
\end{equation}

\subsection{Conceptual Fuzzing Model of \textsc{Clover}}
\label{sec: seed_equivalent_fuzzing}
\subsubsection{Overview}
Like FGSM, \textsc{Clover} does not impose additional requirements on the optimization objective, denoted by $\textit{obj}$, and uses the same objective function as FGSM when computing the gradients. 
It accepts a set of seeds $X$ as input and generates test cases from the seed occurrence sequence $Z \in X^*$ (where $X^*$ is the Kleene closure of $X$).

The conceptual model of \textsc{Clover} includes five main steps: 
\textsc{Clover} tracks the test case with the highest CC value among the test cases generated for each seed occurrence, which is referred to as the $\alpha$-representative test case of the seed occurrence (see Section~\ref{sec: representative test case}). 
In each fuzzing round on an occurrence of a seed, it retrieves the seed label of the seed and the prediction label of a test case of the seed, where this test case is the one that achieves the highest CC value among all generated $\alpha$-representative test cases of all seed occurrences of that seed.
(We refer to these two labels as the label pair of the seed at the fuzzing round.)
It finds the piece of difference between the seed and this test case, referred to as the $\beta$-AFO for the seed at the fuzzing round (see Section \ref{sec: adversarial front objects}). 

On fuzzing the DL model with $z$ (a seed occurrence of seed $x$) as the input seed at a fuzzing round, it determines the set of seeds that contains these seeds sharing the same label pair with $x$ (referred to as seed equivalence among these seeds in Section \ref{sec: seed equivalence}).
It generates a test case for $z$ using the collected perturbation information, including the current $\beta$-AFO of $x$ and the $\beta$-AFO of a seed in this set of seeds, and generates more such test cases for $z$ if the fuzzing budget for $z$ allows (see Section \ref{sec: sq_to_sq_af}).
Finally, \textsc{Clover} constructs a test suite (see Section~\ref{sec: selection strategy}).

We note that $\alpha$-representative test case is a notional concept to ease our presentation of two other main concepts ($\beta$-adversarial front object and seed equivalence) in \textsc{Clover}.

The next five subsections present these steps.
For brevity, we \textbf{overload the term ``seed''} as the terminology to refer to the term ``seed occurrence'' in $Z$ to condense the presentation.
Moreover, unless stated otherwise, a seed refers to a seed occurrence in $Z$.

\subsubsection{Track representative test case}
\label{sec: representative test case}
Like other fuzzing techniques, \textsc{Clover} aims to generate a set of test cases, denoted by $Q(z_i)$, for each given seed $z_i \in Z$ (where $z_i = Z[i]$ for all $i)$.
We refer to the test case with the highest CC value among the test cases in $Q(z_i)$ as the \textbf{\textit{$\boldsymbol{\alpha}$-representative test case}} for $z_i$, denoted as $\boldsymbol{\overline{z_i}}$.
(We use the symbol $\bot$ to stand for the bottom element, which indicates all applied fuzzing attempts on $z_i$ are unsuccessful, and no test case is generated from $z_i$ yet. As such, we choose $-\infty$ as its CC value.)
\begin{align*}
 \overline{z_i} & = 
 \begin{cases}
     t' = \textit{argmax}_{t} \{CC(t) \mid t \in Q(z_i) \land \textsc{isAdversarial}(z_i, t) = true \} & \text{if } Q(z_i) \neq \emptyset  \\
     \bot & \text{otherwise}
 \end{cases} 
\end{align*}

\begin{figure} [t]
\includegraphics[width=0.6\textwidth]{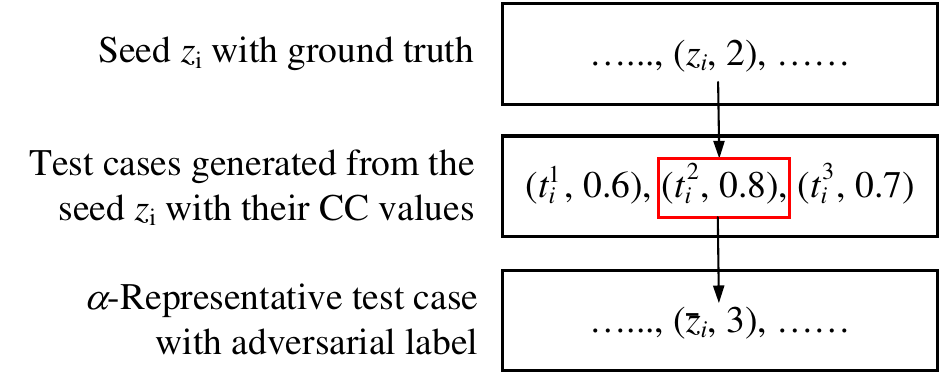}
\caption{Relationship between the Seed $z_i$ in Z and its Test Cases and $\alpha$-Representative Test Case}
\label{fig: most_rep_test_case}
\end{figure}

\begin{figure} [t]
\includegraphics[width=0.85\textwidth]{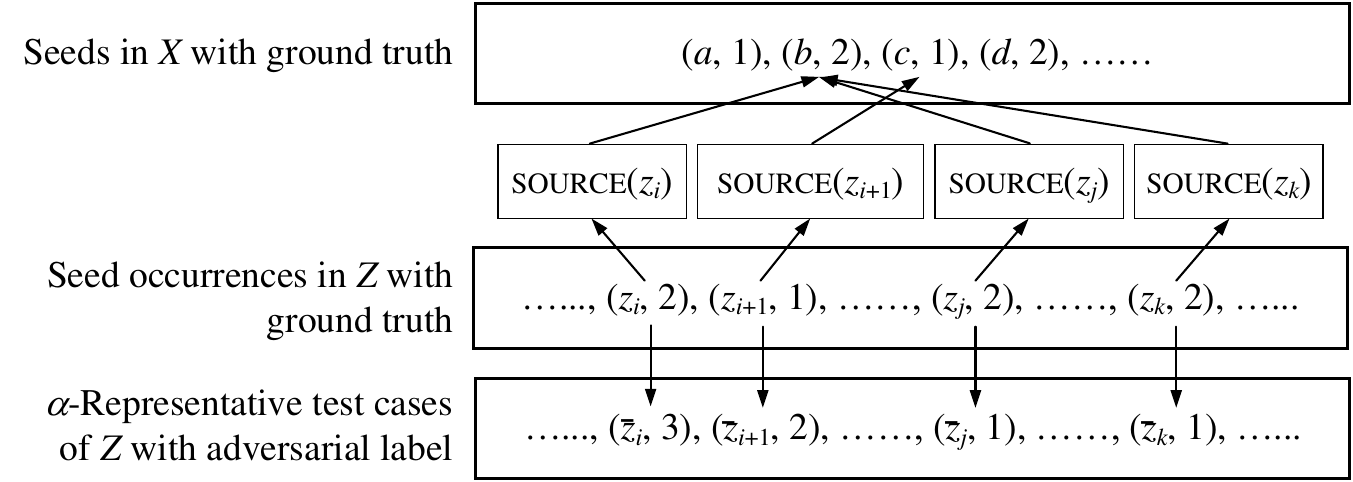}
\caption{Relationship among Seeds in Z, Seeds in X, and $\alpha$-Representative Test Cases}
\label{fig: LINK}
\end{figure}

Fig. \ref{fig: most_rep_test_case} depicts that  $t_i^1$, $t_i^2$, and $t_i^3$ are three test cases of the seed (occurrence) $z_i$.
Their CC values are 0.6, 0.8, and 0.7, respectively. 
The test case $t_i^2$ is depicted as the $\alpha$-representative test case of $z_i$, i.e., $\overline{z_i}$, because its CC value is highest among the test cases of $z_i$. The diagram also shows that $t_i^2$ has an adversarial label of 3.
Fig. \ref{fig: LINK} depicts the relationships between the seeds in $Z$ and $X$ where the three seed occurrences $z_i$, $z_j$, and $z_k$ map to the same seed $b \in X$ through the function \textsc{source}(.).

\subsubsection{Track the representative adversarial front object}
\label{sec: adversarial front objects}
After knowing $\overline{z_i}$ for the seed $z_i$, we can compute their difference (i.e., $\overline{z_i} - z_i$), which captures the overall change that leads $z_i$ to become its $\alpha$-representative test case.
We refer to this piece of difference as the \textit{\textbf{\boldsymbol{$\alpha$}-adversarial front object}} (\textit{\textbf{$\boldsymbol{\alpha}$-AFO}}) for $z_i$.

As such, we annotate the $\alpha$-AFO for $z_i$ with three indexes ($i, u, v$), denoted as $\Delta_i^{(u, v)}$, where $i$ is the index of the seed $z_i$ in $Z$ (as the seed under test in the $i^{th}$ fuzzing round), i.e., $z_i = Z[i]$, $u$ is the seed label returned by $\textsc{semantic}(z_i)$, and $v$ is the adversarial label $f(\overline{z_i})$.
Accordingly, the matching sequence of $\alpha$-AFOs for the seed sequence $Z$, denoted as $\Pi$, is defined below.
\[ 
\Pi = \langle \Delta_i^{(u, v)} \mid z_i = Z[i] \land  
u = \textsc{semantic}(z_i) \land v = f(\overline{z_i}) \land \Delta_i^{(u, v)} = \overline{z_i} - z_i 
\rangle_{i = 0}^{|Z|-1}
\]
A seed in $X$ may have several occurrences in $Z$.
We thus raise the level of adversarial front objects from the seed occurrence level (the level for $Z$) to the seed level (the level for $X$).
\textsc{Clover} tracks the \textit{\textbf{\boldsymbol{$\beta$}-representative adversarial front object}} for each seed $x$ in $X$ (\textit{\textbf{$\boldsymbol{\beta}$-AFO}} for short) at each fuzzing round, where Eq. (\ref{def: most_af}) summarizes how to find the $\beta$-AFO, denoted as $_{x}\Delta_i^{(u, v)}$, for a seed $x \in X$ at the $i^{th}$ fuzzing round.%
\footnote{
If the set $\Gamma(x, i)$ in Eq. (\ref{def: most_af}) is empty (i.e., $x$ has not generated any test cases), we set $_{x}\Delta_i^{(u, v)}$ to zero, and we denote the adversarial label of this special case by the label $a_{\bot}$, i.e., the retrieved $\beta$-AFO for $x$ by Eq. (\ref{def: most_af}) is $_{x}\Delta_i^{(\textsc{semantic}(x), a_\bot)}$, and its value is 0.
}
Specifically, Eq. (\ref{def: most_af}) retrieves the $\alpha$-AFO among these in $\Pi$ that are produced in all fuzzing rounds up to and including the $i^{th}$ fuzzing round and share the same seed as $x$ such that $x$ added with the retrieved $\alpha$-AFO achieves the highest CC value among these test cases formed by individually adding these $\alpha$-AFOs to $x$.
Moreover, we define the helper function $\phi(x, i)$ to return the $\beta$-AFO of  $x$ at the $i^{th}$ fuzzing round.

\begin{equation}
\begin{split}
_{x}\Delta_i^{(u, v)} = & \,\phi(x,i) = \mathop{argmax}\limits_{\Delta_j^{(u, v)} \in \Gamma(x,i)} \ 
CC(x + \Delta_j^{(u, v)}) \\
&
\begin{split}
\text{such that } 
\Gamma(x,i) = \{\Delta_j^{(u, v)} \mid 
& 
0 {\leq} j {\leq} i \land
\Delta_j^{(u, v)} = \Pi[j] \land  
z_j = Z[j] \land x = \textsc{source}(z_j) 
\}
\end{split}
\label{def: most_af}
\end{split}
\end{equation}

We note that we use the two similar notations $_{x}\Delta_i^{(u, v)}$ and $\Delta_i^{(u, v)}$  to refer to the adversarial front objects for the seed $x$ in $X$ and the seed occurrence of $z_i$ in $Z$, respectively, where $z_i$ is a seed occurrence of $x$ at the $i^{th}$ fuzzing round.

We refer to the set $\{ \phi(x,i)\}_{x \in X}$ as the \textit{adversarial front} at the $(i+1)^{th}$ fuzzing round. Intuitively, it captures the best configuration to make $X$ adversarial and attain the highest possible CC values.

\begin{figure}[t]
\includegraphics[width=0.9\textwidth]{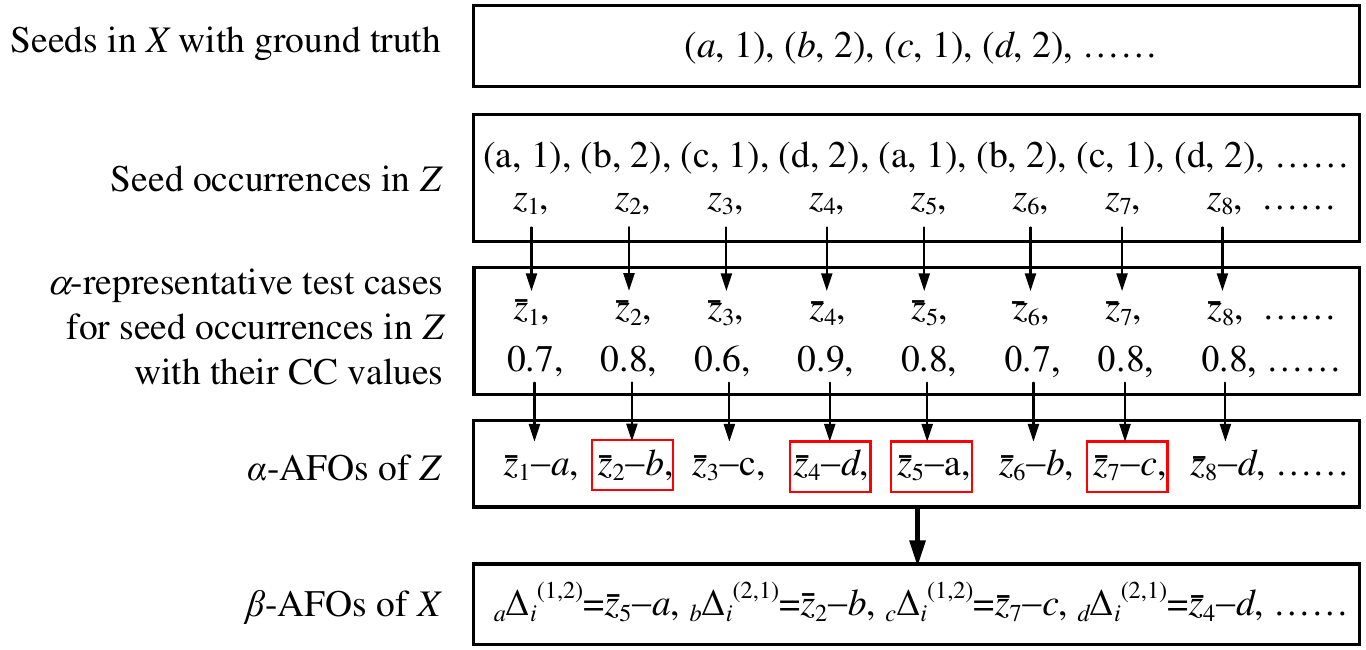}
\caption{Relationship between $\alpha$-AFO and $\beta$-AFO at the $\text{(}i+1\text{)}^{th}$ Fuzzing Round}
\label{fig: alpha_beta_AFO}
\end{figure}

Fig. \ref{fig: alpha_beta_AFO} depicts the relationships between $\alpha$-AFOs of an exemplified seed occurrence sequence $Z$ and $\beta$-AFOs of an exemplified seed list $X$ at the $\text{(}i+1\text{)}^{th}$ fuzzing round.
$X$ contains four seeds $\{a, b, c, d \}$ and $Z$ contains eight seeds from $z_1$ to $z_8$.
The $\alpha$-representative test cases of the seeds in $Z$, denoted as $\overline{z_1}$ to $\overline{z_8}$, and their CC values are depicted in the third row. 
For simplicity, if the ground truth label of a seed in $Z$ is 1, then the adversarial label of its $\alpha$-representative test case is 2, and vice versa.
The corresponding $\alpha$-AOFs of $z_1$ to $z_8$ (i.e., the pieces of differences are $\overline{z_i} - z_i$ for $i$ = 1 to 8) are shown in the fourth row.
Finally, the $\beta$-AFOs for the seeds in $X$ are shown in the last row. 
Take the $\beta$-AFO for the seed $a$ as an example. 
There are two $\alpha$-AFOs ($\overline{z_1}-a$ and $\overline{z_5}-a$) generated from the seed occurrences of this seed. 
The CC value of the $\alpha$-representative test case for $z_5$ is higher than that of $z_1$. 
So, the $\beta$-AFO $_{a}\Delta_i^{(1,2)}$ is computed as $\overline{z_5}-a$. The $\beta$-AFOs for the other three seeds in $X$ are computed in the same manner.
The set of $\beta$-AFOs in the last row in Fig. \ref{fig: alpha_beta_AFO} also depicts an adversarial front.
Intuitively, as the fuzzing campaign continues, the adversarial front moves toward the end of higher contextual confidence.

\subsubsection{Find seed equivalence and equivalence class}
\label{sec: seed equivalence}
We further raise the above concept of adversarial front objects at the $i^{th}$ fuzzing round from the seed level to the seed equivalence level.
Intuitively, these adversarial front objects associated with the equivalent seeds in $X$ capture the most successful discovery of the fuzzing technique to produce test cases with the highest CC values for label pairs.

\begin{definition}[\textbf{\textit{Seed Equivalence and Equicalence Class of Seeds}}]
\label{def: seed_equivalence}
Suppose $x$ and $x'$ are two seeds in $X$ having $_{x}\Delta_i^{(u, v)}$ and $_{x'}\Delta_i^{(u', v')}$ as their $\beta$-AFOs at the $i^{th}$ fuzzing round, respectively.
We call $x$ and $x'$ seed-equivalent at the $i^{th}$ fuzzing round, denoted by $x \sim_i x'$, if and only if $u=u'$ and $v=v'$.
The equivalence class for the seed $x$ at the $i^{th}$ fuzzing round is defined as
$[x]_i = \{ x' \in X \mid x \sim_i x' \}$.
\end{definition}

Recall that the behavior of a DL model for a class 
(i.e., a label in $\mathbb{C}$) 
is generalized from a set of samples for that class.
Definition \ref{def: seed_equivalence} presents the notions of seed equivalence and the equivalence class of seeds. 
In short, to quantify a seed $x \in X$ to be seed-equivalent to other seeds at the $i^{th}$ fuzzing round, we look for semantic (seed label) equivalence among these seeds and the adversarial label equivalence 
---
Those seeds in the same equivalence class are predicted by the DL model under test to the same adversarial label end after perturbation from the same seed label end.
Moreover, moving from one end to another end for a seed is exactly captured by the $\beta$-AFO of the seed.

\begin{figure}[t]
\includegraphics[width=0.7\textwidth]{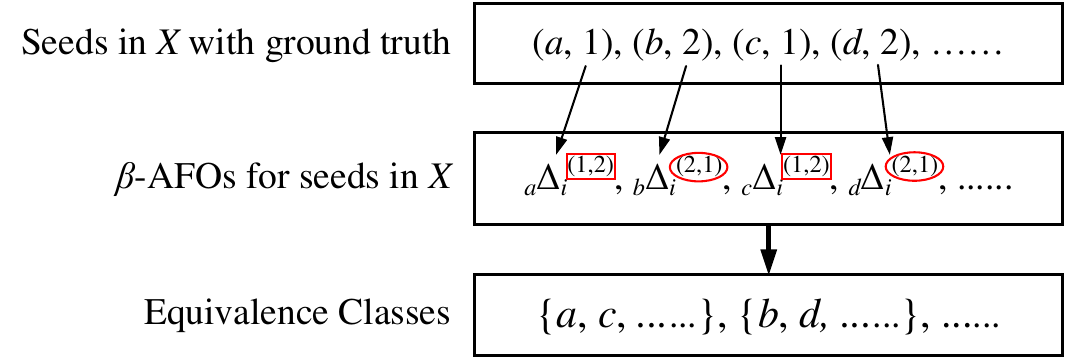}
\caption{Equivalence Classes at the $\text{(}i+1\text{)}^{th}$ Fuzzing Round.}
\label{fig: equivalence_classes}
\end{figure}

Fig. \ref{fig: equivalence_classes} depicts how equivalent classes look like for the scenario shown in Fig. \ref{fig: alpha_beta_AFO}.
By grouping these seeds by their pairs of seed labels and adversarial labels, the two pairs of seeds ($a$, $c$) and ($b$, $d$) are grouped into two different equivalence classes.

Thus, based on seed equivalence, we exploit the $\beta$-AFOs of the seeds in the same equivalence class to assist the generation of test cases for individual seeds in the corresponding equivalence classes at respective fuzzing rounds.
To aid our presentation, we denote the set of $\beta$-AFOs for the seeds in the equivalence class $[x]_i$ that excludes the one for $x$ by $AC(x, i) = \{ \phi(x',i) \mid x' \neq x \land x' \in [x]_i\}$.

The next section presents our idea to generate test cases and track the $\beta$-AFOs for their seeds.

\subsubsection{Generate $\beta$-AFO and test cases for the current seed under fuzzing}
\label{sec: sq_to_sq_af} 
In this section, we use the concepts presented in the last two subsections to illustrate how to generate test cases for a seed.
We present the main principle below and leave the details to the next section when we present the \textit{ContextTranslate} algorithm.

Suppose the sample $z_{i+1}$ is the fuzzing seed at the ${(i{+}1)}^{th}$ fuzzing round (i.e., $z_{i+1} = Z[i+1]$). 
We denote the seed for this seed occurrence $z_{i+1}$ by $x''$ (= $\textsc{source}(z_{i+1})$).

In the ${(i{+}1)}^{th}$ fuzzing round, to make use of information captured in the set $\cup_{x \in X} AC(x, i)$, our idea is to add one of these closely relevant $\beta$-AFOs (modeled as these $\beta$-AFOs in the set $AC(x'', i)$) as a perturbation to the current test case of $z_{i+1}$ (i.e., the test case corresponding to the current $\beta$-AFO for the seed $x''$) to produce the next test case candidate in each iteration, thereby generating a sequence of test case candidates (denoted by $T$ here) iteratively.%
\footnote{
We leave the presentation of the other details of the addition process to Section \ref{sec: algorithms}.
}

Nonetheless, not every such test case candidate could be useful.
So, during the generation of $T$, if a test case candidate $t \in T$ is adversarial (i.e., $\textsc{isAdversarial}(z_{i+1}, t)$ returns true), it will be placed into the set of generated test cases $Q(z_{i+1})$ for $z_{i+1}$. Moreover, there are two cases to consider. Suppose the current iteration to process $T$ is $w$ and the current test case candidate $T[w]$.
\textit{Case (1):}
Suppose $T[w]$ achieves a higher CC value than the existing test cases of the seed $x''$. In this case, the $\beta$-AFO for $x''$ is updated to $T[w] - z_{i+1}$ before proceeding to the next iteration. This test case will be used in the addition process in the next iteration.
\textit{Case (2):} On the other hand, suppose $T[w]$ cannot achieve a higher CC value than the existing test cases of the seed $x''$. We will discard the candidate $T[w]$ from the sequence $T$ and continue the addition process in the next iteration based on the preceding test case candidate in $T$ that is not discarded (i.e., the latest test case that is found in \textit{Case (1)} above).

Like other typical fuzzing techniques \cite{robot, adaptfuzz, quote_robot_extend}, \textsc{Clover} assigns a test budget to each seed to control the number of fuzzing attempts applicable to the seed, which is modeled as the length of the sequence of $\beta$-AFOs applied to generate the sequence $T$.

\begin{figure}[t]
\includegraphics[width=\textwidth]{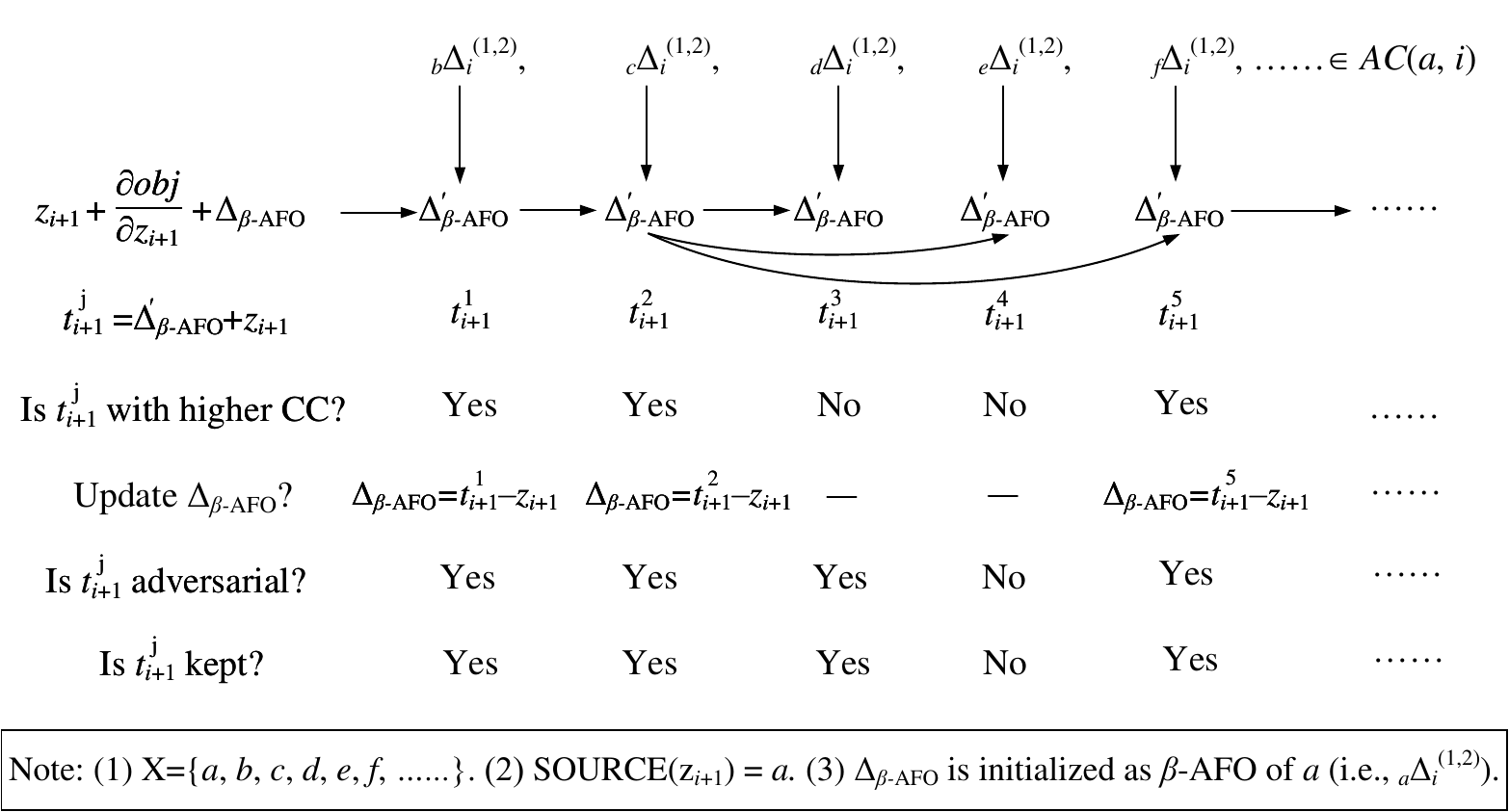}
\caption{Incremental Generation of Test Cases of Seed $z_{i+1} (=Z[i+1])$ using the Set of $\beta$-AFOs $AC(a, i)$ of These Seeds in the Equivalence Class $[a]_i$ where $a = \textsc{source}(z_{i+1})$.}
\label{fig: evolve}
\end{figure}

We use the following example to illustrate the test case generation process.

Fig. \ref{fig: evolve} depicts how \textsc{Clover} evolves $\beta$-AFOs for the seed $a$ in the following set $X$, and generates test cases in the same fuzzing round. Suppose we are in the $(i+1)^{th}$ fuzzing round, in which the current seed occurrence under fuzzing is $z_{i+1}$, which is an occurrence of the seed $a$ in the set $X = \{a, b, c, $ $d, e, f, \dots \}$.
Suppose further the $\beta$-AFO for the seed $a$ at the $i^{th}$ fuzzing round is $_{a}\Delta_i^{(1,2)}$, and the seeds $\{b, c, d, e, f, \dots \}$ are in the equivalence class $[a]_i$.
Let $\Delta_{\beta-\text{AFO}}$ be a variable, which is initialized to $_{a}\Delta_i^{(1,2)}$.
Let us focus on the evolution of the $\beta$-AFO for the seed $a$ as well as the generation of test cases for the current seed occurrence $z_{i+1}$.
\textsc{Clover} firstly constructs the set $AC(a, i)$, which contains the $\beta$-AFOs for the seeds in $[a]_i$.
In the first attempt, it initializes a working sample $z_{i+1}$ + $\Delta_{\beta-\text{AFO}}'$ as $z_{i+1}$ + $\partial \textit{obj} / \partial z_{i+1}$ + $\Delta_{\beta-\text{AFO}}$, where the idea is to slightly perturb the current $\alpha$-representative test case of $z_{i+1}$ (i.e., $z_{i+1} + \Delta_{\beta-\text{AFO}}$) along the gradient of $z_{i+1}$ to produce this working sample (see Alg. \ref{alg: context_translate} for the details). 
After attempting to add the first $\beta$-AFO (i.e., depicted as $_{b}\Delta_i^{(1,2)}$ in the figure) taken from $AC(a, i)$ to the working sample and then evolve it into a resulting test case candidate $t_{i+1}^1$, \textsc{Clover} detects $t_{i+1}^1$ having a higher CC value than all existing test cases of the seed $a$ (also see Alg. \ref{alg: context_translate} for the details).
(We note that in the above evolution step, the idea is to perturb the sample along the gradient, which is inspired by an observed common phenomenon of many existing attacker techniques: they often generate test cases with very high prediction confidence (close to 1).)
Thus, \textsc{Clover} updates $\Delta_{\beta-\text{AFO}}$ to the piece of difference represented by $t_{i+1}^1 - z_{i+1}$ (i.e., assign $t_{i+1}^1 - z_{i+1}$ to the variable $\Delta_{\beta-\text{AFO}}$) and keeps $t_{i+1}^1$ in the test set $Q(z_{i+1})$ since $t_{i+1}^1$ is adversarial. 
In the second attempt, it repeats the process of the first attempt (but adds with another $\beta$-AFO taken from $AC(a, i)$ instead of $_{b}\Delta_i^{(1,2)}$) and produces the test case candidate $t_{i+1}^2$.
It updates the variable $\Delta_{\beta-\text{AFO}}$ to the piece of difference represented by $t_{i+1}^2 - z_{i+1}$ and keeps $t_{i+1}^2$ in the test set $Q(z_{i+1})$ for the same reason as the first attempt above.
Similarly, in the third attempt, \textsc{Clover} generates the test case candidate $t_{i+1}^3$ and detects it being adversarial but not with a higher CC value than the existing test cases of $a$.
Thus, it only keeps $t_{i+1}^3$ in the test set $Q(z_{i+1})$ without updating $\Delta_{\beta-\text{AFO}}$. 
Therefore, the fourth attempt continues based on the results of the second attempt. It detects the generated test case candidate $t_{i+1}^4$ in that attempt not having a higher CC value and not adversarial.
Thus, it neither updates $\Delta_{\beta-\text{AFO}}$ nor keeps $t_{i+1}^4$. 
Similarly, the fifth attempt also continues from the second attempt. 

We note that the membership of an equivalence class $[\textsc{source}(z_{j+1})]_j$ depends on the latest adversarial label and the highest CC value among the test cases of the seed, which may vary as the index $j$ varies.
Thus, two seeds $x$ and $x'$ in $X$ may belong to the same equivalence class for some $j$ but different equivalence classes for some other $j$.
This achieves an adaptive partitioning scheme on $X$ while processing $Z$ over different fuzzing rounds.

\subsubsection{Construct a test suite for robustness improvement}
\label{sec: selection strategy}
\begin{figure}[t]
\includegraphics[width=0.9\textwidth]{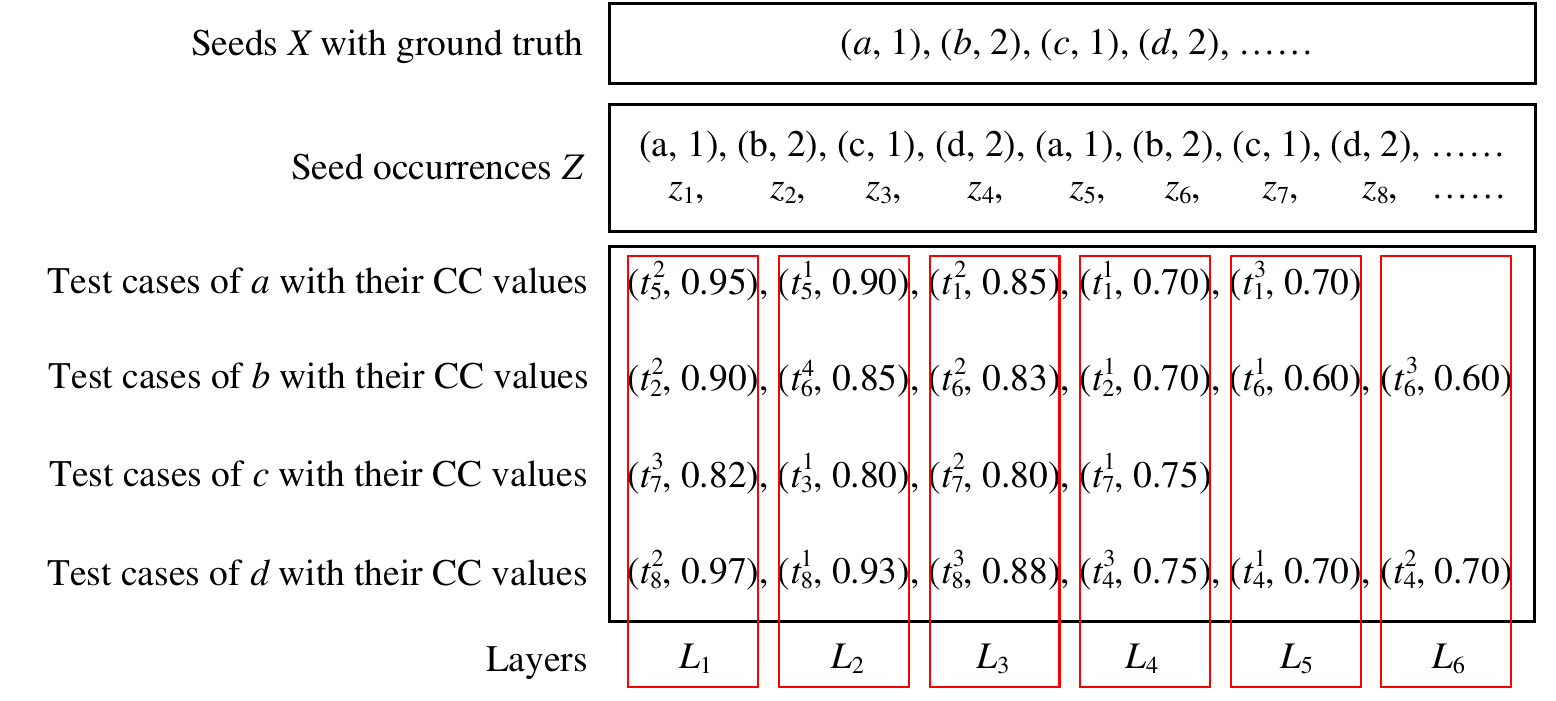}
\caption{Layering Test Cases with Their CC Values for Test Case Selection.}
\label{fig: selection}
\end{figure}

After fuzzing the sequence $Z$, \textsc{Clover} selects test cases among groups of test cases, one group per seed in $X$, as even as possible at the seed (of $X$) level and with a preference for selecting test cases with higher CC values.
After constructing the test suite, the testing-retraining pipeline that is configured with \textsc{Clover} retrains the DL model under test.

Fig. \ref{fig: selection} depicts how \textsc{Clover} selects the test cases into the constructing test suite $A$ in concept. 
It illustrates that test cases are reordered in the style of layering (i.e., depicted as layers $L_i$ for $1 \leq i \leq 6$) in the descending order of the CC values of the test cases, and \textsc{Clover} selects test cases layer by layer.
Suppose the required size of a test suite is 10. In this case, all eight test cases in layers $L_1$ and $L_2$ are selected. 
The remaining two test cases are selected from $L_3$ with the descending order of CC values. Thus, the two test cases $t_8^3$ and $t_1^2$ are selected.

\subsection{Algorithms}
\label{sec: algorithms}
This section presents the detailed algorithms of \textsc{Clover}. 

\begin{algorithm}[t]
\caption{\textsc{Clover}: Context-Aware Robustness Fuzzing}
\label{alg: colver overview}
\SetKwInOut{KwIn}{Input}
\SetKwInOut{KwOut}{Output}
\KwIn{$f \gets$ model under test  \\
    $X \gets $ seeds \\
    $m \gets$ average number of fuzzing attempts for a seed \\
    $n \gets$ size of test suite \\
    }
\KwOut{test suite $A$}
$P$ = ContextFuzz($f, X, m$, ContextTranslate) \Comment{generate test cases} \\
$A$ = ContextSelect($P$, $n$) \Comment{construct a test suite} \\
\textbf{return} $A$ \\
\end{algorithm}

\begin{algorithm}[t]
\caption{ContextFuzz: Seed-Equivalent Sequence-to-Sequence Fuzzing}
\label{alg: context_fuzz}
\SetKwInOut{KwIn}{Input}
\SetKwInOut{KwOut}{Output}
\KwIn{$f \gets$ model under test \\
    $X \gets $ seeds \\
    $m \gets$ average number of fuzzing attempts per seed \\
    ContextTranslate $\gets$ seed-equivalent sequence-to-sequence translator \\
    }
\KwOut{test cases $P$}

$P = \emptyset$, $\mathcal{F} = \emptyset$, $\mathcal{I} = \emptyset$, $i = 0$
\Comment{$\mathcal{F}$ is the adversarial front}\\
\For{$x \in X$} {
    $\mathcal{F}(x) = \bm{0}$, $\mathcal{I}(x)$ = 1 \\
    $x'$ = $x$ + \textsc{processing}($x$,$\partial{obj} / \partial{x}$) \\
    $f$.predict$(x')$ \\
    \If{\rm{\textsc{isAdversarial}}($x, x'$)  }{
        $P(x)$.add($x', CC(x')$)   \\
        $\mathcal{F}(x) = \partial{obj} / \partial{x}$ 
        \Comment{initialize a $\beta$-AFO for the seed}
        \\
        $\mathcal{I}(x) = \mathcal{I}(x) + 1$ 
        \\
    }
}

\While{\rm{budget is not exhausted}}{
    \If{i \rm{= 0}}{
        $energy$ = $\langle 
        {\lceil} \frac{m \mathcal{I}(x)}{\sum_{x \in X} \mathcal{I}(x) / |X|} {\rceil} \rangle_{x \in X}$ \Comment{compute the energy for every seed} \\
    }
    $x_i = X[i]$ \\
    $AC(x_i)$ = ${\langle}\mathcal{F}(x')\rangle_{x' \in [x_i] - \{x_i\} \text{ such that } |AC(x_i)| = min(energy[i], |[x_i]|-1)}$ \\
    $\Delta_{\beta\text{-AFO}}, Q(x_i)$, \#\textit{evolved} = ContextTranslate($f, x_i, AC(x_i), \mathcal{F}(x_i)$, $\partial \textit{obj} / \partial x_i$, $\text{CC}_{max}$($P(x_i)$)) \\
    $\mathcal{F}(x_i) = \Delta_{\beta\text{-AFO}}$  \Comment{maintain the $\beta$-AFO of the seed} \\
    $P(x_i) = P(x_i)$  $\cup$ $ Q(x_i)$ \Comment{keep the test cases} \\
    $\mathcal{I}(x_i) = \mathcal{I}(x_i) + $ \#\textit{evolved} \\
    $i = (i+1) $ mod $|X|$
}
\textbf{return} $P$
\end{algorithm}

\subsubsection{The \textsc{Clover} Algorithm}
\label{sec: Clover algorithm}
Algorithm \ref{alg: colver overview} presents the overall algorithm of \textsc{Clover}. 
It accepts four parameters: 
$f$ is the model under test, 
$X$ is a set of given seeds, 
$m$ is the average number of fuzzing attempts per seed, and $n$ is the size of the test suite to be constructed.
It generates a test suite $P$ via Algorithms \ref{alg: context_fuzz} and \ref{alg: context_translate}, and returns a subset $A$ of $P$ with $n$ test cases via Algorithm \ref{alg: context_selection}.

\subsubsection{The ContextFuzz Algorithm}
\label{sec: contextFuzz_algorithm}

Algorithm \ref{alg: context_fuzz} presents the \textit{ContextFuzz} algorithm.
It accepts four inputs: 
$f$ is the model under test, 
$X$ is the seeds, 
$m$ is the mean number of fuzzing attempts per seed, and 
\textit{ContextTranslate} is our Algorithm \ref{alg: context_translate}.
It returns all generated test cases with their CC values.

\textit{ContextFuzz} sets three maps $P$, $\mathcal{F}$, and $\mathcal{I}$ to empty and the counter $i$ to 0 (line 1).
$P$ keeps the generated test cases.
The entry $\mathcal{F}(x)$ in the map $\mathcal{F}$ keeps the current $\beta$-AFO for the seed $x$.
The map $\mathcal{I}$ keeps the total number of $\beta$-AFO discovered for each seed in $X$. 

Since each $\beta$-AFO represents a successful seed evolution resulting in a test case with the highest CC value up to the corresponding fuzzing attempt, we use the entry $\mathcal{I}(x)$ to bookkeep the number of successful evolutions experienced on $x$. 
As the potentials of different seeds to produce successful evolutions in the fuzzing campaign could be different, in lines 13--15, the algorithm sets up a power schedule \cite{duo_differential_fuzzing}, computing the energy of each seed (line 14) by the formula ${\lceil} \frac{m \textit{$\mathcal{I}$}(x)}{\sum_{x \in X} \mathcal{I}(x) / |X|} {\rceil}$, which scales the average number of fuzzing attempts $m$ by the relative number of successful evolutions experienced on $x$. 

\textit{ContextFuzz} iterates over $X$ to initialize the above three maps (lines 2--11).
Like other perturbation techniques (e.g., FGSM and \textsc{RobOT}), it computes a clipped gradient for a seed $x$ (denoted by \textsc{processing}($x$, $r$)) and adds it to $x$ to generate a test case $x'$, i.e., $x' = clip(x + \epsilon r)$ and ${\parallel}x - x' {\parallel_p} < {\epsilon}$ (line 4), where $r = \nicefrac{\partial \textit{obj}}{\partial x}$ in line 4.
It also clips $x'$ to ensure the generated test cases are within the lower and upper bound, such as 0 and 1 for normalized images.
If $x'$ is a failing test case (lines 5--6), \textit{ContextFuzz} adds $x'$ with its CC value $CC(x')$ to $P(x)$,  sets $\mathcal{F}(x)$ to the gradient of $x$, and increments $\mathcal{I}(x)$ (lines 7--9).

In lines 12--24, the algorithm conducts fuzzing until exhausting the overall fuzzing budget.
It picks the current seed $x_i$ from $X$ (line 16).
It captures a set of $\beta$-AFOs for $x_i$ by the object $AC(x_i)$ (line 17), where $AC(x_i)$ in the $j^{th}$ iteration over lines 12--23 means $AC(x_i, j{-}1)$ in Section \ref{sec: seed equivalence}.
As presented in Section~\ref{sec: sq_to_sq_af}, each $\beta$-AFO in $AC(x_i)$ is a perturbation that results in one representative test case of a seed that is seed-equivalent to $x_i$. 
\textsc{Clover} controls the fuzzing budget of this seed $x_i$ to be $|AC(x_i)|$ to cut off the long tail.
Specifically, $AC(x_i)$ is modeled as an object that conceptually contains the set of $\beta$-AFOs of these seeds in the equivalence class $[x_i]_{j-1}$ (at the immediately past fuzzing round $j-1$, where line 17 simply uses the notation $[x_i]$ to stand for the equivalence class $[x_i]_{j-1}$) maximally satisfying the condition $|AC(x_i)|$ $\leq$ the energy of $x_i$ and the set does not contain the $\beta$-AFO of $x_i$ (line 17). Moreover, if $AC(x_i, j{-}1)$ is empty, the object $AC(x_i)$ is initialized to contain the $\beta$-AFO of $x_i$ as its single element. 
As long as the fuzzing budget on $x_i$ in the current fuzzing round has not been exhausted, if an element is requested from this object $AC(x_i)$ (which is modeled by an invocation of the $pop$(.) operation in line 3 of Algorithm \ref{alg: context_translate}), $AC(x_i)$.pop(.) returns the $\beta$-AFO of a seed in the equivalent class that the object has never returned before. 

In line 18, the algorithm calls \textit{ContextTranslate} to generate test cases for the seed $x_i$.
It passes the set $AC(x_i)$, the current $\beta$-AFO of $x_i$, and the highest CC value among all generated test cases of $x_i$ kept in $P(x_i)$ (denoted by the function $\text{CC}_{max}$(.)) to  \textit{ContextTranslate}.
\textit{ContextTranslate} returns three results: $\Delta_{\beta\text{-AFO}}$ (which is updated to $\mathcal{F}(x_i)$), a sequence of test cases $Q(x_i)$ (which is appended to $P(x_i)$), and the number of times (\#\textit{evolved}) that the evolution of the seed $x_i$ in the invocation to \textit{ContextTranslate} is regarded as successful (which is added to \textit{$\mathcal{I}$}($x_i$) (lines 19--22).
Line 23 resets the variable $i$ to 0 if the whole $X$ has been processed.
Thus, the number of resets on $i$ at line 22 (denoted by \textit{v} here) and the current value of $i$ maps to the fuzzing round $j$ on the seed sequence $Z$ presented in Section \ref{sec: seed_equivalent_fuzzing} are:  $j = v|X| + i$ and $\textsc{source}(z_j) = X[i]$, where $X$ refer to the input to this algorithm.

\begin{algorithm}[t]
\caption{ContextTranslate: Seed-Equivalent Sequence-to-Sequence Translator}
\label{alg: context_translate}
\SetKwInOut{KwIn}{Input}
\SetKwInOut{KwOut}{Output}
\KwIn{
    $f \gets$ model under test\\
    $x \gets $ seed for fuzzing \\
    $AC(x) \gets$ set of $\beta$-AFOs for seeds that are seed-equivalent to $x$ \\
    $\Delta_{\beta\text{-AFO}} \gets$ current $\beta$-AFO of $x$\\
    $\partial \textit{obj} / \partial x \gets$ gradient of $x$ \\
    $CC^* \gets$ the highest CC value among the generated test cases from $x$ \\
    }
\KwOut{updated version of $\Delta_{\beta\text{-AFO}}$ \\ 
        test cases $Q(x)$ for seed $x$ \\ 
        number of representative evolutions $\#\textit{evolved}$}

$Q(x) = \emptyset$, $\#\textit{evolved} = 0$ \\
\While{$|AC(x)| > 0$ \rm{and the overall budget is not exhausted}} 
{
    $\textit{dir} = AC(x)$.pop() \Comment{a $\beta$-AFO due to seed equivalence} \\
    $\Delta_{\beta\text{-AFO}}' = \textit{dir} + \Delta_{\beta\text{-AFO}} +  \partial{obj} / \partial{x} $ \\
    $x'$ = $x$ + \textsc{processing}($x$, $\Delta_{\beta\text{-AFO}}'$) \\
    \For{r= \rm{1} \rm{to} ${\lceil}\epsilon/\delta{\rceil}$} {
        $x'$ = $x'$ + \textsc{processing}($x'$, $\partial{obj} / \partial{x'}$) \Comment{move $x'$ along its gradient}\\
    }
    $f$.predict$(x')$ \\
    \If{\rm{\textsc{isAdversarial}}($x, x'$)}{
        $Q(x)$.add($x', CC(x')$) \Comment{keep the test case} \\
        \If{$CC(x') > CC^*$} {
            $\Delta_{\beta\text{-AFO}} = x' - x$ \Comment{maintain the $\beta$-AFO for $x$} \\
            $CC^* = CC(x')$ \Comment{maintain the highest CC value for $x'$} \\
            $ \#\textit{evolved} = \#\textit{evolved} + 1$ \Comment{deem the seed evolution as successful}\\
        }
    }
}
\textbf{return} $\Delta_{\beta\text{-AFO}}, Q(x), \#\textit{evolved} $
\end{algorithm}

\subsubsection{The ContextTranslate Algorithm}
\label{sec: contextTranslate}

Algorithm \ref{alg: context_translate} presents the (\textit{ContextTranslate}) algorithm of \textsc{Clover}.
It accepts six parameters: 
$f$ is the model under test.
$x$ is the current seed under fuzzing.
$AC(x)$ is the set of $\beta$-AFOs for $x$.
$\Delta_{\beta\text{-AFO}}$ is the $\beta$-AFO of $x$ obtained from the previous fuzzing rounds on $x$.
$\partial \textit{obj}/\partial x$ is the gradient of $x$. 
$CC^*$ is the highest CC value among the test cases generated from $x$ in all previous fuzzing rounds on $x$.
The algorithm aims to generate a sequence of test case candidates for $x$ with increasingly higher CC values, one such test case (unless discarded) for a $\beta$-AFO in $AC(x)$ if the budget allows.

\textit{ContextTranslate} optimizes the procedure to track the $\beta$-AFO for $x$ at the current fuzzing round.
It iteratively generates test cases of $x$ (lines 3--10) and places them into the set $Q(x)$ in line 11. 
The $\alpha$-representative test case $\overline{x}$ at the current fuzzing round is the one in $Q(x)$ with the highest CC value, and the $\alpha$-AFO is thus $\overline{x} - x$.
Specifically, suppose a test case $x'$ of $x$ is added to $Q(x)$. 
If $x'$ also has a higher CC value than the CC value of the resulting sample of $x$ + $\beta$-AFO (line 12), it implies that the piece of difference represented by $x' - x$ is a newly discovered $\beta$-AFO for $x$ at the current fuzzing round (lines 13--14).
On the other hand, if the condition in line 12 is not satisfied, this $\alpha$-AFO for $x'$ is no longer tracked because it will not affect how the $\beta$-AFO for $x$ is computed.
To save effort, the algorithm also does not track which test case is the $\alpha$-representative test case of the current seed occurrence.

In the rest of this subsection, we present the details of \textit{ContextTranslate}.

Algorithm \ref{alg: context_translate} initializes the map $Q(x)$ to empty.
It also sets the counter \#\textit{evolved} to 0,
which counts the number of $\beta$-AFOs for the seed $x$ generated by the current invocation of the algorithm.
The map and the counter record the set of test cases and the number of $\beta$-AFOs generated by the current invocation of the algorithm for the seed $x$, respectively.

The algorithm iteratively applies the sequence of $\beta$-AFOs for $x$ to produce a sequence of test cases (lines 2--18).
In line 4, it combines the $\beta$-AFO of an equivalent seed of $x$ (i.e., $\textit{dir}$), 
the current $\beta$-AFO of $x$ (i.e., $\Delta_{\beta\text{-AFO}}$), and the gradient of $x$ (i.e., $\partial \textit{obj}/\partial x$) to pinpoint a sample $x'$ in the vicinity of $x$.
Like PGD, in lines 6--8, it iteratively moves $x'$ along its gradient (for the reasons of achieving a high prediction on $x$).
However, unlike PGD, it uses a typical cyclic learning rate schedule, which scales the clipped gradient by a cyclic learning rate $\lambda(r) = |max \sin(\pi \cdot \nicefrac{r}{({\lceil}\epsilon/\delta{\rceil}+1)})|$, where $sin(.)$ is the sine function and $max$ is the maximum cyclic learning rate
\footnote{
Previous works (e.g., \cite{fast_better_free, adversarial_training_3, cyclic_learning}) inspire us that using a uniform step size in training could be ineffective for stepping into certain regions due to the regularity in stepping. 
The design of this cyclic learning rate schedule is to complete one cycle by walking through all the steps from $i = 1$ to ${\lceil}\epsilon/\delta{\rceil}$, which will lead to a series of periodic values changing from 0 to 1 then 0. 
Using the term ${\lceil}\epsilon/\delta{\rceil}+1$ in the denominator follows a typical design in activation functions to avoid the division-by-zero error. 
The number of steps (${\lceil}\epsilon/\delta{\rceil}$) corresponds to walking through one cycle of the cyclic learning rate schedule.
}.
Specifically, in the $j^{th}$ iteration, the algorithm computes a fractional clipped gradient, 
i.e., $x' = clip(x' + \epsilon \lambda(j) \nicefrac{\partial \textit{obj}}{\partial x'})$ and ${\parallel}x - x' {\parallel_p} < {\epsilon}$.

After having perturbed $x'$, it applies $f$ to infer $x'$ for prediction label inspection (line 9).

In lines 10--17, the algorithm maintains its data structure for tracking the generated test cases and adversarial front objects for $x$.
Specifically, if  $x'$ is adversarial (line 10), it adds $x'$ and its CC value (which is $CC(x')$) to the map $Q(x)$ (line 11).
Moreover, if $x'$ is higher than all historical test cases of $x$ measured in CC (line 12), the algorithm updates $\Delta_{\beta\text{-AFO}}$ to $x' - x$ (line 13), updates $CC^*$ to $CC(x)$ (line 14), and increments the counter $\#\textit{evolved}$ (line 15). 

The algorithm finally returns $\Delta_{\beta\text{-AFO}}$, $Q(x)$, and $ \#\textit{evolved} $ (line 19).

\begin{algorithm}[t]
\caption{ContextSelect: Context-Aware Test Suite Construction}
\label{alg: context_selection}
\SetKwInOut{KwIn}{Input}
\SetKwInOut{KwOut}{Output}
\KwIn{$P \gets$ pool of test cases \\
    $n \gets$ the required size of the test suite}
\KwOut{list of $n$ test cases}
$X$ = $P$.keys() \\
\If{P \emph{has no CC}}{
    compute $CC(t)$ for each test case $t$ in $P$
}

$\forall x \in X$, Sort($P(x)$) in descending order of CC \\
$A$ = ${\langle}{\rangle}$, $i$ = 1 \\
\While{$|A| < n$} { 
    $\textit{layer} = \{P(x)[i] \mid x \in X \}$ \\
    \If {$|\textit{layer}| + |A| > n $} {
        Sort(\textit{layer}) in descending order of CC \\
        \textit{layer} = $\langle \textit{layer}[j] \rangle_{j \in [1, n{-}|A|]}$ \Comment{get the top $n - |A|$ elements} \\
    }
    $A$ = $A + \textit{layer}$ \Comment{extend $A$ with the current \textit{layer}} \\ 
    $i = i + 1$
}
\textbf{return} $A$
\end{algorithm}

\subsubsection{The ContextSelect Algorithm}
\label{sec: context_selection}
Algorithm \ref{alg: context_selection} presents the \textit{ContextSelect} test case selection strategy.
It accepts two parameters: a pool of test cases $P$ and the required number $n$ of test cases to be selected. 
In the map $P$, the key is a seed, and the value is a set of test cases with their CC values.

Algorithm \ref{alg: context_selection} collects the seeds stored in $P$ into a set $X$ (line 1).
It computes the CC values for these test cases if missing (lines 2--4).
In line 5, it sorts the test cases in $P(x)$ for each seed $x \in X$ in descending order of CC.
Line 6 initializes an empty list $A$ and sets the counter $i$ as 1.
Lines 7--15 iteratively include test cases with higher CC values of each seed into $A$.
Each iteration performs the following:
The algorithm collects the $i^{th}$ test case of each seed $x \in X$ into \textit{layer} (line 8).
Then, it checks whether the size of \textit{layer} is larger than the number of samples remaining to be selected (line 9). 
If this is the case, it sorts \textit{layer} in descending order of CC (line 10) and trims \textit{layer} by dropping the samples beyond the number of samples to be selected (line 11).
Next, it appends \textit{layer} to $A$ (line 13) and increments the counter $i$ by 1 (line 14).
Line 16 returns the prioritized test suite $A$.

\subsection{Tracking the Equivalence Classes for Seed Equivalence }  
\label{sec: eqv_class_computation}
The \textit{ContextFuzz} algorithm requires knowing an equivalence class in line 17 of Algorithm~\ref{alg: context_fuzz} at each fuzzing round, which can be computed efficiently: linear to the number of seeds in $X$ plus the number of test cases generated by the \textit{ContextTranslate} algorithm.
We recall that if the current iteration in the conceptual model of \textsc{Clover} is $j$, then the notation $[x]$ in line 17 refers to the equivalence class $[x]_{j-1}$ in Definition \ref{def: seed_equivalence}.

Suppose $\mathcal{G}$ is a $p \times q$ matrix, and $N$ is a vector of size $p$.
The number of rows $p$ of $\mathcal{G}$ equals the number of class labels, representing the possible seed labels of a sample returned by \mbox{\textsc{semantic}}(.).
The number of columns $q$ of $\mathcal{G}$ equals the number of class labels plus 1, representing the possible adversarial labels of a sample. 
The additional column, denoted by $\bot$, indicates that the seed is not marked with any adversarial label.

The matrix is first initialized at line 3 of  Algorithm~\ref{alg: context_fuzz}.
It places each seed $x \in X$ to the cell $\mathcal{G}(c, \bot)$ where $c$ is the label returned by \mbox{\textsc{semantic}}($x$) and sets  $N[x] = \bot$.
Since $\mathcal{F}(x)$ is initialized as 0 (i.e., no difference), the $\beta$-AFO for $x$ at this moment is 0.

Ckecking the condition \textsc{isAdversarial}($x, x'$) in line 6 of Algorithm~\ref{alg: context_fuzz} further triggers the updates of $\mathcal{G}$ and $N$.
If the condition is satisfied, we move the corresponding seed $x$ from the cell $\mathcal{G}(c, \bot)$ to the cell $\mathcal{G}(c, v)$, where $c$ is the label returned by \mbox{\textsc{semantic}}($x$) and $v$ is the adversarial label of $x'$ obtained from this line 6.
It also updates the cell $N[x]$ to $v$.

In each invocation of the \textit{ContextTranslate} algorithm for the seed $x$ as the value for input parameter $x$, whenever line 11 of Algorithm~\ref{alg: context_translate} executes, we get the adversarial label (denoted by $v$) from line 9 of Algorithm~\ref{alg: context_translate}. 
The sample $x$ is thus removed from the cell $\mathcal{G}(c, N[x])$ and added to the cell $\mathcal{G}(c, v)$. 
The cell $N[x]$ is updated to $v$ as well.

The seeds in the equivalence class needed in line 17 of Algorithm~\ref{alg: context_fuzz} is the full set of seeds in the cell $\mathcal{G}(c, N[x])$.

\section{Experiment}
\label{sec: experiment}
This section presents the experiment to evaluate \textsc{Clover} in the scope of the testing-retraining pipeline that is configured with \textsc{Clover} as the DL testing technique to improve the robustness of DL models.
The tool, the source code, the model weights, the results, and the data sets of the experiments are available at \cite{clover_github}.

\subsection{Research Questions}
As presented in Section \ref{sec: AM methodology}, the AM methodology includes a testing-retraining pipeline. Like the experiments presented in \cite{robot, quote_robot_extend}, the pipeline can be configured into two configurations.
In Configuration A, the pipeline constructs a test suite by selecting test cases from a given test pool via a configured DL testing technique. This configuration refers to the scenario for applying a test case selection technique (e.g., DeepGini \cite{deepgini}), where users provide a test pool to the overall workflow before the testing-retraining pipeline in the workflow is executed.
In Configuration B, the pipeline generates a test suite via a configured DL fuzzing technique, which is a scenario of test case generation.
Corresponding to these two configurations, we refer to \textsc{Clover} with $P$ in line 2 of Algorithm \ref{alg: colver overview} input from an external source and line 1 of Algorithm \ref{alg: colver overview} deleted as \textbf{\textsc{Clover} in Configuration \textit{A}}, and the original \textsc{Clover} as \textbf{\textsc{Clover} in Configuration \textit{B}}.

To align with the setting in these experiments of DL testing techniques for robustness improvement delivered through the testing-retraining pipeline in the literature \cite{deepgini, robot, tensorfuzz, deepxplore, dlfuzz, sensei, seed_select_dl_test, quote_robot_extend, ATS2022, drfuzz, test_selection_tosem}, in our main experiment (for answering research questions RQ1--RQ5 below), we evaluate \textsc{Clover} in both configurations by studying the robustness-oriented quality of the constructed test suites and the robust accuracy improvement%
\footnote{
We note that the literature on DL model fuzzing \cite{sensei, robot, quote_robot_extend, deepxplore, fuzzing_roadmap} has reiterated that this measurement metric is very important in studying the effectiveness of fuzzing techniques for DL model testing.}
achieved by \textsc{Clover} on clean models.
Additionally, we evaluate the performance of \textsc{Clover} on adversarially trained DL models by an exploratory study through the experiment for answering research question RQ6.

Specifically, we aim to answer the following six research questions.

\begin{enumerate}
\item[RQ1:]
To what extent is \textsc{Clover} in Configuration \textit{A} effective in constructing test suites for robustness improvement delivered through the testing-retraining pipeline?

\item[RQ2:]
Is there any difference in robustness improvement by using test suites with different CC values constructed by \textsc{Clover} in Configuration \textit{A} delivered through the testing-retraining pipeline?

\item[RQ3:]
To what extent is \textsc{Clover} in Configuration \textit{B} effective in robustness improvement delivered 
through the testing-retraining pipeline? How about the robustness-oriented quality of the constructed test suites?

\item[RQ4:]
To what extent is \textsc{Clover} in Configuration \textit{B} effective in robustness improvement delivered through the testing-retraining pipeline if the fuzzing technique is configured with an existing state-of-the-art testing metric instead of the testing metric CC to identify test cases? How about the robustness-oriented quality of the constructed test suites?

\item[RQ5:]
To what extent do the major design decisions and hyperparameters to configure \textsc{Clover} in Configuration \textit{B} affect the performance of \textsc{Clover} in robustness improvement delivered through the testing-retraining pipeline?

\item[RQ6:]
To what extent is \textsc{Clover} effective in test case generation on adversarially trained models for the model validation purpose? 
\end{enumerate}

RQ1 and RQ2 evaluate the effectiveness of \textsc{Clover} in Configuration \textit{A}, and RQ3 to RQ4 evaluate \textsc{Clover} in Configuration \textit{B}.
Like the previous experiments~\cite{deepgini, robot, quote_robot_extend, sensei, seed_select_dl_test, surprise_adequacy}, the pair RQ1 and RQ3 assess the effectiveness of \textsc{Clover} in robust accuracy improvement delivered through the testing-retraining pipeline.
The pair RQ2 and RQ4 assess the effects of configuring \textsc{Clover} with the metric CC and other peer metrics on the corresponding robust accuracy improvement.
Furthermore, on the one hand, \textsc{Clover} in Configuration \textit{A} does not generate test cases (and thus RQ1 and RQ2 do not study the ability of \textsc{Clover} to generate test cases).
On the other hand, \textsc{Clover} in Configuration \textit{B} generates test cases.
In RQ3 and RQ4, we additionally evaluate \textsc{Clover} about the robustness-oriented quality of the constructed test suites produced by the technique. 
RQ5 explores the key design decisions and hyperparameters in formulating a fuzzing attempt in the \textsc{Clover} algorithm to study the extent these factors affect \textsc{Clover} in achieving robustness improvement delivered through the testing-retraining pipeline.
To explore the effect of fuzzing on adversarially trained models, we further evaluate \textsc{Clover} on these models about its robustness-oriented quality of the generated test cases through RQ6.

We should note again that robustness improvement is achieved through applying the test suite constructed by a DL testing technique to a retraining step in the pipeline. Thus, through the comparison between different DL testing techniques to construct test suites and apply their test suites to the same retraining step with the same DL model under retraining, one can compare the relative robustness improvement of these DL testing techniques in a controlled manner.

\subsection{Experimental Setup}
This subsection details the experimental setup to answer the research questions.

\subsubsection{\textbf{\textit{Implementation}}} 
We implement our test framework in tensorflow-gpu 2.4.0 on a computer with Ubuntu 20.04 equipped with an Intel Xeon Gold 6136 processor, 256GB RAM, and an NVIDIA GeForce RTX 2080Ti GPU card with 12GB VRAM. 
We implement all techniques on it after porting \textsc{Adapt} and \textsc{RobOT} from their publicly-available code repositories \cite{adapt_github, robot_github}.
The \textsc{RobOT} code repository includes the implementations of PGD and FGSM.
We also reuse their implementations after porting.
We adopt the retraining script \cite{robot_github} of \textsc{RobOT} for all techniques to retrain each model under test 40 epochs and refer to it as \textsc{Retrain}. 
The retraining script \textsc{Retrain}($f, A$) follows the typical finetuning training procedure on a test suite $A$ (which is the set of selected test cases) and the original training dataset of the model under test $f$ to reduce the training error. 
We configure the learning rate of the retraining script with the learning rate value appearing at the end of the training process for each corresponding model under test, i.e., 0.0001 for cases \circled{1}, \circled{3}, \circled{4} and 0.001 for \circled{2} in Table \ref{tab: clean_model}.

We have ensured all models converged in the training/retraining processes of our experiments. 
We enable early stopping in the scripts to reduce overfitting when training the DL models under test by calling the ``EarlyStopping'' API in the callback and restoring the best weights of models.\footnote{
In our pre-experiment, we have also attempted to use another popular approach to early stopping: Keeping each model after each training/retraining epoch and selecting the one right before the change of validation accuracy turns direction after reaching the maximum (despite the small variations) when studying \textsc{Clover} on cases \circled{1}--\circled{4}. From our data analyses on their results, we do not spot any noticeable difference from the present results reported in RQ1--RQ5 in this paper. 
}

The testing-retraining pipeline is to apply a DL testing technique on a model $f$ to construct a test suite $A$, followed by applying \textsc{Retrain}($f, A$) to obtain a retrained model $f'$. We adopt the pipeline script provided by \cite{robot_github}.
The robust dataset to assess DL models $f$ and its retrained version $f'$ will be introduced in Section~\ref{sec: robustness_assessment_universe}.

\subsubsection{\textbf{\textit{Datasets}}} 
We adopt the following four datasets as our benchmark datasets: FashionMnist \cite{fashion_mnist}, SVHN \cite{svhn}, CIFAR10 \cite{cifar}, and CIFAR100 \cite{cifar}. 
The FashionMnist and CIFAR10 datasets each contain ten classes with 5000 training and 1000 test images. 
The SVHN dataset contains 73257 digits for training and 26032 digits for testing, divided into ten classes.
The CIFAR100 dataset contains 60000 samples with 500 training and 100 test images, divided into 100 classes.
For each benchmark dataset, we construct the validation dataset by randomly sampling 5000 images from the downloaded training dataset, where the remaining images serve as the training samples, and we use the downloaded test datasets as our (clean) test datasets.

\begin{table}[t]
\caption{Benchmark Cases of Clean Models (Baselines)}
\label{tab: clean_model}
\begin{tabu}{|c||c|c|c|c|c|c|}
    \hline
    \multirow{2}{*}{Case} & 
    \multirow{2}{*}{Dataset} &
    \multirow{2}{*}{Model} &
    \multirow{2}{*}{Parameters} &
    \multirow{2}{*}{\begin{tabu}[c]{@{}c@{}} Training \\ Accuracy \end{tabu}} &
    \multirow{2}{*}{\begin{tabu}[c]{@{}c@{}} Validation \\ Accuracy \end{tabu}} &
    \multirow{2}{*}{\begin{tabu}[c]{@{}c@{}} Test \\ Accuracy \end{tabu}} \\
    & & & & & & \\ \hline
    \circled{1} & FashionMnist & VGG16 & 33,624,202 & 97.24 & 94.18 & 93.77 \\ \hline
    \circled{2} & SVHN & LeNet5 & 136,886 & 90.14 & 89.38 & 88.98 \\ \hline
    \circled{3} & CIFAR10 & ResNet20 & 273,066 & 92.16 & 89.38 & 88.45 \\ \hline
    \circled{4} & CIFAR100 & ResNet56 & 867,620 & 72.78 & 60.52 & 61.49 \\ \hline
\end{tabu}
\end{table}

\begin{table}[t]
\caption{Benchmark Cases of Adversarially Trained Models (Baselines)\tablefootnote{``Ratio'' refers to the proportion of samples in each batch to be replaced with their adversarial counterparts in the function \textit{BasicIterativeMethod}~\cite{adversarial_robustness_toolbox} for data augmentation.}}
\label{tab: adv_trained_model}
\begin{tabular}{|c||c|c|c|c|c|c|c|}
    \hline
    \multirow{2}{*}{Case} & 
    \multirow{2}{*}{Dataset} &
    \multirow{2}{*}{Model} &
    \multirow{2}{*}{Ratio} &
    \multirow{2}{*}{Parameters} &
    \multirow{2}{*}{\begin{tabu}[c]{@{}c@{}} Training \\ Accuracy \end{tabu}} &
    \multirow{2}{*}{\begin{tabu}[c]{@{}c@{}} Validation \\ Accuracy \end{tabu}} &
    \multirow{2}{*}{\begin{tabu}[c]{@{}c@{}} Test \\ Accuracy \end{tabu}} \\
    & & & & & & & \\ \hline
    \circled{5} & CIFAR10 & ResNet20 & 50\% & 273,066 & 90.98 & 84.72 & 84.26 \\ \hline
    \circled{6} & CIFAR10 & ResNet20 & 100\% & 273,066 & 86.93 & 82.04 & 81.58 \\ \hline
    \circled{7} & CIFAR100 & ResNet56 & 50\% & 867,620 & 77.22 & 58.26 & 59.20 \\ \hline
    \circled{8} & CIFAR100 & ResNet56 & 100\% & 867,620 & 70.92 & 55.72 & 56.74 \\ \hline
\end{tabular}
\end{table}

\subsubsection{\textbf{\textit{DL Models for Benchmark}}}
Like the experiments presented in existing work \cite{adaptfuzz, robot, quote_robot_extend, deephunter, deepmutation++, deepxplore, tensorfuzz, dlfuzz, npc}, to evaluate \textsc{Clover}, we select serveral DL models for this purpose: VGG16 \cite{very_deep_cov}, LeNet5 \cite{lenet5}, ResNet20 \cite{resnet20}, and ResNet56 \cite{resnet20}. 
We include an earlier model, LeNet5 \cite{lenet5}, because \textsc{RobOT} \cite{robot} is the state-of-the-art robustness-oriented fuzzing technique to generate test cases for robustness testing, and its experiment uses LeNet5 extensively. 
DL testing and maintenance research \cite{robot, quote_robot_extend, deephunter, dlfuzz, tensorfuzz, ood_detect, npc, deeppatch} widely use the other three DL models for experiments. 

We adopt their existing implementations and training scripts of model architectures from the official TensorFlow library and train them from scratch on the clean training and validation datasets for each benchmark. 
We evaluate the resulting models on their test datasets, where we observe that the test accuracy of each model matches the published benchmark values in the literature \cite{fashionmnist_benchmark, robot_github, cifar10_benchmark, cifar100_benchmark}.
Table \ref{tab: clean_model} summarizes the number of parameters, training accuracy, validation accuracy, and test accuracy of each model under test.
We index these four models as cases \circled{1}, \circled{2}, \circled{3}, and \circled{4}, respectively.

Apart from applying \textsc{Clover} and peer fuzzing techniques as a DL testing technique in a testing-retraining pipeline, we also explore an alternate scenario to use \textsc{Clover} and peer fuzzing techniques to quality assure the retrained models (i.e., adversarially trained models) that are produced by the pipeline of pure adversarial training.
Specifically, we start from the above-mentioned trained models as the DL models under test and further adversarially retrain them to produce four adversarially trained models, summarized in Table \ref{tab: adv_trained_model}.
We index these four models as cases \circled{5}--\circled{8}.%
\footnote{
We note that the experiments reported in a vast majority of existing DL fuzzing literature \cite{adaptfuzz, robot, deephunter, deepmutation++, deepxplore, tensorfuzz, dlfuzz, sensei, ATS2022, drfuzz} did not report the results on fuzzing adversarially trained DL models.
}
We will apply each of \textsc{Clover} and peer fuzzing techniques on these adversarially trained models to generate test cases. To facilitate the comparison, we use test suites of the same size to compare the robustness-oriented quality of the respective test suites, where the measurement metrics will be introduced in Section~\ref{sec: evaluation metrics}.

Specifically, we follow \cite{arch_search_adv_robust, adversarial_robustness_toolbox, test_selection_tosem} to adopt the widely-used IBM adversarial robustness toolbox (ART) version 1.17 ~\cite{adversarial_robustness_toolbox} to adversarially retrain the models in cases \circled{3} and \circled{4} to produce models indexed as cases \circled{5}--\circled{8} with the function \textit{AdversarialTrainer} with the default hyperparameters and the function \textit{BasicIterativeMethod}~\cite{adversarial_robustness_toolbox} for data augmentation with the parameter ``\textit{ratio}'' (which is a parameter to specify the proportion of samples in each batch to be replaced with their adversarial example counterparts%
)~\footnote{
The article~\cite{arch_search_adv_robust} is a supplementary document of the article~\cite{neuro_cell_search} to present the experiments of the latter article. ART \cite{adversarial_robustness_toolbox} is a popular automated adversarial robustness toolbox and does not require users to provide hyperparameters such as the learning rate and the early stopping rule.}. 
The models in cases \circled{5} and \circled{6} are retrained from case \circled{3}, and the models in cases \circled{7} and \circled{8} are retrained from case \circled{4}, both with the values of ratio set to 0.5 and 1.0, respectively. 

We note that our infrastructure is built on top of TensorFlow.
To identify adversarially \emph{pretrained} models on our datasets and architecture, we have attempted to port those models built on Pytorch to Tensorflow via ONNX \cite{onnx}, but we have found that ONNX nonetheless has produced converted models with severe losses in accuracy. So, we do not adopt these converted models.
We have also attempted to search on the web about the well-maintained official repository of pretrained models implemented on Tensorflow for our datasets but failed.
Therefore, we choose to adopt the state-of-the-art framework ART, which automatically trains a model without user interventions to provide hyperparameters such as the learning rate and early stopping criteria to the ART.
We have also attempted to produce adversarially trained models of cases \circled{1} and \circled{2}.
However, we have observed that the application of adversarial training using \textit{BasicIterativeMethod} and \textit{AdversarialTrainer}~\cite{adversarial_robustness_toolbox} function on cases \circled{1} and \circled{2} led to a significant decrease in test accuracy.
For instance, ART reduces the test accuracy of the resulting models by more than $70\%$ when setting \textit{ratio} to 1.0 for \textit{BasicIterativeMethod} on SVHN with LeNet5. 
Thus, we exclude their adversarially trained versions as subjects in our experiments.

\subsubsection{\textbf{\textit{Selection Universes for Test Case Selection}}}
\label{sec: selection_universes}
In Configuration A, a DL testing technique selects test cases from a test pool, referred to as the \textit{selection universe} in this paper, to construct test suites.
To conduct a fair controlled experiment, the previous work \cite{robot, quote_robot_extend} prepared one or more common baselines for each dataset for all techniques to select test cases from it to construct test suites.
We follow this practice in our experiment.
Specifically, we prepare three common baselines for each dataset for all test case selection techniques to select samples from it for test case construction and check the robustness improvement effects of the constructed test suites through the testing-retraining pipeline. 

The first selection universe follows the setting presented in \cite{robot}. 
It uses FGSM \cite{fgsm} and PGD \cite{pgd} (see Section~\ref{sec: related_methods_limitations} for the introduction of these two techniques) to generate the test pool because they are frequently used as baselines to benchmark techniques in many experiments and easily produce many hard adversarial examples.
Following \cite{robot, quote_robot_extend}, we configure each of FGSM \cite{fgsm} and PGD \cite{pgd} in Configuration \textit{A} to run on the training dataset of each model under test to generate 50000 test cases.
The first selection universe $P_{train}^{\textsc{FGSM+PGD}}$ is the union of these two sets of test cases (100000 in total).

The second and the third selection universes $P_{train}^{\textsc{Adapt}}$ and $P_{train}^{\textsc{RobOT}}$ are generated by \textsc{Adapt} and \textsc{RobOT} on the training dataset of each model under test, respectively.
We configure \textsc{Adapt} and \textsc{RobOT}, respectively, to run 18000 seconds, the maximum fuzzing budget used for our whole experiment.
The descriptive statistics of $P_{train}^{\textsc{Adapt}}$ and $P_{train}^{\textsc{RobOT}}$ are shown in Table \ref{tab: throughout} and discuss in Section \ref{sec: threats_to_validity}.

\subsubsection{\textbf{\textit{Peer Techniques}}}
\textsc{Clover} in Configuration \textit{A} does not generate any test cases. Rather, it selects test cases from test pools. As such, we compare it with four peer test case selection techniques: {\textsc{Random}}, {\textsc{DeepGini}} \cite{deepgini}, {\textsc{be-st}} \cite{robot}, and {\textsc{km-st}} \cite{robot}.
\textsc{Random} is the widely used baseline in the experiments of software engineering research. It randomly selects $n$ samples from a given set of test cases. 
The next three techniques have been reviewed in Section~\ref{sec: related_methods_limitations}, which, in brief, select test cases in descending order of the Gini index, among the top-performing test cases and worse-performing test cases in terms of FOL values, and among test cases from each equally-divided subrange of FOL values, respectively. 

We compare \textsc{Clover} in Configuration \textit{B} (the original Algorithms \ref{alg: colver overview}--\ref{alg: context_selection}) with \textsc{Adapt} \cite{adaptfuzz} and \textsc{RobOT} \cite{robot}.
Both peer techniques have been reviewed in Section~\ref{sec: related_methods_limitations}.
They represent state-of-the-art coverage-based and loss-based fuzzing techniques, respectively. 
We configure each of the two fuzzing techniques to apply \textsc{km-st} to construct a subset $A$ of the set of test cases generated by the fuzzing technique. 
We adopt \textsc{km-st} rather than \textsc{be-st} because the effect of robustness improvement for both \textsc{be-st} and \textsc{km-st} are similar in \textsc{RobOT}'s original experiment (their Fig.~7) \cite{robot}, and \textsc{km-st} is more effective than \textsc{be-st} in most cases in our Experiment 1 as summarized in Fig. \ref{fig: rq1_1} of the present paper.

\subsubsection{\textbf{\textit{Hyperparameters}}} 
\label{sec: hyperparameters}
We use the default setting of Adam in the official TensorFlow library for training the clean models, where Adam uses 0.001 as the initial learning rate. It reduces the rate after 50 epochs by a factor of 10 at every 30 epochs. We observed the learning rates stopped at 0.0001 for cases \circled{1}, \circled{3}, and \circled{4}, and 0.001 for \circled{2} when the training process finished. 
FGSM and PGD are set to have the step size of 0.03 with a single step, and 0.03/6 with ten steps each, respectively, for cases \circled{1} and \circled{2}, and of 0.01 with a single step, and 0.01/6 with ten steps each, respectively, for cases \circled{3} and \circled{4} by following \cite{robot, robot_github}.
The step size and single step size of cases \circled{3} and \circled{4} are also used for cases \circled{5}--\circled{8} in \textit{BasicIterativeMethod} \cite{adversarial_robustness_toolbox} to generate adversarially trained models.

We follow \cite{adaptfuzz, robot} to set up \textsc{Adapt}: the activation threshold for the neuron coverage \cite{deepxplore} is 0.5. 
The time budget for each seed is 10 seconds for cases \circled{1}--\circled{3} and 20 seconds for cases \circled{4}.
We follow \cite{robot} to set the hyperparameters of \textsc{RobOT} (presented in the original symbols in \cite{robot}): $\xi = 10^{-18}$, $k = 5$, $\lambda = 1$, and $iters = 3$. 
Other hyperparameters, including the number of sections for \textsc{km-st} to be 4, are set according to its source code \cite{robot_github}.

For \textsc{Clover}, we set $p$-norm = $L_{\infty}$-norm, $k = 20$, $m = 5$, $\delta = 0.01$, and $max = 0.2$. 
We set $\epsilon$=0.05 for cases \circled{1}--\circled{4} and $\epsilon$=0.025 for cases \circled{5}--\circled{8}.
We choose these values for $\epsilon$ to ensure that the distance between each seed and their test cases on average is comparable to peer techniques for the purpose of comparison.\footnote{
Initially, we chose $\epsilon = 0.05$ for cases \circled{5}--\circled{8}. 
However, after fuzzing these models, we found that the corresponding mean $L_2$-norm distances of the selected test cases are larger than those for \textsc{Adapt} and \textsc{RobOT}, which are listed in the paragraph after the present paragraph in the main text of the paper. Therefore, we reduced it by half and observed that the resulting mean $L_2$-norm distances achieved by \textsc{Clover} became comparable with these for \textsc{Adapt} and \textsc{RobOT}.
}
The current values of $k$ and $m$ are based on some experimental trials (see Section \ref{sec: experimental_procedure}). We set $\delta$ to 0.01 because it is the smallest value with two decimal places (where a number with two decimal places is popularly stated as the value for measuring a perturbation bound in the literature on DL model testing\cite{deepxplore, deepmutation++}), much smaller than $\epsilon$. The number of steps ${\lceil}\epsilon/\delta{\rceil}$ in Algorithm \ref{alg: context_translate} is 5. The current value for $max$ is chosen so that the learning rate schedule can complete one cycle. Therefore, the maximum cyclic learning rate $max$ for each step is set to $1/5=0.2$.

\textsc{Adapt} and \textsc{RobOT} perform test case generations while measuring the $L_2$-norm distance between a seed and its test cases in their experiments \cite{adaptfuzz, robot}, but their papers do not present how to find the specific value for the fuzzing boundary $\epsilon$ in their experiments.
Our experiment follows the default parameters specified in their repositories and follows \textsc{Adapt} to measure the mean $L_2$-norm distances for \textsc{Adapt}, \textsc{RobOT}, and \textsc{Clover} between the seeds and their test cases, respectively.
The results for \textsc{Adapt}, \textsc{RobOT}, and \textsc{Clover} are summarized below, from left to right in each case.
\begin{multicols}{2}
\begin{itemize}
    \item Case \circled{1}: 0.57, 0.34, 0.08.
    \item Case \circled{2}: 0.06, 0.06, 0.05.
    \item Case \circled{3}: 0.05, 0.12, 0.05.
    \item Case \circled{4}: 0.07, 0.08, 0.07.
\end{itemize}
\begin{itemize}
    \item Case \circled{5}: 0.03, 0.03, 0.03.
    \item Case \circled{6}: 0.03, 0.03, 0.03.
    \item Case \circled{7}: 0.04, 0.03, 0.03.
    \item Case \circled{8}: 0.03, 0.03, 0.03.
\end{itemize}
\end{multicols}
Although \textsc{Clover} uses the $L_{\infty}$-norm distance, the mean $L_2$-norm distance of the test cases generated by \textsc{Clover} is similar to these generated by \textsc{Adapt} or \textsc{RobOT} --- the test cases generated by \textsc{Adapt} and \textsc{RobOT} in our experiments are not more restrictive than these generated by \textsc{Clover}.

\subsubsection{\textbf{\textit{Test Dataset and Robustness Assessment Universe}}}
\label{sec: robustness_assessment_universe}
We adopt the downloaded test dataset of each model under test (\circled{1} to \circled{4}) as our test dataset. 
We follow the methodology in \cite{robot, quote_robot_extend} to construct our robustness assessment universe, named $P_{test}$.
Specifically, we repeat the procedure that generates $P_{train}^{\textsc{FGSM+PGD}}$ presented in Section~\ref{sec: selection_universes} to generate 20000 test cases for each of $P_{test}$ from the samples in the corresponding test dataset instead of 100000 test cases for $P_{train}^{\textsc{FGSM+PGD}}$ from the samples in the corresponding training dataset.

\subsubsection{\textbf{\textit{Evaluation Metrics}}}
\label{sec: evaluation metrics}
Apart from measuring the CC values, we measure the robustness-oriented quality of the constructed test suites and the robustness improvement achieved by a technique delivered through the testing-retraining pipeline.

Suppose a fuzzing technique generates a test suite $A$ from a seed list $X$ or constructs a test suite $A$ from a selection universe when testing a DL model $f$.

We measure the mean CC value of $A$, denoted by \textbf{\textit{\textit{\#CC}}}, defined as $\sum_{t \in A} CC(t) \div |A|$.

The metrics {\textit{\textit{\#AdvLabel}}} \cite{adaptfuzz, ATS2022, drfuzz} and {\textit{\textit{\#Category}}} \cite{deephunter, adaptfuzz, ATS2022, drfuzz} measure the number of unique adversarial labels per seed at the test suite level and the number of seeds that the technique \textit{can} produce test cases (kept in $A$) from them, respectively.
We simply refer to these two metrics as the number of unique adversarial labels and the number of unique categories, respectively.
It is important to ensure the diversity of the generated test suite and test cases, as failures exposed from the same seed can indicate the existence of a common defect in the model \cite{ATS2022, drfuzz}. 
The inclusion of diverse test cases provides valuable feedback to developers, which helps them better understand the problem at hand and improves the overall robustness of the model \cite{deephunter, mode, adaptfuzz, ATS2022, drfuzz}.

\begin{description}
\item[\textit{\textit{\#AdvLabel}} \cite{adaptfuzz, ATS2022, drfuzz}] achieved by a technique on testing $f$ by generating a test suite $A$ is the total number of unique pairs of a seed in $X$ and an adversarial label of any test cases (in $A$) of the seed, defined as $|\{(x,f(t)) \mid x \in X \land t \in A \land \text{$x$ is the seed for the test case $t$}\}|$.
It measures the diversity and magnitude of the misclassification patterns captured by the test suites produced by the technique.

\item[\textit{\textit{\#Category}} \cite{adaptfuzz, deephunter, ATS2022, drfuzz}] for a technique is the number of seeds having at least one test case in $A$, defined as $|\{x \mid x \in X \land t \in A \land \text{$x$ is the seed for the test case $t$}\}$.
It measures the scope of coverage achieved by the technique to produce adversarial examples successfully from the pool $X$.
\end{description}
A higher \textit{\textit{\#AdvLabel}} or a higher \textit{\textit{\#Category}} achieved by a test suite indicates that the test suite successfully contains more misclassification patterns, thereby the technique constructing the test suite considered to be more effective in test suite construction \cite{adaptfuzz, deephunter, ATS2022, drfuzz}.

To evaluate a DL testing technique on robustness improvement on clean models delivered through the testing-retraining pipeline, we measure the resulting robust accuracy improvement following the experiments presented in the most recent research in the field \cite{quote_robot_extend, seed_select_dl_test, robot, deepgini, sensei, surprise_adequacy, deeptest, test_selection_tosem, ATS2022, drfuzz}. 
Suppose the DL model $f$ is retrained on $A$ and the original training dataset to produce a retrained model $f'$.
We refer to the proportion of samples in the robustness assessment universe $P_{test}$ such that $f'$ predicts their labels to the ground truth labels in top-1 prediction as the \textit{\textbf{robust accuracy}} achieved by the technique.
We compute the \textbf{\textit{robust accuracy improvement achieved by the technique}} by subtracting the robust accuracy of $f$ on $P_{test}$ from the robust accuracy of $f'$ on $P_{test}$.

We measure the CC values of the test cases in $A$ on each of $f$ and $f'$ and reorder $A$ in ascending order of CC values to produce a reordered list (denoted as $U$ for $f$ and $V$ for $f'$).
The \textbf{\textit{mean CC reduction}} achieved by the technique is the total change in CC values of the samples in the same index positions from $f$ to $f'$, which is computed as $\sum_{1 \leq i \leq |A|} (CC(U[i]) - CC(V[i]))) \div |A|$.

\subsubsection{\textbf{\textit{Experimental Procedure}}} 
\label{sec: experimental_procedure}
We conduct the following experiments to answer the RQs. 

\textbf{Experiment 1 (for Answering RQ1):} 
In the first sub-experiment (Experiment 1a), for each model (cases \circled{1}--\circled{4}), we run \textsc{Clover} in Configuration \textit{A} to select $n_1$ test cases from the selection universe $P_{train}^{\textsc{FGSM+PGD}}$ of the model, which outputs a test suite for each $n_1$ in $N_1 = \{1000, 2000, 4000, \\ 6000, 8000$, $10000, 20000 \}$. 
We apply the testing-retraining pipeline to get a retrained model for each such test suite.
We measure the robust accuracy of each retrained model and compute the robust accuracy improvement.
We repeat Experiment 1a except that we use each of $P^{\textsc{Adapt}}_{train}$ and $P^{\textsc{RobOT}}_{train}$ instead of $P^{\textsc{FGSM+PGD}}_{train}$, referred to as Experiment 1b and 1c. 
(We choose $n_1$ up to $20000$ because the original experiments \cite{robot} of \textsc{km-st} and \textsc{be-st} use a similar range.)
We repeat Experiments 1a, 1b, and 1c except for using each of \textsc{Random}, \textsc{DeepGini}, \textsc{be-st}, and \textsc{km-st} instead of \textit{ContextSelect}.
As such, each technique constructs 84 test suites and 84 retrained models.

We conduct the Wilcoxon signed-rank test \cite{wilcoxon} at the 5\% significance level with Bonferroni correction on the robust accuracy improvements achieved by each of \textsc{DeepGini}, \textsc{be-st}, \textsc{km-st}, and \textsc{Clover} compared to those achieved by \textsc{Random} on each selection universe over all four cases as a whole.
(We pair the robust accuracy improvements of all techniques by the same combination of test suite size and selection universe.)
We also calculate the effect size by Cohen's $d$ \cite{cohen_d} to check whether the difference in robust accuracy improvement of each pair of techniques is observable. 
The effect size measures the strength of the relationship between two variables in a population divided into several magnitudes, including 0.01, 0.20, 0.50, 0.80, 1.20, and 2.0, corresponding to very small, small, medium, large, very large, and huge strength levels \cite{cohen_d}.
If the effect size is at a low strength level (e.g., small and very small), the difference between the two lists is negligible, even if the $p$-value indicates a significant difference.

\textbf{Experiment 2 (for Answering RQ2):} 
We reorder the selection universe $P_{train}^{\textsc{FGSM+PGD}}$ in ascending order of CC value.
(Note that a CC value ranges over [0, 1].)
We divide the reordered list into five consecutive sections with an equal range of CC values, i.e., 
$N_2$ = \{[0, 0.2], (0.2, 0.4], (0.4, 0.6], (0.6, 0.8], (0.8, 1]\} so that each section only keeps the test cases with CC values within its range.
We then run \textsc{Clover} in Configuration \textit{A} to output a test suite $A$ containing $n_3$ test cases for each possible value $n_3 \in N_3 = \{1000, 2000, 4000, 6000\}$\footnote{
This range is chosen because in the original experiments of \textsc{Adapt} and \textsc{RobOT}, they select no more than 10\% of generated test cases for retraining a model under test. Or, in the case of \textsc{RobOT}, its experiment only sets to fuzz a model with 300 to 1200 seconds and uses all the generated test cases for model retraining.}.
We repeat the experiment using \textsc{Random} instead of \textsc{Clover}. 
It represents a state of the practice of using a random subset of a selection universe to construct a test suite for retraining the model under test.
We measure the robust accuracy improvement achieved by each technique after applying the test suite to get a retrained model delivered through the testing-retraining pipeline.

\textbf{Experiment 3 (for Answering RQ3):} 
We construct a subset $X$, containing 18000 samples of the training dataset of each clean model (cases \circled{1}--\circled{4}).
Same as Experiment 2, we set the number of selected test cases to $n_3 \in N_3 = \{ 1000, 2000, 4000, 6000 \}$.
We then conduct the following experiment on \textsc{Clover} in Configuration \textit{B} (named Experiment 3a): 
We run \textsc{Clover} on each model under test with the seed list $X$ and set $n_4$ seconds as the total fuzzing time budget to generate a test suite, and also keep all the generated test cases as a test pool, denoted as \textit{All}, i.e., without selection.
We conduct the above procedure for each $n_4$ in $N_4 = \{ 1800, 3600, 7200, 18000 \}$.
We measure \textit{\textit{\#AdvLabel}}, \textit{\textit{\#Category}}, and \textit{\textit{\#CC}} of the constructed test suites and the robust accuracy improvement achieved by \textsc{Clover} after applying each such test suite via the testing-retraining pipeline.
We further repeat Experiment 3a except using each of \textsc{Adapt} and \textsc{RobOT} instead of \textsc{Clover} (named Experiments 3b and 3c, respectively).

Moreover, for each pair of a model under test and a respective retrained model produced via the testing-retraining pipeline via \textsc{Clover}, we measure the CC value of each test case in each test suite generated by \textsc{Clover} and the corresponding mean CC reduction under the setting of $n_3 \in N_3$ and $n_4$=18000.

We repeat Experiment 3 three times to alleviate the influence of randomness in the experiment.

\textbf{Experiment 4 (for Answering RQ4)}:
We repeat Experiment 3a with $n_4$=18000, except that we use the test case prioritization metric Gini adopted by \textsc{DeepGini} and the loss-based metric FOL proposed by \textsc{RobOT} to replace our CC metric in \textsc{Clover}, and also use \textsc{DeepGini} and \textsc{km-st} to replace line 2 of Algorithm \ref{alg: colver overview}, named Experiments 4a and 4b, respectively.

\textbf{Experiment 5 (for Anserwing RQ5)}:
In this series of experiments, we explore alternatives to the key design decisions and vary the hyperparameters in the \textsc{Clover} algorithm to study to what extent \textsc{Clover} in Configuration \textit{B} is sensitive to these settings on clean models. Experiment 5 consists of several sub-experiments, which we will present in turn.

Two main ideas implemented in \textsc{Clover} are to find test cases of the same seed with higher CC values and perturb test cases based on seed equivalence. To know the extent of their effects contributing to the robustness improvement effectiveness of \textsc{Clover} delivered through the testing-retraining pipeline, we create two variants of \textsc{Clover}. The first variant of \textsc{Clover}, named \textsc{Clover+Smallest}, finds test cases of the same seed with lower CC values. The second variant, named \textsc{Clover+SingleDir}, ablates the use of the seed equivalence information in perturbing a test case so that the gradient direction provided to perturb the test case only depends on the given seed and the existing test case of the seed that achieves the highest CC values so far for that fuzzing attempt.
The experimental procedures to apply these two variants to generate test cases are presented as Experiments 5a and 5b, respectively.

\textit{Experiment 5a for \textsc{Clover+Smallest}:} We repeat Experiment 3a with $n_4$=18000, except that we modify line 12 in Algorithm \ref{alg: context_translate} from ``{\text{if}} $CC(x') > CC^*$ {\text{then}}'' to ``{\text{if}} $CC(x') < CC^*$ {\text{then}}'', where $CC^*$ is modified to keep the lowest CC value among the generated test case from the same seed (instead of highest one in the original \textsc{Clover}). Also, we modify line 10 in Algorithm \ref{alg: context_selection} from ``Sort(\textit{layer}) in descending order of CC'' to ``Sort(\textit{layer}) in ascending order of CC''. 
We measure the difference in mean robust accuracy improvement achieved by the \textsc{Clover} variant and that achieved by \textsc{Clover} in Experiment 3a.

\textit{Experiment 5b for \textsc{Clover+SingleDir}:} We repeat Experiment 5a, except that we modify line 4 in Algorithm \ref{alg: context_translate} from ``$\Delta_{\beta\text{-AFO}}' {=} \textit{dir} + \Delta_{\beta\text{-AFO}} + \partial{obj} / \partial{x} $'' to ``$\Delta_{\beta\text{-AFO}}' {=} \partial{obj} / \partial{x}$'' instead of the modification stated in Experiment 5a.

In the algorithmic design of \textsc{Clover}, there are several hyperparameters that contribute to the formulation of a fuzzing attempt. While the design itself does not prioritize one fuzzing attempt over another in terms of time spent or sensitivity to the model type, certain hyperparameters, i.e., $m, k, \delta, p\text{-norm}$, and $\epsilon$ may potentially impact the performance of fuzzing attempts according to intuition and empirical evidence. Therefore, we conduct a study to examine the influence of these hyperparameters on cases \circled{1}--\circled{4} within the robust accuracy improvement achieved through the testing-retraining pipeline.

First, \textsc{Clover} requires users to provide an estimate of the average resources allocated to fuzz a seed, which is represented by the hyperparameter $m$. Intuitively, using a larger $m$, more fuzzing attempts on a seed will be conducted, which may discover test cases with higher CC values than using a smaller $m$. Nonetheless, fuzzing is a time-constrained test activity. Spending more fuzzing attempts on one seed implies allocating fewer fuzzing attempts on the other seed if the two fuzzing campaigns spend the same amount of total time. 
What we are unclear is whether the difference is small, and if this is the case, using a smaller $m$ to allow the fuzzing technique to fuzz on more seeds within the same time budget seems to be a viable choice. The corresponding experimental procedure is presented as Experiment 5c.

\textit{Experiment 5c:} We repeat Experiment 3a with $n_4$=18000, except that we use each $m \in \{3, 5, 7, 9\}$ instead of keeping $m$ to 5.

Second, in each fuzzing attempt, \textsc{Clover} makes a ballpark estimation of the contextual confidence of each test case, which requires collecting the prediction probability of $k$ data points in the surrounding of the test case. Intuitively, having a larger $k$ produces a more accurate estimation, and yet evolutionary algorithms generally do not require accurate estimation to evolve its elements in each evolution attempt.  Furthermore, \textsc{Clover} discovers test cases by seed equivalence where its algorithmic design is to apply the $\beta$-AFOs achieved by some other test cases to assist the evolution of the current test case. 
Thus, it is unclear the extent of benefits brought by obtaining a more accurate estimation to \textsc{Clover}. We thus vary $k$ to study the effect of $k$, where its experimental procedure is summarized as Experiment 5d.

\textit{Experiment 5d:} We repeat Experiment 3a with $n_4$=18000, except that we use each $k \in \{5, 10, 20, 40\}$ instead of keeping $k$ to 5.

Third, in estimating the contextual confidence in each fuzzing attempt, \textsc{Clover} selects several data points within a perturbation bound $\delta$. 
As presented in Section \ref{sec: Clover}, \textsc{Clover} is developed in the background of $\delta \ll \epsilon$. 
We vary  $\delta$ to evaluate the sensitivity of \textsc{Clover} on this hyperparameter, which is presented in Experiment 5e.
Intuitively, even $\delta$ becomes smaller, as adversarial examples widely exist around samples within many different fuzzing bounds in the literature, we tend to believe that the differences in effect could be small.
On the other hand, as $\delta$ becomes close to  $\epsilon$, more perturbed samples are seriously clipped, which in essence, makes the resulting sample use the information of $\Delta_{\beta\text{-AFO}}$ less effectively. 

\textit{Experiment 5e:} We repeat Experiment 3a with $n_4$=18000, except that we use each $\delta \in \{ 0.001, 0.005,$ $0.01, 0.05 \}$ instead of keeping $\delta$ to 0.01.

Fourth, in measuring the distance between two data points, \textsc{Clover} uses the notion of $p\text{-norm}$ distance, which is popular in machine learning. 
We design the \textsc{Clover} algorithm to compute the probability values of the adversarial label of the test case observed on the surrounding data points of that test case.
This design naturally leads to the use of $L_{\infty}$-norm, which is to find the maximum value in the prediction vector of each such surrounding data point.
Changing $L_{\infty}$-norm to another $p$-norm, such as $L_2$-norm, will unavoidably destroy the notion of contextual confidence. 
Therefore, to keep the essence of \textsc{Clover}, we do not change how CC is computed.
On the other hand, to apply the other $p$-norm distance in generating perturbed data, the \textsc{processing}(.) function called in lines 5 and 7 of Algorithm \ref{alg: context_translate} is modified to skip clipping $x'$ with $\epsilon$.
We also modify line 10 from ``if ${\textsc{isAdversarial}}(x, x')$ then'' to ``if ${\textsc{isAdversarial}}(x, x')$ and $||x-x'||_2 \leq \epsilon$ then'' in Algorithm \ref{alg: context_translate} to filter out $x'$ if its $L_2\text{-norm}$ distance to $x$ is larger than $\epsilon$.
Experiment 5f below describes the experimental procedure for this variant of \textsc{Clover}.

\textit{Experiment 5f:} We repeat Experiment 3a with $n_4$=18000, except that we use each type of $p\text{-norm}$, where $p \in \{L_{\infty}, L_2\}$.
To make the mean $L_2$-norm distance of generated test cases similar to those generated by \textsc{Clover} in Experiment 3a (see Section \ref{sec: hyperparameters}), after several trials, we adjust and set $\epsilon$ to 0.10, 0.06, 0.10, and 0.08 for cases \circled{1} to \circled{4}, respectively.

Lastly, any fuzzing technique is configured with a fuzzing bound. Varying the fuzzing bound is a typical ablation study in many experiments. Following the results in the literature \cite{adversarial_training_2, adversarial_training_3, ae_train_free, fast_ae_train, fast_better_free}, in general, we expect that using a larger bound can produce a larger robust accuracy improvement.

\textit{Experiment 5g:} We repeat Experiment 3a with $n_4$=18000, except that we use each $\epsilon \in \{0.01, 0.03, $ $ 0.05, 0.07\}$ instead of keeping $\epsilon$ to 0.05. 

\textbf{Experiment 6 (for Anserwing RQ6)}:
To explore an alternative use of \textsc{Clover} for quality assurance of DL models, we use the test dataset of each adversarially trained model as the seed list $X$\footnote{
we randomly sample 10000 samples from the SVHN test dataset.
}, which contains 10000 test cases, to attack the model.
Same as Experiment 3a, we set the number of selected test cases to $n_3 \in N_3 = \{ 1000, 2000, 4000, 6000 \}$.

We conduct the following experiment on \textsc{Clover} in Configuration \textit{B} (named Experiment 6a): 
we run \textsc{Clover} on each adversarially trained model with the seed list $X$ and set $n_4=18000$ seconds as the total fuzzing time budget to generate a test suite, and keep all the generated test cases into a test pool, denoted as \textit{All}, i.e., without selection.
We measure \textit{\textit{\#AdvLabel}}, \textit{\textit{\#Category}}, and \textit{\textit{\#CC}} on these test suites and the test pool.
We repeat Experiment 6a three times to alleviate the influence of randomness during the experiment. 

We further repeat Experiment 6a except using each of \textsc{Adapt} and \textsc{RobOT} instead of \textsc{Clover} (named Experiments 6b and 6c, respectively).

We also repeat Experiment 4 except that we adopt cases \circled{5}--\circled{8} instead of cases \circled{1}--\circled{4} as the DL models under fuzzing (as Experiment 6d).

\subsubsection{Discussion on Fuzzing Budgets}
\label{sec:fuzzing-budget-and-model-limitation}
The fuzzing process on DL models with modern model architecture could be slow.
The downloaded \textsc{RobOT} source code specifies the size of the seed list as 1000. 
In the original experiment \cite{robot}, \textsc{RobOT} only fuzzes a DL model for a short period (5, 10, and 20 minutes) on models such as LetNet-5 on MNIST, Fashion-MNIST, and SVHN and ResNet20 on CIFAR10 in the evaluation for robustness improvement delivered through the testing-retraining pipeline.
In cases \circled{1} to \circled{4}, using merely 5 minutes for fuzzing, on average, \textsc{Adapt} can only process 30, 30, 30, and 15 seeds, and \textsc{RobOT} can only process 205, 183, 120, and 60 seeds, which correspond to only 0.03\%--0.46\% of the training dataset of the four cases and is insignificant.
They can only generate 350, 3417, 1023, and 404 test cases (for \textsc{Adapt}) and 756, 1767, 763, and 310 test cases (for \textsc{RobOT}), respectively, which correspond to 0.69\%--7.59\% of the training datasets.
Our preliminary experiment shows that such a small test suite cannot make meaningful changes in robust accuracy.
We thus choose a larger time budget to cater to models with more complicated model architecture in our experiments. 

Apart from setting $n_4=18000$, we have also attempted to set $n_4$ to 1800, 3600, and 7200 for Experiment 6a--6c, but the numbers of generated test cases produced by \textsc{Adapt} and \textsc{RobOT} are often too small to make a reasonable comparison with \textsc{Clover}. 
For instance, \textsc{RobOT} only generates 2977 and 2772 test cases in cases \circled{7} and \circled{8} with $n_4=7200$, respectively, which are insufficient to construct test suites for $n_3 = 4000$ and $=6000$. 
If $n_4 = 1800$, it generates much fewer than the number of test cases we require to construct any test suite in our data analysis (where the minimum is $n_3 = 1000$).

\subsubsection{Discussion on Adversarially Trained Models as Models under Fuzzing}
\label{sec:discussion-adv-models}
To train an adversarially trained model, the training scheme has used the training dataset to produce adversarial examples, where in each training epoch, a round of search for an adversarial example of a seed is conducted. Since there are many training epochs, the number of searches on each training sample is already extensive. Configuring a fuzzing technique to fuzz on the same training sample for that adversarially trained model could not meaningfully validate the robustness quality of the model.
Thus, following \cite{deephunter, dlfuzz, adaptfuzz}, our experiments configure a fuzzing technique to fuzz with a test dataset of the model as the seed list. 
With this experimental setting, the generated test cases will be adversarial examples of the test dataset.
If one trains the adversarially trained model under fuzzing with these adversarial examples, one effectively instructs the model to learn from the test dataset. 
So, our experiment does not retrain these adversarially trained models.
Therefore, we do not measure the robust accuracy improvement achieved by a fuzzing technique on these models delivered through the testing-retraining pipeline.

We have attempted to include more adversarially trained DL models, such as the \textit{pretrained} models provided in \cite{adversarial_robustness_toolbox, arch_search_adv_robust} on CIFAR10 using the evolutionary neural network architecture search technique \cite{neuro_cell_search} combined with adversarial training, including using 50\% and 100\% of the samples in each mini-batch to be replaced by their adversarial examples, to facilitate the comparison between \textsc{Clover} and the peer techniques\footnote{Reference \cite{arch_search_adv_robust} reports that, on CIFAR10 with ratio = 0.5 and 1.0, the best synthetic models achieve the test accuracy of 93.21\% and 82.88\% on the test dataset, and the robust accuracy of 48.99\% and 50.60\% on a set of adversarial samples generated by PGD, respectively. 
The architectures of the synthetic models are complicated and contain 20,435,714 and 7,705,130 parameters, respectively.
}%
.
However, both \textsc{RobOT} and \textsc{Adapt} \textbf{cannot} generate \textbf{any} test cases from the synthetic CIFAR10 model with the ratio 1.0 and the time budget of $n_4=18000$ seconds.
We have attempted to increase the time budget $n_4$ further to 24 hours and increase the value of fuzzing bound $\epsilon$ by 100\%, both separately and in combination.
Still, both \textsc{RobOT} and \textsc{Adapt} cannot generate any test cases.
We have profiled their executions for examination. We observed that they iterated over seeds, mutated test case candidates, and checked whether their prediction labels were adversarial as usual. 
On the one hand, the situation may indicate that the adversarially trained models are robust against the attacks of \textsc{RobOT} and \textsc{Adapt}.
On the other hand, we cannot compare the results of \textsc{Clover} with them.
Thus, we do not include these models in the data analysis of the present paper because we are not sure whether the inability of \textsc{RobOT} and \textsc{Adapt} in test case generation is due to the insufficient fuzzing budget issue or the limitation of their algorithms or any unknown factors. 
As a reference, \textsc{Clover} generates more than 5000 test cases on fuzzing each of these two models with $n_4$=18000.

\section{Results and Data Analysis for \textsc{Clover} in Configuration \textit{A}} 
\label{sec: result_in_configuration_A}
This section presents the results and data analysis for answering RQ 1 and RQ2 through Experiments 1 and 2 for \textsc{Clover} in Configuration \textit{A} configured in the testing-retraining pipeline.
We use the term \textsc{Clover} instead of the full qualification in the respective section for brevity.
We also use the term \textit{robustness improvement} achieved by \textsc{Clover} (or a peer technique, respectively) to mean the robustness improvement achieved by the retrained model output by the testing-retraining pipeline when applying \textsc{Clover} (or the peer technique, respectively) as the DL testing technique in the pipeline. 
The results and data analyses for RQ3--RQ4, RQ5, and RQ6 can be found in the next three sections.

\begin{figure}[]
  \centering
  \includegraphics[width=0.30\textwidth]{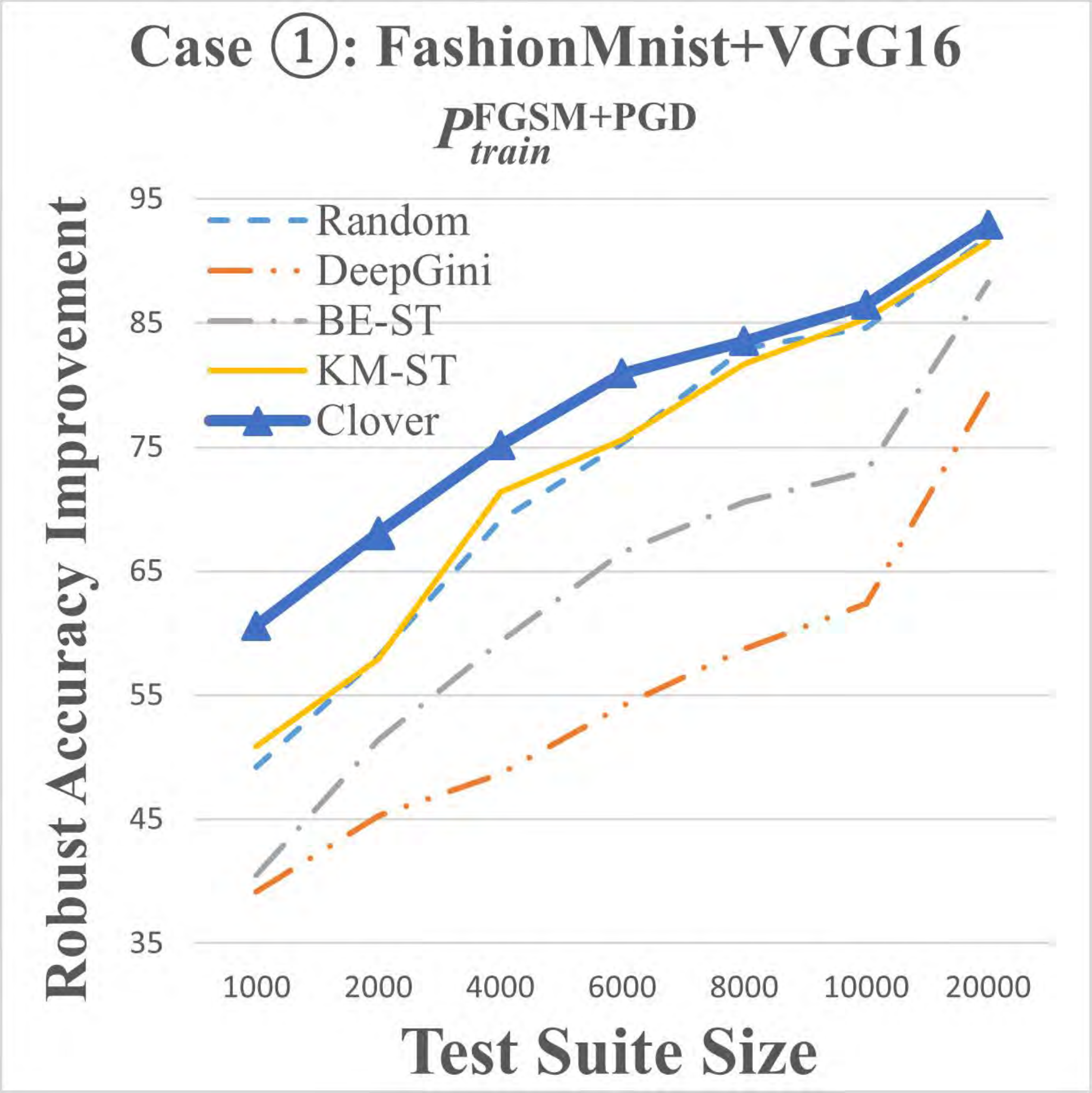}
  \includegraphics[width=0.30\textwidth]{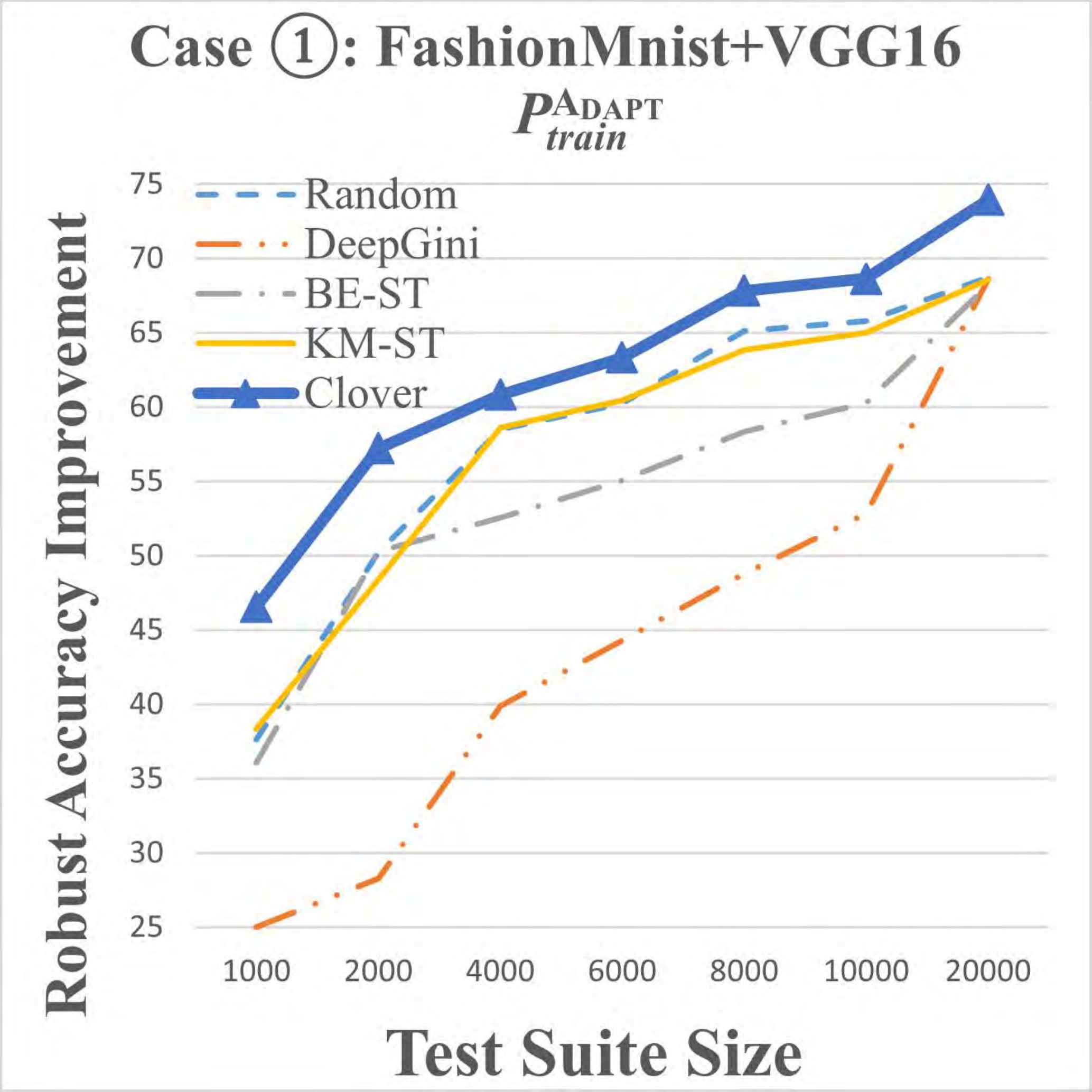}
  \includegraphics[width=0.30\textwidth]{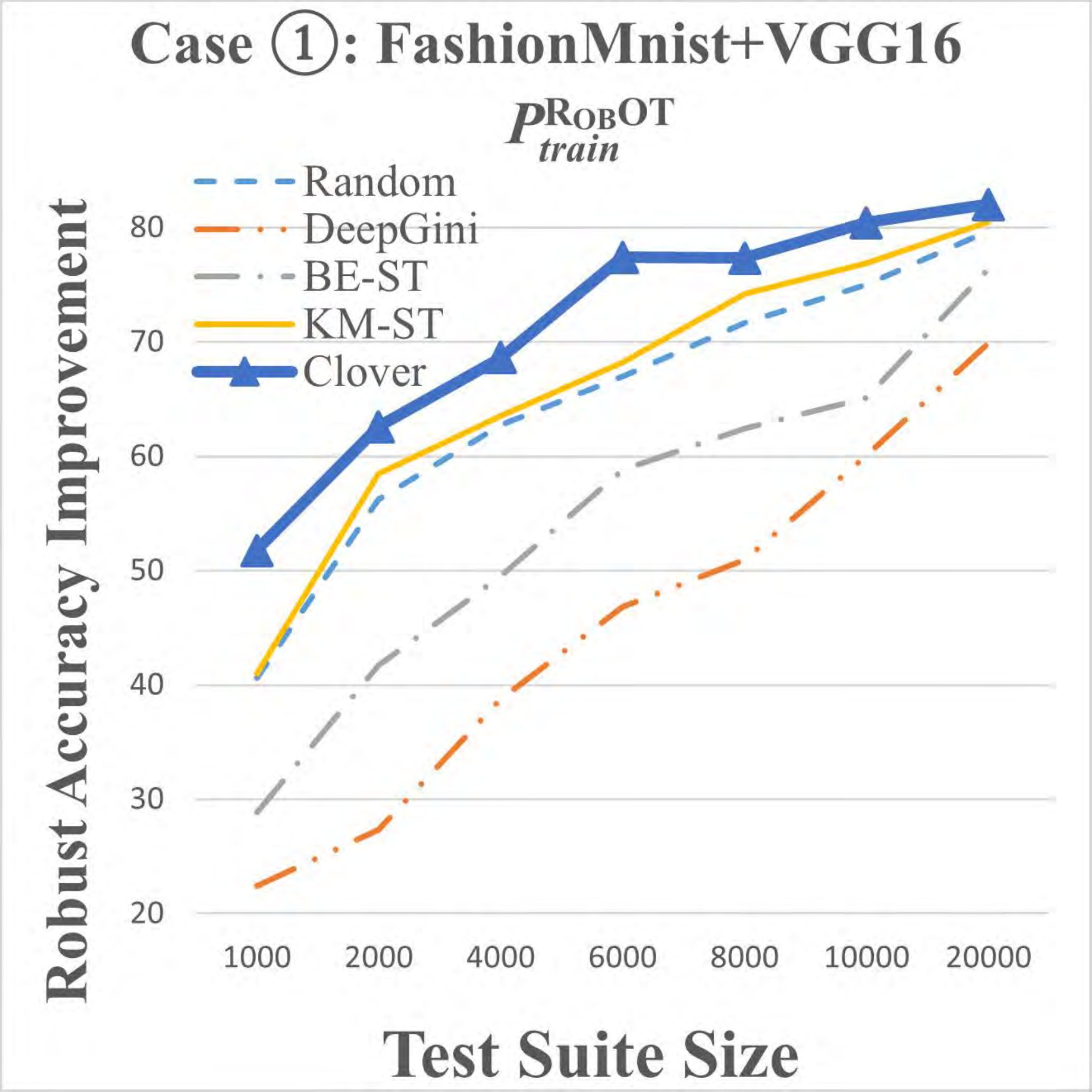} \\
  {\small (a) Case \circled{1} on the three selection universes $P_{train}^{\textsc{FGSM+PGD}}$, $P_{train}^{\textsc{Adapt}}$, and $P_{train}^{\textsc{RobOT}}$, respectively}

  \includegraphics[width=0.30\textwidth]{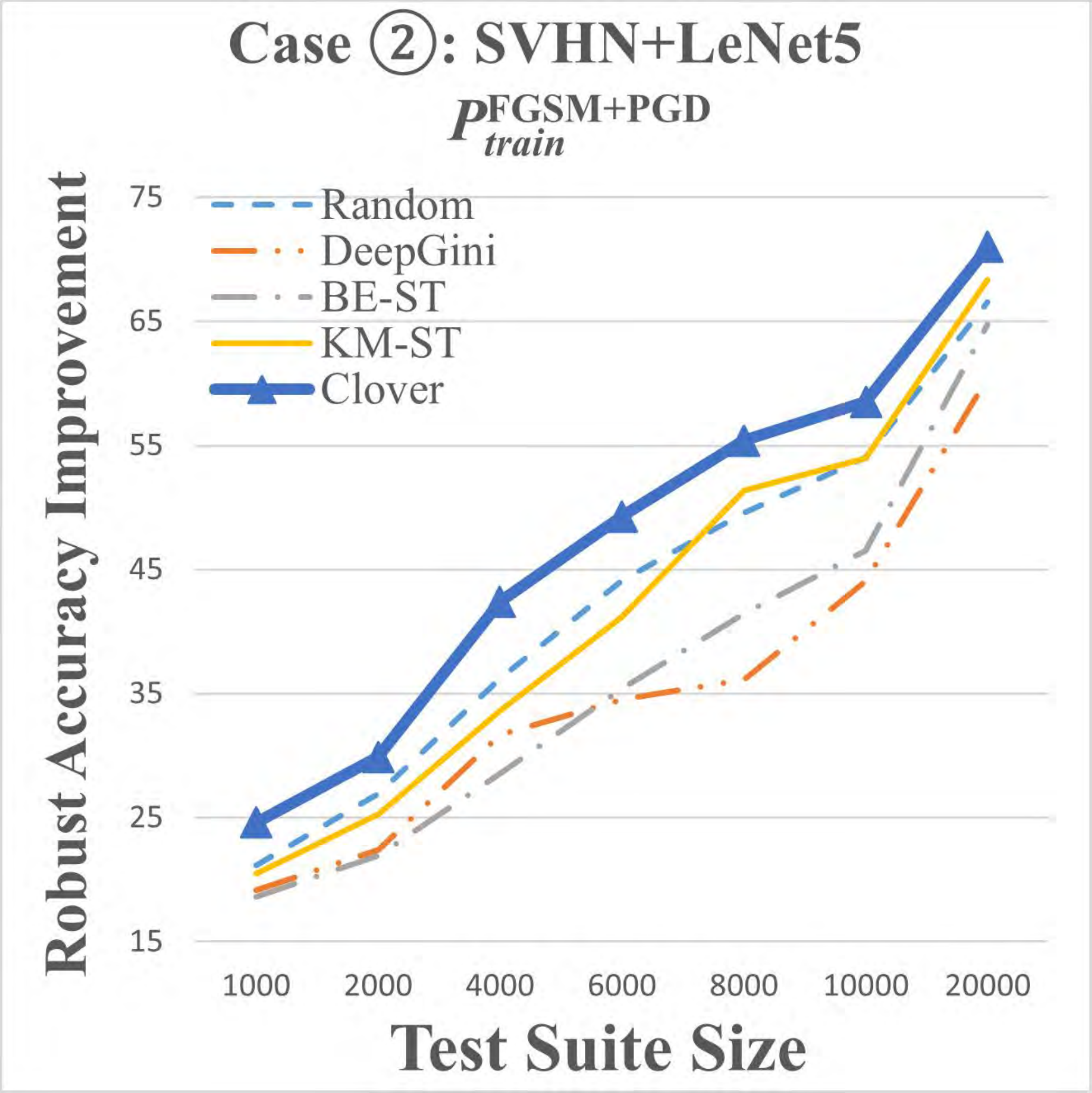}  
  \includegraphics[width=0.30\textwidth]{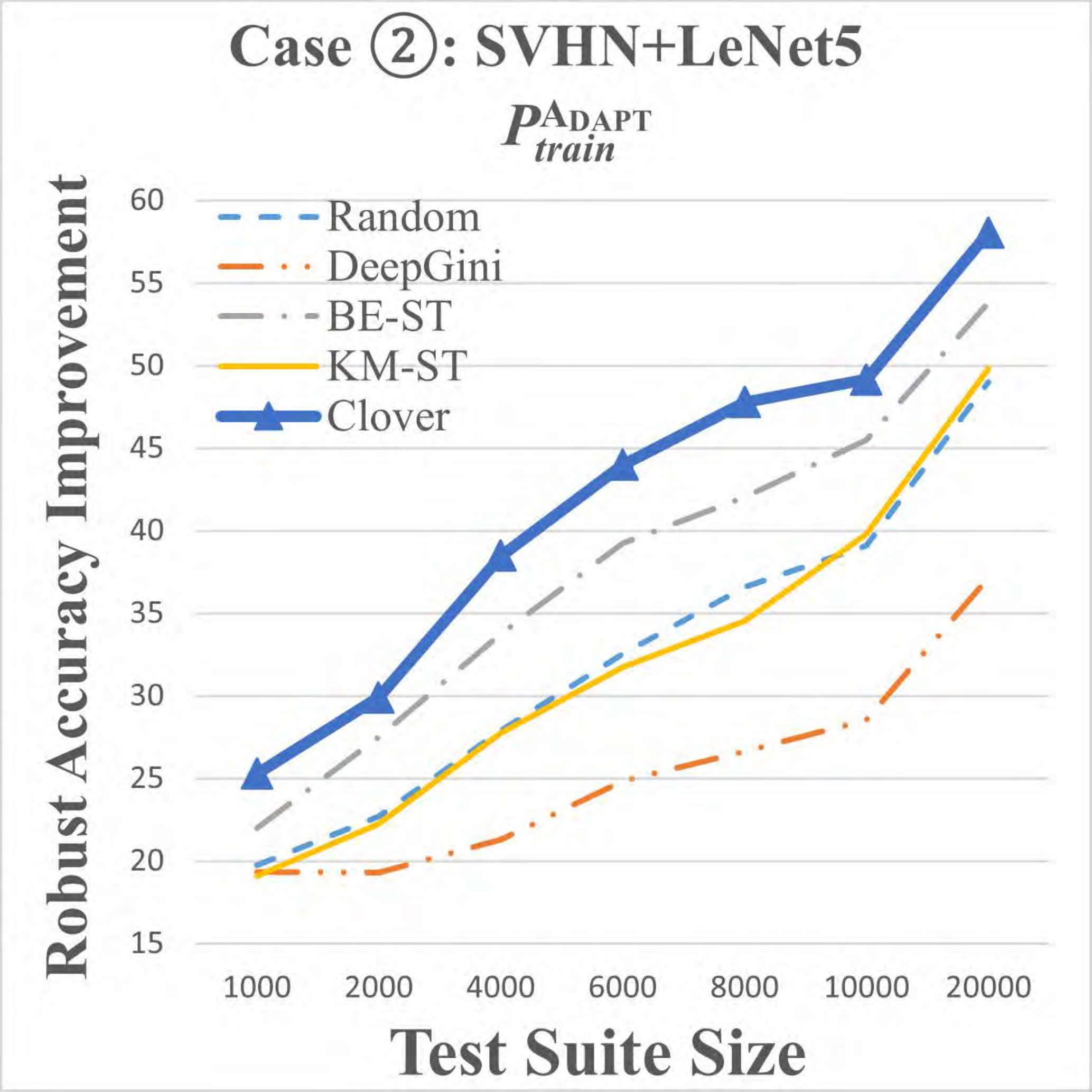}  
  \includegraphics[width=0.30\textwidth]{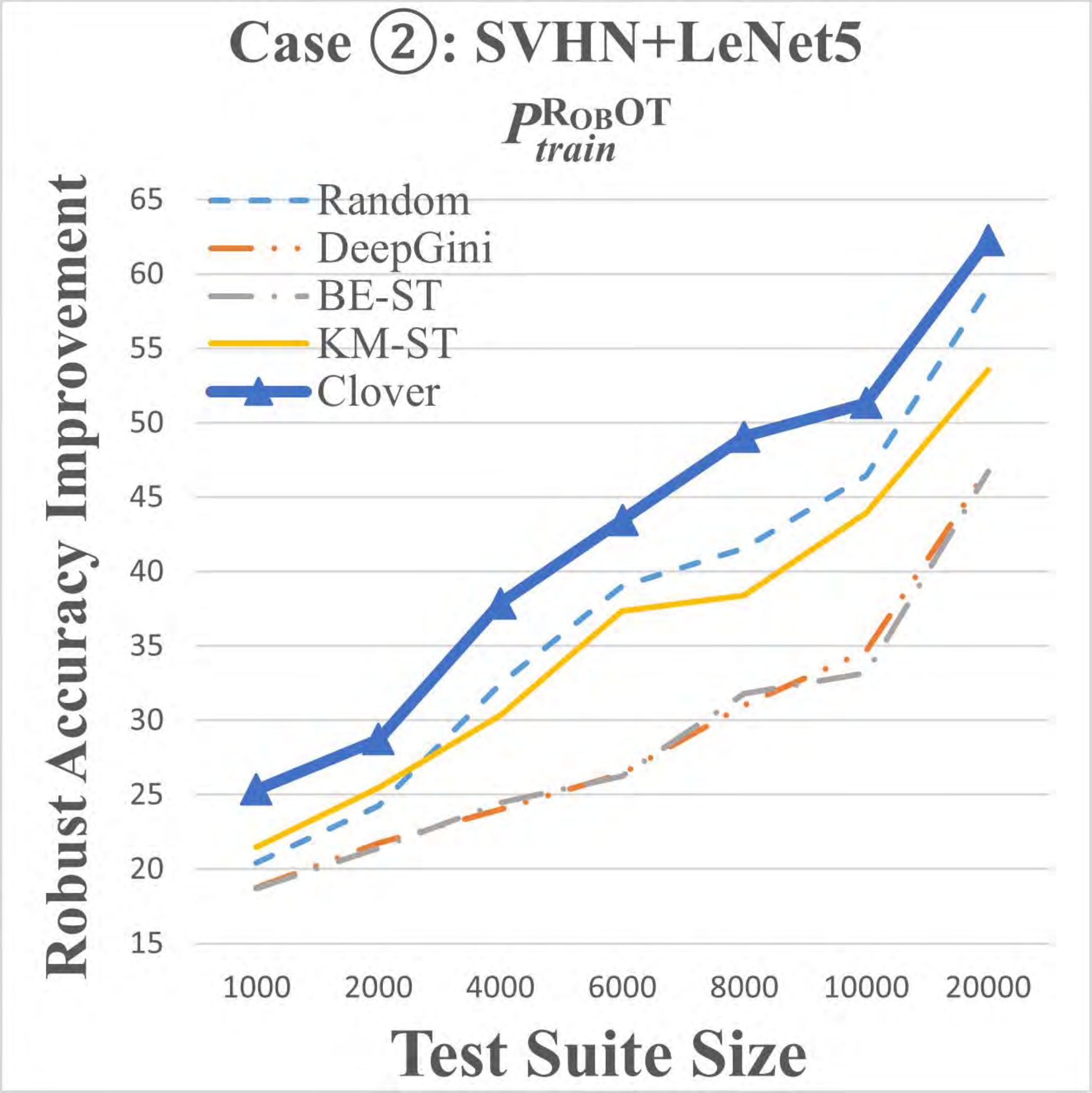} \\
  {\small (b) Case \circled{2} on the three selection universes $P_{train}^{\textsc{FGSM+PGD}}$, $P_{train}^{\textsc{Adapt}}$, and $P_{train}^{\textsc{RobOT}}$, respectively}
   
  \includegraphics[width=0.30\textwidth]{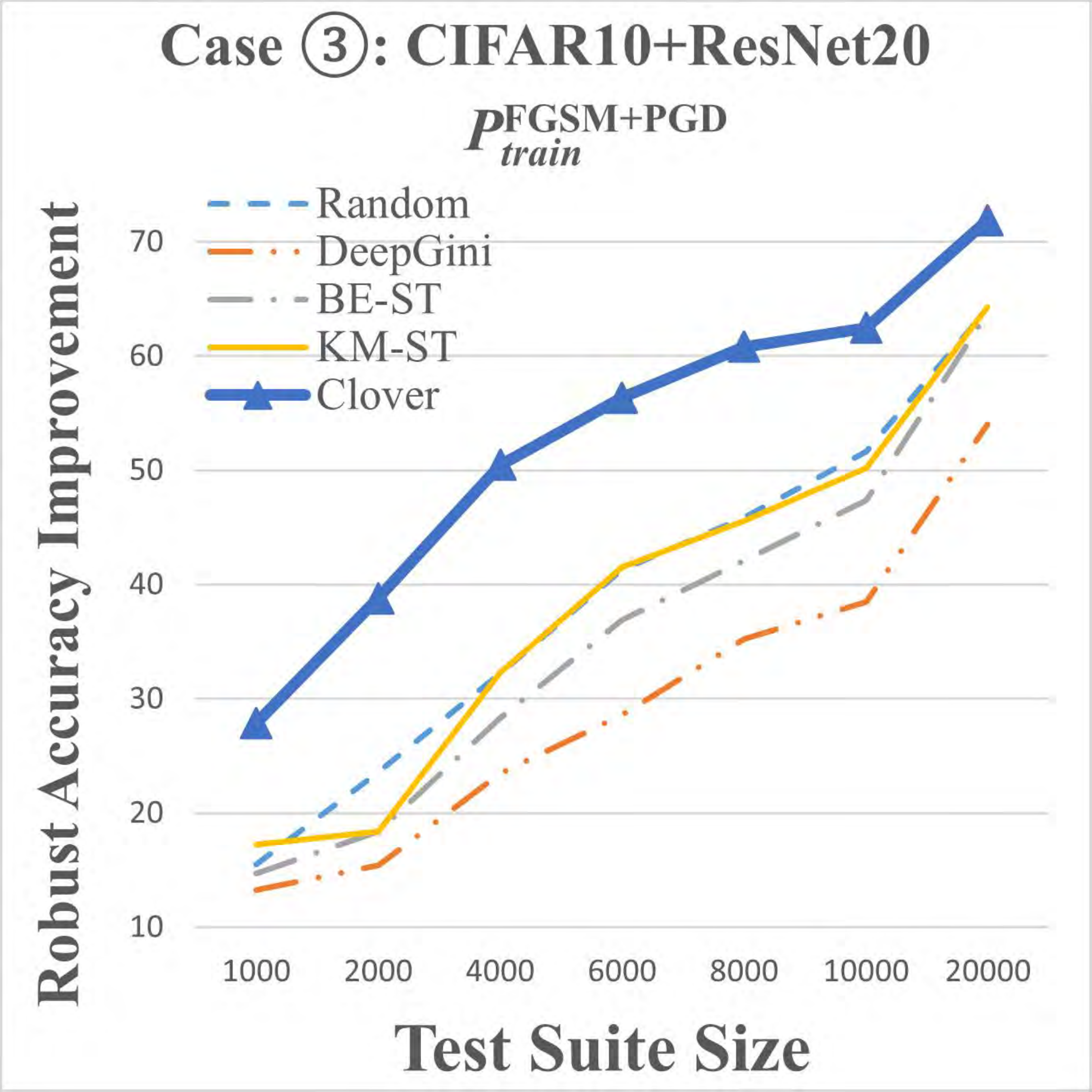}
  \includegraphics[width=0.30\textwidth]{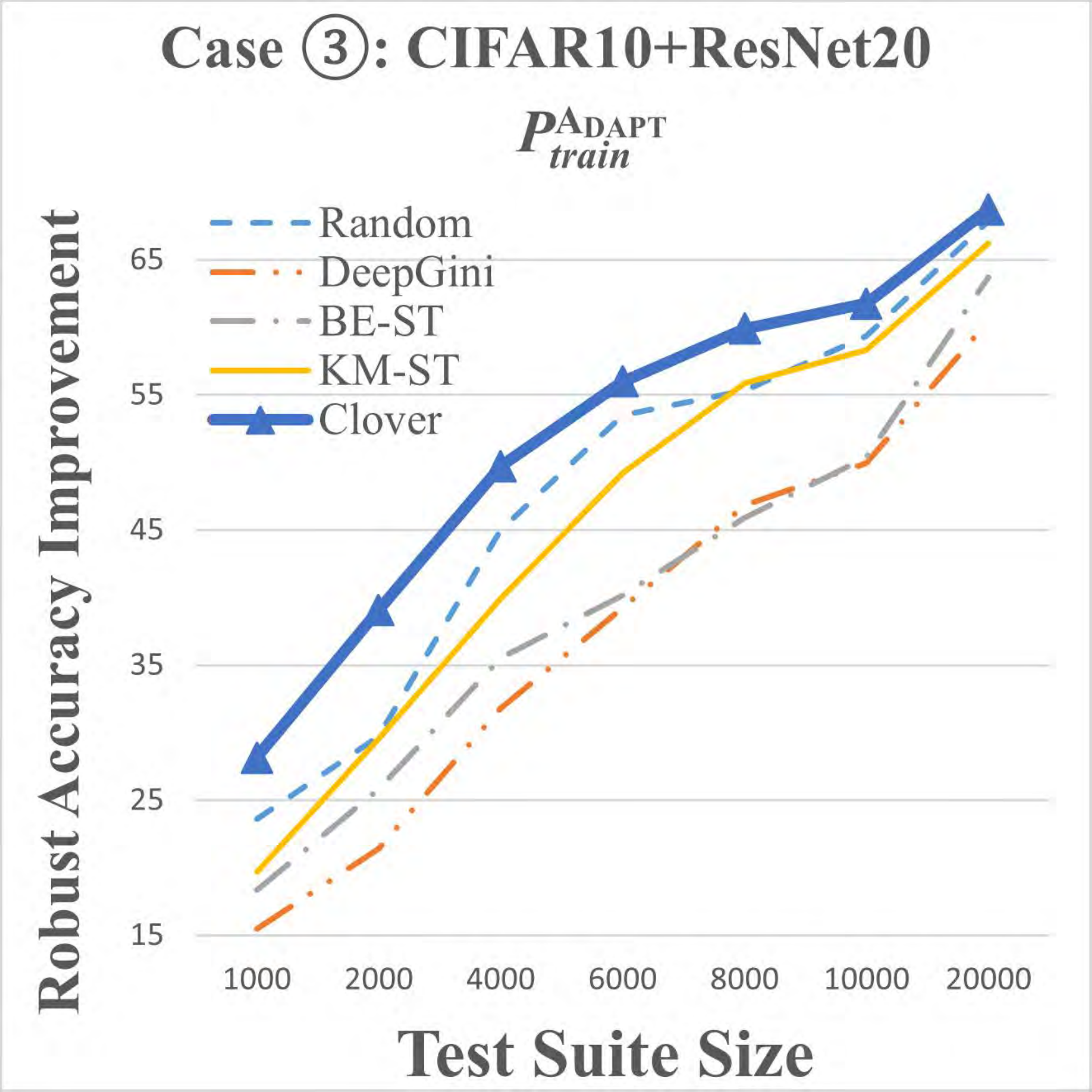}
  \includegraphics[width=0.30\textwidth]{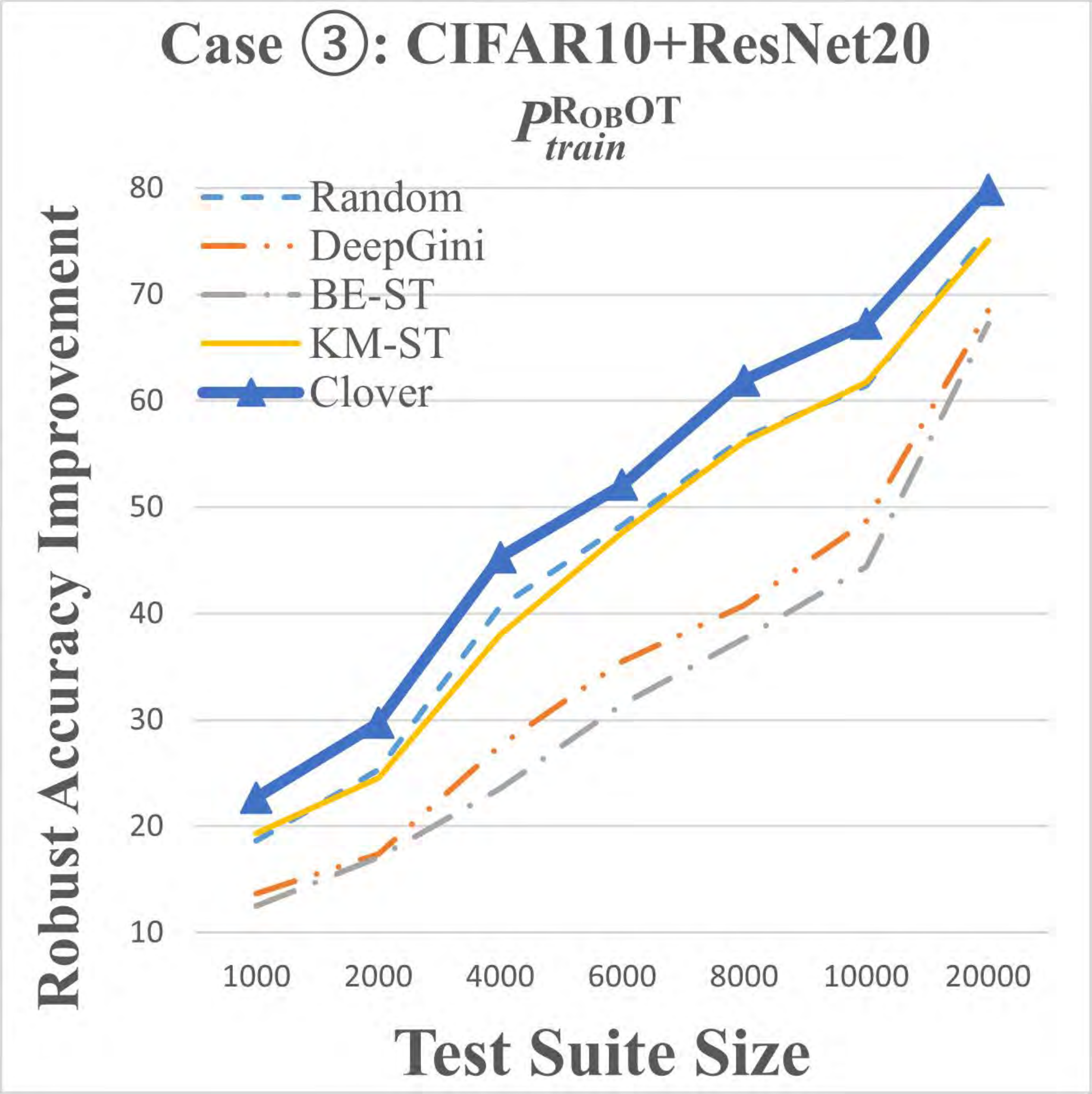} \\
  {\small (c) Case \circled{3} on the three selection universes $P_{train}^{\textsc{FGSM+PGD}}$, $P_{train}^{\textsc{Adapt}}$, and $P_{train}^{\textsc{RobOT}}$, respectively}
  
  \includegraphics[width=0.30\textwidth]{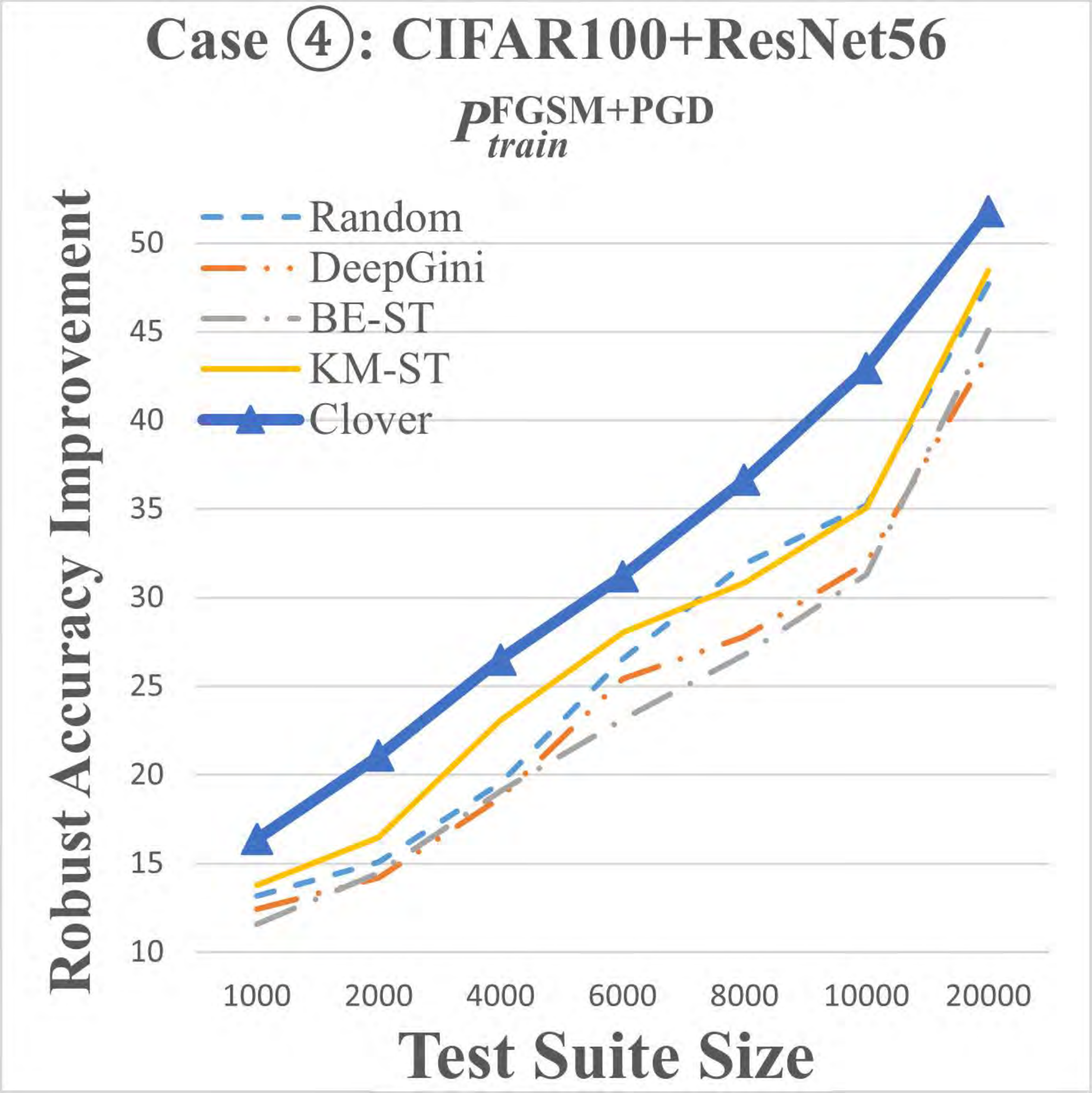}
  \includegraphics[width=0.30\textwidth]{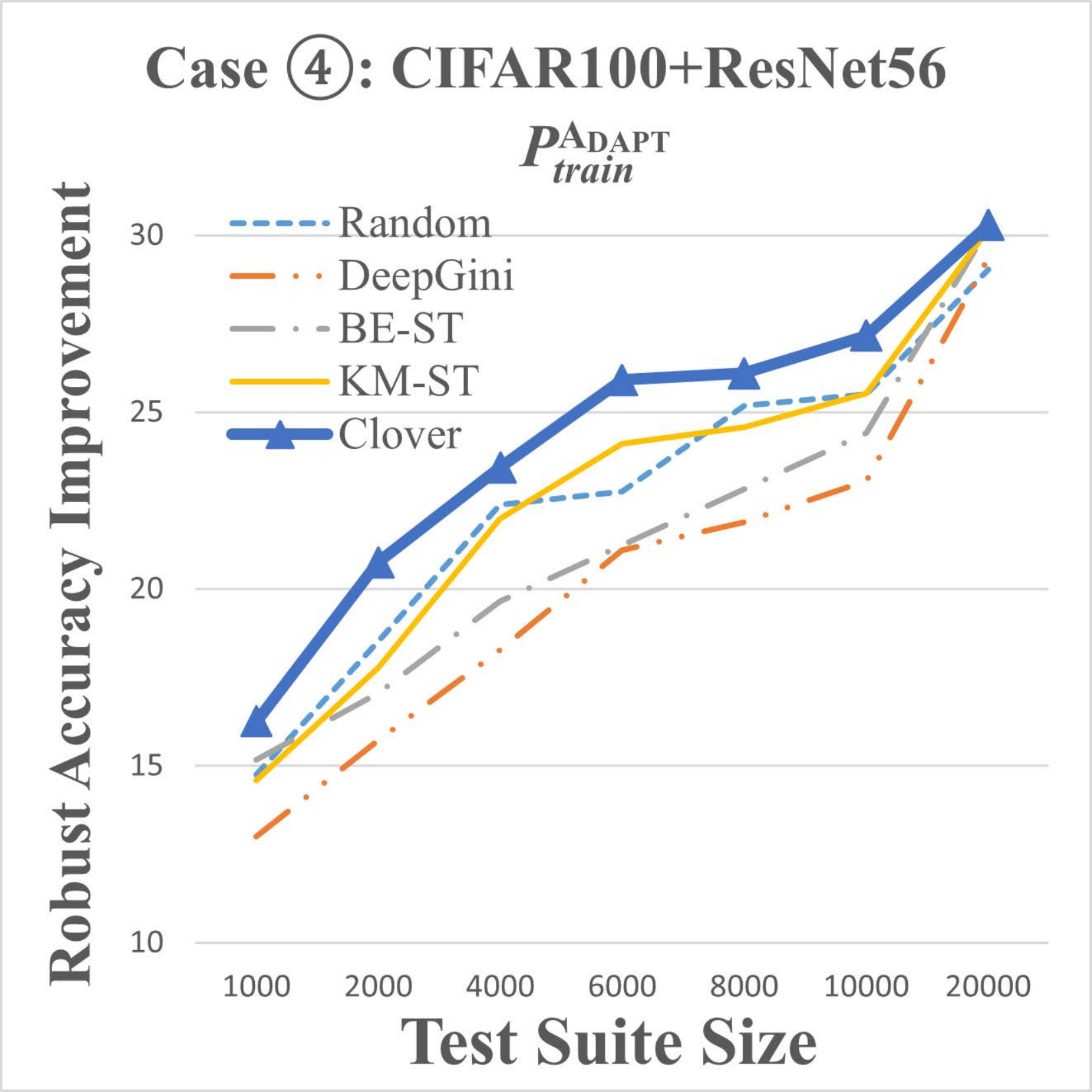} 
  \includegraphics[width=0.30\textwidth]{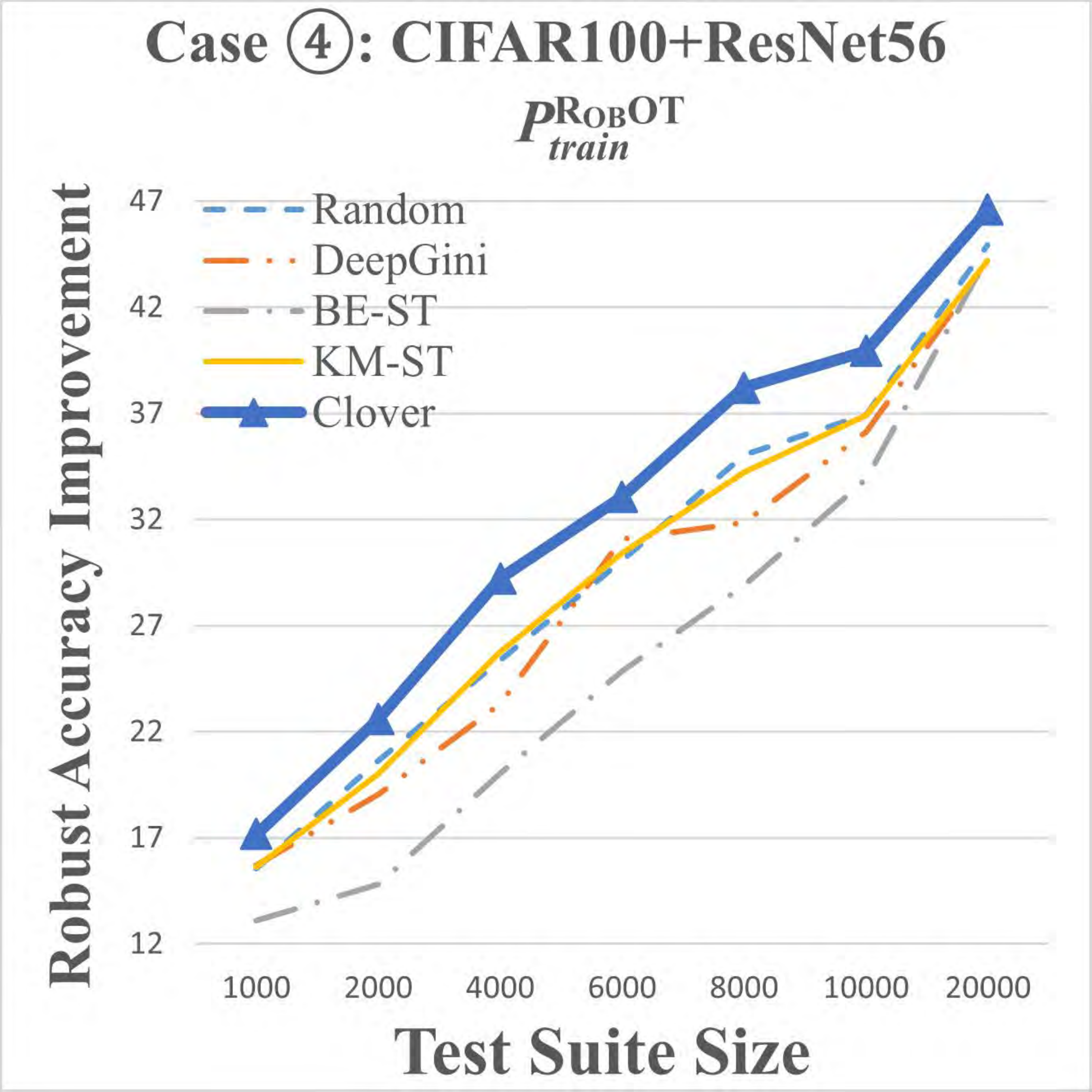} \\
  {\small (d) Case \circled{4} on the three selection universes $P_{train}^{\textsc{FGSM+PGD}}$, $P_{train}^{\textsc{Adapt}}$, and $P_{train}^{\textsc{RobOT}}$, respectively}
  \caption{Robust Accuracy Improvements Achieved by Technique in Configuration \textit{A} on Selection Universes}
\label{fig: rq1_1}
\end{figure}

\begin{figure}[]
  \centering
  \includegraphics[width=0.30\textwidth]{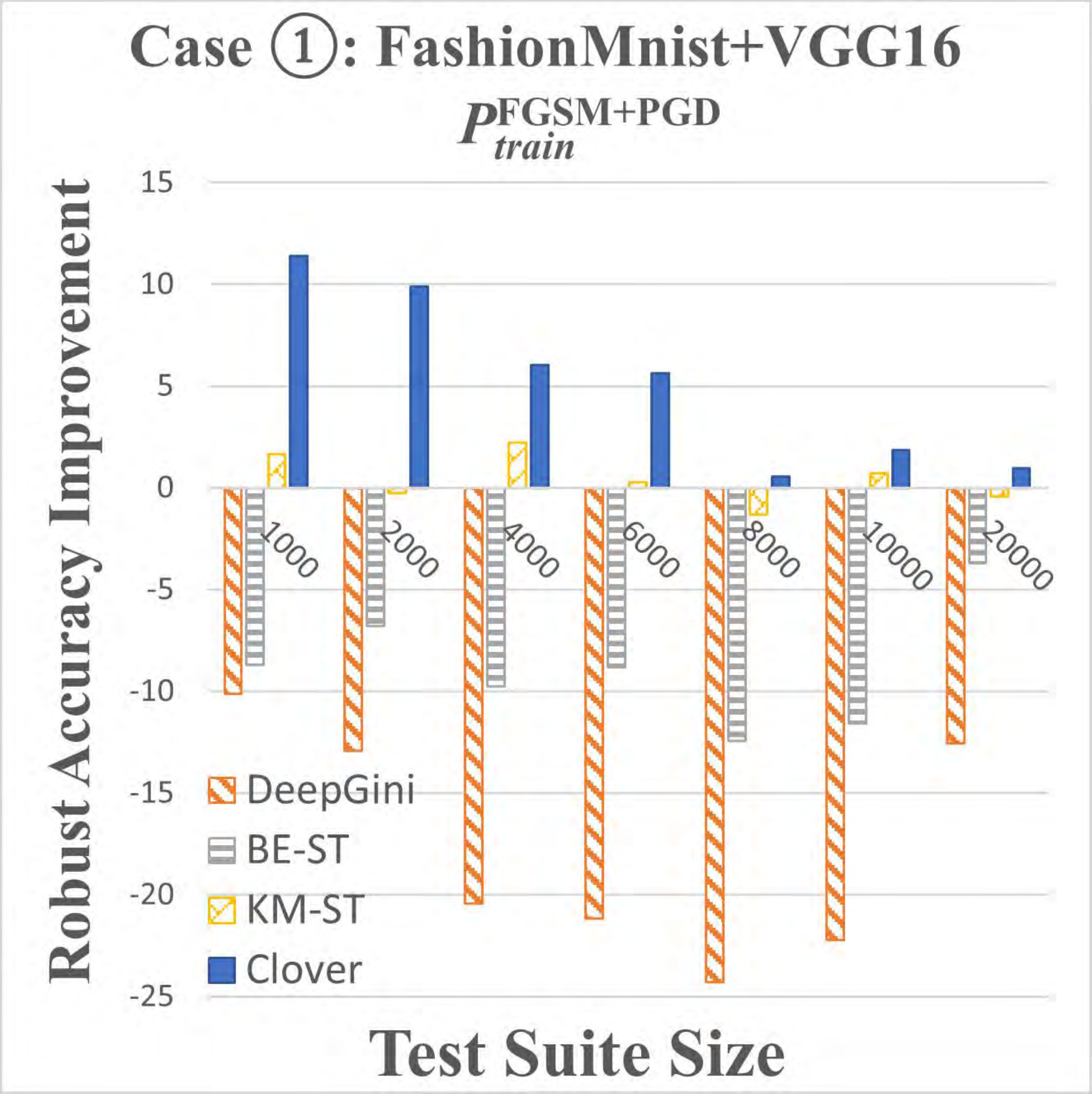}
  \includegraphics[width=0.30\textwidth]{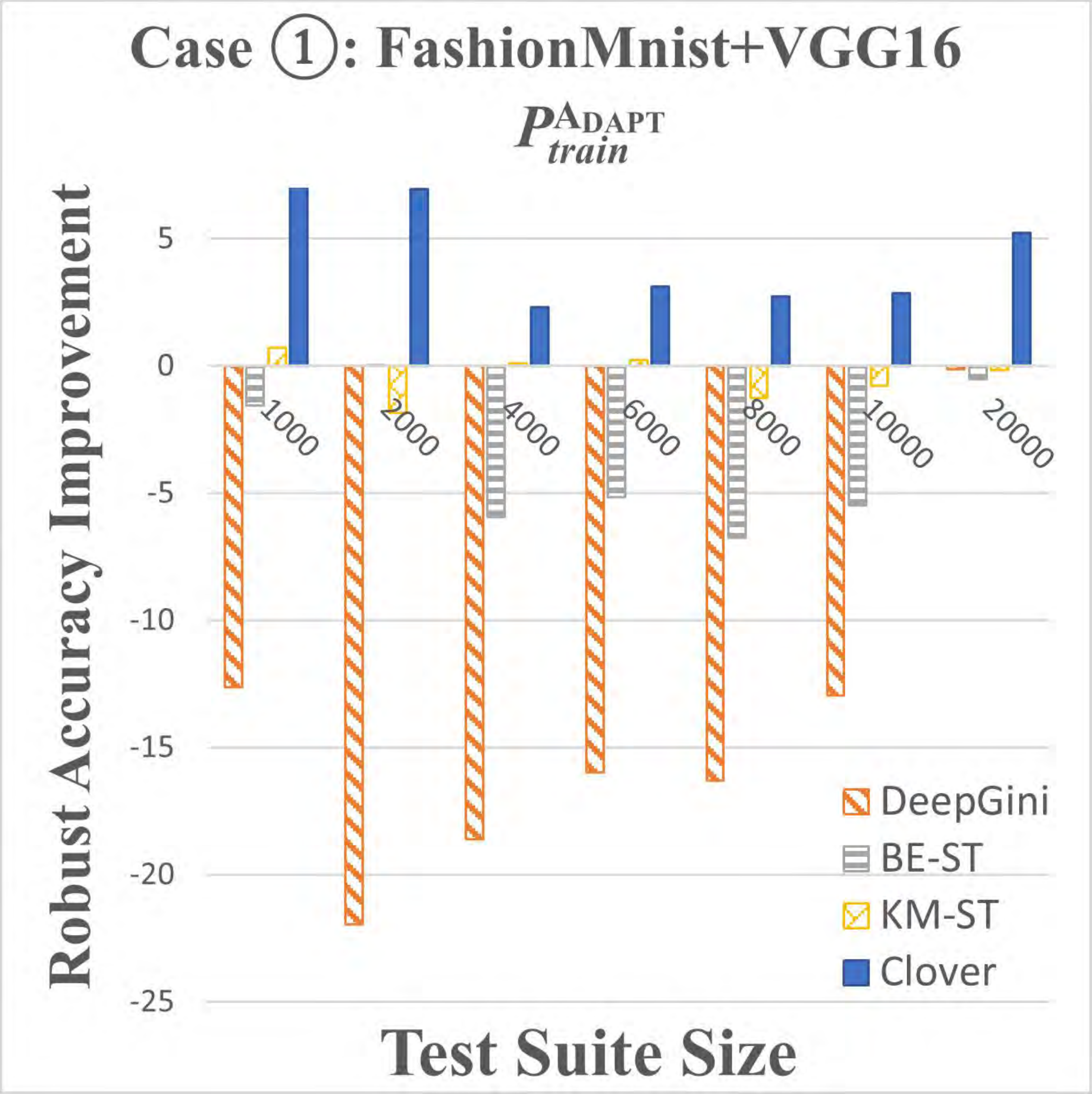}
  \includegraphics[width=0.30\textwidth]{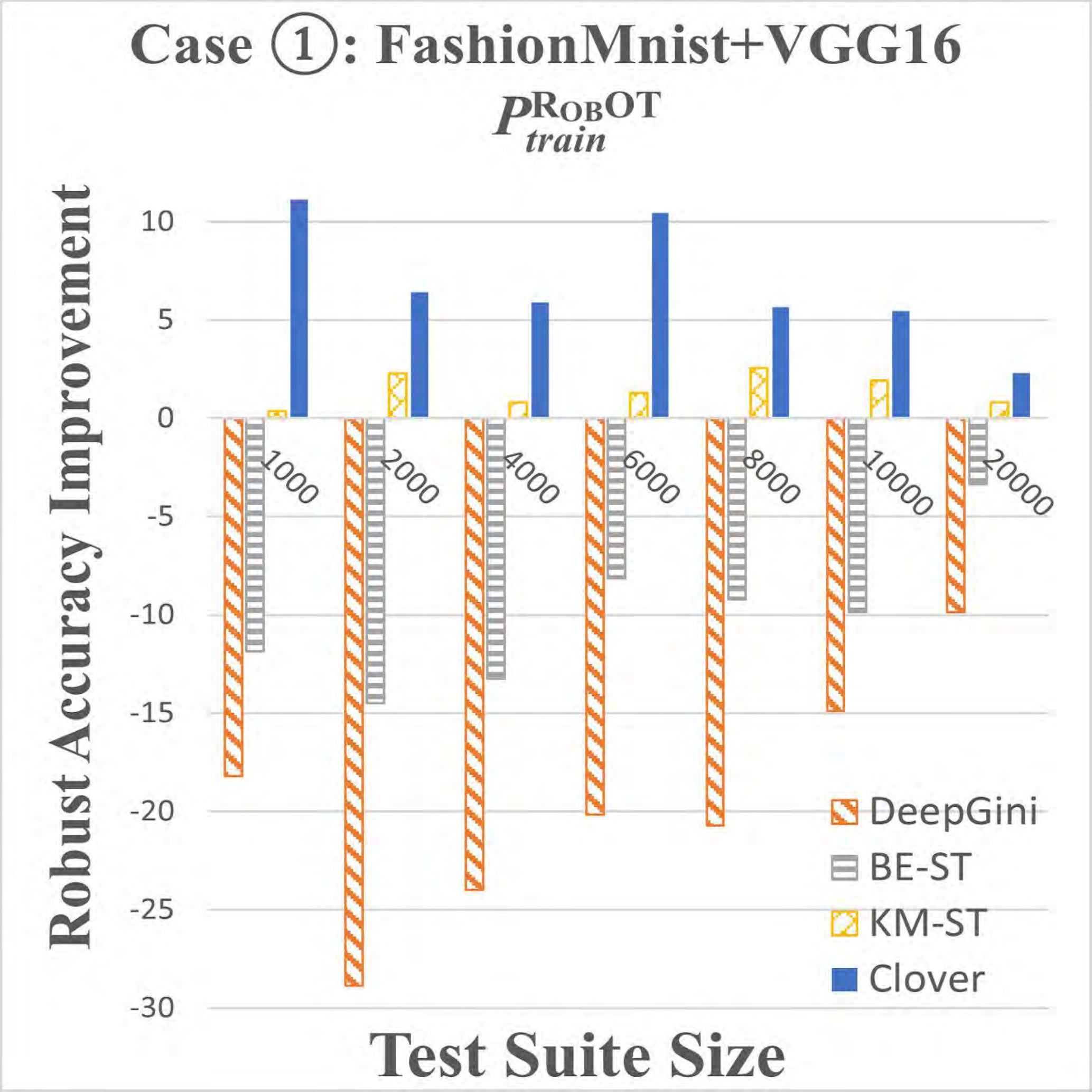} \\
  {\small (a)  Case \circled{1} on the three selection universes in Fig. \ref{fig: rq1_1} (a)}

  \includegraphics[width=0.30\textwidth]{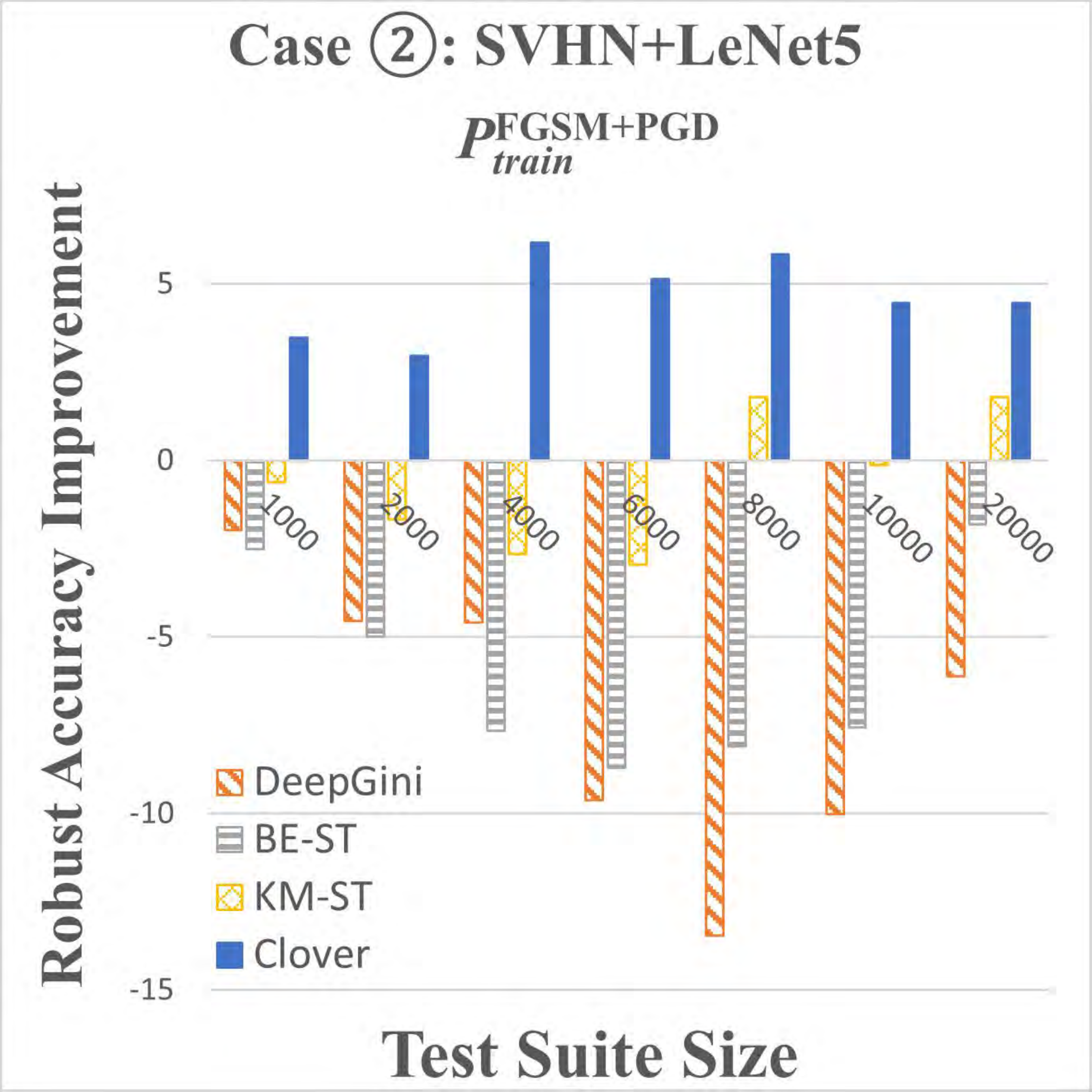}
  \includegraphics[width=0.30\textwidth]{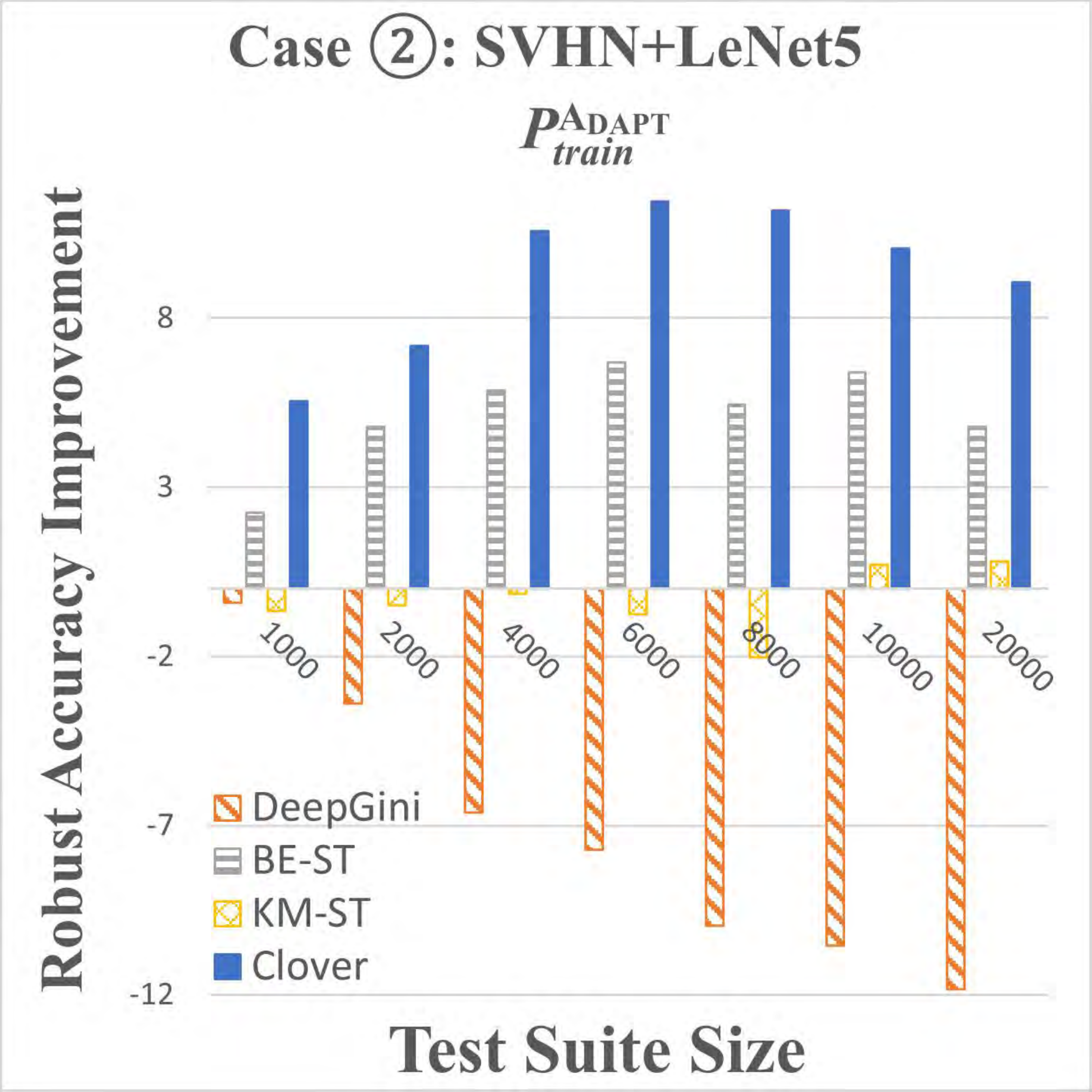}
  \includegraphics[width=0.30\textwidth]{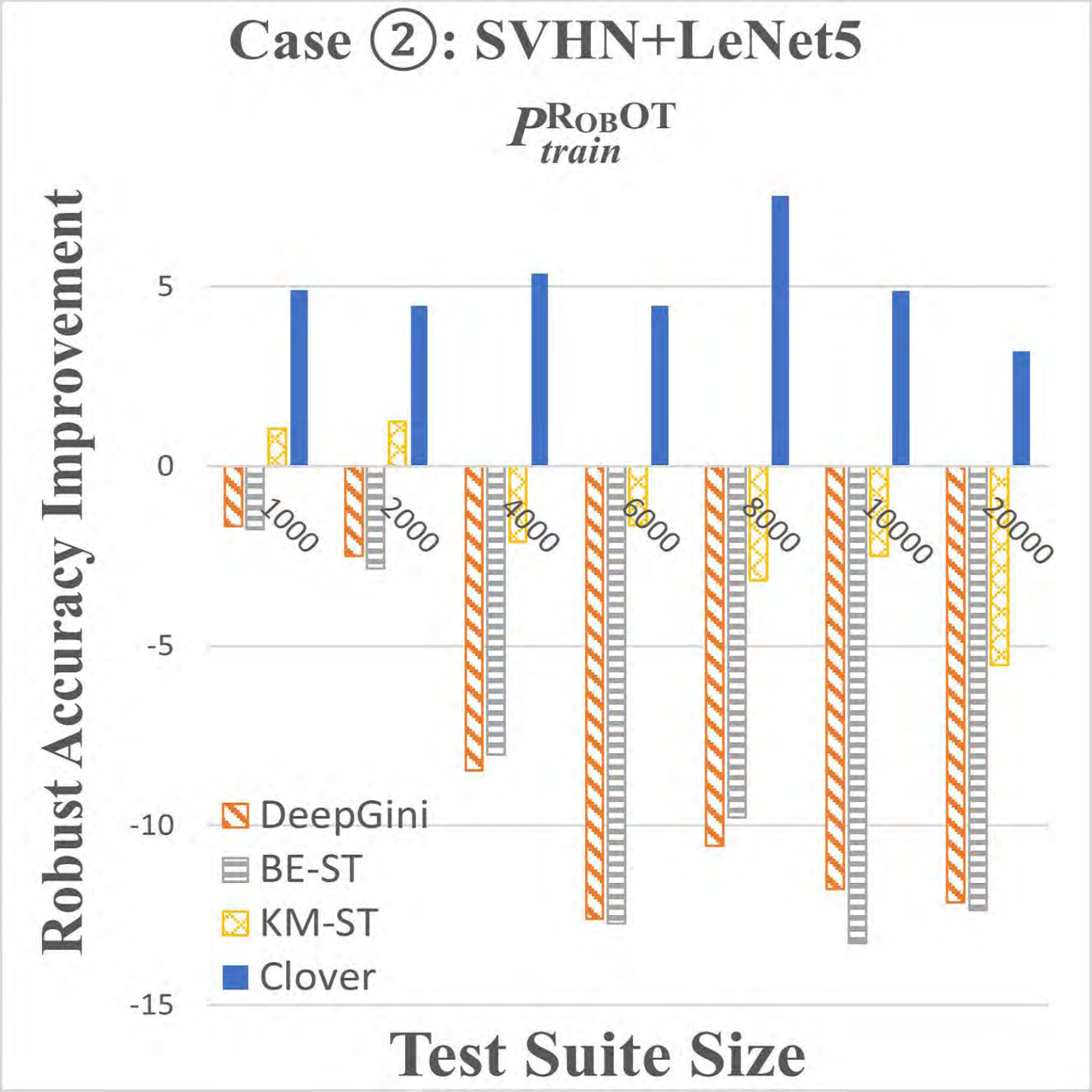} \\
  {\small (b)  Case \circled{2} on the three selection universes in Fig. \ref{fig: rq1_1} (b)}

  \includegraphics[width=0.30\textwidth]{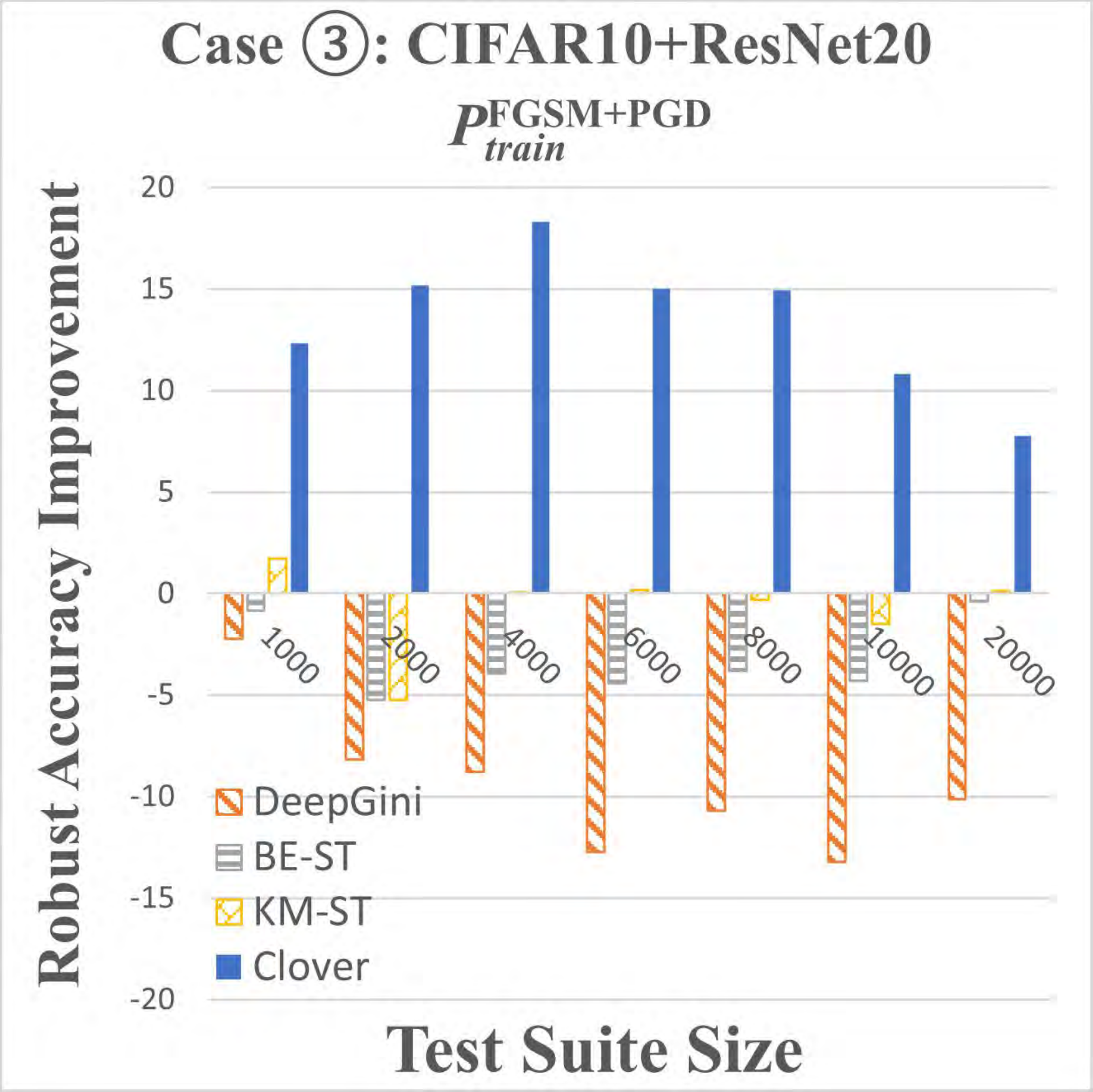}
  \includegraphics[width=0.30\textwidth]{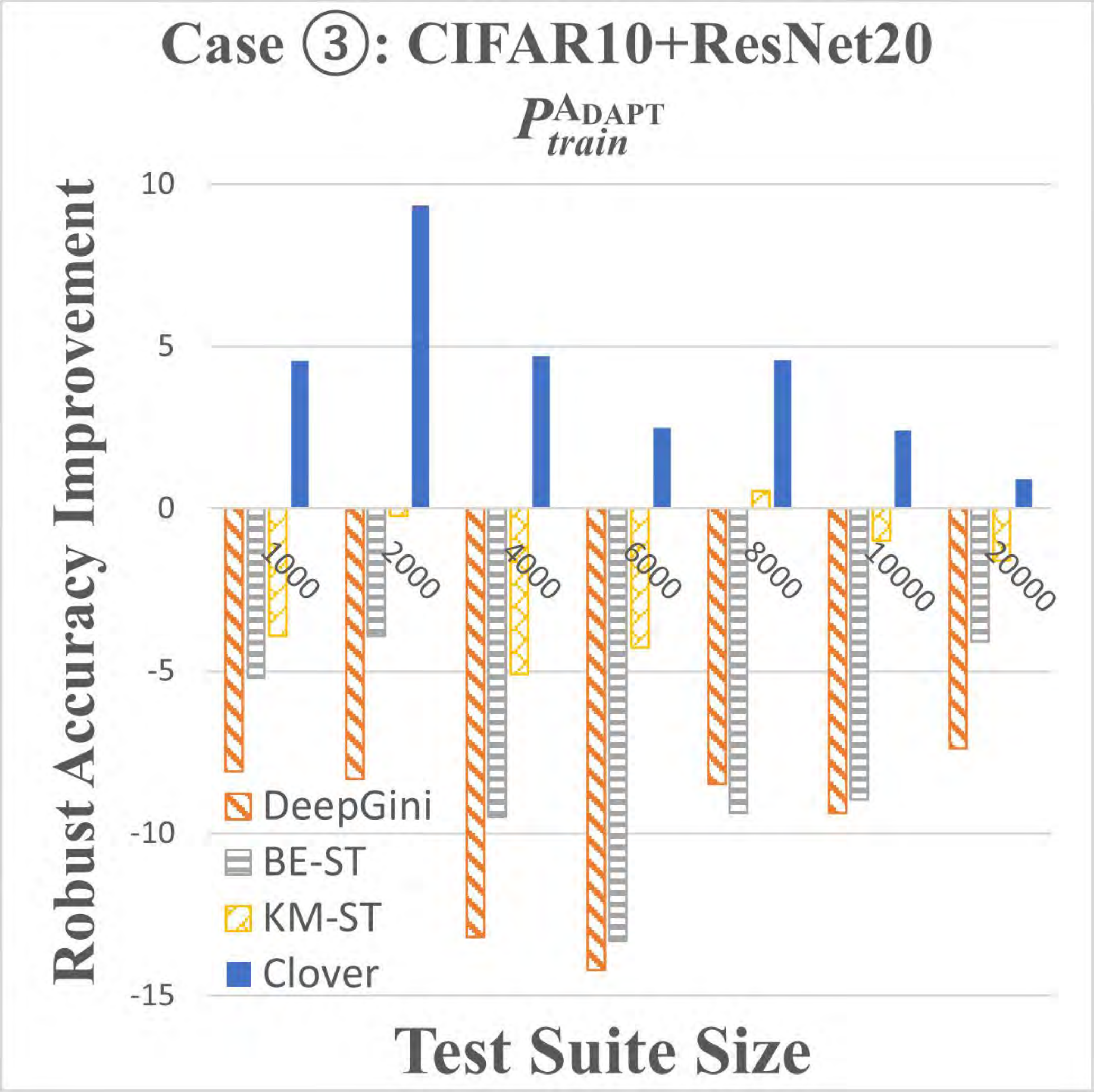}
  \includegraphics[width=0.30\textwidth]{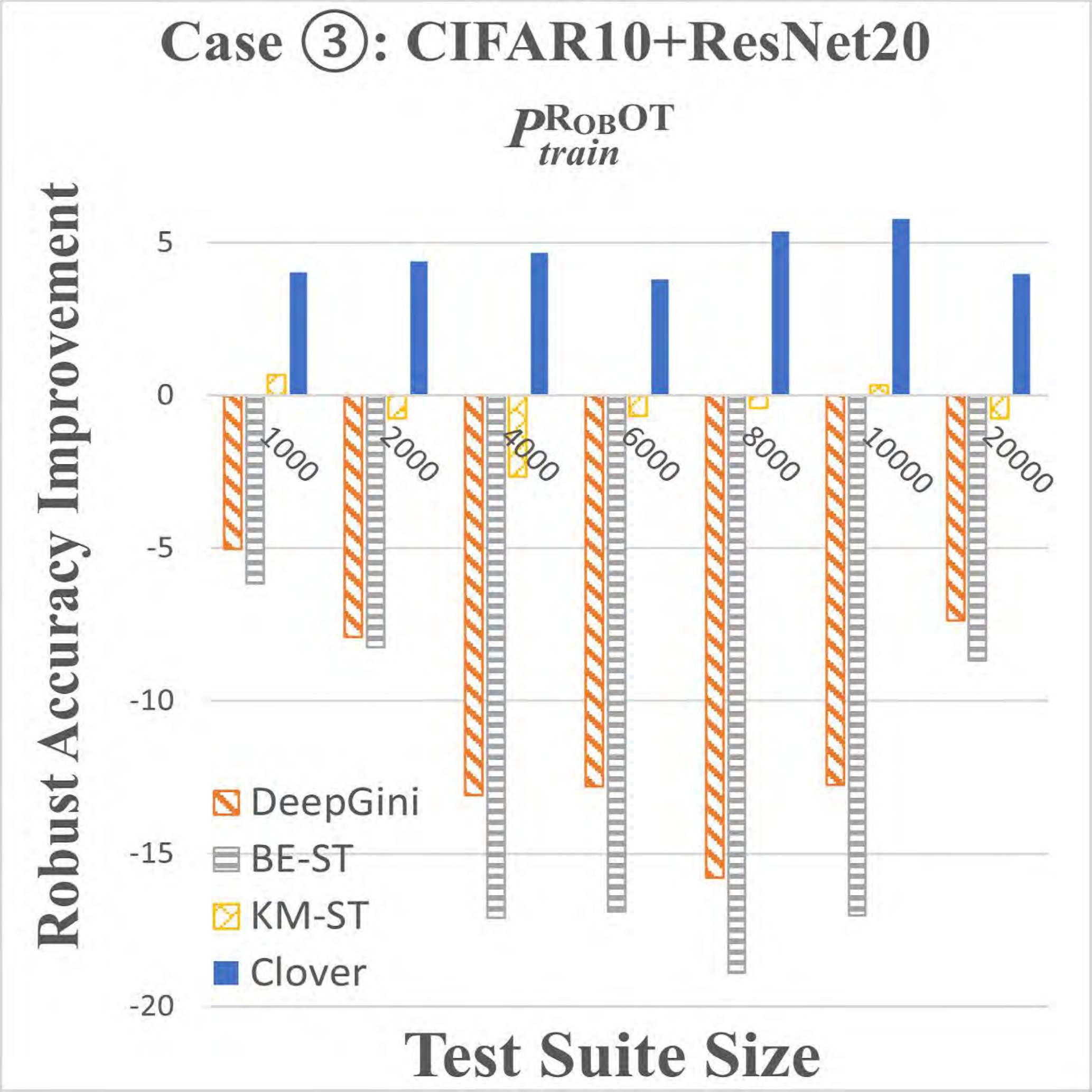} \\
  {\small (c)  Case \circled{3} on the three selection universes in Fig. \ref{fig: rq1_1} (c)}

  \includegraphics[width=0.30\textwidth]{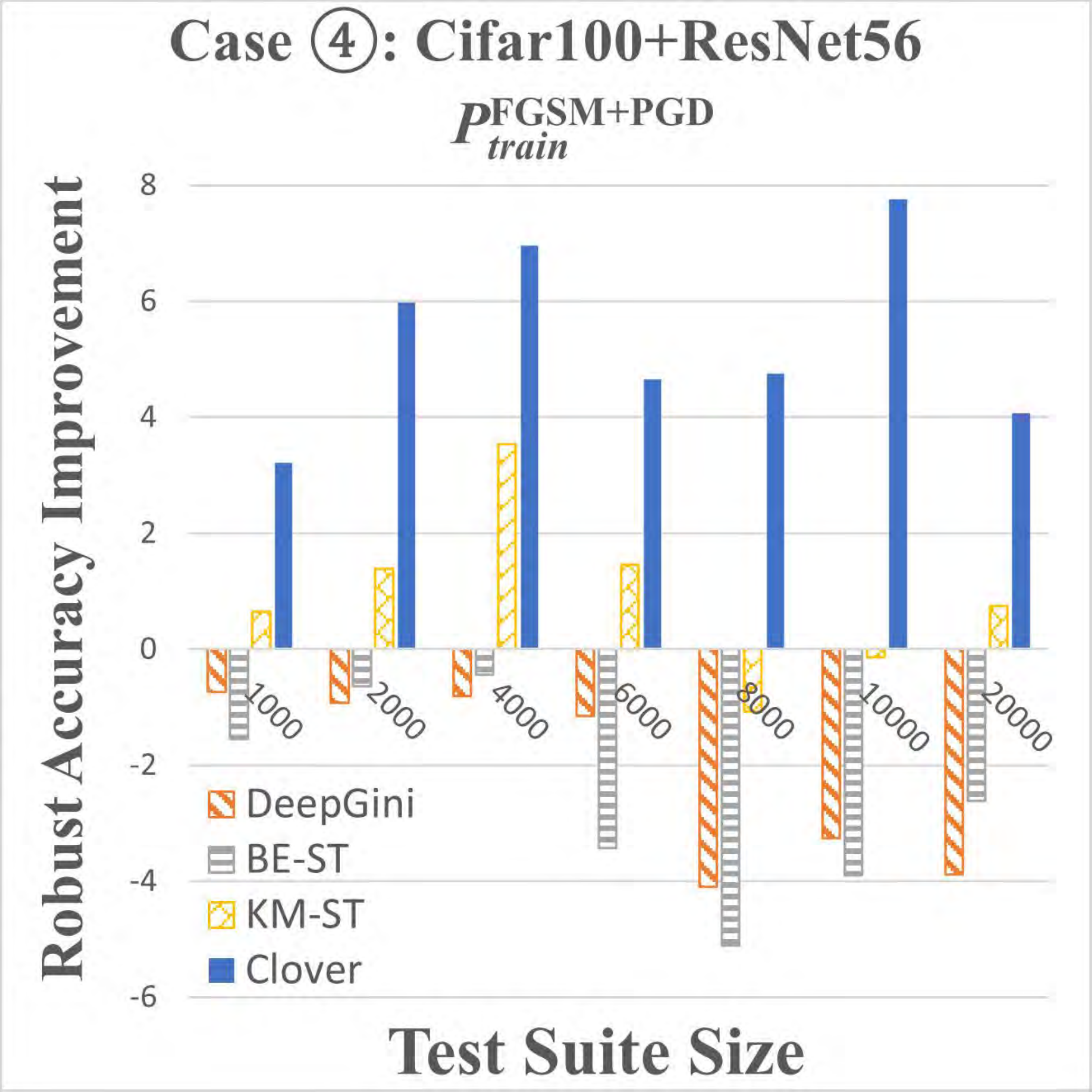}
  \includegraphics[width=0.30\textwidth]{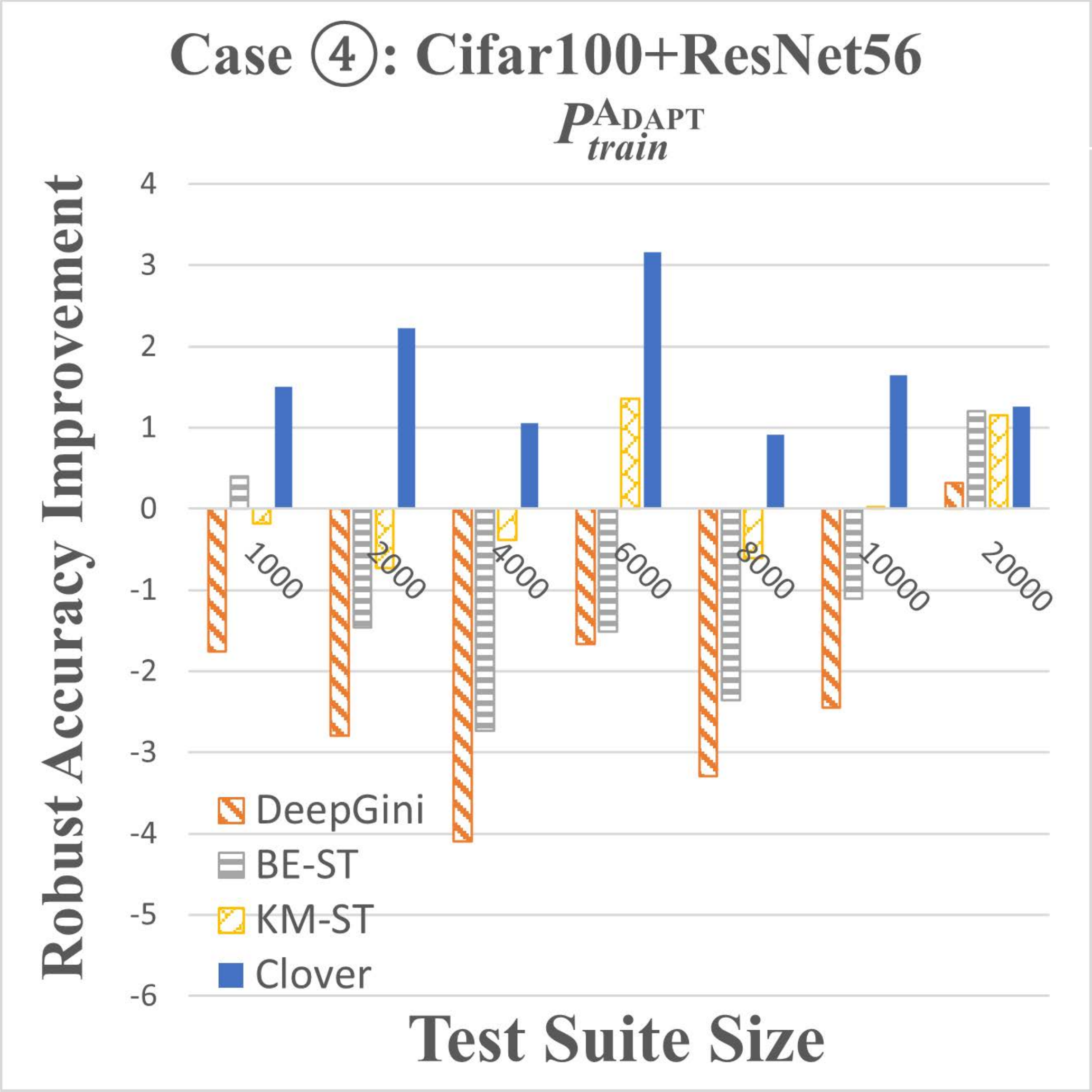}
  \includegraphics[width=0.30\textwidth]{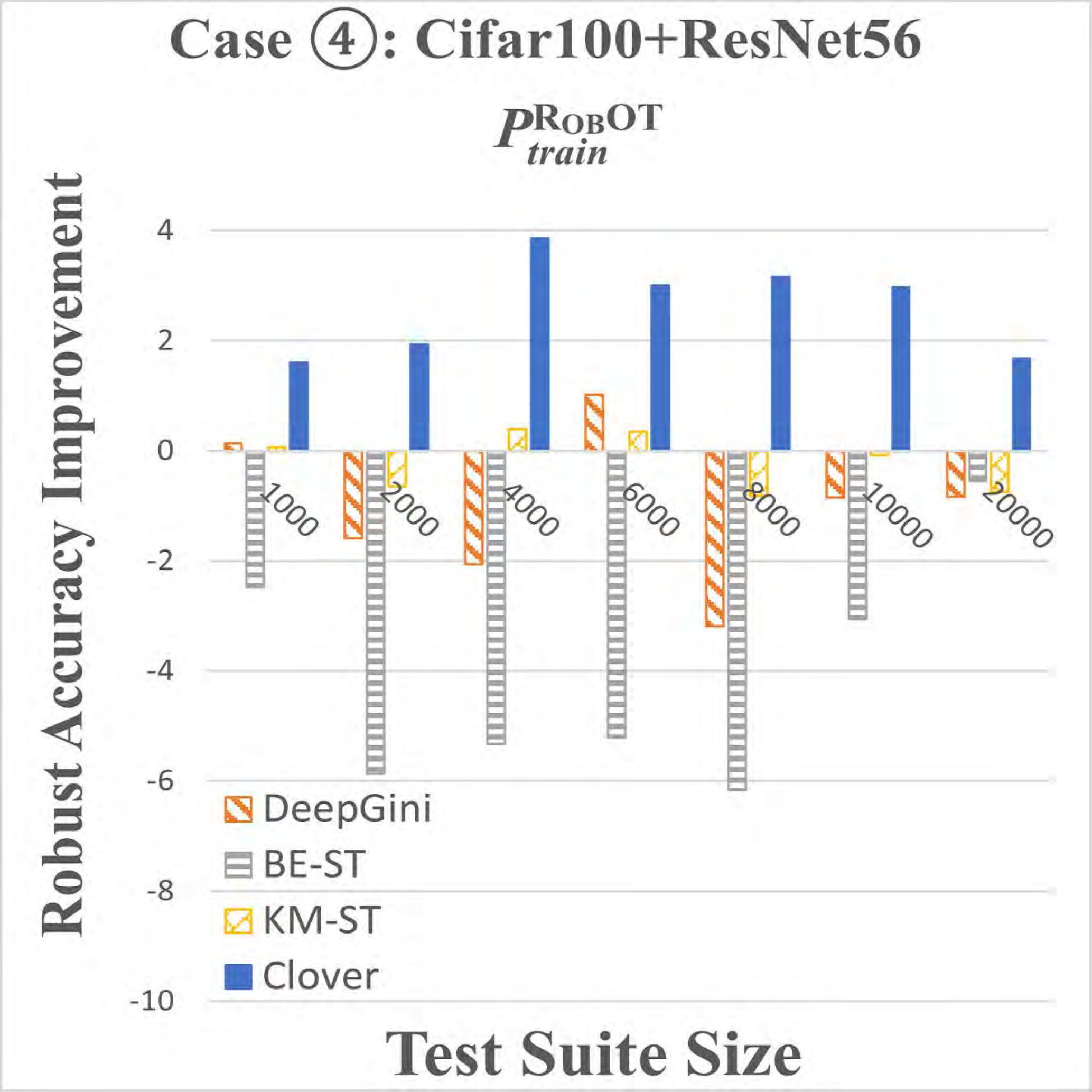} \\
  {\small (d)  Case \circled{4} on the three selection universes in Fig. \ref{fig: rq1_1} (d)}

  \caption{Robust Accuracy Improvement in Fig. \ref{fig: rq1_1} Relative to \textsc{Random} as the Baseline ($x = 0$)}
\label{fig: rq1_2}
\end{figure}

\subsection{Answering RQ1 (Comparison with Peer Test Suite Construction Techniques)} 
\label{subsection: answering_rq1}
Fig. \ref{fig: rq1_1} summarizes the results of Experiment 1. 
It has four subfigures (a)--(d), one for each case from \circled{1} to \circled{4}. 
In each subfigure, the three plots from left to right correspond to the robust accuracy improvements achieved on the three selection universes $P_{train}^{\textsc{FGSM+PGD}}$, $P_{train}^{\textsc{Adapt}}$, and $P_{train}^{\textsc{RobOT}}$, respectively. 
The five series of seven points in the five different colors in each plot corresponds to the five techniques in Experiment 1, which are depicted in the legend, from top to bottom as \textsc{Random}, \textsc{DeepGini}, \textsc{be-st}, \textsc{km-st}, and \textsc{Clover}. 
In each plot, the $x$-axis is the number of test cases in the test suite constructed by a technique (i.e., the possible values in $N_1$). 
The $y$-axis is the robust accuracy improvement. 
A higher $y$ value indicates a more effective technique. 
In Fig. \ref{fig: rq1_1}, in total, there are 420 points representing the robust accuracy improvement results of the 420 combinations of five techniques, seven possible values for $n_1$, four dataset+model cases, and three selection universes ($420 = 5 \times 7 \times 4 \times 3$).

In Fig. \ref{fig: rq1_1}, \textsc{Clover} locates higher than \textsc{Random}, \textsc{DeepGini}, \textsc{be-st}, and \textsc{km-st} in all 84 cases. \textsc{DeepGini} performs less effective than \textsc{be-st} and \textsc{km-st}. 
\textsc{km-st} performs slightly more effective than \textsc{be-st} in 72 out of 84 ($86\%$) cases. The differences in robust accuracy improvement between \textsc{Clover} and \textsc{km-st} are as noticeable as the differences between \textsc{km-st} and \textsc{DeepGini}. 
The difference in case \circled{1} is smaller, and the lines for \textsc{km-st} and \textsc{Random} cross each other and are very close. 

For the Wilcoxon signed-rank test \cite{wilcoxon} at the 5\% significance level with Bonferroni correction, the $p$-values for \textsc{DeepGini}, \textsc{be-st}, \textsc{km-st}, and \textsc{Clover} compared to \textsc{Random} are 
1.00, 1.00, 0.33, and $\leq 1e^{-5}$ on $P_{train}^{\textsc{FGSM+PGD}}$, 
1.00, 0.96, 0.98, and $\leq 1e^{-5}$ on $P_{train}^{\textsc{Adapt}}$, and 
1.00, 1.00, 0.70, and $\leq 1e^{-5}$ on $P_{train}^{\textsc{RobOT}}$, respectively.
The respective effect sizes are
2.06, 1.12, 0.01, and 1.55 on $P_{train}^{\textsc{FGSM+PGD}}$, 
2.10, 0.45, 0.18, and 1.19 on $P_{train}^{\textsc{Adpat}}$, and 
2.42, 2.12, 0.07, and 1.08 on $P_{train}^{\textsc{RobOT}}$.
Only \textsc{Clover} and \textsc{Random} are significantly different at the 5\% significance level with a large effect size in all four cases.

\textsc{km-st} and \textsc{Random} closely compete with each other. Fig. \ref{fig: rq1_2} uses \textsc{Random} in Fig. \ref{fig: rq1_1} as the baseline (i.e., $x$=0) to show the relative performance among techniques. Positive and negative bars indicate a technique more effective and less effective than \textsc{Random}, respectively. For instance, in the leftmost plot (for $P_{train}^{\textsc{FGSM+PGD}}$) in Fig. \ref{fig: rq1_1}(c), when $n_1 = 10000$, \textsc{Clover} and \textsc{Random} are $62.49\%$ and $51.66\%$, respectively, in $y$-values. 
In the leftmost plot of Fig. \ref{fig: rq1_2}(c), the respective bar height is 10.83\%.

In Fig. \ref{fig: rq1_2}, \textsc{Clover} outperforms \textsc{Random} in all 84 cases ($100\%$). Almost all bars for \textsc{Clover} are longer than the others in the same plot. 
On average, \textsc{Clover} is more effective than \textsc{Random} in robust accuracy improvement by $5.51\%$, $6.30\%$, $7.40\%$, and $3.21\%$ in cases \circled{1} to \circled{4}, respectively. 
\textsc{km-st} wins \textsc{Random} in only 39 out of 84 cases ($46\%$) but underperforms \textsc{Random} in the remaining $54\%$. The bars for \textsc{km-st} are all short (compared to the bars for \textsc{Clover}). The average change in robust accuracy improvement achieved by \textsc{km-st} atop \textsc{Random} for the four models are $0.47\%$, $-0.93\%$, $-1.18\%$, and $0.27\%$ only (i.e., less effective than \textsc{Random} in two out of four cases). The differences in robust accuracy improvement between \textsc{Clover} and \textsc{km-st} are large and consistent.

\textsc{be-st} wins \textsc{Random} in 12 cases ($14\%$ only), but the differences are all small (less than $3.79\%$ on average). For the remaining 72 cases, most bars for \textsc{be-st} are much longer than the bars for \textsc{km-st}. 

\textsc{DeepGini} always underperforms \textsc{Random} by a large extent. It is the least effective among the techniques in 49 cases ($58\%$), and if \textsc{DeepGini} is not the least effective one, then \textsc{be-st} or \textsc{km-st} is.

Across all four cases \circled{1} to \circled{4}, \textsc{Clover} outperforms \textsc{Random} by 0.56\%--11.39\%, 2.96\%--11.45\%, 2.42\%--18.30\%, 0.91\%--6.96\%, respectively.
Across all three selection universes, the corresponding mean differences in robust accuracy improvements achieved by \textsc{Clover} are 7.17\%, 4.93\%, and 4.73\%.

We also conduct the Wilcoxon signed-rank test at the 5\% significance level with Bonferroni correction on the robust accuracy improvements achieved by \textsc{Clover} compared to those achieved by each of \textsc{DeepGini}, \textsc{be-st} and \textsc{km-st} by pairing based on test suite size and selection universe. Their $p$-values are all $\leq 1e^{-5}$, and the effect sizes are all at a large level on $P_{train}^{\textsc{FGSM+PGD}}$, $P_{train}^{\textsc{Adapt}}$ and $P_{train}^{\textsc{RobOT}}$, meaning that \textsc{Clover} is significantly different from \textsc{DeepGini}, \textsc{be-st} and \textsc{km-st} at the 5\% significance level in statistically meaningful ways.

In summary, the difference between \textsc{Clover} and \textsc{Random} is statistically significant at the 5\% significance level with a large effect size.
\textsc{km-st}, \textsc{be-st}, and \textsc{DeepGini} perform either similarly to \textsc{Random} or worse than it significantly at the 5\% significance level with medium to large effect sizes.

\begin{tcolorbox}[
enhanced, breakable,
attach boxed title to top left = {yshift = -3mm, xshift = 5mm},
boxed title style = {sharp corners},
colback = white, 
title={Answering RQ1}
]
\textsc{Clover} in Configuration \textit{A} is the only technique consistently outperforming \textsc{Random} and all other peer techniques in the experiment in a statistically meaningful way.
\end{tcolorbox}

\subsection{Answering RQ2 (Trend of Robustness Improvement)}
\label{subsection: answering_rq2}

\begin{figure*} [t]
\begin{minipage}[c]{\textwidth}
  \centering
  \includegraphics[width=0.9\linewidth]{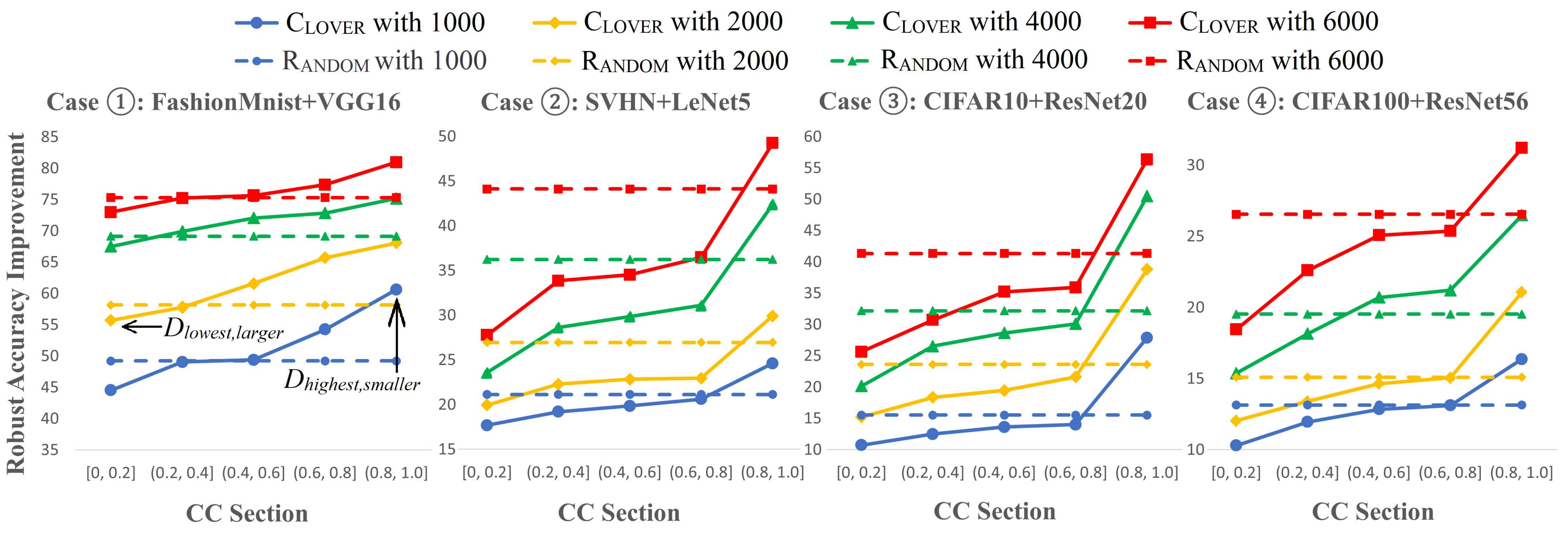}
\end{minipage}%
\caption{Robust Accuracy Improvement Achieved by \textsc{Clover} in Configuration \textit{A} on Test Suites Concentrated with Test Cases in Different Sections of CC (\textsc{Random} as the Baseline)}
\label{fig: rq2}
\end{figure*}

Fig. \ref{fig: rq2} shows the result of Experiment 2. 
The four plots from left to right correspond to cases \circled{1} to \circled{4}, respectively.
In each plot, the $x$-axis is the five pools of test cases with CC falling into the section indicated by the $x$-value accepted by \textsc{Clover} to produce test suites.
These sections are the possible values of $n_2$ in the set $N_2$ (i.e., from $n_2 = [0, 0.2]$ to $n_2 = (0.8, 1]$).
We refer to the five test suites as Pool-1 to Pool-5, respectively.
The $y$-axis is the robust accuracy improvement achieved by \textsc{Clover}.
The four solid curves and dashed lines from bottom to top correspond to the four possible sizes (values in $N_3$) of the test suite constructed by \textsc{Clover} and \textsc{Random}, respectively. 

Across all four cases \circled{1} to \circled{4} and all sizes of the constructed test suites (points along the curve in the same line style), as the test suite accepted by \textsc{Clover} is replaced by the one filled with samples with higher CC (i.e., as the $x$-value increases), 
\textsc{Clover} achieves an increasingly higher robust accuracy improvement.
The cumulative increases in robust accuracy improvement from using Pool-1 (i.e., the test suite with the section of the lowest CC) to using Pool-5 (i.e., the test suite with the section of the highest CC) for all curves are large. 
For instance, when the test suite size ($n_3$) is 6000, these cumulative increases in cases \circled{1} to \circled{4} are $7.98\%$, $21.48\%$, $30.74\%$, and $12.75\%$, (or 10.93\%, 77.35\%, 119.94\%, 69.11\% in percentage by normalizing the robust accuracy improvement for the one with the lowest CC ($n_2 = [0, 0.2]$) to 1) respectively.
For each case from \circled{1} to \circled{4}, the mean cumulative increases in robust accuracy improvements from using Pool-1 to Pool-5 achieved by \textsc{Clover} over the five test suites are $11.02\%$, $14.30\%$, $25.46\%$, and $9.74\%$, respectively. 
The effect of higher CC is significant.

Across the four cases and every consecutive pair of curves in the same plot, we compare two particular points in the curve pair:
The point with the highest $y$-value in the lower curve (referred to as configuration $D_{\textit{highest, smaller}}$) and the point with the lowest $y$-value in the upper curve (referred to as configuration $D_{\textit{lowerest, larger}}$). 
In Fig. \ref{fig: rq2}, we mark these two points in a curve pair in case \circled{1} to ease readers to follow. 
They represent \textsc{Clover} using Pool-5 to construct a \textit{smaller} test suite and \textsc{Clover} using Pool-1 to construct a \textit{larger} test suite, respectively.
Interestingly, the former configuration ($D_{\textit{highest, smaller}}$) even achieves a significantly higher robustness improvement than the configuration ($D_{\textit{lowerest, larger}}$).
\begin{enumerate}
\item
    From top to bottom in each plot, there are three pairs of such points. 
    We compute the gain in robust accuracy improvement achieved by the point $D_{\textit{highest, smaller}}$ on top of $D_{\textit{lowerest, larger}}$ in each such pair.
    For instance, the gain for the two points marked in case \circled{1} in Fig. \ref{fig: rq2} is 4.92\% (= 60.62\% $-$ 55.70\%).
    The mean gains of the four cases for the three pairs of points from top to bottom are 2.56\%, 8.54\%, 18.72\%, and 6.02\%, respectively.
\item
    Also, from top to down in every plot,
    the reductions in selection universe size for the three pairs of consecutive curves in percentage are 
    33.3\% (= $\frac{6000-4000}{6000}$), 
    50.0\% (= $\frac{4000-2000}{4000}$), 
    50.0\% (= $\frac{2000-1000}{2000}$), respectively.
    The reductions in ratio and absolute number are large while achieving large gains in robust accuracy improvements.
\end{enumerate}

Fig. \ref{fig: rq2} also shows that \textsc{Clover} consistently outperforms \textsc{Random} of the same test suite size when using Pool-5.
The overall trend is clear, moving upward by large steps in percentage, and consistent across all four cases.
There are large upward increases in most curves when the $x$-value changes from Pool-4 to Pool-5 (i.e., from the section (0.6, 0.8] to the section (0.8, 1]).
The results show that a test suite with high CC (Pool-5) can potentially outperform an observably larger test suite with low CC (Pool-1).

We further compute the change in robust accuracy improvement achieved by each test suite of \textsc{Clover} in each section (Pool-1 to Pool-5) compared to \textsc{Random} for the same test suite size in each of the four plots.
We measure Spearman's correlation coefficient \cite{spearman_correlation} between this change in robust accuracy improvement and the CC value of the corresponding test suite of \textsc{Clover} for each of the four cases.
The results for cases \circled{1} to \circled{4} are 0.94, 0.68, 0.84, and 0.93, respectively. 
They are all strong correlations \cite{correlation_measure}.

We summarize that, under the same test case selection budget (i.e., the test suite size is the same), \textsc{Clover} is more effective than \textsc{Random}.

\begin{tcolorbox}[
enhanced, breakable,
attach boxed title to top left = {yshift = -2mm, xshift = 5mm},
boxed title style = {sharp corners},
colback = white, 
title={Answering RQ2}
]
\textsc{Clover} in Configuration \textit{A}, when constructing test suites with higher CC (i.e., a section closer to 1), achieves a higher robust accuracy improvement.
A test suite with test cases taken from Pool-5 (a high CC section) significantly outperforms (in terms of the number of test cases) both a random test suite of the same size and an observably larger test suite that contains test cases taken from Pool-1 (a low CC section) in terms of the robust accuracy improvement.
\end{tcolorbox}

\section{Results and Data Analysis for \textsc{Clover} in Configuration \textit{B}} 
\label{sec: results_in_configuration_B}
This section presents the results and data analysis for answering RQ3 and RQ4 through Experiments 3 and 4 for \textsc{Clover} in Configuration \textit{B}.
Like what we have clarified in Section~\ref{sec: result_in_configuration_A}, we use the term robustness improvement achieved by a DL testing technique to mean the robustness improvement achieved by the retrained model output by the testing-retraining pipeline with the technique as the DL testing in the pipeline.

\subsection{Answering RQ3 (Overall Effect of \textsc{Clover})}
\label{sec: rq3_result}

\begin{figure}[t]
  \centering
  \includegraphics[width=0.35\textwidth]{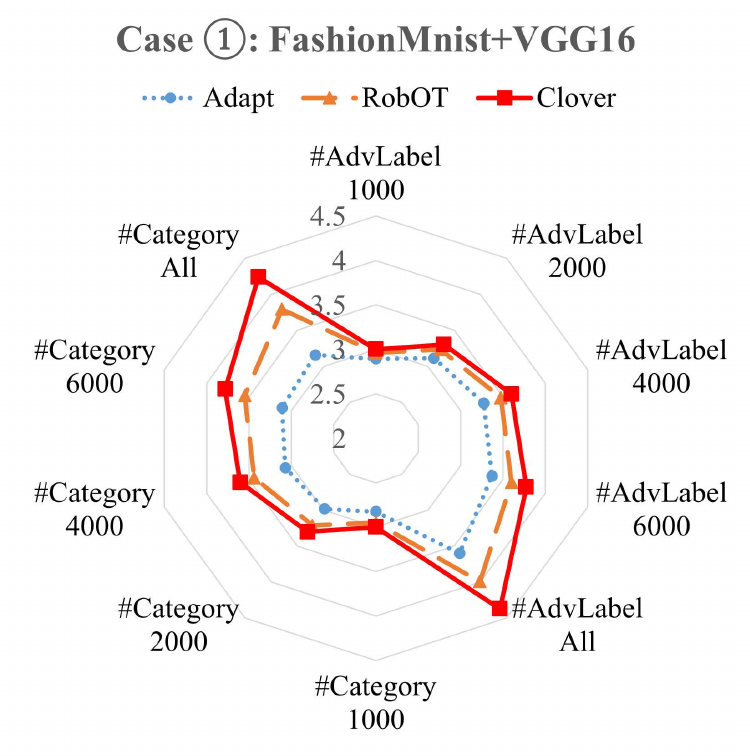}
  \includegraphics[width=0.35\textwidth]{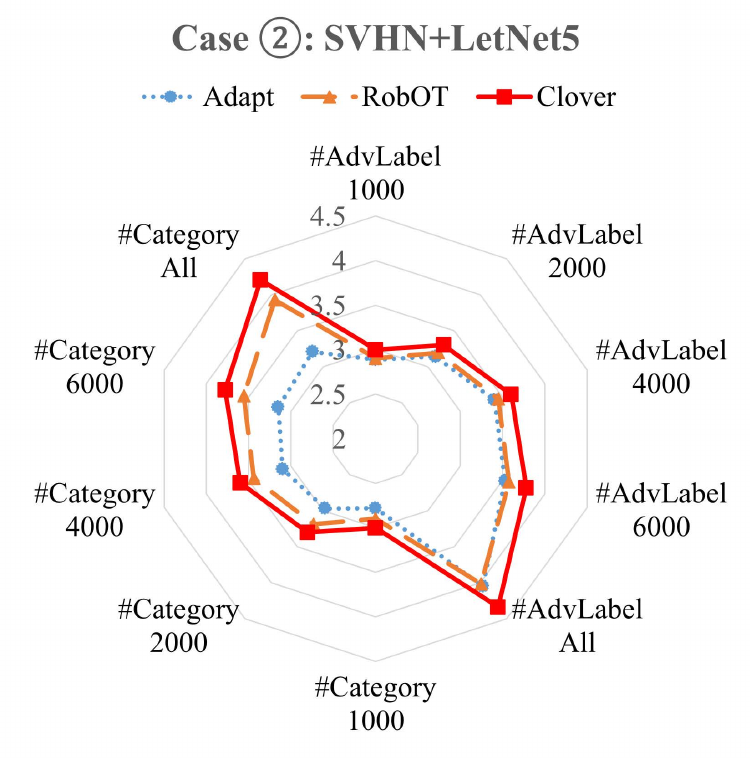} \\
  \includegraphics[width=0.35\textwidth]{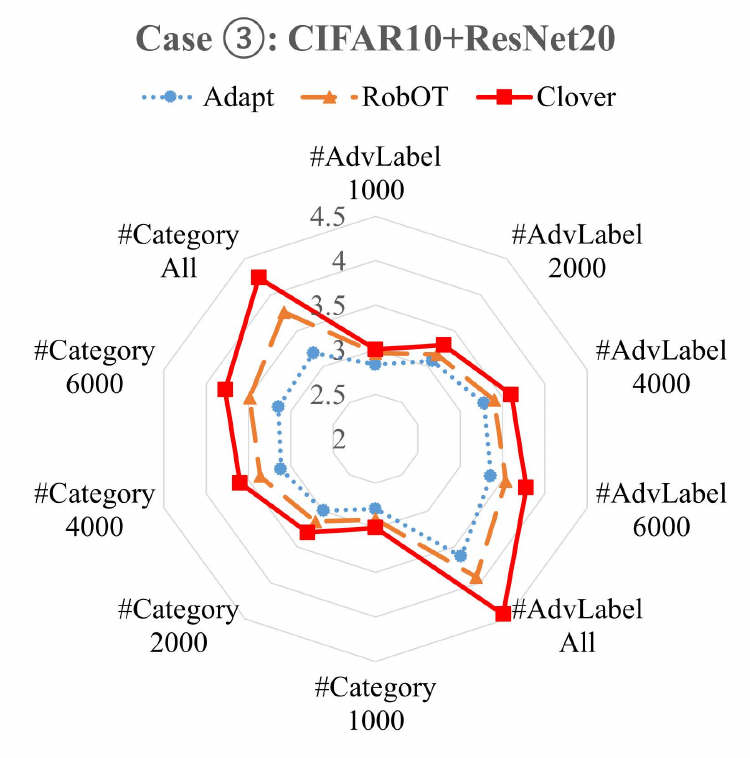}
  \includegraphics[width=0.35\textwidth]{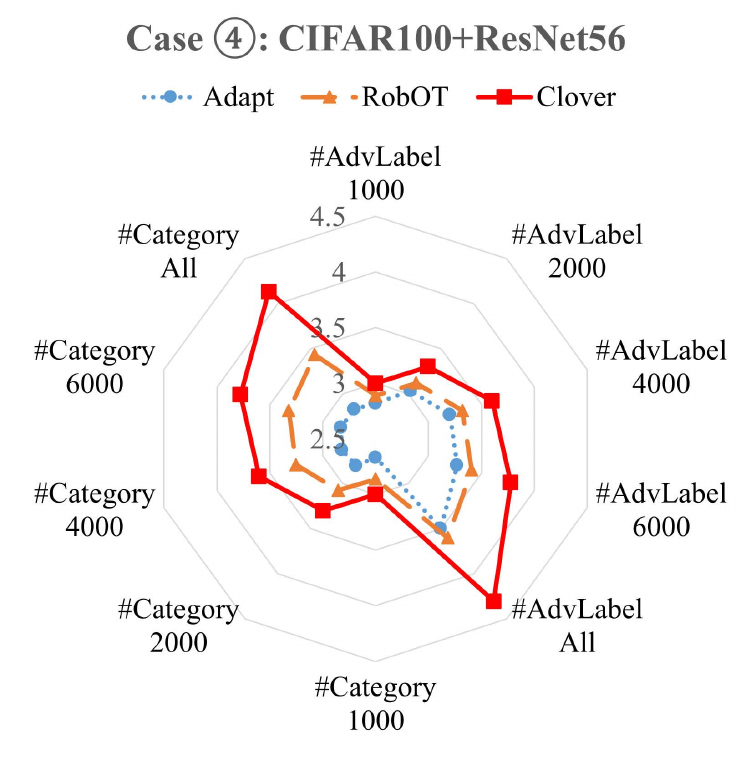} \\
  \caption{\textit{\#AdvLabel} and \textit{\#Category} for Different Techniques with $n_4=18000$ (in $log_{10}$ scale)}
\label{fig: rq3_statistics_advlabel_category}
\end{figure}

\begin{figure}[t]
  \centering
  \includegraphics[width=0.24\textwidth]{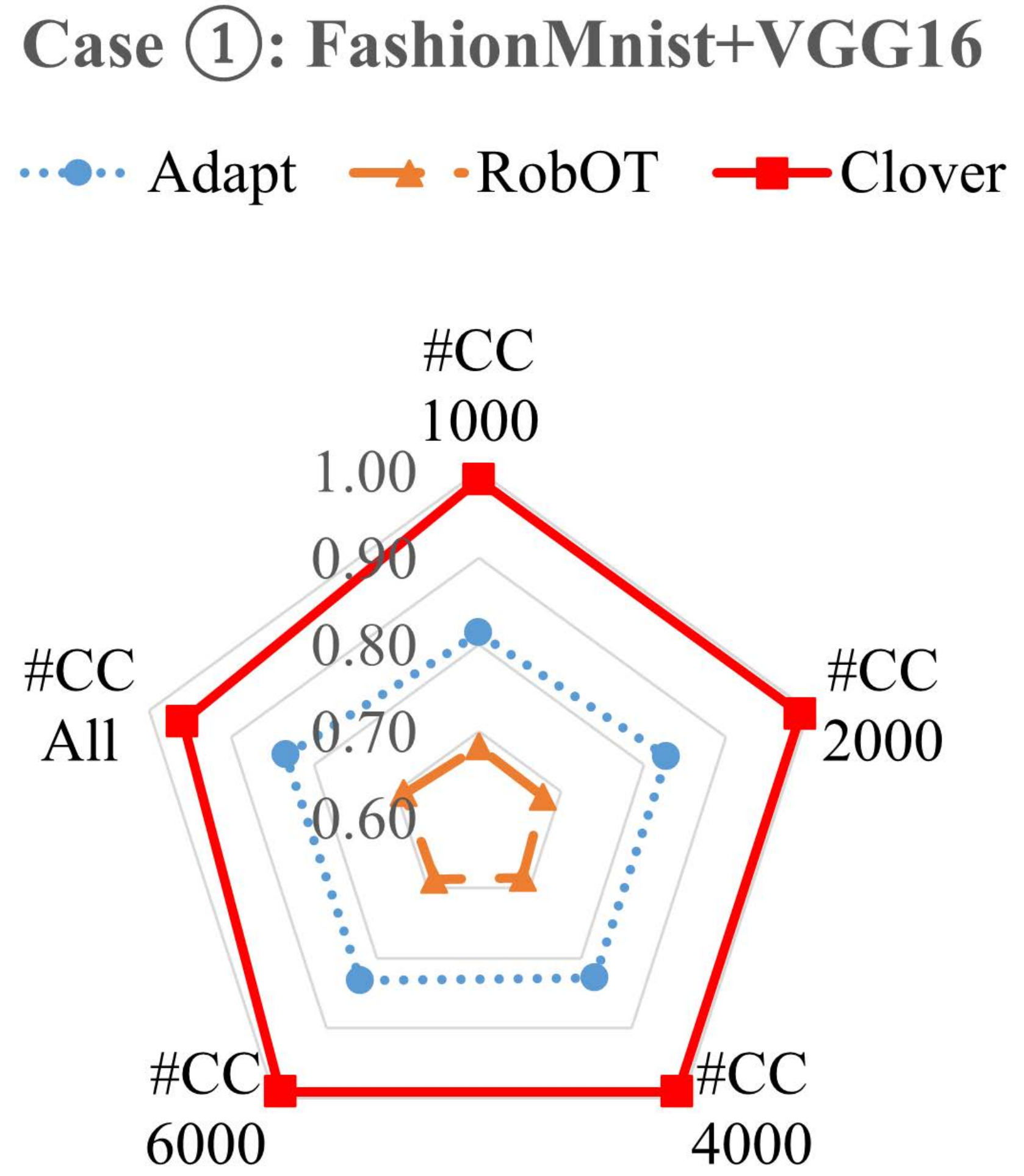}
  \includegraphics[width=0.24\textwidth]{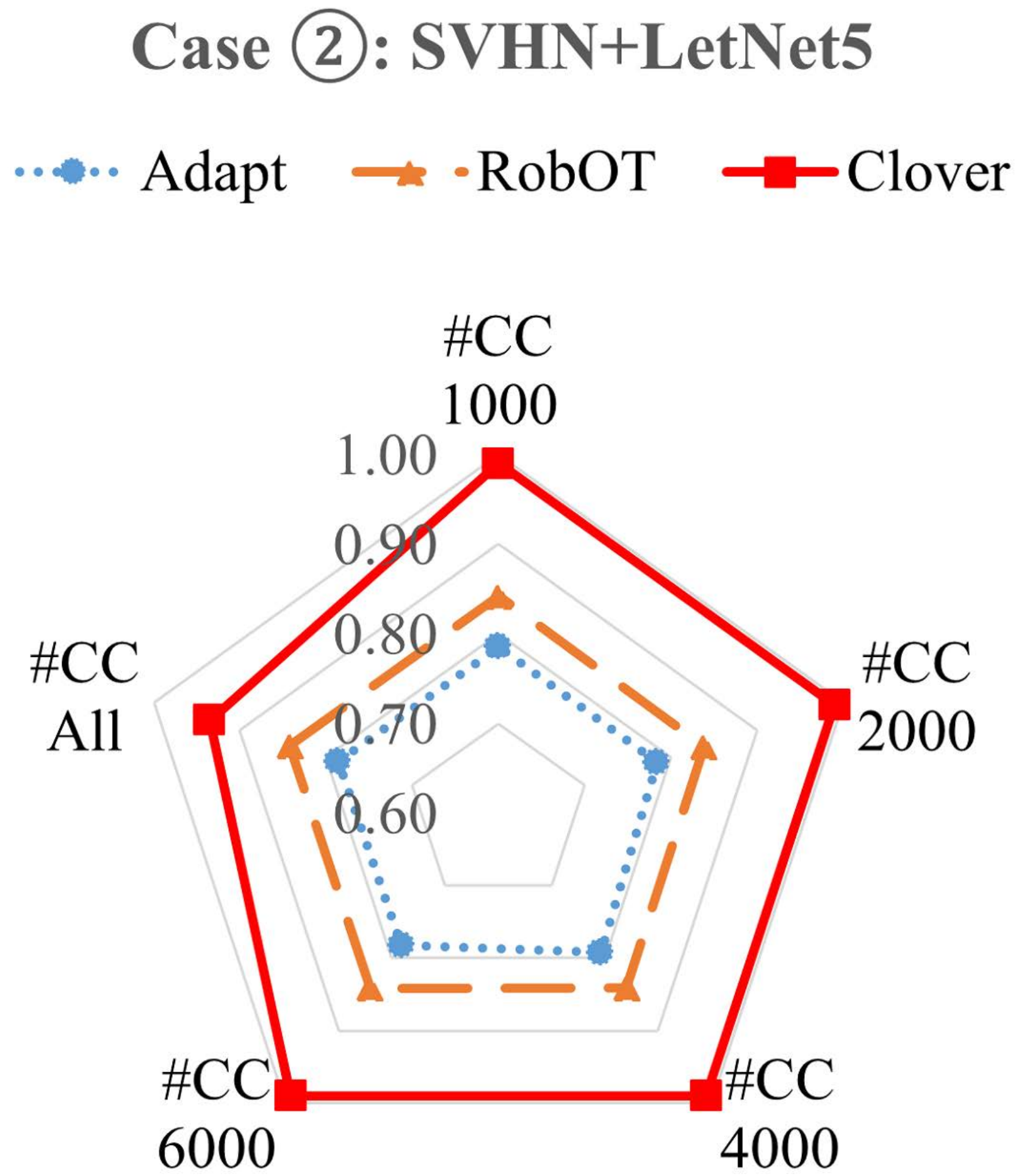}
  \includegraphics[width=0.24\textwidth]{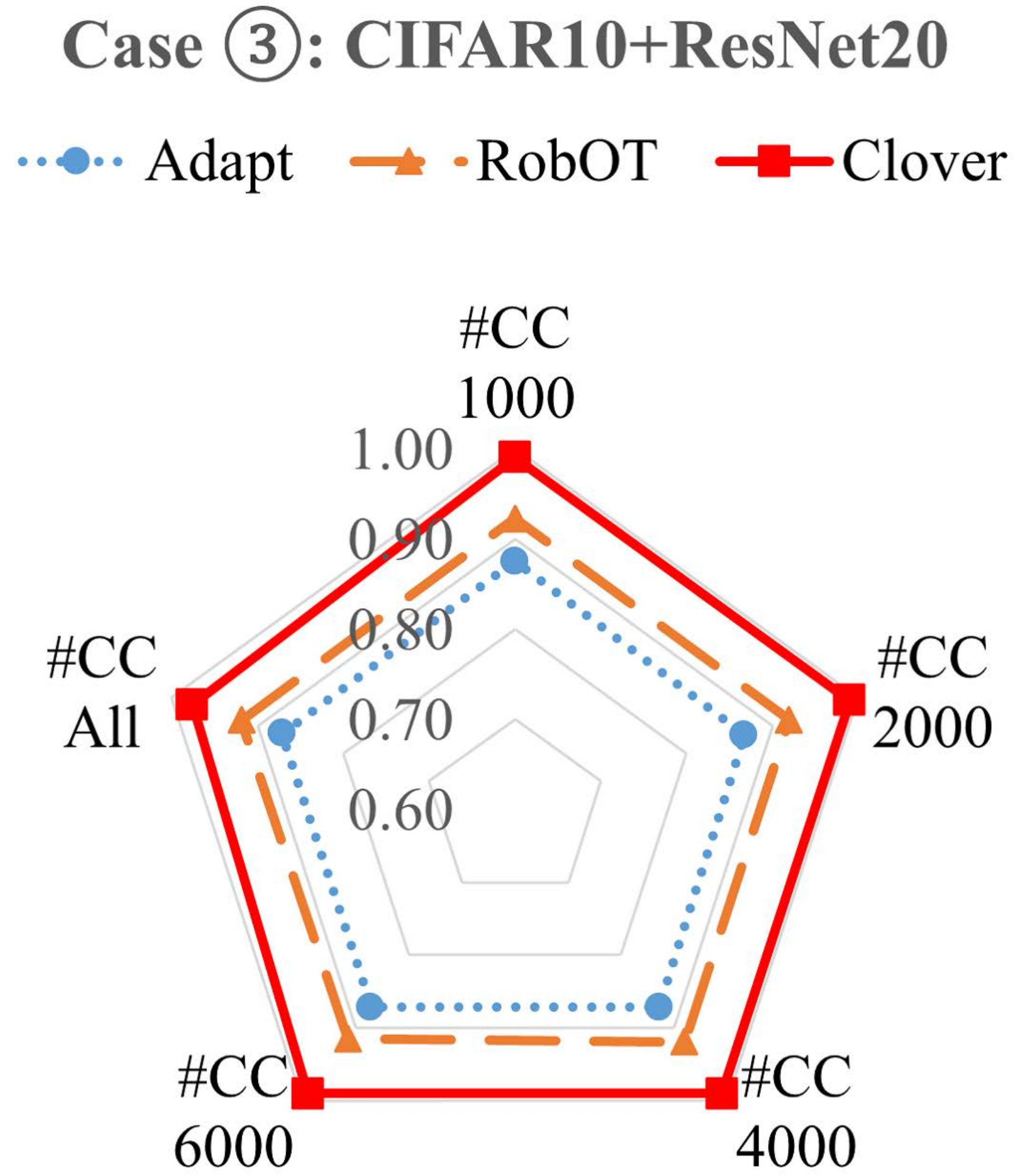}
  \includegraphics[width=0.24\textwidth]{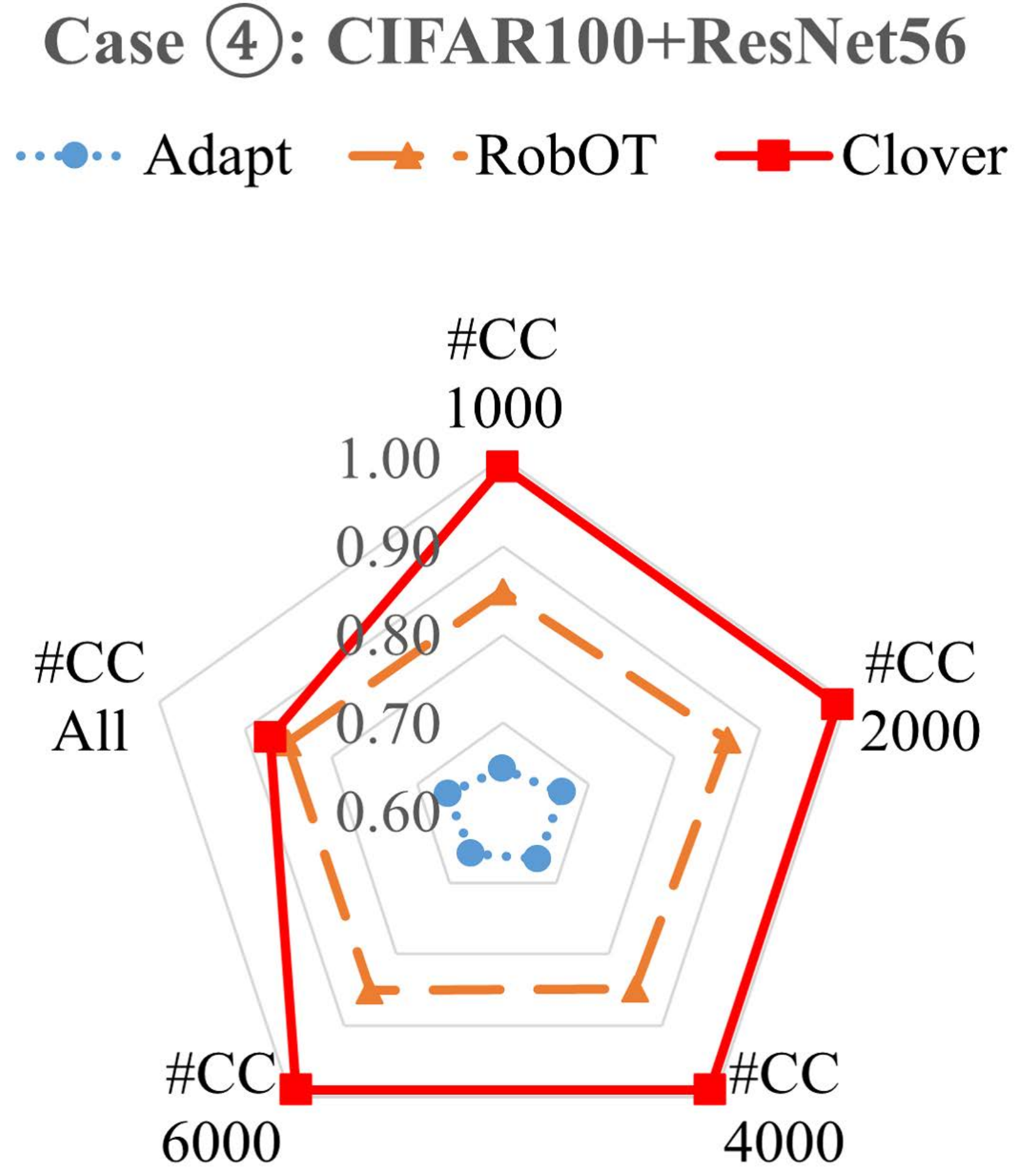} \\
  \caption{\textit{\#CC} for Different Techniques with $n_4 = 18000$}
\label{fig: rq3_statistics_cc}
\end{figure}

\paragraph{\textbf{Effectiveness in test case generation:}}
Figures \ref{fig: rq3_statistics_advlabel_category} and \ref{fig: rq3_statistics_cc} summarize the results of Experiments 3a, 3b, and 3c on the fuzzing effectiveness in producing samples in terms of the number of unique adversarial labels (\textit{\#AdvLabel}) and the number of unique categories (\textit{\#Category}) as well as the mean CC values achieved by the test suites (\textit{\#CC}).
Each of the two figures consists of four radar charts for cases \circled{1}--\circled{4}, respectively.
We note that the values achieved by each technique in each axis in each radar chart can be found in Tables \ref{tab: rq3_fuzz_clean_model_test_suites} and \ref{tab: rq3_fuzz_clean_model_universe} in Appendix \ref{appendix: rq3_fuzz_clean_model_detail}. 

In Fig. \ref{fig: rq3_statistics_advlabel_category}, each radar chart has ten radial axes, five for  \textit{\#AdvLabel} and five for \textit{\#Category}.
The five radial axes for \textit{\#AdvLabel} and these for \textit{\#Category} starting from 12 o'clock and 6 o'clock in clock position and then clockwise, respectively, are the results for $n_3$ = 1000, 2000, 4000, 6000, and \textit{All}, respectively, where \textit{All} refers to the test pool that contains all test cases generated by the technique (i.e., without selection criteria to filter test cases), with the fuzzing budget of $n_4 = 18000$ seconds. 
Similarly, in Fig. \ref{fig: rq3_statistics_cc}, the five axes starting from 12 o'clock in clock position and then clockwise are \textit{\#CC} for $n_3$ = 1000, 2000, 4000, 6000, and \textit{All}.
In each radar chart, the results for \textsc{Adapt}, \textsc{RobOT}, and \textsc{Clover} are shown in different line styles and markers.

The result of each technique shown in each radar axis in each radar chart is the result for the value of $n_3$ as indicated by the axis label, where we show the results in the base-10 log scale (i.e., $log_{10}(.)$) in Fig. \ref{fig: rq3_statistics_advlabel_category}. 
We use the base-10 log scale to show the results for the metrics \textit{\#AdvLabel} and \textit{\#Category} because the results among the three techniques vary by \emph{more than one order of magnitude} at some values of $n_3$.
For instance, the \textit{\#Category} values of the test suites generated with $n_3=6000$ and $n_4=18000$ for \textsc{Clover} versus \textsc{RobOT} versus \textsc{Adapt} in case \circled{4} are 6000 versus 2104 versus 672, respectively, or 3.78 versus 3.32 versus 2.83 in the base-10 log scale, respectively.
Besides, although, visually, the difference of 0.46 between 3.78 and 3.32 in a chart looks small, the actual difference in the original number is 3896 = 6000 $-$ 2104, or 185\% (=3896/2104).

We observe that in every radar chart in the two figures except the chart for cases \circled{1} in Fig. \ref{fig: rq3_statistics_cc}, the region enclosed by the data points for \textsc{Clover} is the largest, that for \textsc{Adapt} is the smallest, and that for \textsc{RobOT} is in between the former two.
For the remaining chart (shown in Fig. \ref{fig: rq3_statistics_cc}), the enclosed region for \textsc{Clover} is still the largest: the enclosed regions in cases \circled{1} for \textsc{RobOT} is smaller than these for \textsc{Adapt}. 
We also observe that in each chart, the difference in the region size between any two pairs of techniques appears large (note that Fig. \ref{fig: rq3_statistics_advlabel_category} is in the log scale).

In terms of individual axes, \textsc{Clover} achieves a larger value than both \textsc{Adapt} and \textsc{RobOT} in each radar chart.
The results indicate that in terms of \textit{\#AdvLabel}, \textit{\#Category}, and \textit{\#CC}, \textsc{Clover} consistently outperforms \textsc{Adapt} and \textsc{RobOT}.
As a summary of these radar charts (before taking $log_{10}$), the mean \textit{\#AdvLabel}, the mean \textit{\#Category}, and the mean \textit{\#CC}, each across all combinations of the four cases (\circled{1}--\circled{4}) and $n_3 \in N_3$ are 
3250.00, 3250.00, and 0.99 for \textsc{Clover}, 
1603.44, 941.38, and 0.79 for \textsc{Adapt}, and 
2100.63, 1931.19, and 0.82 for \textsc{RobOT}, respectively.

They show that \textsc{Clover} produces test cases with much wider ranges of unique adversarial labels
and unique categories by $2.03\times$ and $3.45\times$ compared to \textsc{Adapt} and by $1.55\times$ and $1.68\times$ compared to \textsc{RobOT} across the benchmarks on average, indicating that \textsc{Clover} can produce more diverse robustness-oriented test suites than the two state-of-the-art techniques.
We observe from the data analysis that the test suites produced by \textsc{Clover} also exhibit higher mean CC values, which aligns with the higher performance in \textit{\#AdvLabel} and \textit{\#Category} achieved by \textsc{Clover} compared to the two peer techniques.

To further compare the difference in terms of \textit{\#AdvLabel}, \textit{\#Category}, and \textit{\#CC} between \textsc{Clover} and the other two peer techniques, we conduct the Wilcoxon signed-rank test and calculate Cohen's $d$ to measure the $p$-value and effect size over all combinations of Cases \circled{1} to \circled{4} and the size $n_3$ of the constructed test suites with $n_4=18000$.
The $p$-values are all $\le 1e^5$.
The effect sizes for the test between \textsc{Cloer} and \textsc{Adapt} are 
0.80, 1.24, and 2.41, for \textit{\#AdvLabel}, \textit{\#Category}, and \textit{\#CC}, repsectively.
The effect sizes for the test between \textsc{Cloer} and \textsc{RobOT} are 
0.64, 0.76, and 1.68, for \textit{\#AdvLabel}, \textit{\#Category}, and \textit{\#CC}, repsectively.
They are all at the medium, large, and huge levels, respectively, showing that the higher effectiveness of \textsc{Clover} than both \textsc{Adapt} and \textsc{RobOT} in Experiment 3 is statistically meaningful in terms of the three measurement metrics with observable differences.

\begin{table}[]
\caption{Robust Accuracy Improvement Achieved by Different DL Testing Techniques in Configuration \textit{B}}
\label{tab: rq3_fuzzing_compare_by_time}
\resizebox{\textwidth}{!}{
\begin{tabu}{|l|c|c|c|c|c|c|c|c|c|}
\hline
\multirow{3}{*}{Benchmark Case} & \multirow{3}{*}{Technique} & \multicolumn{4}{c|}{$n_4$ = 1800} & \multicolumn{4}{c|}{$n_4$ = 3600} \\ \cline{3-10}
 &  & \multicolumn{4}{c|}{$n_3$} & \multicolumn{4}{c|}{$n_3$} \\ \cline{3-10} 
 &  & 1000 & 2000 & 4000 & 6000 & 1000 & 2000 & 4000 & 6000\\ \hline 
\multirow{3}{*}{\begin{tabu}[l]{@{}c@{}}\circled{1}: FashionMnist+VGG16\end{tabu}}
 & \textsc{Adapt} & 31.72 & 35.76 & 38.36 & 40.18 & 35.68 & 38.85 & 45.90 & 46.21 \\ 
 & \textsc{RobOT} & 37.10 & 44.66 & 51.18 & 53.91 & 40.55 & 49.16 & 57.64 & 60.82 \\ 
 &\textsc{Clover} & 61.12 & 70.86 & 75.06 & 75.13 & 60.26 & 70.76 & 77.85 & 80.05 \\ \tabucline[1.2pt]{1-10} 
\multirow{3}{*}{\circled{2}: SVHN+LeNet5} 
 & \textsc{Adapt} & 19.98 & 22.22 & 25.91 & 27.62 & 20.36 & 23.43 & 25.99 & 30.50 \\ 
 & \textsc{RobOT} & 20.62 & 23.75 & 28.34 & 32.82 & 20.36 & 23.25 & 29.31 & 34.07 \\ 
 &\textsc{Clover} & 23.29 & 29.37 & 38.20 & 44.26 & 24.47 & 30.30 & 39.63 & 48.22 \\ \tabucline[1.2pt]{1-10} 
\multirow{3}{*}{\circled{3}: CIFAR10+ResNet20} 
 & \textsc{Adapt} & 16.83 & 19.95 & 23.38 & 24.90 & 18.91 & 24.22 & 28.64 & 32.38 \\ 
 & \textsc{RobOT} & 16.03 & 20.39 & 28.11 & 32.39 & 17.13 & 23.57 & 33.09 & 39.81 \\ 
 &\textsc{Clover} & 55.43 & 67.31 & 73.71 & 75.88 & 54.57 & 68.65 & 76.71 & 79.75 \\ \tabucline[1.2pt]{1-10} 
\multirow{3}{*}{\circled{4}: CIFAR100+ResNet56} 
 & \textsc{Adapt} & 11.83 & 13.18 & 13.36 & 13.80 & 13.03 & 14.12 & 15.29 & 15.81 \\ 
 & \textsc{RobOT} & 12.80 & 14.12 & 15.40 & 16.55 & 14.70 & 16.33 & 19.77 & 20.73 \\ 
 &\textsc{Clover} & 28.91 & 37.89 & 44.11 & 47.93 & 28.60 & 39.31 & 51.64 & 57.87 \\ \tabucline[1.2pt]{1-10} 
\multirow{5}{*}{\textbf{\begin{tabu}[l]{@{}l@{}}Mean  \\ Robust Accuracy \\ Improvement \end{tabu}}} 
 & \textsc{Adapt} & 20.09 & 22.78 & 25.25 & 26.63 & 22.00 & 25.16 & 28.96 & 31.23 \\ 
 & \textsc{RobOT} & 21.64 & 25.73 & 30.76 & 33.92 & 23.19 & 28.08 & 34.95 & 38.86 \\ 
 & \textbf{\textsc{Clover}} & \textbf{42.19} & \textbf{51.36} & \textbf{57.77} & \textbf{60.80} & \textbf{41.97} & \textbf{52.26} & \textbf{61.46} & \textbf{66.47} \\ \cline{2-10}
 & \textbf{\textsc{Clover}$\div$\textsc{Adapt}} & \textbf{2.10} & \textbf{2.25} & \textbf{2.29} & \textbf{2.28} & \textbf{1.91} & \textbf{2.08} & \textbf{2.12} & \textbf{2.13} \\ 
 & \textbf{\textsc{Clover}$\div$\textsc{RobOT}} & \textbf{1.95} & \textbf{2.00} & \textbf{1.88} & \textbf{1.79} & \textbf{1.81} & \textbf{1.86} & \textbf{1.76} & \textbf{1.71} \\ \hline \hline

\multirow{3}{*}{Benchmark Case} & \multirow{3}{*}{Technique} & \multicolumn{4}{c|}{$n_4$ = 7200} & \multicolumn{4}{c|}{$n_4$ = 18000} \\ \cline{3-10}
 &  & \multicolumn{4}{c|}{$n_3$} & \multicolumn{4}{c|}{$n_3$} \\ \cline{3-10} 
 &  & 1000 & 2000 & 4000 & 6000 & 1000 & 2000 & 4000 & 6000 \\ \hline
\multirow{3}{*}{\begin{tabu}[c]{@{}c@{}}\circled{1}: FashionMnist+VGG16\end{tabu}} 
 & \textsc{Adapt} & 37.10 & 44.75 & 49.70 & 51.53 & 38.85 & 49.25 & 57.36 & 60.59 \\ 
 & \textsc{RobOT} & 39.19 & 52.60 & 63.77 & 68.04 & 43.24 & 53.87 & 65.33 & 71.68 \\ 
 &\textsc{Clover} & 58.65 & 67.88 & 72.93 & 80.67 & 55.76 & 63.03 & 78.98 & 83.83 \\ \tabucline[1.2pt]{1-10} 
\multirow{3}{*}{\circled{2}: SVHN+LeNet5} 
 & \textsc{Adapt} & 20.58 & 23.18 & 28.65 & 30.70 & 19.53 & 22.56 & 27.50 & 31.34 \\ 
 & \textsc{RobOT} & 20.48 & 23.86 & 30.11 & 35.74 & 20.03 & 24.98 & 29.77 & 36.11 \\ 
 &\textsc{Clover} & 24.90 & 31.68 & 41.57 & 47.82 & 23.73 & 29.96 & 42.08 & 50.07 \\ \tabucline[1.2pt]{1-10} 
\multirow{3}{*}{\circled{3}: CIFAR10+ResNet20}
 & \textsc{Adapt} & 19.85 & 26.73 & 36.06 & 40.73 & 19.71 & 29.61 & 43.94 & 50.72 \\ 
 & \textsc{RobOT} & 17.94 & 23.03 & 35.66 & 44.19 & 18.91 & 26.12 & 37.93 & 48.21 \\ 
 &\textsc{Clover} & 54.83 & 68.29 & 77.59 & 81.31 & 54.45 & 68.99 & 77.51 & 81.22 \\ \tabucline[1.2pt]{1-10} 
\multirow{3}{*}{\circled{4}: CIFAR100+ResNet56}
 & \textsc{Adapt} & 13.96 & 15.61 & 17.84 & 18.29 & 15.12 & 17.90 & 21.48 & 24.38 \\ 
 & \textsc{RobOT} & 15.73 & 18.78 & 23.23 & 24.86 & 16.57 & 20.68 & 27.20 & 31.59 \\ 
 &\textsc{Clover} & 27.71 & 39.09 & 53.53 & 61.10 & 27.34 & 39.05 & 54.75 & 62.98 \\ \tabucline[1.2pt]{1-10} 
\multirow{5}{*}{\textbf{\begin{tabu}[l]{@{}l@{}}Mean \\ Robust Accuracy \\ Improvement \end{tabu}}}
 & \textsc{Adapt} & 22.87 & 27.57 & 33.06 & 35.32 & 23.30 & 29.83 & 37.57 & 41.76 \\ 
 & \textsc{RobOT} & 23.34 & 29.57 & 38.19 & 43.21 & 24.69 & 31.41 & 40.06 & 46.90 \\ 
 & \textbf{\textsc{Clover}} & \textbf{41.52} & \textbf{51.73} & \textbf{61.41} & \textbf{67.73} & \textbf{40.32} & \textbf{50.26} & \textbf{63.33} & \textbf{69.53} \\ \cline{2-10}
 & \textbf{\textsc{Clover}$\div$\textsc{Adapt}} & \textbf{1.82} & \textbf{1.88} & \textbf{1.86} & \textbf{1.92} & \textbf{1.73} & \textbf{1.68} & \textbf{1.69} & \textbf{1.67} \\ 
 & \textbf{\textsc{Clover}$\div$\textsc{RobOT}} & \textbf{1.78} & \textbf{1.75} & \textbf{1.61} & \textbf{1.57} & \textbf{1.63} & \textbf{1.60} & \textbf{1.58} & \textbf{1.48} \\ \hline
\end{tabu}
}
\end{table}

\paragraph{\textbf{Effectiveness in robustness improvement:}}
Table \ref{tab: rq3_fuzzing_compare_by_time} summarizes the results of Experiments 3a, 3b, and 3c in robust accuracy improvement.
It contains two sub-tables due to page size limitations. 
In each sub-table, from top to bottom, there are five sections. The first four sections are for cases \circled{1} to \circled{4}. 
The last section summarizes the statistics for the same column. 
The section for each case (\circled{1} to \circled{4}) shows the robust accuracy improvement of a technique for the combinations of $n_3$ and $n_4$ as specified by the column headings, which is the mean result of three repeated trials.
In the last section, the first three rows show the mean robust accuracy improvement achieved by \textsc{Adapt}, \textsc{RobOT}, and \textsc{Clover} in the four cases, respectively.
The next two rows show the mean robust accuracy improvement of \textsc{Clover} to that of each of \textsc{Adapt} and \textsc{RobOT} in ratio, respectively.
There are 16 combinations of $n_3$ and $n_4$ for each case in Table \ref{tab: rq3_fuzzing_compare_by_time}, so there are 64 values in total.

\textsc{Clover} achieves higher robust accuracy improvements than both \textsc{Adapt} and \textsc{RobOT} in all 64 combinations (100\%). 
Taking all 64 combinations for each technique as a whole, the mean robust accuracy improvements for \textsc{Adapt}, \textsc{RobOT} and \textsc{Clover} are 28.33\%, 32.15\%, and 55.01\%, respectively. 
In 41 out of 64 combinations (64\%), \textsc{Clover} is more effective than \textsc{Adapt} by at least 60\% in ratio.
For instance, in case \circled{1} with $n_4$=1800 and $n_3$=1000, the robust accuracy improvements of \textsc{Clover} and \textsc{Adapt} are 61.12 and 31.72, respectively. The improvement ratio is $(61.12 - 31.72)/31.72 = 92\%$.
Compared to \textsc{RobOT}, in 51 out of 64 cases (80\%), \textsc{Clover} is more effective by at least 30\%. 
Across these 64 combinations of $n_3$ and $n_4$, \textsc{Clover} is more effective than \textsc{Adapt} and \textsc{RobOT} up to 266\% and 246\%, respectively. 
The largest differences between \textsc{Clover} and each of \textsc{Adapt} and \textsc{RobOT} appear in case \circled{4} with $n_4$=3600 and $n_3$=6000 and case \circled{3} with $n_4$=1800 and $n_3$=1000, respectively. 
Across the 16 combinations of $n_4$ and $n_3$, in terms of mean robust accuracy improvement ratio, on average, \textsc{Clover} is more effective than \textsc{Adapt} and \textsc{RobOT} by 67\% to 129\% and 48\% to 100\%, respectively. 

The table also shows that if $n_3$ is relatively small (e.g., $n_3$ = 1000), increasing the time budget $n_4$ from 1800 to 18000 has little (and sometimes negative) effect on the robustness improvement on these four models achieved by \textsc{Clover}. 
On the other hand, when $n_3$ is relatively large (e.g., $n_3$ = 6000), the robustness improvement becomes more obvious and always positive by increasing the fuzzing budget.
The result shows that using a longer time budget (i.e., a higher cost) does not warrant getting a higher robustness improvement (i.e., a higher effectiveness) if the constructed test suite could be small.
However, suppose the constructed test suite size is affordable to be relatively large. In that case, spending more time on fuzzing the DL models can generally make the pipeline produce retrained DL models with higher robustness improvement.

\begin{figure}[t]
\begin{minipage}[c]{\textwidth}
  \centering
  \includegraphics[width=\linewidth]{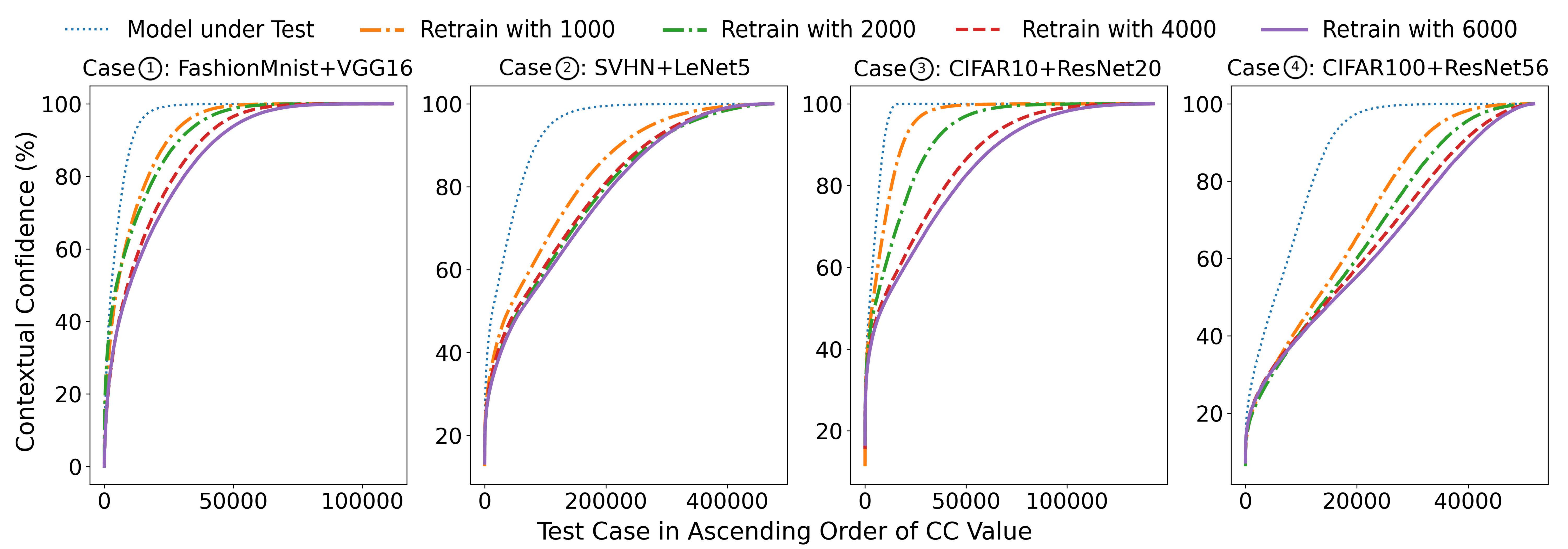}
\end{minipage}
\caption{CC Values of Test Cases Generated by \textsc{Clover} in Configuration \textit{B}}
\label{fig: rq3_contextual_diff}
\end{figure}

We compute the CC value of each test case generated by \textsc{Clover} with the fuzzing budget of 18000 seconds on each pair of the model under test and the respective retrained model.
Fig. \ref{fig: rq3_contextual_diff} shows the CC values of all generated test cases corresponding to models retrained with different test suite sizes (containing 1000 to 6000 test cases, indicated by the legends ``Retrain with 1000'' to ``Retrain with 6000'', respectively), where the $x$-axis shows the test cases in the corresponding test suite sorted in the ascending order of CC value. The $y$-axis is the CC value of the corresponding test case.

We observe a clear and large reduction in CC achieved by \textsc{Clover} in each plot in the figure.
We conduct the Mann-Whitney U test and calculate Cohen's $d$ to measure the $p$-value and effect size between each such pair of CC value sequences achieved by the same test suite.
The $p$-values are all smaller than $1e^{-5}$, indicating they are all significantly different at the 5\% significance level with Bonferroni correction.
The effect sizes corresponding to the legends of the figure from left to right are: case \circled{1}: 0.29, 0.35, 0.53, 0.62; case \circled{2}: 0.63, 0.85, 0.80, 0.87; case \circled{3}: 0.21, 0.46, 0.81, 0.93; and case \circled{4}: 0.62, 0.77, 0.90, 0.98, respectively.
The effect sizes in 12 out of these 16 cases (75\%) are medium or large with four small effect sizes (in case \circled{1} and \circled{3} for $n_3 = 1000$ and $n_3 = 2000$).
The result shows that the reduction achieved by \textsc{Clover} is statistically meaningful.

For each such set of test cases, we further compute the increment of robust accuracy improvement (denoted by $I_{\textit{acc}}$) and the decrement of mean CC (denoted by $D_{cc}$) achieved on this set of test cases after retraining the model under test with the constructed test suite. 
In each case, we measure the Spearman's correlation coefficient \cite{spearman_correlation} between $I_{\textit{acc}}$ and $D_{cc}$ on the set of all test cases generated by \textsc{Clover}.
The results for cases \circled{1} to \circled{4} are 1.00, 0.80, 1.00, and 1.00, respectively. 
They are all strong correlations \cite{correlation_measure}.

We recall that FGSM/PGD and two fuzzing techniques were used to generate the selection universes in Experiment 1. In contrast, the test suites in Experiment 3 are generated by \textsc{Clover} in  Configuration $B$.
We compare each of \textsc{Random} and \textsc{Clover} in Experiment 1 with \textsc{Clover} in Experiment 3 to study the effect of using the fuzzing and selection components of \textsc{Clover} and that of using only the fuzzing component of \textsc{Clover}, respectively.

We also recall that the mean robust accuracy improvements of \textsc{Random} and \textsc{Clover} in Experiment 1 are 34.88\% and 41.09\%, respectively, and that of \textsc{Clover} in Experiment 3 is 55.01\%, where the differences are large.
To further compare the robust accuracy improvement between \textsc{Clover} in Experiments 1 and 3 and between \textsc{Random} in Experiment 1 and \textsc{Clover} in Experiment 3, we conduct the Wilcoxon signed-rank test (paired by test suite size) and calculate Cohen's $d$ to measure the $p$-value and effect size over all combinations of Cases \circled{1} to \circled{4} and the size $n_3$ of the constructed test suites.
The $p$-values are both $\le 1e^5$, and the effect sizes are large (i.e., 0.87 and 0.79 in the two comparisons, respectively).
The result shows that the higher effectiveness of \textsc{Clover} in Experiment 3 than both \textsc{Random} and \textsc{Clover} in Experiment 1 is statistically meaningful.

\begin{tcolorbox}[
enhanced, breakable,
attach boxed title to top left = {yshift = -2mm, xshift = 5mm},
boxed title style = {sharp corners},
colback = white, 
title={Answering RQ3}
]
\textsc{Clover} outperforms the current state-of-the-art coverage-based and loss-based techniques (\textsc{Adapt} and \textsc{RobOT}) in generating more diverse test suites by $2.03\times$ and $1.55\times$ regarding the number of unique adversarial labels, and by $3.45\times$ and $1.68\times$ regarding the number of unique categories, respectively.
\textsc{Clover} achieves higher mean CC values than \textsc{Adapt} and \textsc{RobOT} by 26\% and 20\%, respectively.
 It also outperforms \textsc{Adapt} and \textsc{RobOT} in generating test suites for robust accuracy improvements by 67\%--129\% and 48\%--100\%, respectively.
In addition, \textsc{Clover} achieves a strong correlation of 0.80--1.00 between the increment of robust accuracy improvement and the decrement of mean CC of its generated test cases in Spearman's correlation coefficient. 
\end{tcolorbox}

\subsection{Answering RQ4 (Effects of \textsc{Clover} Variants)}
\begin{figure}[t]
  \centering
  \includegraphics[width=0.35\textwidth]{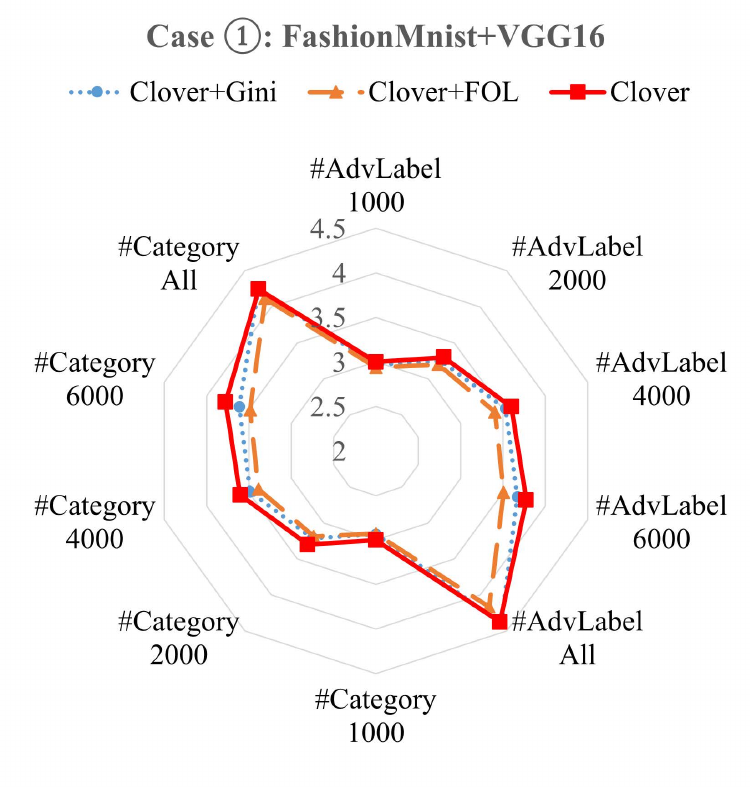}
  \includegraphics[width=0.35\textwidth]{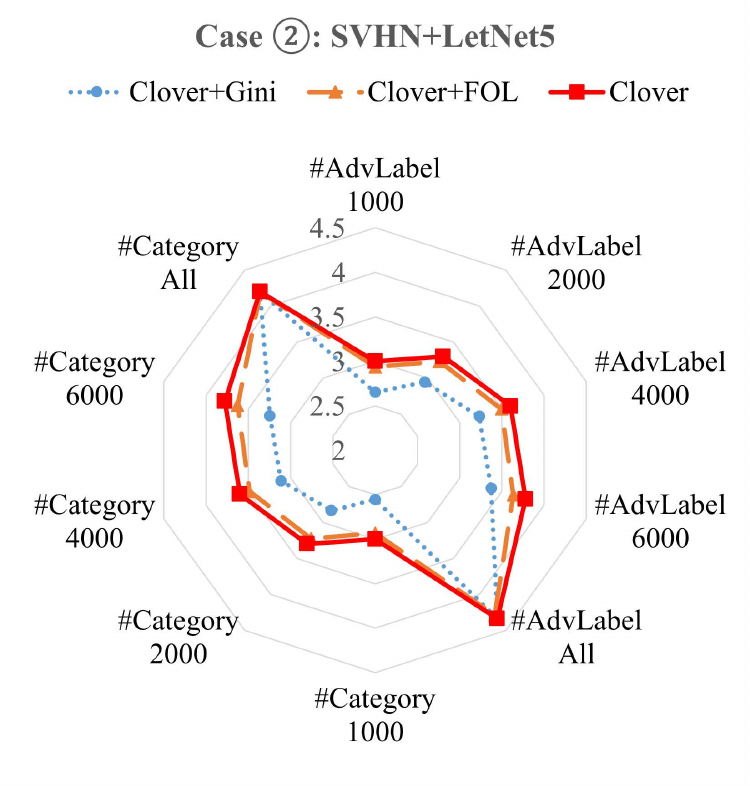} \\
  \includegraphics[width=0.35\textwidth]{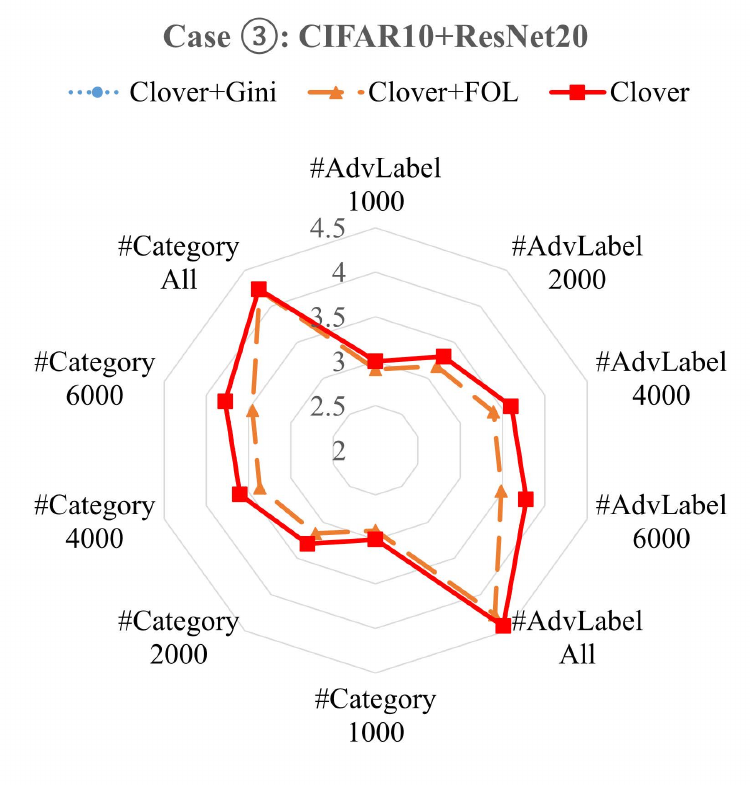}
  \includegraphics[width=0.35\textwidth]{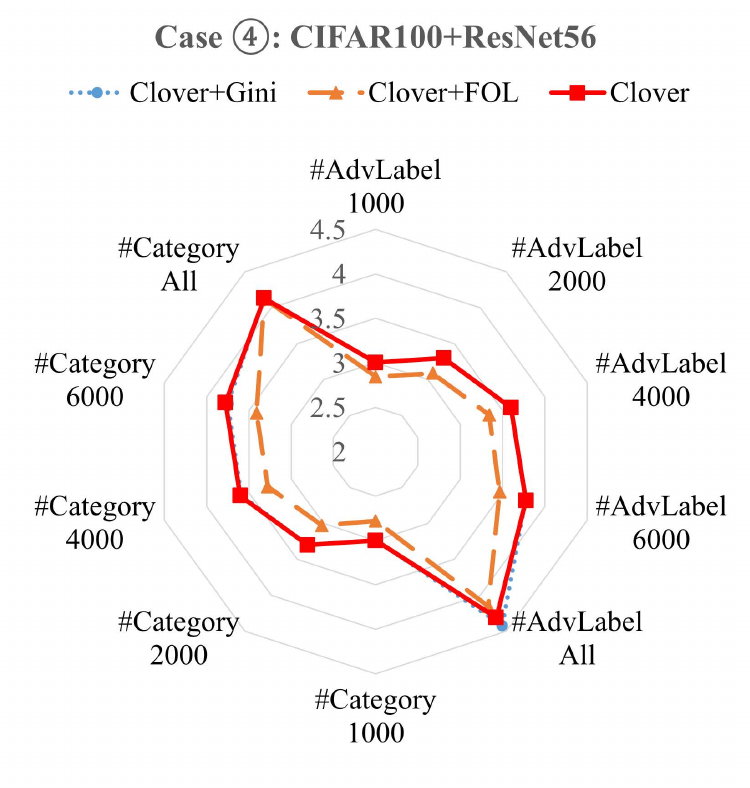} \\
  \caption{\textit{\#AdvLabel} and \textit{\#Category} for \textsc{Clover} and its Variants (in $log_{10}$ scale)}
\label{fig: rq4_statistics_advlabel_category}
\end{figure}

\begin{figure}[t]
  \centering
  \includegraphics[width=0.24\textwidth]{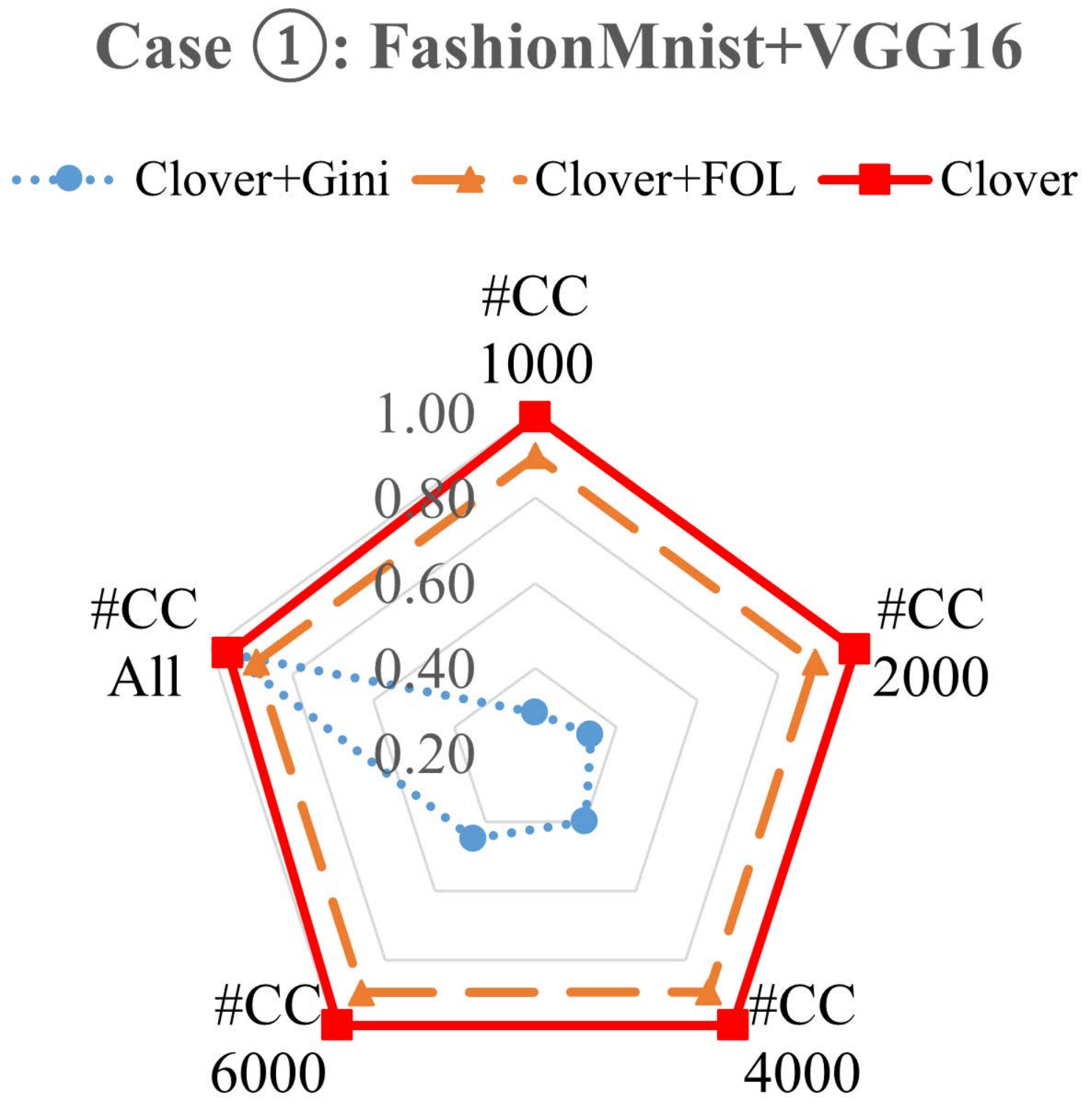}
  \includegraphics[width=0.24\textwidth]{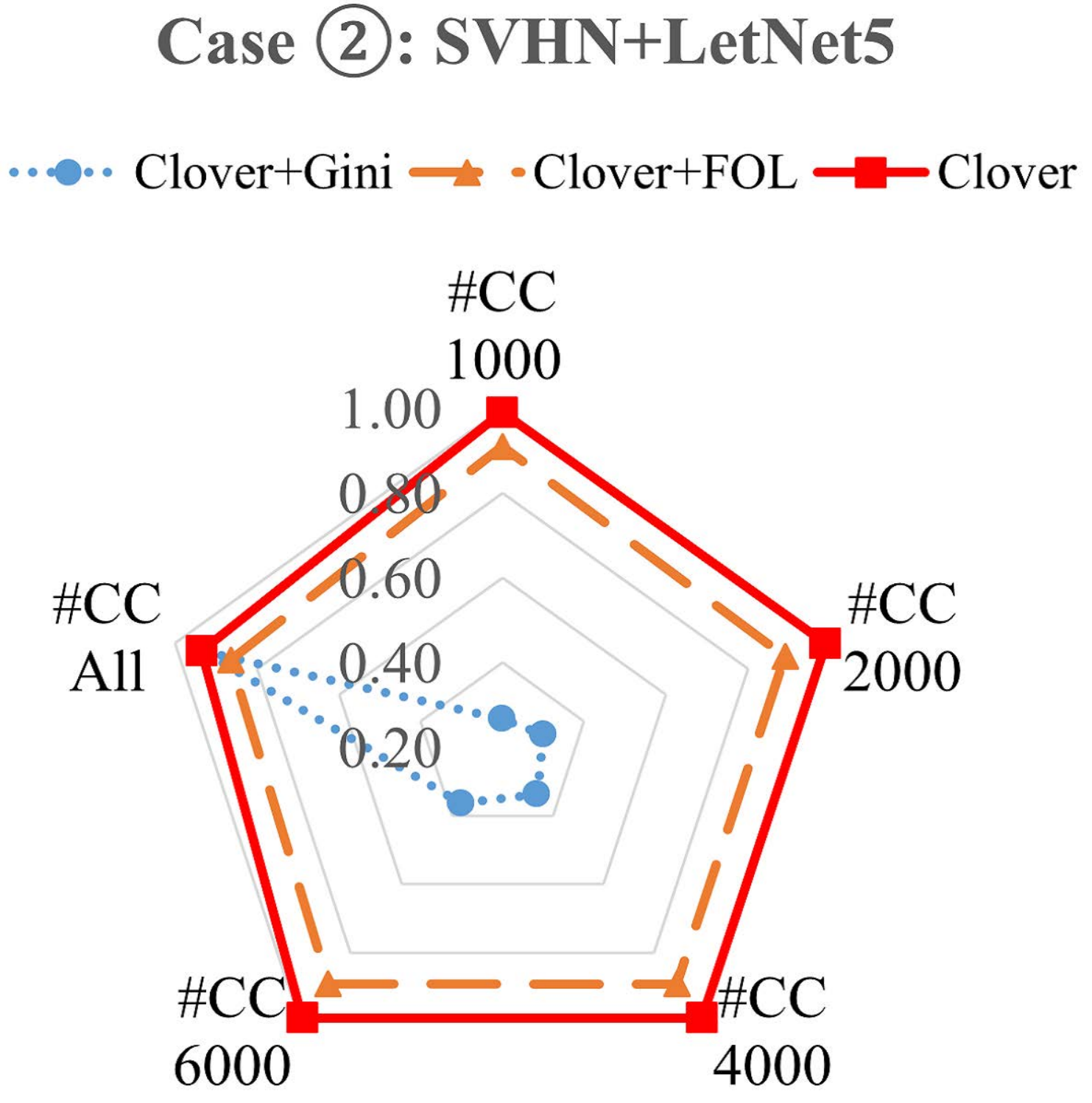}
  \includegraphics[width=0.24\textwidth]{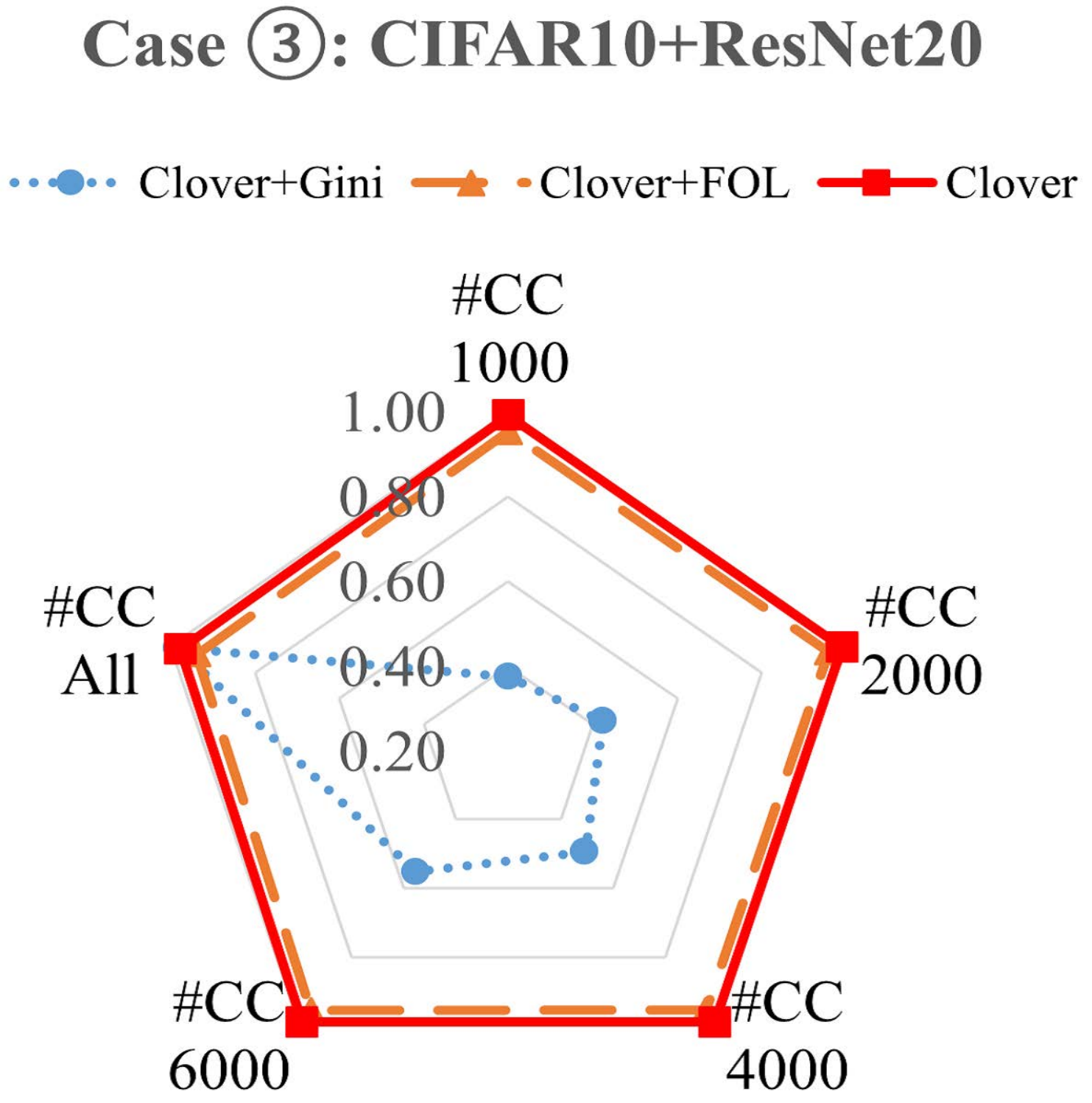}
  \includegraphics[width=0.24\textwidth]{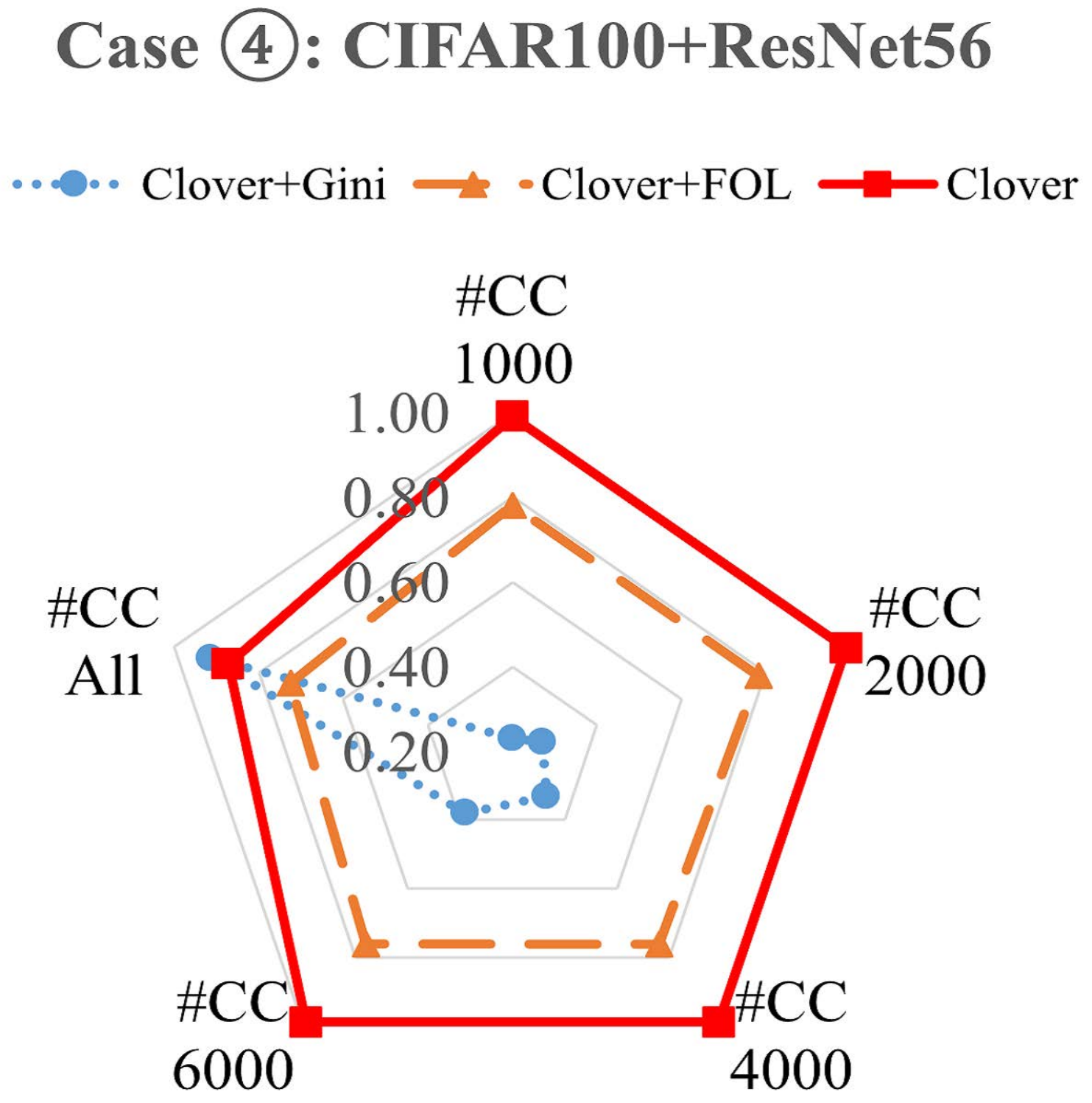} \\
  \caption{\textit{\#CC} for \textsc{Clover} and its Variants}
\label{fig: rq4_statistics_cc}
\end{figure}

\paragraph{\textbf{Effectiveness in test case generation:}}
Figures \ref{fig: rq4_statistics_advlabel_category} and \ref{fig: rq4_statistics_cc} summarize the results of Experiment 4a and 4b measured in \textit{\#AdvLabel}, \textit{\#Category}, and \textit{\#CC} for \textsc{Clover} and its two variants, i.e., \textsc{Clover+Gini}, and \textsc{Clover+FOL}.
Readers can interpret the axes of the charts in these two figures like these in Figures \ref{fig: rq3_statistics_advlabel_category} and \ref{fig: rq3_statistics_cc}, respectively.
We also copy \textsc{Clover}'s results from Figures \ref{fig: rq3_statistics_advlabel_category} and \ref{fig: rq3_statistics_cc} to ease the comparison between Experiments 3 and 4.

We observe that the enclosed regions for \textsc{Clover} are either the largest or almost completely overlapping with the largest regions in the two charts for cases \circled{3} and \circled{4} in Fig. \ref{fig: rq4_statistics_advlabel_category}.
The regions enclosed by the data points for \textsc{Clover+FOL} are always the smallest in Fig. \ref{fig: rq4_statistics_advlabel_category}, and smaller than or close to these for \textsc{Clover} in Fig. \ref{fig: rq4_statistics_cc}.
Moreover, in Fig. \ref{fig: rq4_statistics_cc}, the regions for \textsc{Clover+Gini} are much smaller than those for the other two techniques.
The axes for $n_3$=\textit{All} in the four charts in Fig. \ref{fig: rq4_statistics_cc} further show that \textsc{Clover+Gini} achieves similar \textit{\#CC} values compared to \textsc{Clover} or larger than \textsc{Clover} a bit, but its values are much smaller than \textsc{Clover} with $n_3 \in \{1000, 2000, 4000, 6000\}$.
The result indicates that the test case prioritization metric Gini is effective in guiding test case generation with higher CC values but quite ineffective in selecting those with higher CC values, which is consistent with the results of robust accuracy improvement we observed in Experiment 1 (see Section \ref{subsection: answering_rq1}).

As a summary of the results presented in these radar charts (before taking a $log$ operation), the mean \textit{\#AdvLabel}, the mean \textit{\#Category}, and the mean \textit{\#CC}, each across all combinations of cases \circled{1}--\circled{4} and $n_3 {\in} N_3$ are 
3250.00, 3250.00, and 0.99 for \textsc{Clover}, 
2630.44, 2449.88, and 0.36 for \textsc{Clover+Gini}, and 
2069.44, 1954.88, and 0.88 for \textsc{Clover+FOL}, respectively.
\textsc{Clover} produces test suites with wider ranges of unique adversarial labels and unique categories by $1.24\times$ and $1.33\times$ compared to \textsc{Clover+Gini} and by $1.57\times$ and $1.66\times$ compared to \textsc{Clover+FOL} across the benchmarks on average.
Comparing the results of \textsc{Clover}'s two variants with the results of \textsc{Adapt} and \textsc{RobOT} presented in answering RQ3 (see Section \ref{sec: rq3_result}), both \textsc{Clover+Gini} and \textsc{Clover+FOL}  outperform \textsc{Adapt} and \textsc{RobOT} in terms of \textit{\#AdvLabel},\textit{\#Category}, and \textit{\#CC} except \textit{\#CC} for \textsc{Clover+Gini}.

\begin{table}[t]
\caption{Robust Accuracy Improvements Achieved by \textsc{Clover} Variants in Configuration \textit{B}}
\label{tab: rq4_robustness_improvement}
\resizebox{0.79\textwidth}{!}{
\begin{tabu}{|l|c|c|c|c|c|}
\hline
\multirow{3}{*}{Benchmark Case} & \multirow{3}{*}{Technique} & \multicolumn{4}{c|}{$n_4$ = 18000} \\ \cline{3-6}
 &  & \multicolumn{4}{c|}{$n_3$} \\ \cline{3-6} 
 &  & 1000 & 2000 & 4000 & 6000 \\ \hline
\multirow{3}{*}{\begin{tabular}[c]{@{}c@{}}\circled{1}: FashionMnist+VGG16\end{tabular}} 
 &\textsc{Clover} & 55.76 & 63.03 & 78.98 & 83.83 \\
 &\textsc{Clover+Gini} & 37.01 & 43.27 & 50.79 & 56.73 \\
 &\textsc{Clover+FOL} & 49.35 & 59.03 & 73.09 & 80.05 \\ \tabucline[1.2pt]{1-6} 
\multirow{3}{*}{\begin{tabular}[c]{@{}c@{}}\circled{2}: SVHN+LeNet5\end{tabular}} 
 &\textsc{Clover} & 23.73 & 29.96 & 42.08 & 50.07 \\
 &\textsc{Clover+Gini} & 17.67 & 20.23 & 24.68 & 28.03 \\
 &\textsc{Clover+FOL} & 20.64 & 25.28 & 31.51 & 36.37 \\ \tabucline[1.2pt]{1-6} 
\multirow{3}{*}{\begin{tabular}[c]{@{}c@{}}\circled{3}: CIFAR10+ResNet20\end{tabular}}
 &\textsc{Clover} & 54.45 & 68.99 & 77.51 & 81.22 \\
 &\textsc{Clover+Gini} & 16.62 & 21.32 & 33.68 & 46.24 \\
 &\textsc{Clover+FOL} & 39.77 & 54.33 & 68.37 & 72.83 \\ \tabucline[1.2pt]{1-6} 
\multirow{3}{*}{\begin{tabular}[c]{@{}c@{}}\circled{4}: CIFAR100+ResNet56\end{tabular}} 
 &\textsc{Clover} & 27.34 & 39.05 & 54.75 & 62.98 \\
 &\textsc{Clover+Gini} & 16.12 & 24.68 & 34.85 & 41.95 \\
 &\textsc{Clover+FOL} & 20.25 & 26.55 & 35.30 & 40.76 \\ \tabucline[1.2pt]{1-6} 

\multirow{5}{*}{\textbf{\begin{tabular}[l]{@{}l@{}} Mean\\ Robust Accuracy\\ Improvement \end{tabular}}} & 
\textsc{Clover} & 40.32 & 50.26 & 63.33 & 69.53 \\ &
\textsc{Clover+Gini} & 21.86 & 27.38 & 36.00 & 43.24 \\ &
\textsc{Clover+FOL} & 32.50 & 41.30 & 52.07 & 57.50 \\ \cline{2-6} & 
\textsc{Clover} $\div$ (\textsc{Clover+Gini}) & 1.84 & 1.84 & 1.76 & 1.61 \\ & 
\textsc{Clover} $\div$ (\textsc{Clover+FOL}) & 1.24 & 1.22 & 1.22 & 1.21 \\ \hline
\end{tabu}
}
\end{table}

\paragraph{\textbf{Effectiveness in robustness improvement:}}
Table \ref{tab: rq4_robustness_improvement} summarizes the results of Experiments 4a and 4b in robust accuracy improvement.
In the table, there are five sections from top to bottom.
The first four sections are for cases \circled{1} to \circled{4}, and the last section summarizes the statistics for the same column. 
For cases \circled{1} to \circled{4}, the corresponding section presents the robust accuracy improvement of a technique in the experimental setting of $n_3$ (values in $N_3$) specified by the column heading with the time budget of 18000 seconds.
In the last section, the first three rows show the mean robust accuracy improvement achieved by the original \textsc{Clover}, \textsc{Clover+Gini}, and \textsc{Clover+FOL} in the four cases, respectively. 
The next two rows show the mean robust accuracy improvement of \textsc{Clover} compared to each of \textsc{Clover+Gini} and \textsc{Clover+FOL} in ratio.

From cases \circled{1} to \circled{4}, we observe the original \textsc{Clover} achieves higher robust accuracy improvements than the two variants \textsc{Clover+Gini} and \textsc{Clover+FOL} in all 16 combinations consistently.
Across all four cases, the mean robust accuracy improvements for the original \textsc{Clover}, \textsc{Clover+Gini}, and \textsc{Clover+FOL} over the 16 combinations are 55.86\%, 32.12\%, and 45.84\%, respectively.
Across all values of $n_3$, in terms of mean robust accuracy improvement ratio, the original \textsc{Clover} is significantly more effective than \textsc{Clover+Gini} by 61\%--84\%, which is consistent with the large difference of \textit{\#CC} value between \textsc{Clover} and \textsc{Clover+Gini} in Fig. \ref{fig: rq4_statistics_cc} for answering RQ4.

For each case from \circled{1} to \circled{4}, \textsc{Clover+FOL} in Table \ref{tab: rq4_robustness_improvement} and \textsc{RobOT} in Table \ref{tab: rq3_fuzzing_compare_by_time} with $n_4$=18000 are the results with the same fuzzing time budget of 18000 seconds.
The mean differences in robust accuracy improvement between them (computed as \textsc{Clover+FOL} minus \textsc{RobOT}) are 7.81\%, 9.89\%, 12.01\%, and 10.60\% for the four cases.
They are all positive, and their underlying values before taking the average are also all positive. The result shows that the algorithm of \textsc{Clover} is more effective than that of \textsc{RobOT} when using the same FOL metric to guide their fuzzing processes.

Moreover, if the guiding metric is changed from FOL to CC, \textsc{Clover} further enlarges the difference.
Specifically, across all values of $n_3$, in terms of mean robust accuracy improvement ratio, the original \textsc{Clover} is more effective than \textsc{Clover+FOL} and \textsc{RobOT}, by 21\%--24\% and 48\%--63\%, respectively.

\begin{tcolorbox}[
enhanced, breakable,
attach boxed title to top left = {yshift = -2mm, xshift = 5mm},
boxed title style = {sharp corners},
colback = white, 
title={Answering RQ4}
]
\textsc{Clover} in Configuration \textit{B} is more effective in both test case generation and robustness improvement than its variant configured with Gini or FOL as the guiding metric. 
The two variants of \textsc{Clover} are also more effective than \textsc{Adapt} and \textsc{RobOT}.
\end{tcolorbox}

\section{Results and Data Analysis for Key Design Decision in \textsc{Clover}}
This section reports the data analysis for answering RQ5.

\begin{table}[t]
\caption{Robust Accuracy Improvements Achieved by \textsc{Clover} and Its Variants}
\label{tab: rq5_smallest_single_dir}
\resizebox{0.79\textwidth}{!}{
\begin{tabu}{|l|c|c|c|c|c|}
\hline
\multirow{3}{*}{Benchmark Case} & \multirow{3}{*}{Technique} & \multicolumn{4}{c|}{$n_4$ = 18000} \\ \cline{3-6}
 &  & \multicolumn{4}{c|}{$n_3$} \\ \cline{3-6} 
 &  & 1000 & 2000 & 4000 & 6000 \\ \hline
\multirow{3}{*}{\begin{tabu}[c]{@{}c@{}}\circled{1}: FashionMnist+VGG16\end{tabu}} 
 &\textsc{Clover} & 55.76 & 63.03 & 78.98 & 83.83 \\
 &\textsc{Clover+Smallest} & 41.81 & 47.37 & 57.63 & 64.80 \\
 &\textsc{Clover+SingleDir} & 52.60 & 60.27 & 71.24 & 75.52 \\
 \tabucline[1.2pt]{1-6} 
\multirow{3}{*}{\circled{2}: SVHN+LeNet5} 
 &\textsc{Clover} & 23.73 & 29.96 & 42.08 & 50.07 \\
 &\textsc{Clover+Smallest} & 19.81 & 24.19 & 28.07 & 33.77 \\
 &\textsc{Clover+SingleDir} & 22.47 & 26.38 & 33.21 & 38.87 \\
 \tabucline[1.2pt]{1-6} 
\multirow{3}{*}{\circled{3}: CIFAR10+ResNet20}
 &\textsc{Clover} & 54.45 & 68.99 & 77.51 & 81.22 \\
 &\textsc{Clover+Smallest} & 17.18 & 22.04 & 32.86 & 45.52 \\
 &\textsc{Clover+SingleDir} & 52.93 & 64.90 & 75.86 & 78.93 \\
 \tabucline[1.2pt]{1-6} 
\multirow{3}{*}{\circled{4}: CIFAR100+ResNet56}
 &\textsc{Clover} & 27.34 & 39.05 & 54.75 & 62.98 \\
 &\textsc{Clover+Smallest} & 16.64 & 23.68 & 34.10 & 44.93 \\
 &\textsc{Clover+SingleDir} & 24.56 & 35.06 & 48.85 & 59.75 \\
 \tabucline[1.2pt]{1-6} 
 
\multirow{7}{*}{\textbf{\begin{tabu}[l]{@{}l@{}}Mean \\ Robust Accuracy \\ Improvement \end{tabu}}}
 &\textsc{Clover} & 40.32 & 50.26 & 63.33 & 69.53 \\
 &\textsc{Clover+Smallest} & 23.86 & 29.32 & 38.17 & 47.26 \\
 &\textsc{Clover+SingleDir} & 38.14 & 46.65 & 57.29 & 63.27 \\ \cline{2-6} 
 & \begin{tabu}[l]{@{}l@{}} \textbf{((\textsc{Clover+Smallest})} \\ - \textbf{\textsc{Clover})$\div$Clover} \end{tabu} & $-\textbf{41\%}$ & $-\textbf{42\%}$ & $-\textbf{40\%}$ & $-\textbf{32\%}$ \\ 
 & \begin{tabu}[l]{@{}l@{}} \textbf{((\textsc{Clover+SingleDir})} \\ - \textbf{\textsc{Clover})$\div$\textsc{Clover}} \end{tabu} & $-\textbf{5\%}$ & $-\textbf{7\%}$ & $-\textbf{10\%}$ & $-\textbf{9\%}$ \\ \hline
\end{tabu}
}
\end{table}

\subsection{Major Design Decisions}
\label{sec:data-analysis-small}
Table \ref{tab: rq5_smallest_single_dir} summarizes the results of Experiments 5a--5b. 
We copy \textsc{Clover}'s results from Table \ref{tab: rq3_fuzzing_compare_by_time} to this table to ease the comparison.
It has five sections from top to bottom, one section for a case in cases \circled{1} to \circled{4}, and the last section for the comparison statistics. 
The first four sections show the robust accuracy improvements achieved by \textsc{Clover}, \textsc{Clover+Smallest}, and {\textsc{Clover+SingleDir}}.
There are 4 combinations of $n_3$ and $n_4$ for each case and 16 combinations in total. 
Like the previous two sections, the robustness improvement refers to the robustness improvement exhibited by the retrained model output by the testing-retraining pipeline configured with \textsc{Clover} or its variants.

\subsubsection{The smaller CC, the more effective?}
In all 16 combinations of $n_3$ and $n_4$ for cases \circled{1} to \circled{4} presented in Table \ref{tab: rq5_smallest_single_dir}, \textsc{Clover+Smallest} achieves smaller robust accuracy improvements compared with \textsc{Clover}.
Their average robust accuracy improvements are 55.86\% and 34.65\%, respectively. 
The difference is large, and the direction of change is consistent.

We measure the reduction ratio in robust accuracy improvement for each combination of $n_3$ and $n_4$ for \textsc{Clover+Smallest} compared to \textsc{Clover}. 
For example, in case \circled{1} with $n_3$ = 1000 and $n_4$ = 18000, ratio is (55.76 $-$ 41.81)/55.76 = 25\%.
For cases \circled{1} to \circled{4}, the ranges of the reduction ratios are 23\%--27\%, 17\%--33\%, 44\%--68\%, and 29\%--39\%, respectively, with an average of 39\%.

If designing \textsc{Clover} to use a smaller CC value for selecting test cases, the above result shows that the reduction ratio of mean robust accuracy improvement is consistent, large, and observable, indicating that configuring \textsc{Clover} to favor higher CC values when selecting test cases is a more effective strategy.
Compared to the results in Table \ref{tab: rq3_fuzzing_compare_by_time}, we note that \textsc{Clover+Smallest} still achieves 69\% and 25\% cases with higher robust accuracy improvement than \textsc{Adapt} and \textsc{RobOT}, respectively.

Recall from the main result of Experiment 2 that test suites containing test cases with higher CC values are more effective than those with lower CC values.
The two experiments (Experiment 2 and Experiment 5a) consistently show that configuring \textsc{Clover} to prefer test cases with higher CC values in test suite construction produces greater robust accuracy improvements than configuring it to prefer test cases with lower CC values.

\subsubsection{More effective to fuzz each seed using a single direction?}
\label{sec:data-analysis-direction}
In all 16 combinations in Table \ref{tab: rq5_smallest_single_dir}, \textsc{Clover} outperforms \textsc{Clover+SingleDir} in the robust accuracy improvement.
The average robust accuracy improvement of \textsc{Clover+SingleDir} is 51.34\%, and the average reduction ratio in robust accuracy improvement is 8\%.
From cases \circled{1} to \circled{4}, the ranges of the reduction ratios are 4\%--10\%, 5\%--22\%, 2\%--6\%, and 5\%--11\%, respectively.

After factoring out the perturbations due to adding $\beta$-AFOs to each seed in line 4 of Algorithm \ref{alg: context_translate}, the drops in mean robust accuracy improvement are observable and consistent in direction.
It shows that configuring \textsc{Clover} to fuzz seeds with $\beta$-AFOs is more effective.

In the table, the robust accuracy improvement of \textsc{Clover+SingleDir} is generally higher than that of \textsc{Clover+Smallest}. In some extreme cases, such as $n_3 = 4000$ and $n_4 = 18000$ on case \circled{3}, the difference is 43.00 $(= 75.86 - 32.86)$. 
The result shows that the effect of preferring smaller CC exerts a larger effect on \textsc{Clover} to push down the effectiveness of the generated test suites.

\subsection{Hyperparameters to Control Fuzzing Attempts}
\label{sec:data-analysis-hyperparameters}

\begin{figure} [t]
\centering
    \includegraphics[width=0.55\linewidth]{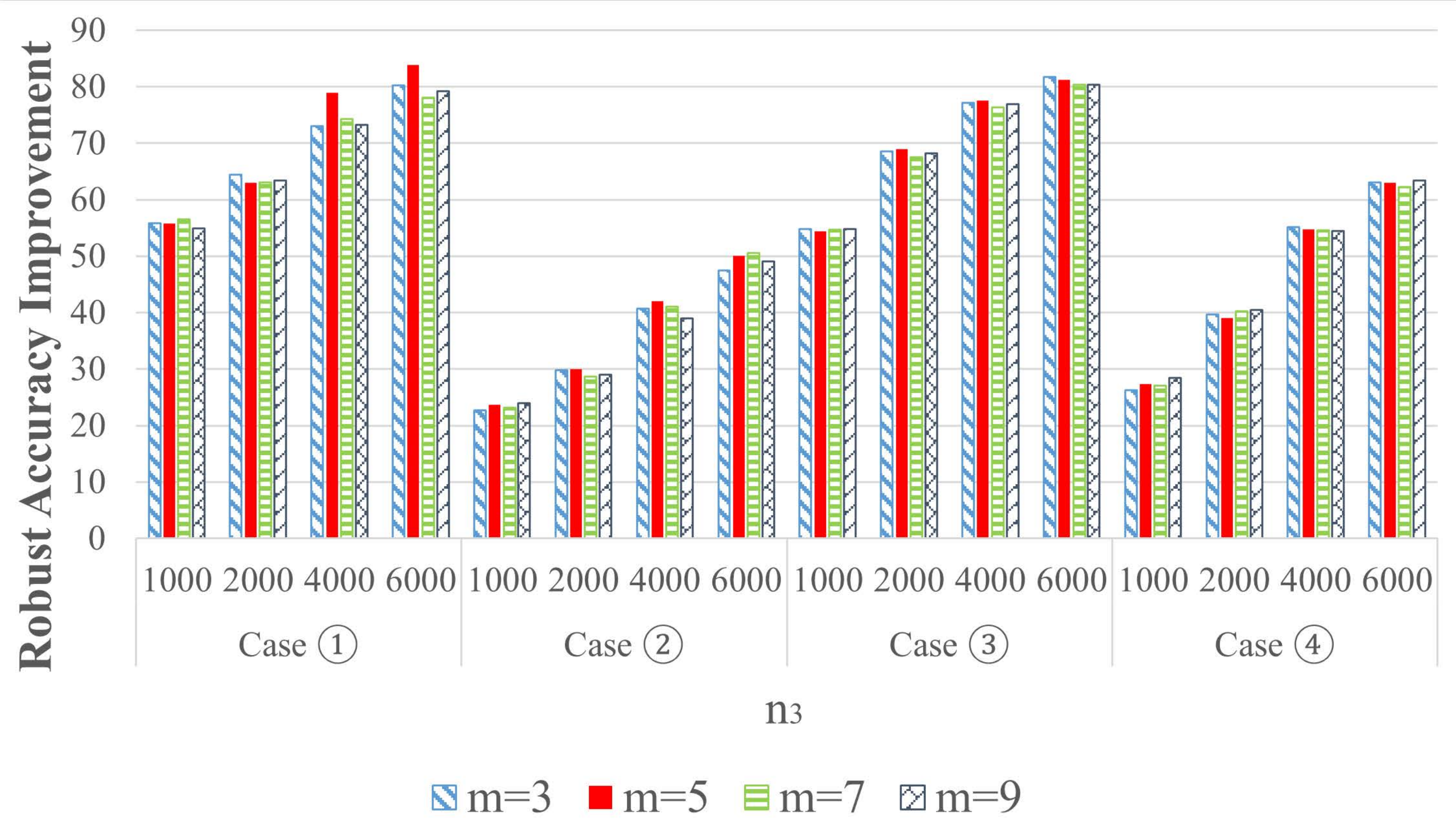}
    \caption{Robust Accuracy Improvement Achieved by \textsc{Clover} Configed with Various Values of $m$}
\label{fig: rq5_various_m}
\end{figure}

\begin{figure} [t]
  \centering
  \includegraphics[width=0.55\linewidth]{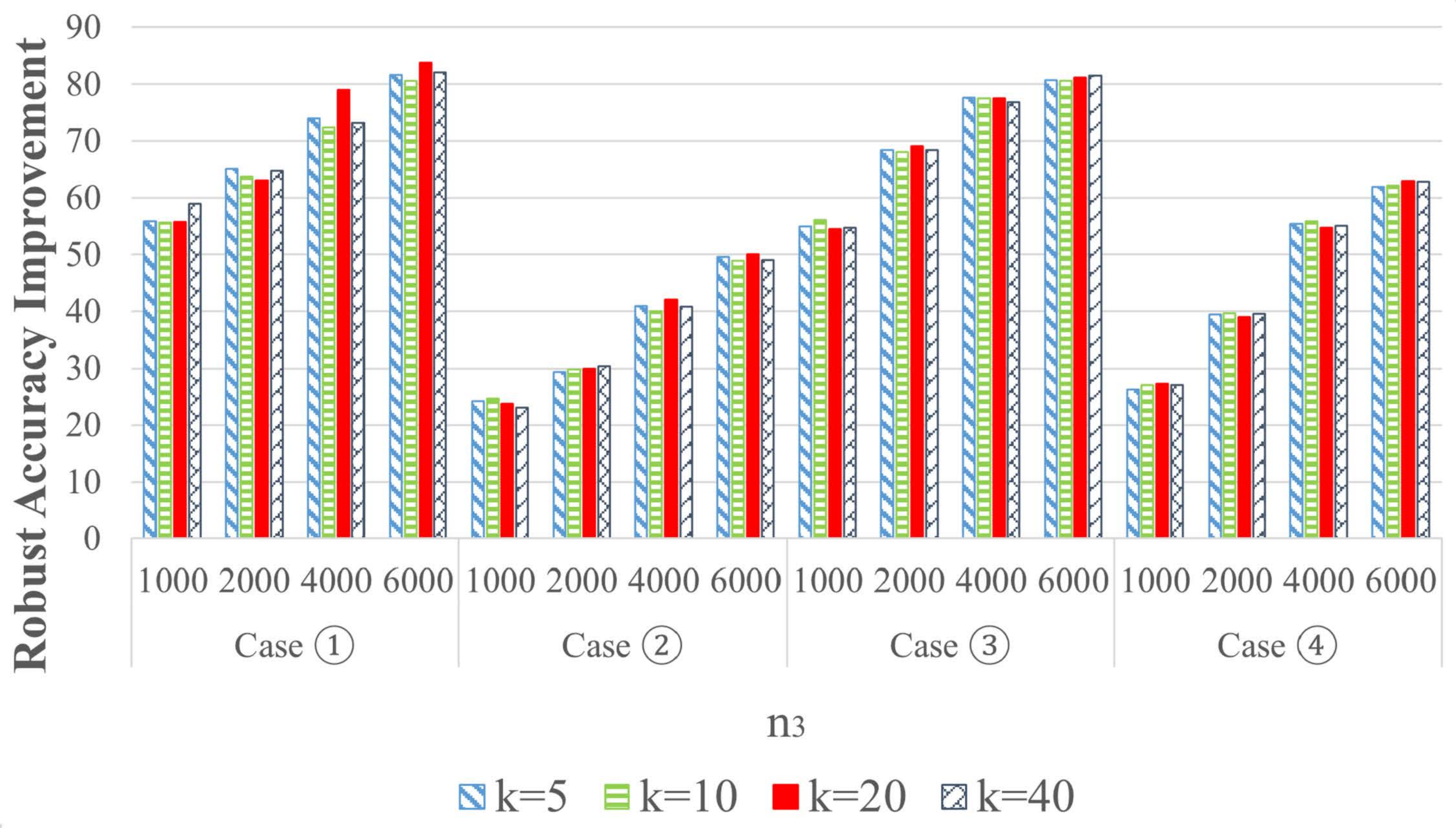}
\caption{Robust Accuracy Improvement Achieved by \textsc{Clover} Configed with Various Values of $k$}
\label{fig: rq5_various_k}
\end{figure}

\begin{figure} [t]
  \centering
  \includegraphics[width=0.55\linewidth]{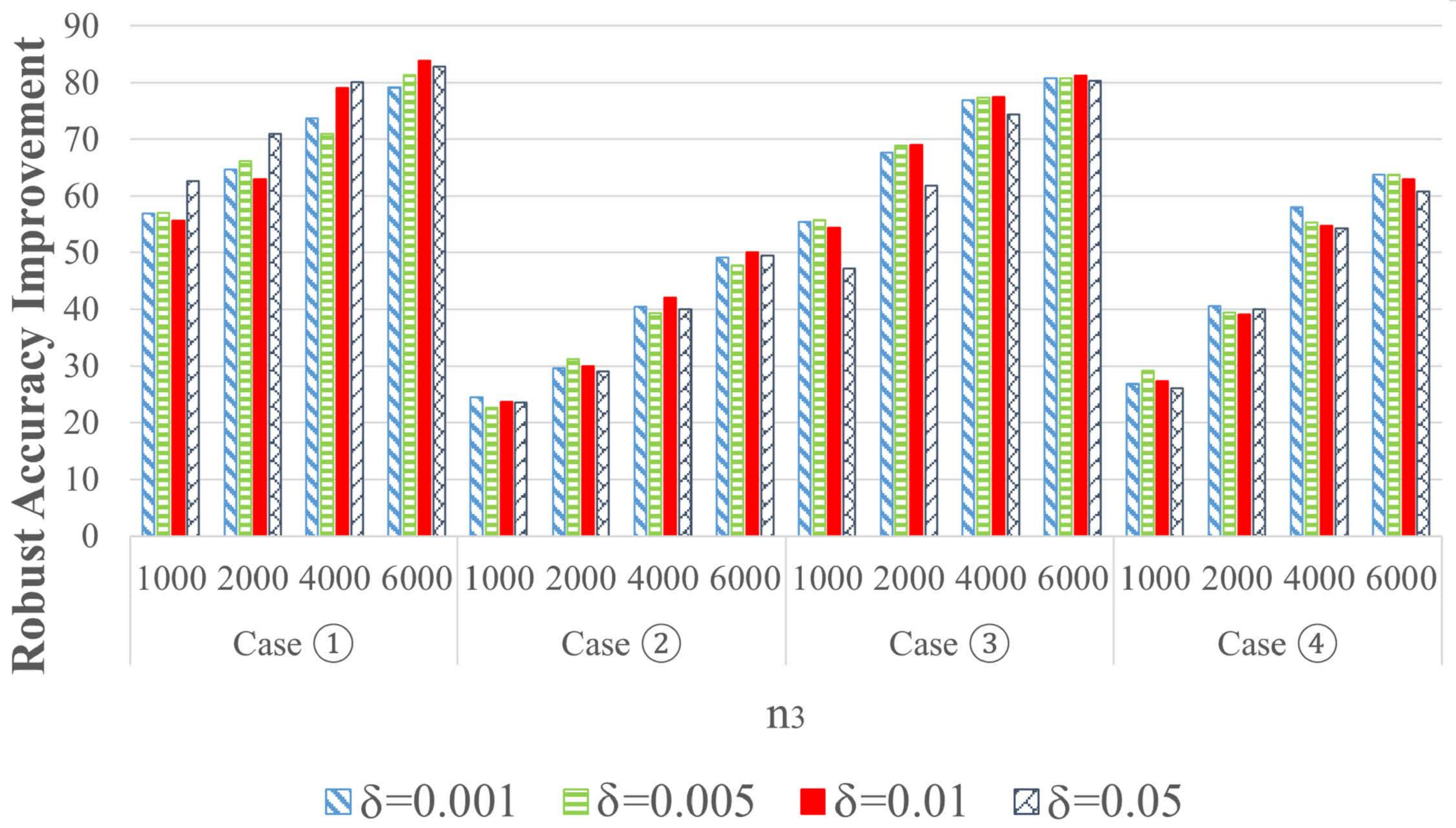}
\caption{Robust Accuracy Improvement Achieved by \textsc{Clover} Configed with Various Values of $\delta$}
\label{fig: rq5_various_delta}
\end{figure}

\begin{figure} [t]
  \centering
  \includegraphics[width=0.55\linewidth]{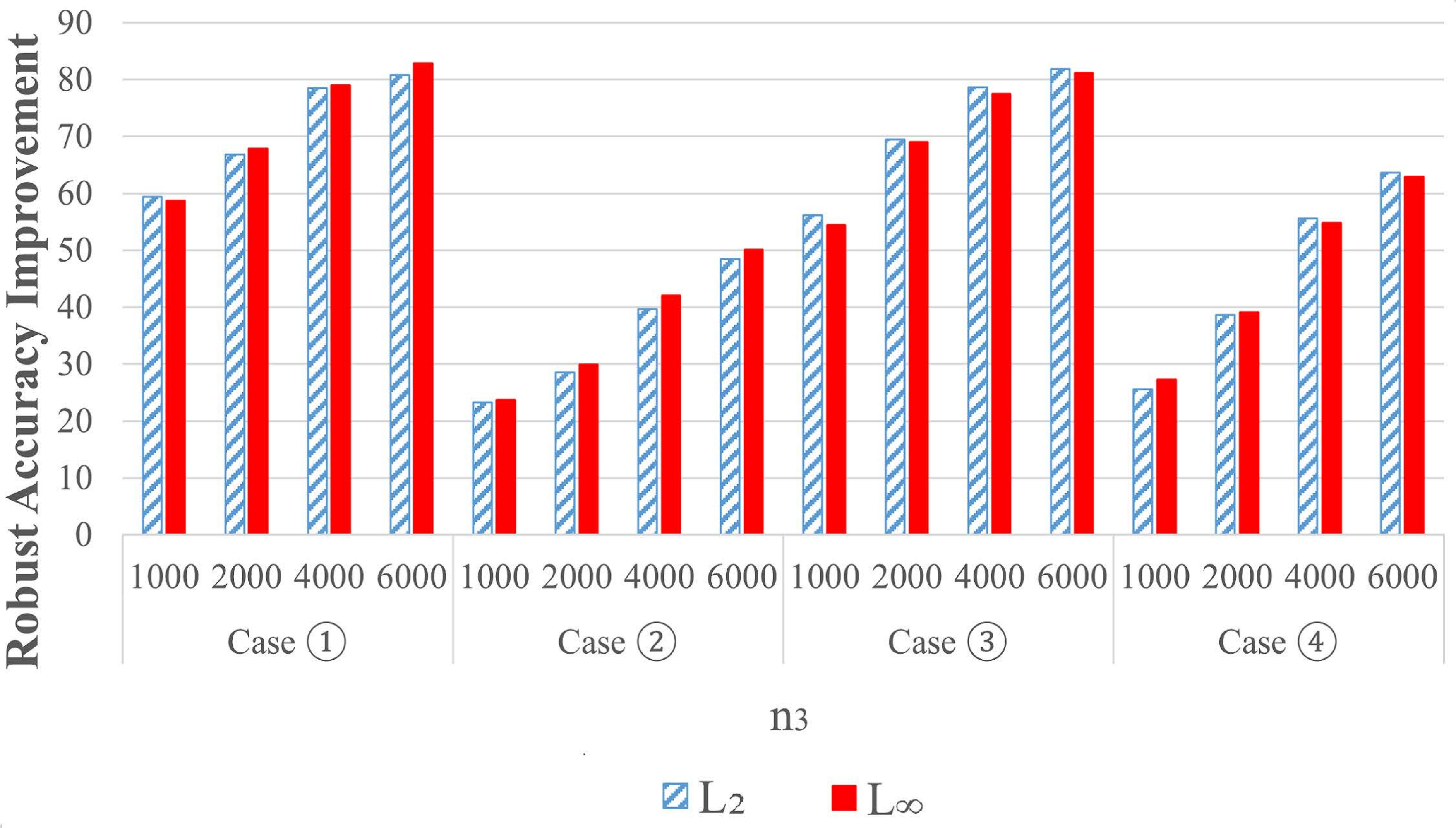}
\caption{Robust Accuracy Improvement Achieved by \textsc{Clover} Configed with Different Types of $p$-norm}
\label{fig: rq5_various_norm}
\end{figure}

\begin{figure} [t]
  \centering
  \includegraphics[width=0.55\linewidth]{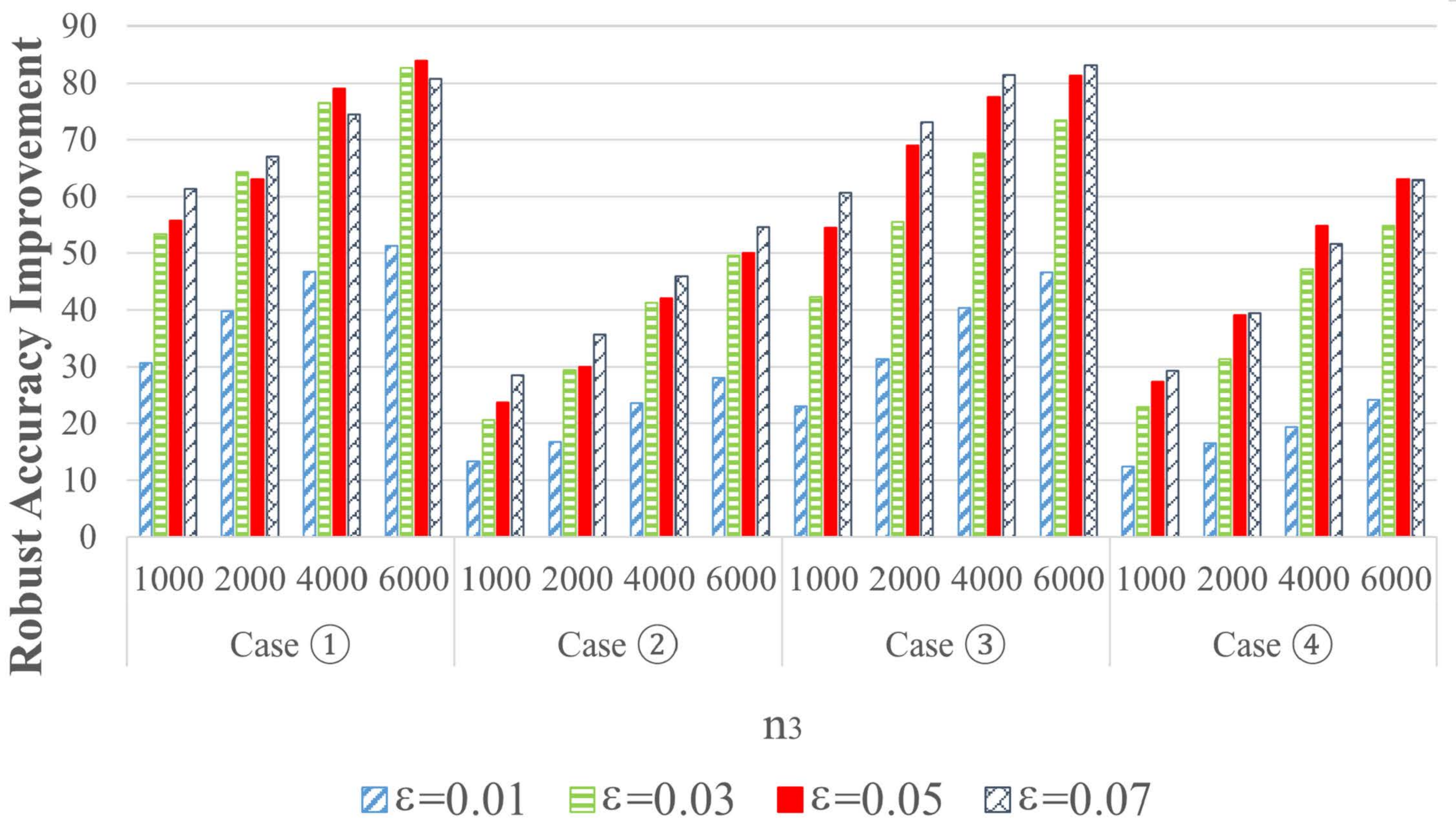}
\caption{Robust Accuracy Improvement Achieved by \textsc{Clover} Configed with Various Values of $\epsilon$}
\label{fig: rq5_various_fuzzing_boundary}
\end{figure}

Figures \ref{fig: rq5_various_m}--\ref{fig: rq5_various_fuzzing_boundary} summarize the results of Experiments 5c--5g, where \textsc{Clover} is configured with different values of $m$, $k$, $\delta$, $\epsilon$, and $p\text{-norm}$, respectively.
Each figure has four sections, one section for each of cases \circled{1}--\circled{4}.
In each section, the $x$-axis is $n_3$, and the $y$-axis is the robust accuracy improvement. 
To ease the comparison between Experiments 3 and 5, we copy the results of the original \textsc{Clover} from Tabel \ref{tab: rq3_fuzzing_compare_by_time} and show them as the solid bars in each figure.
The results for the other settings are shown in the bars filled with different patterns corresponding to the legend.

In Fig. \ref{fig: rq5_various_m} (for varying $m$), in all 16 combinations of $n_3$ and $n_4$ for cases \circled{1} to \circled{4}, \textsc{Clover} configured with different values of $m$ achieve quite similar robust accuracy improvements. Across all four cases, the differences in robust accuracy improvements between the original \textsc{Clover} (with $m=5$) and the other variants, i.e., \textsc{Clover} with $m=3$, $m=7$, and $k=9$ are $-3\%$ to $1\%$, $-3\%$ to $0\%$, and $4\%$ to $1\%$ in ratio (where we scale the original \textsc{Clover} to 1), respectively.

Similarly, in Fig. \ref{fig: rq5_various_k} (for varying $k$), the robust accuracy improvements among the \textsc{Clover} variants are similar. Compared to the original \textsc{Clover} (with $k=20$), the differences range
 from $-2\%$ to $1\%$, $-3\%$ to $1\%$, and $-3\%$ to $2\%$ in ratio (where the original \textsc{Clover} is scaled to 1), respectively, for $k=5$, $k=10$, and $k=40$, respectively, which are also small.

In Fig. \ref{fig: rq5_various_delta} (for varying $\delta$), different \textsc{Clover} variants achieve similar robust accuracy improvements: The differences against the original \textsc{Clover} (which is scaled to 1) are -2\% to $2\%$, $-2\%$ to $4\%$, and $-2\%$ to $0\%$ in ratio for $\delta=0.001, 0.005$, and $0.05$, respectively, which are also small.

In Fig. \ref{fig: rq5_various_norm} (for varying the $p$-norm), the differences in robust accuracy improvement between the original \textsc{Clover} and the variant range from $-0.01\%$ to $0\%$, which is small.

In summary, we observe that the above hyperparameters do not significantly affect the performance of the \textsc{Clover} algorithm.

Fig. \ref{fig: rq5_various_fuzzing_boundary} summarizes the results for Experiment 5g.
In all 16 combinations of $n_3$ and $n_4$, 
As expected, by using different fuzzing bounds  (values for $\epsilon$), \textsc{Clover} achieves different robust accuracy improvements.
Across all four cases, the differences in robust accuracy improvements between \textsc{Clover}'s variants, i.e., \textsc{Clover} with $\epsilon=0.01$, $\epsilon=0.03$, and $\epsilon=0.07$, and the original \textsc{Clover} range from $-51\%$ to $-46\%$, $-14\%$ to $6\%$, and $0\%$ to $12\%$ in ratio, respectively.

In summary, the robust accuracy improvement is noticeably affected by the changes in $\epsilon$ but not by the other studied hyperparameters. It is interesting to explore the underlying reasons as a future work.

\begin{tcolorbox}[
enhanced, breakable,
attach boxed title to top left = {yshift = -2mm, xshift = 5mm},
boxed title style = {sharp corners},
colback = white, 
title={Answering RQ5}
]
Configuring \textsc{Clover} with the preferences of test cases with higher CC values and enabling the use of $\beta$-AFOs in producing test cases improves the effectiveness of \textsc{Clover}.
Varying the key hyperparameters ($m$, $k$, $\delta$, and $p$-norm) in the algorithm does not significantly affect  \textsc{Clover}'s performance.
But, using different fuzzing bounds ($\epsilon$), as expected, affects its performance.
\end{tcolorbox}

\section{Results and Data Analysis for Overall Effect of \textsc{Clover} on Adversarially Trained Models}
This section presents the result and data analysis for answering RQ6 through Experiment 6 on fuzzing adversarially trained models for \textsc{Clover} in Configuration \textit{B}.

\subsection{Answering RQ6 (Effects of \textsc{Clover})}
\label{sec: rq6_results_a_to_c}

\begin{figure}[t]
  \centering
  \includegraphics[width=0.35\textwidth]{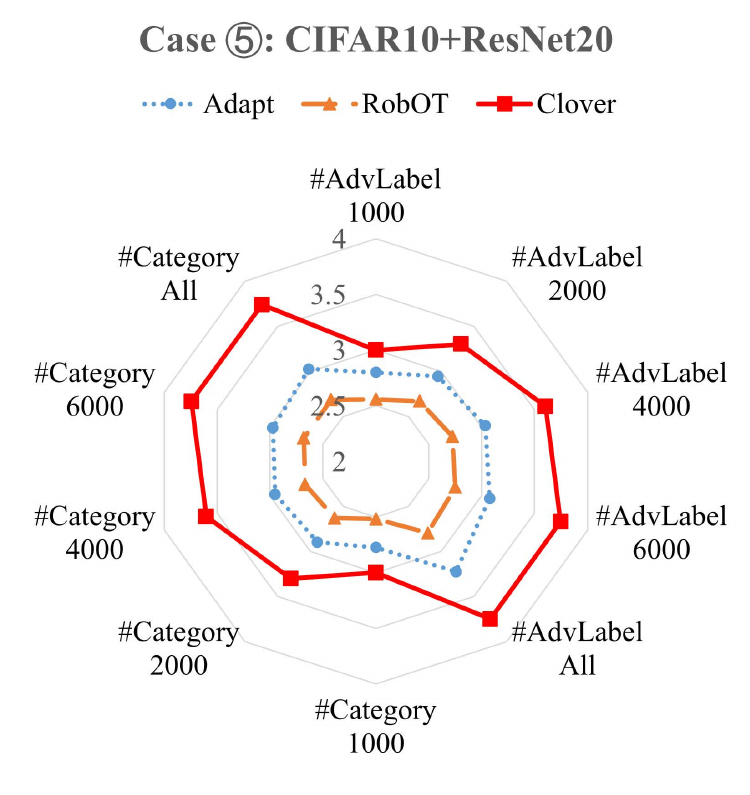}
  \includegraphics[width=0.35\textwidth]{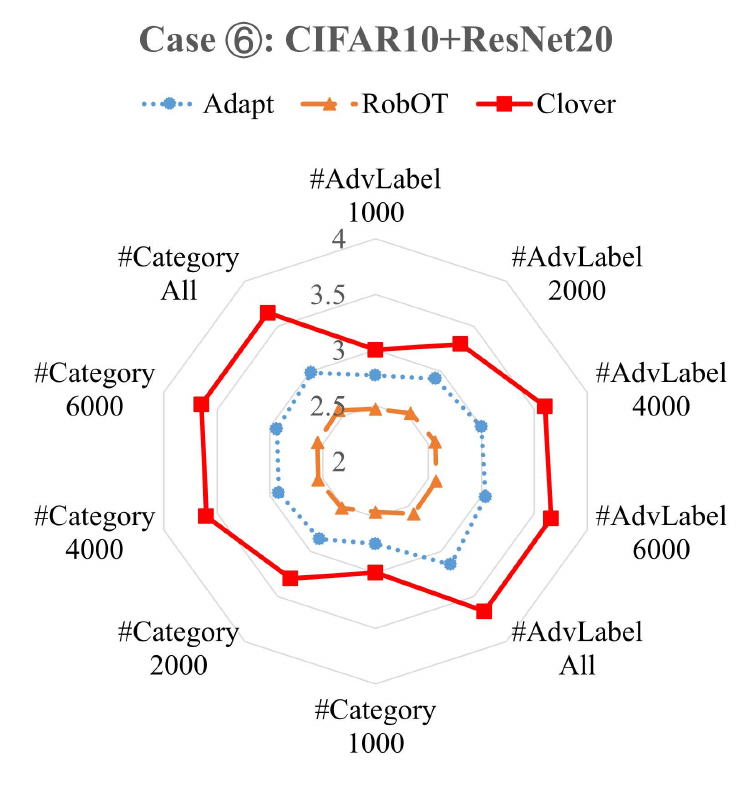} \\
  \includegraphics[width=0.35\textwidth]{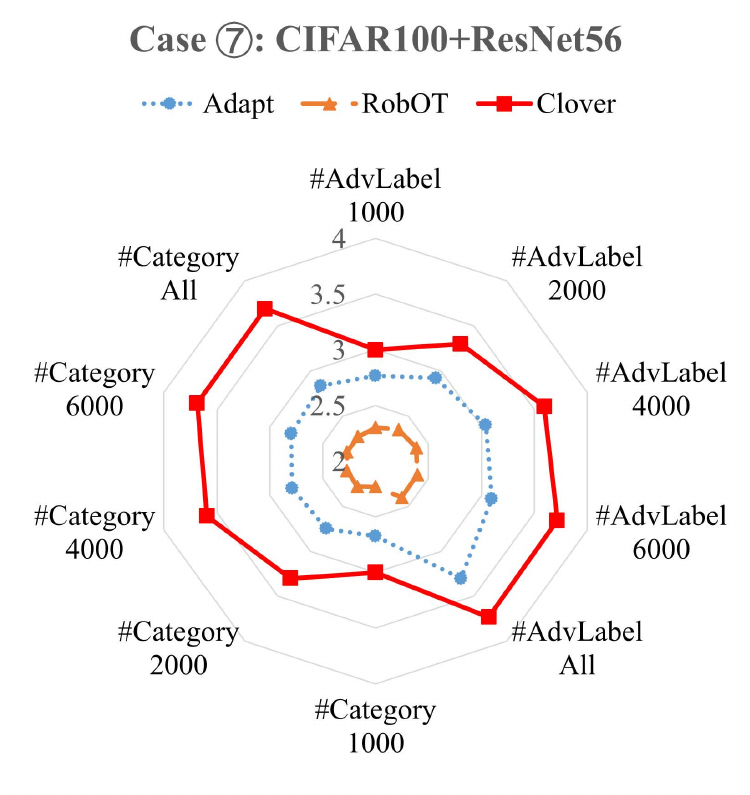}
  \includegraphics[width=0.35\textwidth]{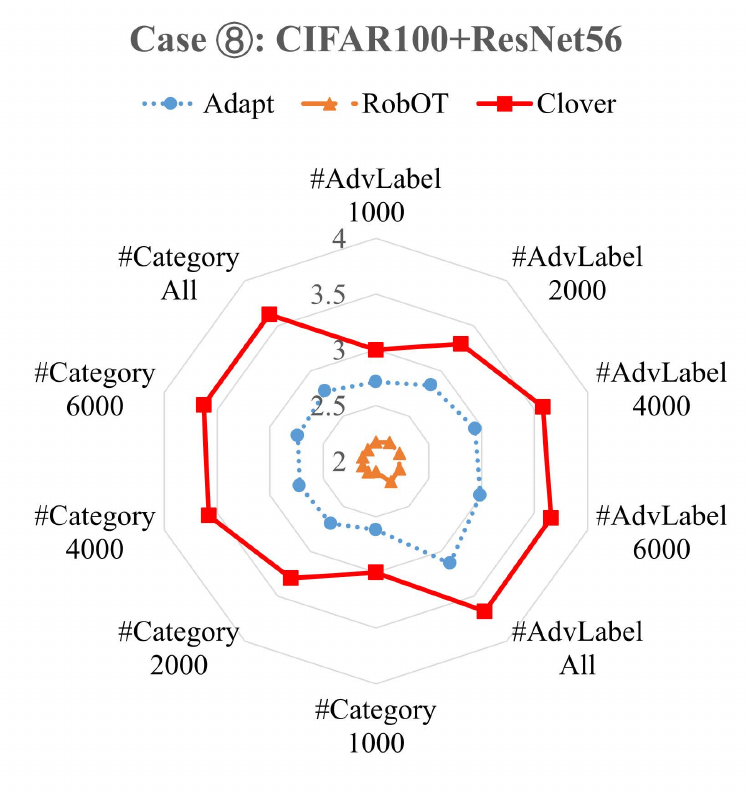} \\
  \caption{\textit{\#AdvLabel} and \textit{\#Category} for Different Techniques on Adversarially Trained Models with $n_4 = 18000$ (in $log_{10}$ scale)}
\label{fig: rq6_statistics_advlabel_category}
\end{figure}

\begin{figure}[t]
  \centering
  \includegraphics[width=0.24\textwidth]{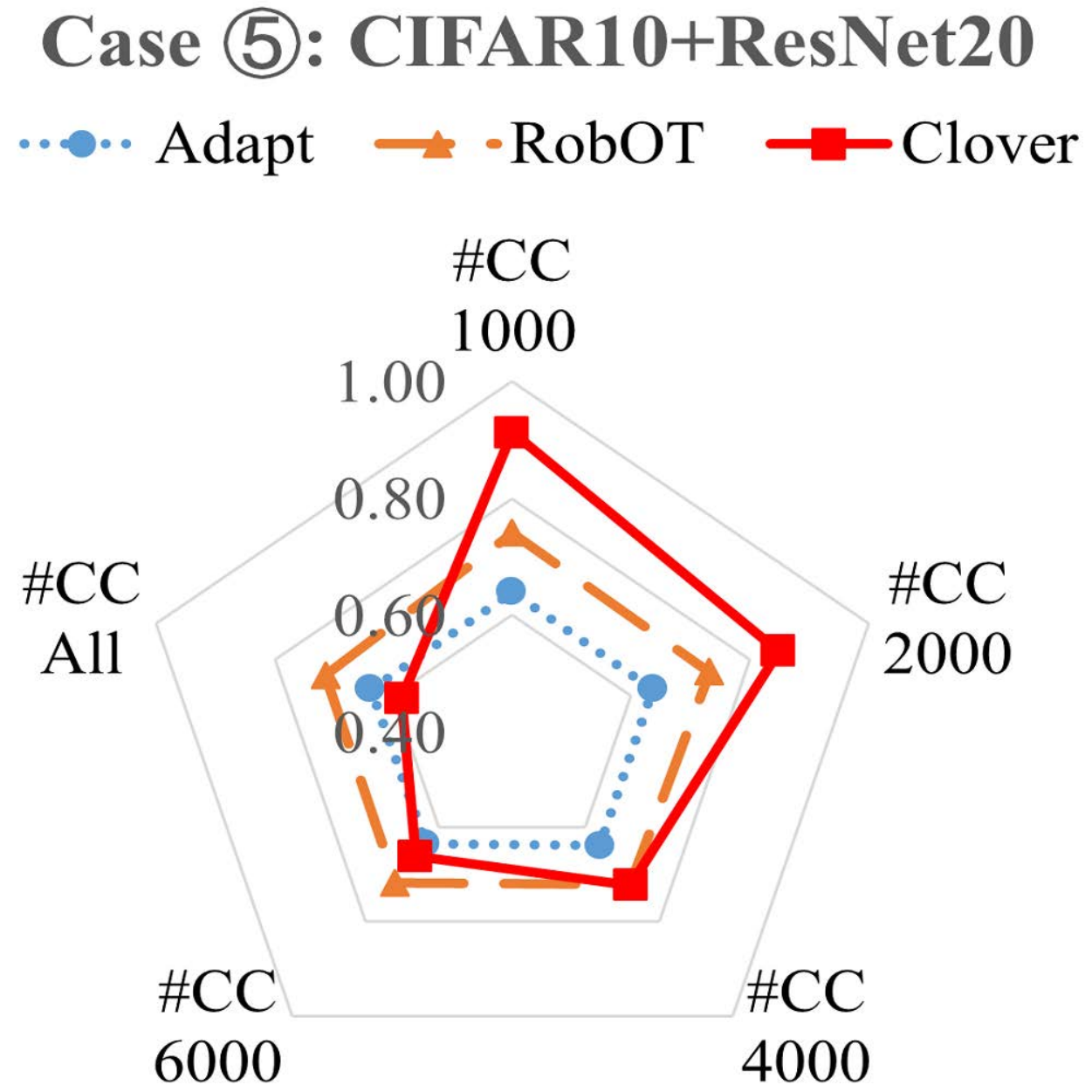}
  \includegraphics[width=0.24\textwidth]{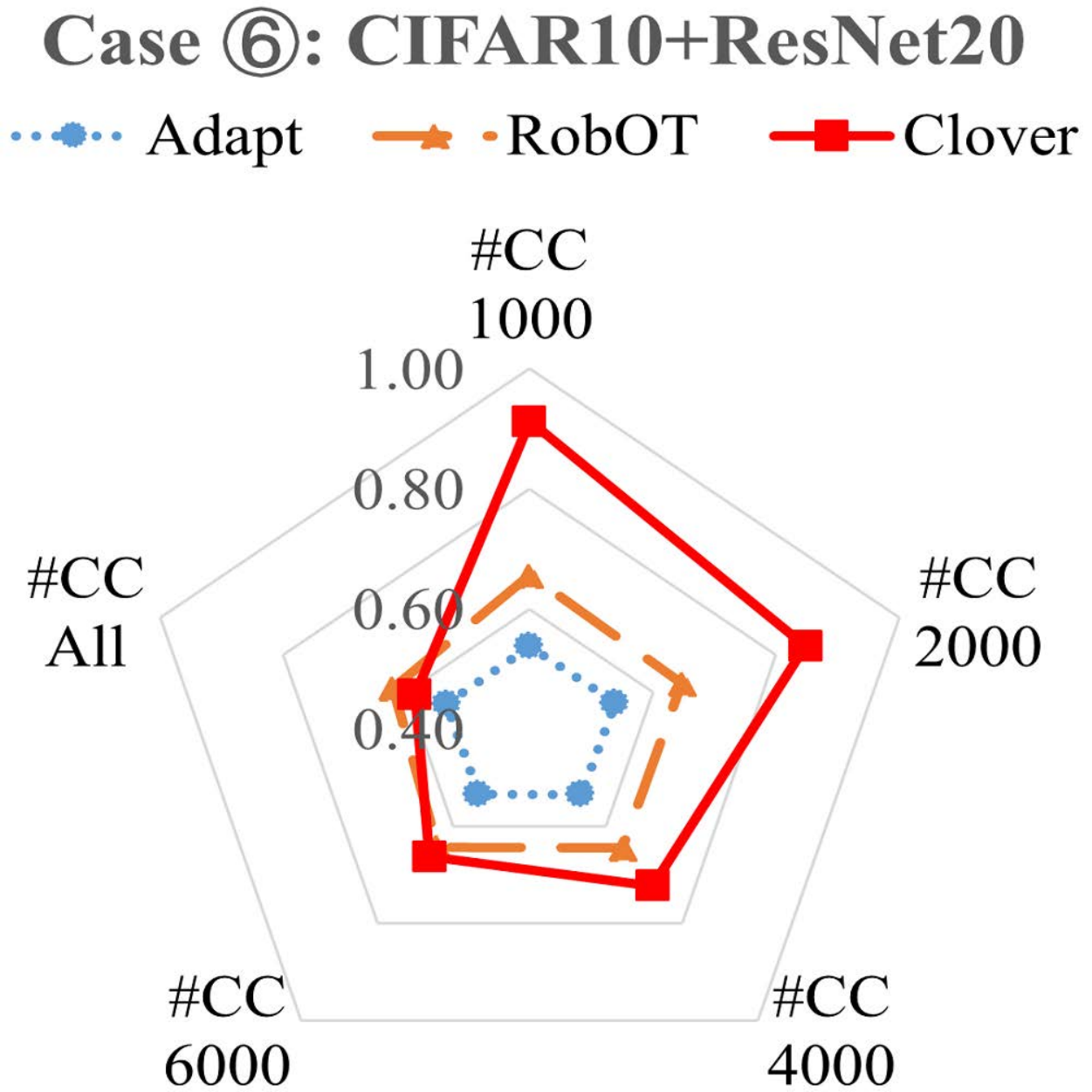}
  \includegraphics[width=0.24\textwidth]{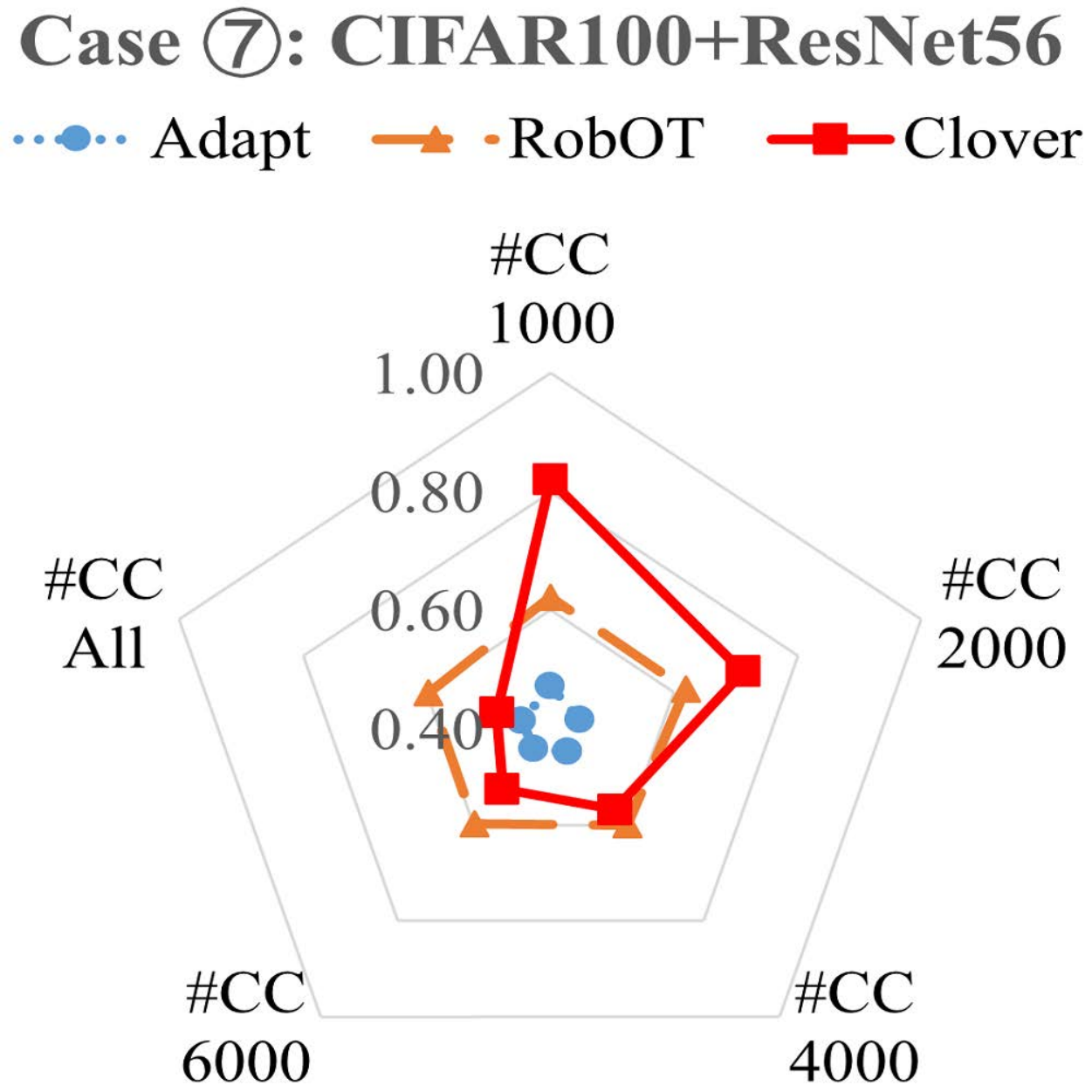}
  \includegraphics[width=0.24\textwidth]{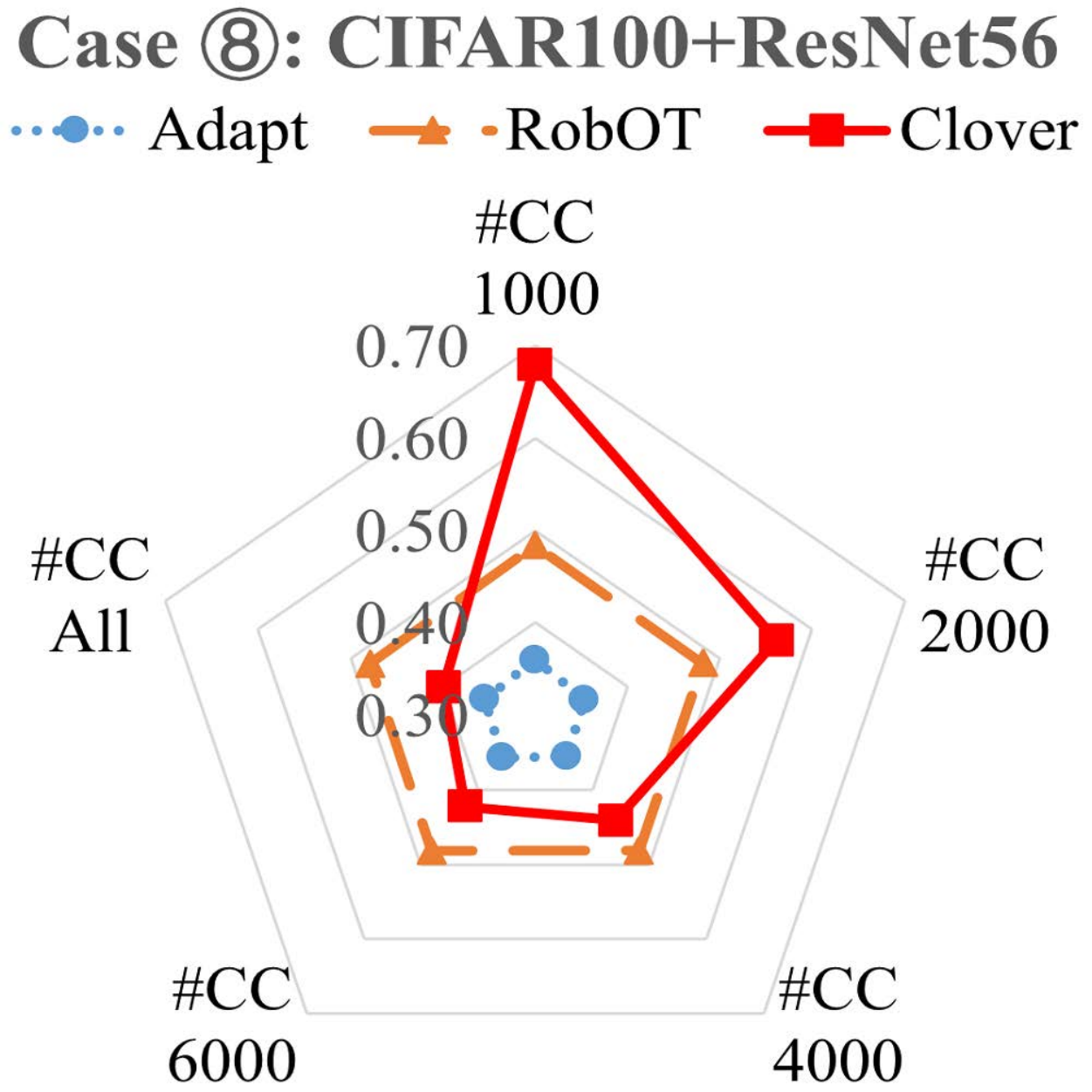} \\
  \caption{\textit{\#CC} for Different Techniques on Adversarially Trained Models with $n_4 = 18000$}
\label{fig: rq6_statistics_cc}
\end{figure}

\begin{table}[thb]
\caption{Mean Results of Test Suites Generated by Different DL Testing Techniques Fuzzing in Configuration \textit{B} on Adversarially Trained Models for 3 Runs with $N_3$ and $n_4=18000$} 
\label{tab: rq6_fuzz_adv_trained_model_test_suites}
\resizebox{\textwidth}{!}{
\begin{tabu}{|l|c|c|c|c|c|c|c|}
\hline
\multirow{2}{*}{Benchmark Case} & \multirow{2}{*}{Technique} & \multicolumn{3}{c|}{$n_3$ = 1000} & \multicolumn{3}{c|}{$n_3$ = 2000} \\ \cline{3-8}
 &  & \textit{\#AdvLabel} & \textit{\#Category} & \textit{\#CC} & \textit{\#AdvLabel} & \textit{\#Category} & \textit{\#CC} \\ \hline 
\multirow{3}{*}{\begin{tabu}[l]{@{}c@{}}\circled{5}: CIFAR10+ResNet20\end{tabu}}
 & \textsc{Adapt} & 629 & 593 & 0.64 & 883 & 786 & 0.64 \\ 
 & \textsc{RobOT} & 362 & 329 & 0.74 & 466 & 425 & 0.73 \\ 
 &\textsc{Clover} & 1000 & 1000 & 0.91 & 2000 & 2000 & 0.85 \\ \tabucline[1.2pt]{1-8} 
\multirow{3}{*}{\circled{6}: CIFAR10+ResNet20}
 & \textsc{Adapt} & 592 & 552 & 0.54 & 826 & 726 & 0.54 \\ 
 & \textsc{RobOT} & 294 & 286 & 0.66 & 343 & 329 & 0.65 \\ 
 &\textsc{Clover} & 1000 & 1000 & 0.79 & 2000 & 2000 & 0.69 \\ \tabucline[1.2pt]{1-8} 
\multirow{3}{*}{\circled{7}: CIFAR100+ResNet56} 
 & \textsc{Adapt} & 584 & 471 & 0.47 & 835 & 563 & 0.45 \\ 
 & \textsc{RobOT} & 200 & 170 & 0.62 & 224 & 191 & 0.62 \\ 
 &\textsc{Clover} & 1000 & 1000 & 0.82 & 2000 & 2000 & 0.71 \\ \tabucline[1.2pt]{1-8} 
\multirow{3}{*}{\circled{8}: CIFAR100+ResNet56} 
 & \textsc{Adapt} & 514 & 415 & 0.36 & 697 & 490 & 0.35 \\ 
 & \textsc{RobOT} & 148 & 125 & 0.48 & 161 & 132 & 0.48 \\ 
 &\textsc{Clover} & 1000 & 1000 & 0.68 & 2000 & 2000 & 0.56 \\ \tabucline[1.2pt]{1-8} 
\multirow{5}{*}{\textbf{\begin{tabu}[l]{@{}l@{}} Mean Results \end{tabu}}} 
 & \textsc{Adapt} & 579.75 & 507.67 & 0.50 & 810.17 & 641.33 & 0.49 \\ 
 & \textsc{RobOT} & 251.00 & 227.42 & 0.63 & 298.75 & 269.17 & 0.62 \\ 
 & \textsc{Clover} & 1000.00 & 1000.00 & 0.80 & 2000.00 & 2000.00 & 0.70 \\ \cline{2-8} 
 & \textsc{Clover}$\div$(\textsc{Adapt}) & 1.72 & 1.97 & 1.59 & 2.47 & 3.12 & 1.42 \\ 
 & \textsc{Clover}$\div$(\textsc{RobOT}) & 3.98 & 4.40 & 1.28 & 6.69 & 7.43 & 1.14 \\ \hline \hline

\multirow{2}{*}{Benchmark Case} & \multirow{2}{*}{Technique} & \multicolumn{3}{c|}{$n_3$ = 4000} & \multicolumn{3}{c|}{$n_3$ = 6000} \\ \cline{3-8}
 &  & \textit{\#AdvLabel} & \textit{\#Category} & \textit{\#CC} & \textit{\#AdvLabel} & \textit{\#Category} & \textit{\#CC} \\ \hline 
\multirow{3}{*}{\begin{tabu}[l]{@{}c@{}}\circled{5}: CIFAR10+ResNet20\end{tabu}}
 & \textsc{Adapt} & 1085 & 895 & 0.64 & 1193 & 943 & 0.64 \\ 
 & \textsc{RobOT} & 528 & 467 & 0.72 & 565 & 480 & 0.72 \\ 
 &\textsc{Clover} & 4000 & 4000 & 0.72 & 5595 & 5550 & 0.66 \\ \tabucline[1.2pt]{1-8} 
\multirow{3}{*}{\circled{5}: CIFAR10+ResNet20}
 & \textsc{Adapt} & 1004 & 826 & 0.54 & 1096 & 860 & 0.53 \\ 
 & \textsc{RobOT} & 367 & 348 & 0.65 & 374 & 354 & 0.64 \\ 
 &\textsc{Clover} & 4000 & 4000 & 0.56 & 4560 & 4431 & 0.56 \\ \tabucline[1.2pt]{1-8} 
\multirow{3}{*}{\circled{7}: CIFAR100+ResNet56} 
 & \textsc{Adapt} & 1101 & 616 & 0.45 & 1251 & 630 & 0.44 \\ 
 & \textsc{RobOT} & 243 & 187 & 0.60 & 248 & 187 & 0.60 \\ 
 &\textsc{Clover} & 4000 & 4000 & 0.57 & 5213 & 4911 & 0.53 \\ \tabucline[1.2pt]{1-8} 
\multirow{3}{*}{\circled{8}: CIFAR100+ResNet56} 
 & \textsc{Adapt} & 868 & 525 & 0.36 & 973 & 547 & 0.36 \\ 
 & \textsc{RobOT} & 167 & 134 & 0.48 & 168 & 134 & 0.48 \\ 
 &\textsc{Clover} & 4000 & 4000 & 0.44 & 4245 & 4245 & 0.42 \\ \tabucline[1.2pt]{1-8} 
\multirow{5}{*}{\textbf{\begin{tabu}[l]{@{}l@{}} Mean Results \end{tabu}}} 
 & \textsc{Adapt} & 1014.50 & 715.58 & 0.49 & 1128.00 & 745.00 & 0.49 \\ 
 & \textsc{RobOT} & 346.42 & 283.83 & 0.61 & 338.58 & 288.42 & 0.61 \\ 
 & \textsc{Clover} & 4000.00 & 4000.00 & 0.57 & 4973.42 & 4784.58 & 0.54 \\ \cline{2-8} 
 & \textsc{Clover}$\div$(\textsc{Adapt}) & 3.94 & 5.59 & 1.16 & 4.41 & 6.42 & 1.10 \\ 
 & \textsc{Clover}$\div$(\textsc{RobOT}) & 12.25 & 14.09 & 0.94 & 14.69 & 16.59 & 0.89 \\ \hline
\end{tabu}
}
\end{table}

\begin{table}[t]
\caption{Mean Results of \textit{All} Test Cases Generated by Different Techniques Fuzzing on Adversarially Trained Models in Configuration \textit{B} for 3 Runs with $n_4=18000$} 
\label{tab: rq6_fuzz_adv_trained_model_universe}
\resizebox{0.7\textwidth}{!}{
\begin{tabu}{|l|c|c|c|c|}
\hline
\multirow{2}{*}{Benchmark Case} & \multirow{2}{*}{Technique} & \multicolumn{3}{c|}{\textit{All}} \\ \cline{3-5}
 &  & \textit{\#AdvLabel} & \textit{\#Category} & \textit{\#CC} \\ \hline 
\multirow{3}{*}{\begin{tabu}[l]{@{}c@{}}\circled{5}: CIFAR10+ResNet20\end{tabu}}
 & \textsc{Adapt} & 1677 & 1062 & 0.64 \\ 
 & \textsc{RobOT} & 623 & 484 & 0.71 \\ 
 &\textsc{Clover} & 5627 & 5550 & 0.59 \\ \tabucline[1.2pt]{1-5} 
\multirow{3}{*}{\circled{6}: CIFAR10+ResNet20}
 & \textsc{Adapt} & 1405 & 960 & 0.53 \\ 
 & \textsc{RobOT} & 382 & 368 & 0.62 \\ 
 &\textsc{Clover} & 4624 & 4431 & 0.49 \\ \tabucline[1.2pt]{1-5} 
\multirow{3}{*}{\circled{7}: CIFAR100+ResNet56} 
 & \textsc{Adapt} & 2012 & 683 & 0.45 \\ 
 & \textsc{RobOT} & 252 & 187 & 0.60 \\ 
 &\textsc{Clover} & 5375 & 4911 & 0.49 \\ \tabucline[1.2pt]{1-5} 
\multirow{3}{*}{\circled{8}: CIFAR100+ResNet56} 
 & \textsc{Adapt} & 1362 & 599 & 0.36 \\ 
 & \textsc{RobOT} & 170 & 134 & 0.48 \\ 
 &\textsc{Clover} & 4646 & 4245 & 0.40 \\ \tabucline[1.2pt]{1-5} 
\multirow{5}{*}{\textbf{\begin{tabu}[l]{@{}l@{}} Mean Results \end{tabu}}} 
 & \textsc{Adapt} & 3.14 & 5.79 & 0.49 \\ 
 & \textsc{RobOT} & 356.83 & 293.00 & 0.60 \\ 
 & \textbf{\textsc{Clover}} & 5067.92 & 4784.58 & 0.49 \\  \cline{2-5}
 & \textbf{\textsc{Clover}$\div$\textsc{Adapt}} & 3.14 & 5.79 & 0.99 \\ 
 & \textbf{\textsc{Clover}$\div$\textsc{RobOT}} & 14.20 & 16.33 & 0.81 \\ \hline
\end{tabu}
}
\end{table}

Figures \ref{fig: rq6_statistics_advlabel_category} and \ref{fig: rq6_statistics_cc} summarize the results of Experiment 6a--6c in terms of \textit{\#AdvLabel}, \textit{\#Category}, and \textit{\#CC} for \textsc{Clover} and the two peer techniques \textsc{Adapt} and \textsc{RobOT} on adversarially trained models.
Readers can interpret the axes of the charts in these two figures like these in Fig. \ref{fig: rq3_statistics_advlabel_category} and \ref{fig: rq3_statistics_cc}, respectively.
The detailed values achieved by each technique in each axis can be found in Tables \ref{tab: rq6_fuzz_adv_trained_model_test_suites} and \ref{tab: rq6_fuzz_adv_trained_model_universe}.

In Fig. \ref{fig: rq6_statistics_advlabel_category}, we observe that the enclosed regions for \textsc{Clover} are always larger than these of \textsc{Adapt} and \textsc{RobOT} in all four charts by large extents, where \textsc{RobOT} always achieves the smallest regions and \textsc{Adapt} is in between \textsc{Clover} and \textsc{RobOT}.

Let us first discuss the results on \textit{\#AdvLabel}. 
Across the axes for different $n_3 \in \{1000, 2000, 4000, \\ 6000\}$, the difference in terms of \textit{\#AdvLabel} between \textsc{Clover} and each of \textsc{Adapt} and \textsc{RobOT} tends to increase as $n_3$ increases. 
Over these four charts, when $n_3 = 1000$, the mean numbers of unique adversarial labels generated by \textsc{Clover} are $1.72\times$ and $3.98\times$ folds of these of \textsc{Adapt} and \textsc{RobOT}, respectively. The corresponding differences increase to $4.41\times$ and $14.69\times$ when $n_3=6000$, respectively. 

When $n_3 = 1000$, in cases \circled{5} and \circled{6}, where the \textit{Ratio} parameters are set to 0.5 and 1.0 for the two CIFAR10 models (see Table \ref{tab: adv_trained_model}) respectively, the mean numbers of unique adversarial labels generated by \textsc{Clover} are $1.59\times$ and $1.69\times$ of those of \textsc{Adapt}, respectively, and $2.76\times$ and $3.40\times$ of those of \textsc{RobOT}. 
The corresponding differences increase to $4.69\times$ and $9.90\times$ (for using 0.5 as the \textit{Ratio} parameter), and $4.16\times$ and $12.19\times$ (for using 1 as the \textit{Ratio} parameter) when $n_3=6000$ for \textsc{Adapt} and \textsc{RobOT}, respectively.

Similarly, on the two CIFAR100 models in cases \circled{7} and \circled{8}, the corresponding diffrences increase from $1.71\times$ and $1.95\times$ when $n_3 = 1000$ to $4.17\times$ and $4.36\times$ when $n_3=6000$ for \textsc{Adapt}, and increase from $5.00\times$ and $6.76\times$ when $n_3 = 1000$ to $21.02\times$ and $25.27\times$ when $n_3=6000$ for \textsc{RobOT}, respectively.

\textsc{RobOT}’s performance in terms of \textit{\#AdvLabel} decreases significantly as the abovementioned \textit{Ratio} parameter for adversarial example generation in each training epoch increases from 0.5 to 1. This is because, in cases \circled{6} and \circled{8}, \textsc{RobOT} generates much fewer test cases (see Table \ref{tab: throughout}). \textsc{Adapt}’s performance also decreases but more moderately. 
From Table \ref{tab: throughout}, we observe that \textsc{Adapt} has generated much more test cases than the largest number of selected test cases ($n_3=6000$) in cases \circled{6} and \circled{8} by $5.00\times$ to $11.42\times$ already, but \textsc{RobOT} only generates test cases by $1.18\times$ to $3.20\times$.

Across cases \circled{5}--\circled{8}, comparing the values for the axes for $n_3=6000$ and $n_3=All$ (i.e., without selection), the corresponding increases in \textit{\#AdvLabel} is relatively gentle, i.e., $1.43\times$, $1.05\times$, and $1.02\times$ for \textsc{Adapt}, \textsc{RobOT} and \textsc{Clover}, respectively. 
We observe that their corresponding pairs of values at $n_3=6000$ and $n_3=All$ are similar between the same model with different parameter values for \textit{Ratio}.
We also observe that the total numbers of unique adversarial labels generated by \textsc{RobOT} and \textsc{Adapt} at $n_3=All$ are smaller than those generated by \textsc{Clover} at $n_3=2000$ and $n_3=4000$ in all four charts, respectively. 

Overall, the results show that \textsc{Clover} has a higher performance in generating test cases with higher diversity in unique adversarial labels than the two peer techniques.

We next discuss the results on \textit{\#Category} in Fig. \ref{fig: rq6_statistics_advlabel_category}.
Similar to the results of \textit{\#Advlabel} discussed above, we observe noticeable differences in \textit{\#Category} among \textsc{Adapt}, \textsc{RobOT} and \textsc{Clover}, and their trends of difference are almost the same as the trends of difference we have discussed on \textit{\#AdvLabel} above from $n_3=1000$ to $n_3=6000$ and from $n_3=6000$ to $n_3=All$.

When $n_3 = 1000$, the mean \textit{\#Category} for \textsc{Clover} is $1.97\times$ and $4.40\times$ of these of \textsc{Adapt} and \textsc{RobOT}, respectively. The corresponding differences increase to $6.42\times$ and $16.59\times$ when $n_3=6000$, respectively. 
When $n_3 = 1000$, in cases \circled{5} and \circled{6}, at the \textit{Ratio} parameter set to 0.5 and 1.0 for the two CIFAR10 models, the mean \textit{\#Category} for \textsc{Clover} are $1.68\times$ and $1.81\times$ of those of \textsc{Adapt}, respectively, and $5.89\times$ and $5.15\times$ of these of \textsc{RobOT}, respectively.
The corresponding differences increase to $3.04\times$ and $3.50\times$ (for \textit{Ratio} = 0.5), and $11.56\times$ and  $12.84\times$ (for \textit{Ratio} = 1) when $n_3=6000$ for \textsc{Adapt} and \textsc{RobOT}, respectively.
Similarly, for the two CIFAR100 models in cases \circled{7} and \circled{8}, the corresponding differences increase from $2.12\times$ and $2.41\times$ when $n_3 = 1000$ to $7.80\times$ and $7.76\times$ when $n_3=6000$ for \textsc{Adapt}, and increase from $5.24\times$ and $8.00\times$ when $n_3 = 1000$ to $26.26\times$ and $31.68\times$ when $n_3=6000$ for \textsc{RobOT}, respectively.

Fig. \ref{fig: rq6_statistics_cc} shows the results of \textit{\#CC}. 
The enclosed regions for \textsc{Clover} are the largest, followed by \textsc{RobOT} and finally \textsc{Adapt}. 
We also observe that \textsc{RobOT} produces higher \textit{\#CC} values than \textsc{Clover} at some $n_3$, nonetheless as we have presented above, \textsc{RobOT} generates an order of magnitude smaller numbers of unique adversarial labels and unique categories than \textsc{Clover}.
We observe that as $n_3$ increases, $\#CC$ of \textsc{Clover} decreases gradually. 
The results indicate that  \textsc{Clover} has to include test cases with lower CC values in the constructing test suites. 
This is because the \textsc{Clover} algorithm iteratively over seeds to select their test cases that have been prioritized in descending order of CC values. 
It further indicates that \textsc{Clover} has not generated test cases with high CC values from some test cases within the given time budget of $n_4 = 18000$.
We recall that \textsc{Clover} generates test cases from the same seed with increasingly higher CC values with the aim of test case diversity through seed equivalence.
Future work of \textsc{Clover} is to improve its cost-effectiveness by allocating a higher time budget to these seeds that have shown a potential to produce test cases yet with relatively low CC values.

To further compare the difference in terms of \textit{\#AdvLabel} and \textit{\#Category} between \textsc{Clover} and the other two peer techniques on adversarially trained models, we conduct the Wilcoxon signed-rank test and calculate Cohen's $d$ to measure the $p$-value and effect size over all combinations of Cases \circled{5} to \circled{8} and the size $n_3$ of the constructed test suites with $n_4=18000$.
The $p$-values are all $\le 1e^5$.
The effect sizes for the test between \textsc{Cloer} and \textsc{Adapt} are 
1.97 and 2.34 for \textit{\#AdvLabel} and \textit{\#Category}, repsectively.
The effect sizes for the test between \textsc{Cloer} and \textsc{RobOT} are 
2.62 and 2.72 for \textit{\#AdvLabel} and \textit{\#Category}, repsectively.
They are all at huge levels, showing that the differences we have presented above are statistically meaningful in these two metrics.

\subsection{Answering RQ6 (Effects of \textsc{Clover} Variants)}
\begin{figure}[t]
  \centering
  \includegraphics[width=0.35\textwidth]{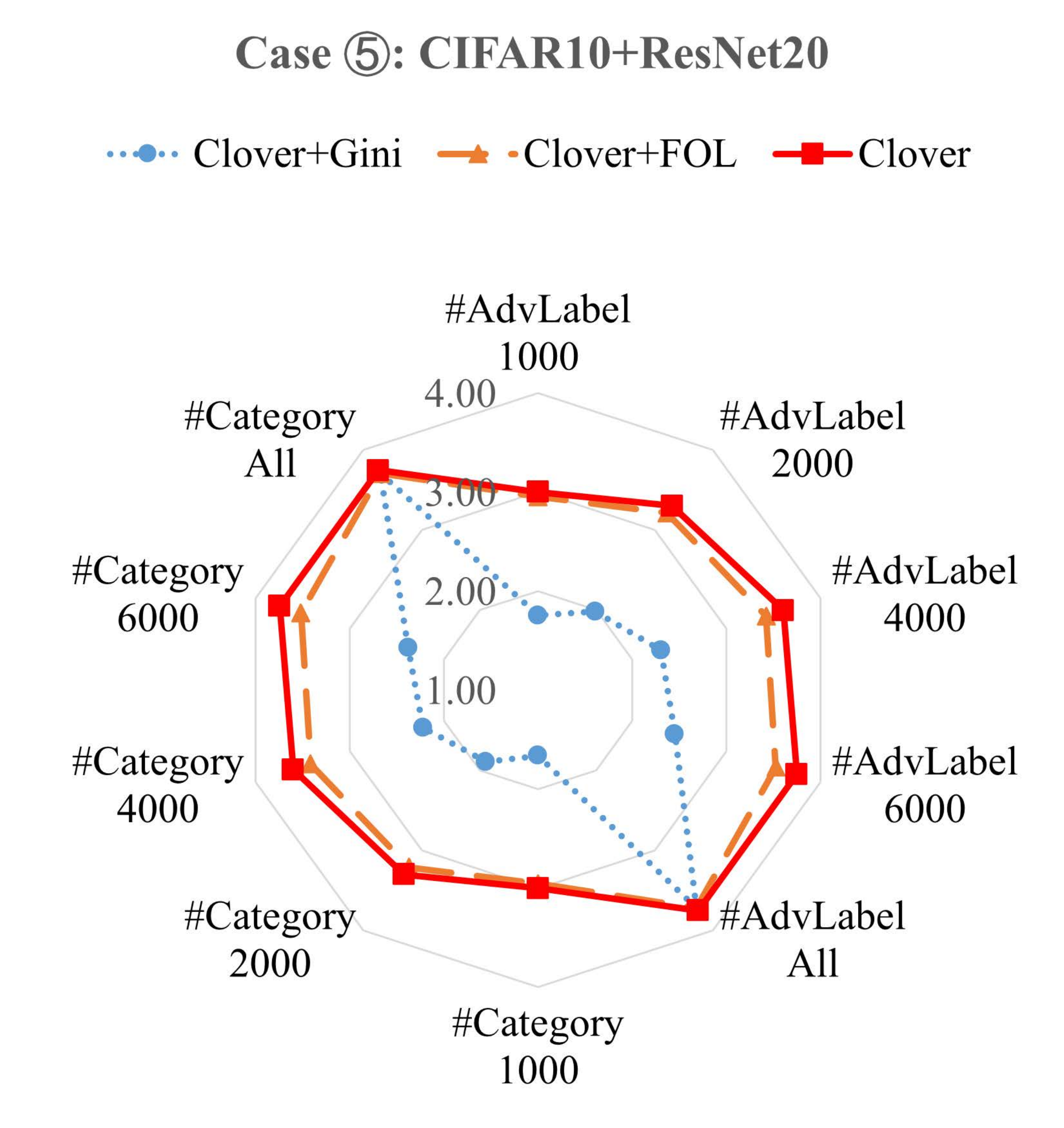}
  \includegraphics[width=0.35\textwidth]{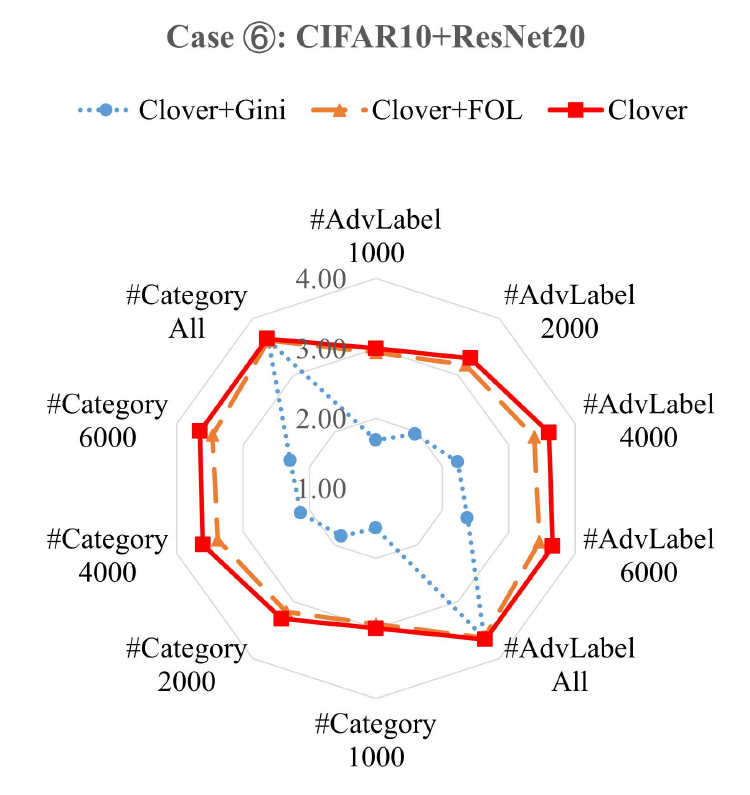} \\
  \includegraphics[width=0.35\textwidth]{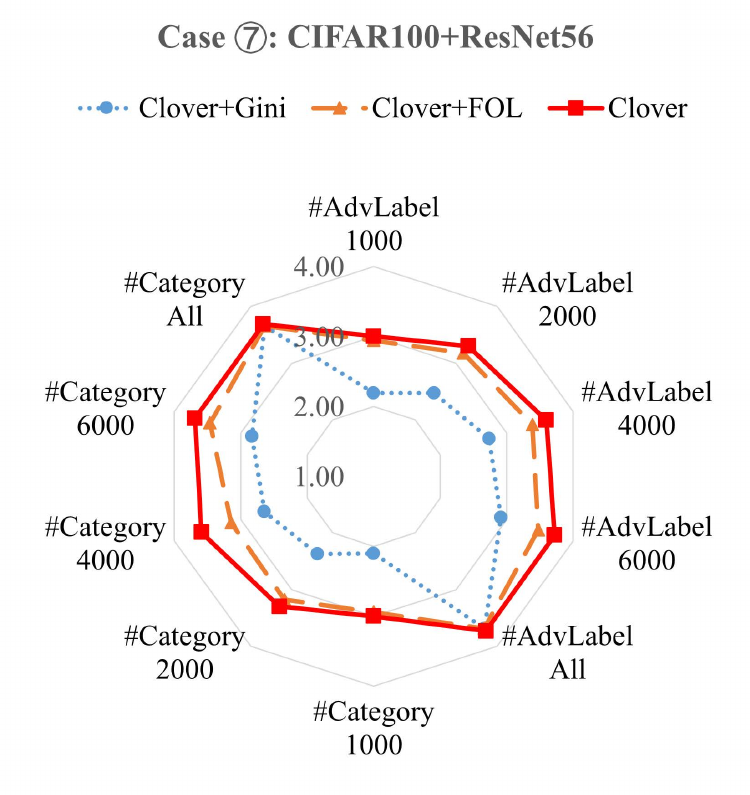}
  \includegraphics[width=0.35\textwidth]{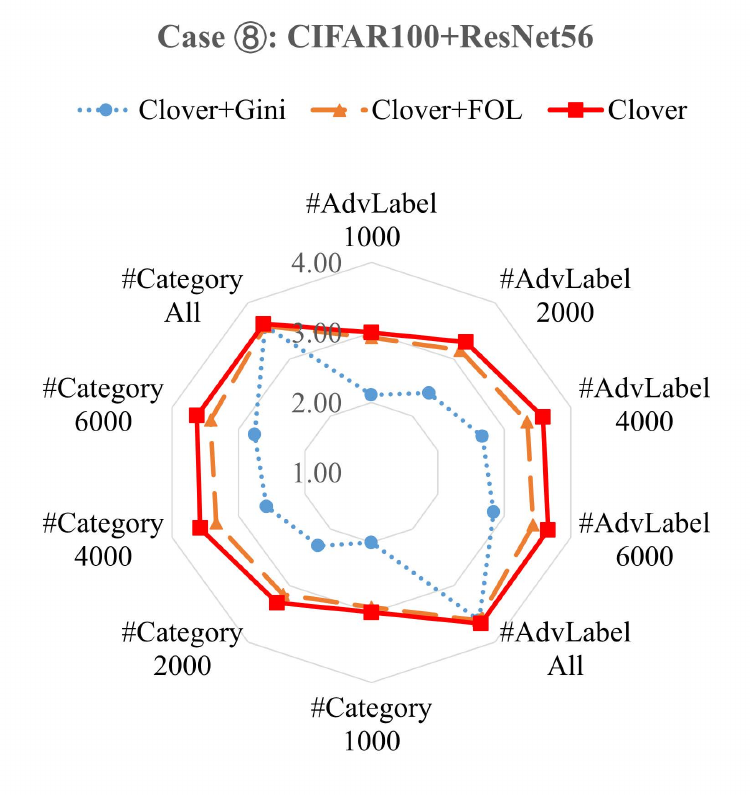} \\
  \caption{\textit{\#AdvLabel} and \textit{\#Category} for \textsc{Clover} and Its Variants on Adversarially Trained Models with $n_4 = 18000$ (in $log_{10}$ scale)}
\label{fig: rq6_statistics_advlabel_category_variants}
\end{figure}

\begin{figure}[t]
  \centering
\includegraphics[width=0.24\textwidth]{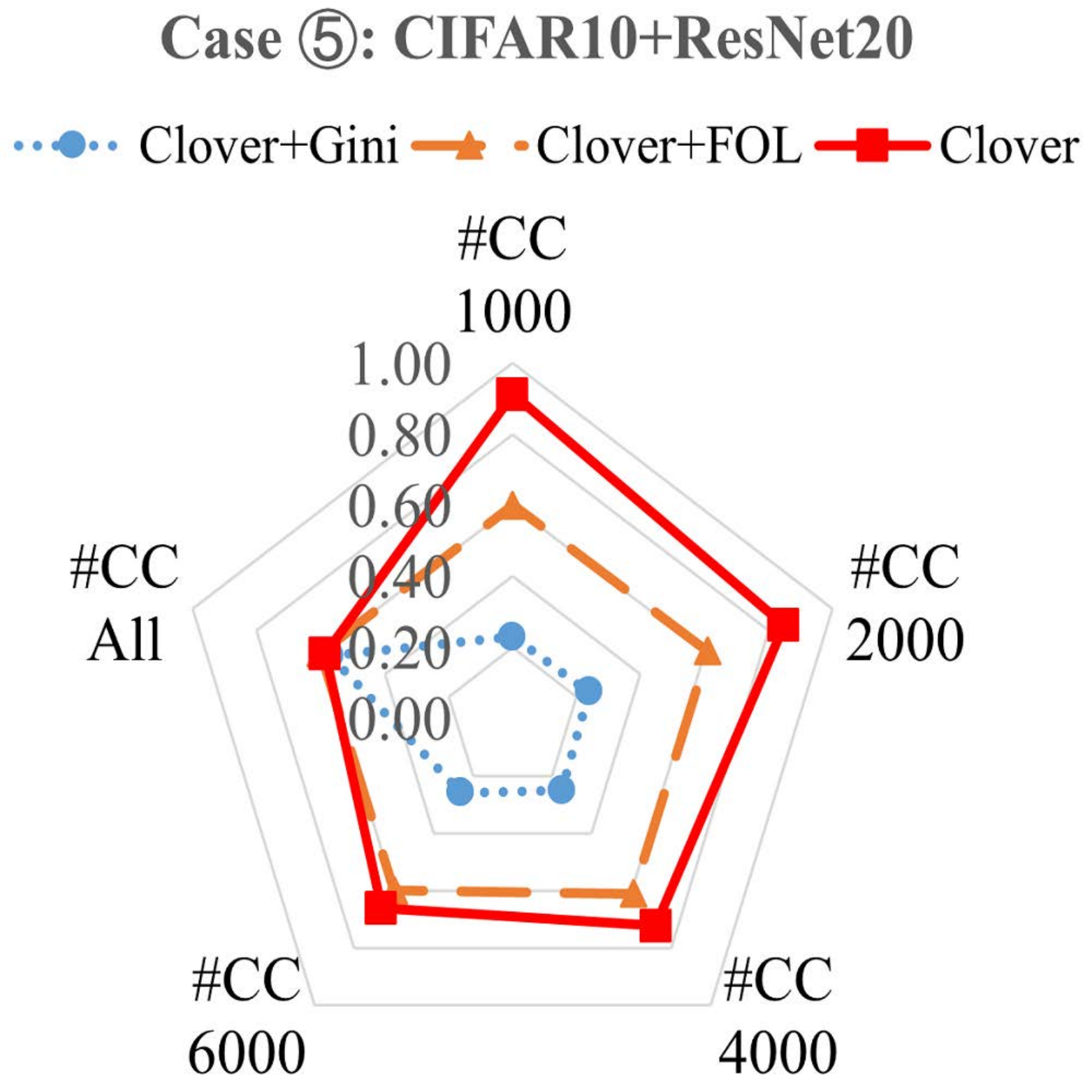}
\includegraphics[width=0.24\textwidth]{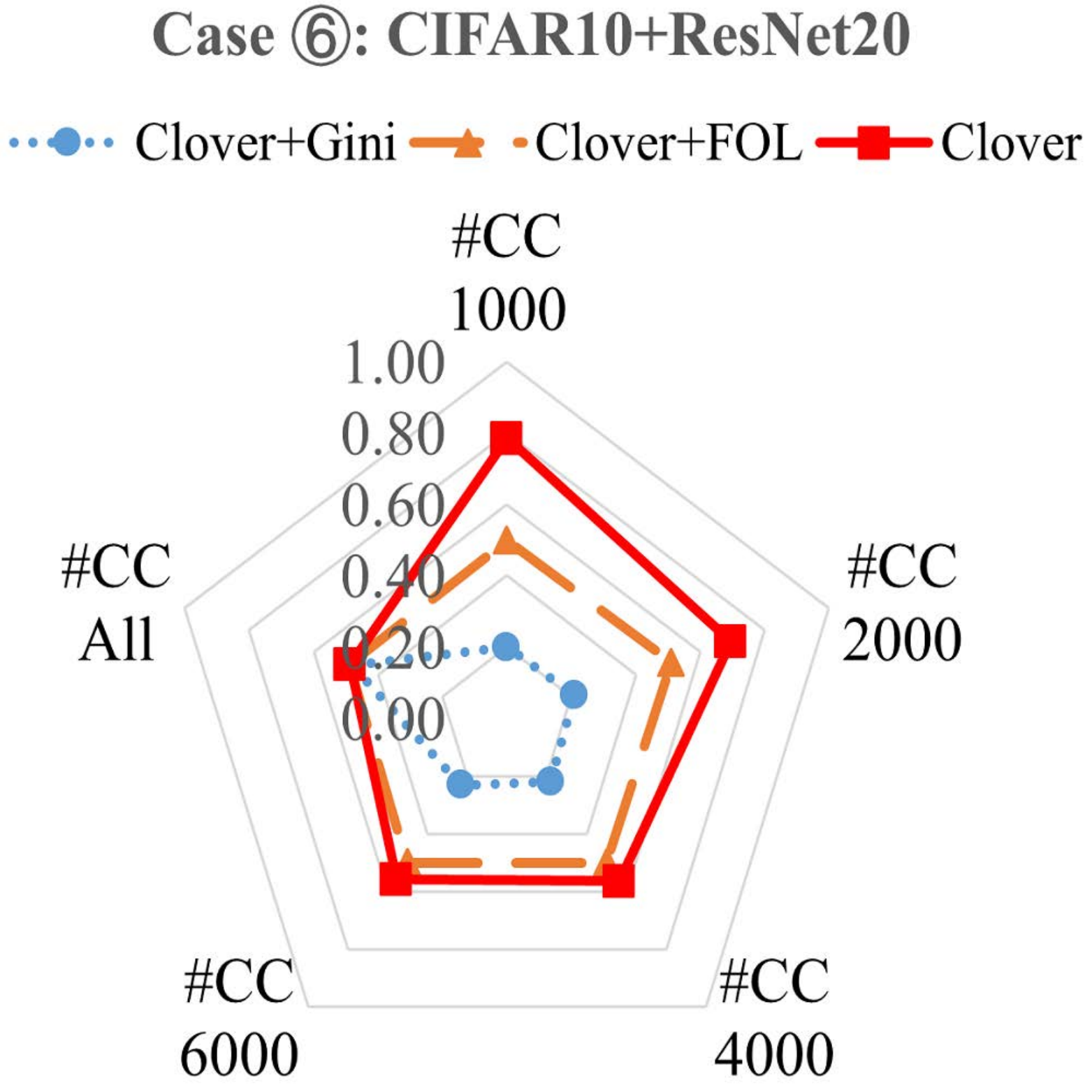}
\includegraphics[width=0.24\textwidth]{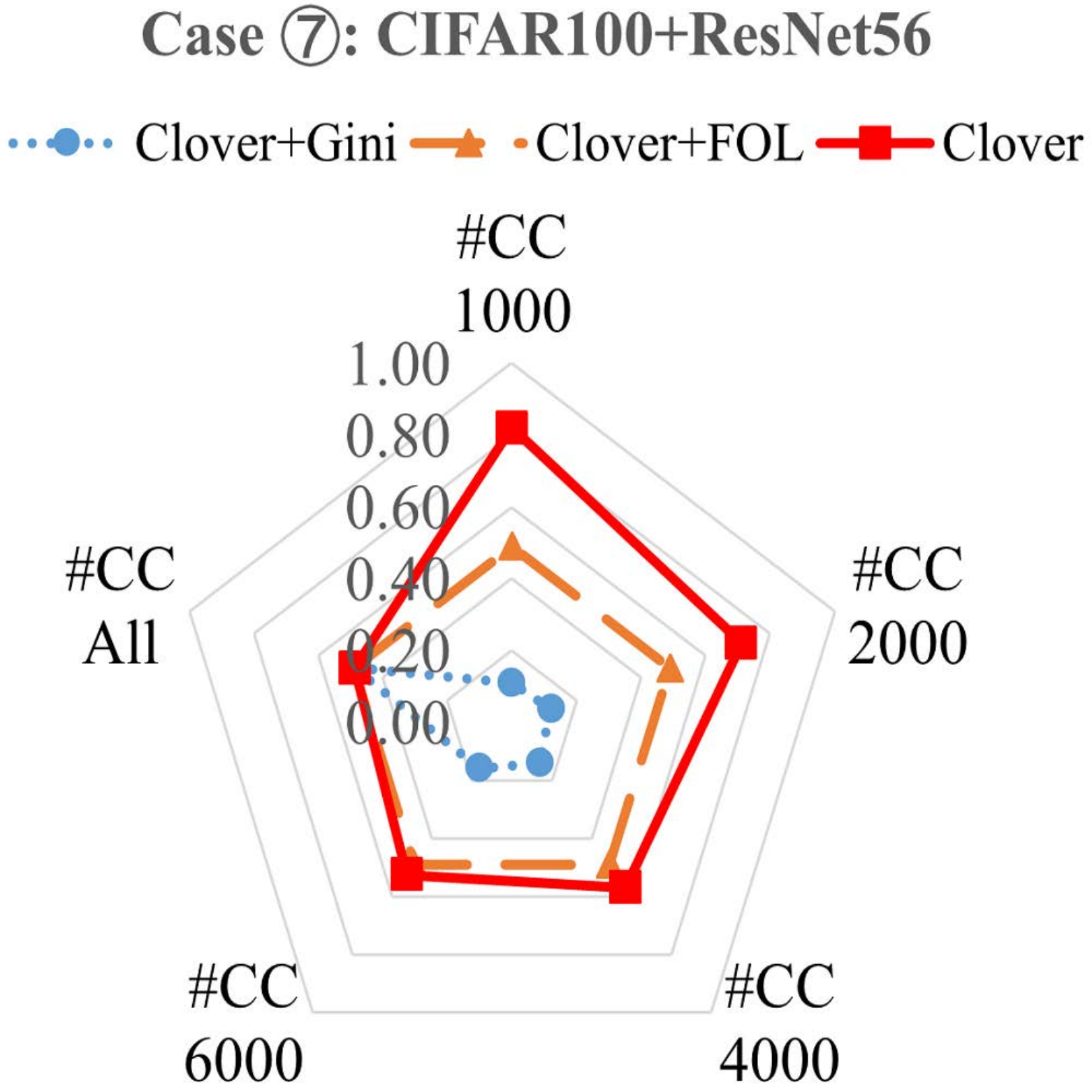}
\includegraphics[width=0.24\textwidth]{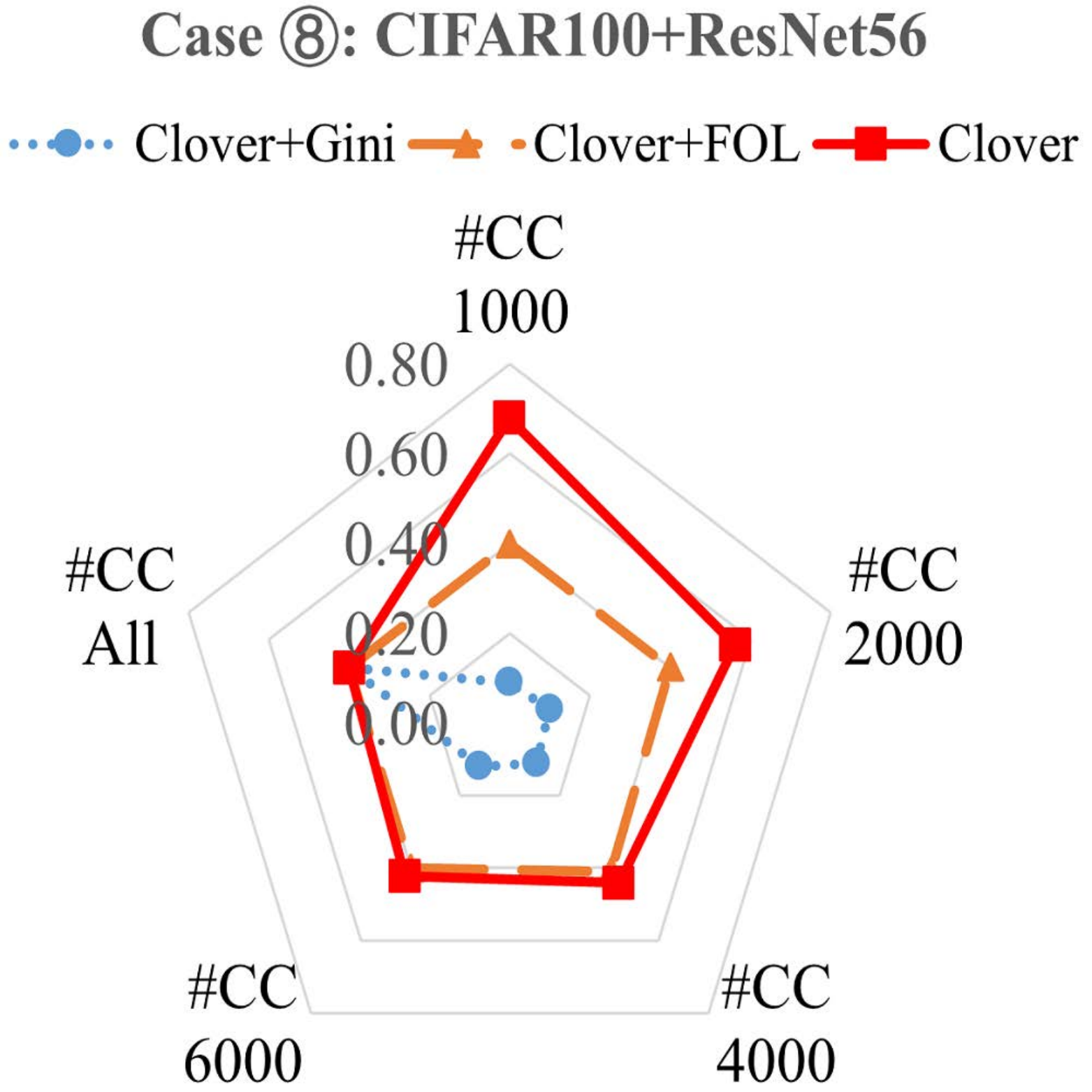} \\
 \caption{\textit{\#CC} for \textsc{Clover} and Its Variants on Adversarially Trained Models with $n_4 = 18000$}
\label{fig: rq6_statistics_cc_variants}
\end{figure}

Fig. \ref{fig: rq6_statistics_advlabel_category_variants} and \ref{fig: rq6_statistics_cc_variants} summarize the results of Experiment 6d in \textit{\#AdvLabel}, \textit{\#Category} and \textit{\#CC} for \textsc{Clover} and its two variants, i.e., \textsc{Clover+Gini} and \textsc{Clover+FOL}. 
Readers can interpret the axes of the charts in these two figures like these in Fig. \ref{fig: rq3_statistics_advlabel_category} and \ref{fig: rq3_statistics_cc}, respectively. 
We also copy \textsc{Clover}'s results from Figures \ref{fig: rq6_statistics_advlabel_category} and \ref{fig: rq6_statistics_cc} to ease the comparison between Experiments 6a--6c and 6d.
We note that the values achieved by \textsc{Clover}'s variants in each axis in each radar chart can be found in Table \ref{tab: rq6_fuzz_adv_trained_model_test_suites_variants} and \ref{tab: rq6_fuzz_adv_trained_model_universe_variants} in Appendix \ref{appendix: rq3_fuzz_clean_model_detail}. 

From Fig. \ref{fig: rq6_statistics_advlabel_category_variants}, we observe that the regions enclosed by the data points of \textsc{Clover} are always the largest in all four charts. The enclosed regions for \textsc{Clover+Gini} are always the smallest, and that for \textsc{Clover+FOL} is always in between the former two.

At each of the axes of $n_3$=1000, 2000, 4000, and 6000, \textsc{Clover+FOL} achieves a smaller value than \textsc{Clover} in each of \textit{\#AdvLabel}, \textit{\#Category}, and \textit{\#CC}.
Specifically, for cases \circled{5} to \circled{8}, in terms of \textit{\#AdvLabel}, \textsc{Clover+FOL} produces fewer unique adversarial labels than the original \textsc{Clover} by 102, 126, 127, and 150 for $n_3$ = 1000, by 368, 477, 490, and 553 for $n_3$ = 2000, by 1370, 1576, 1569, and 1798 for $n_3$ = 4000, and by 2262, 1660, 2240, and 1835 for $n_3$ = 6000, respectively. 
The differences are large. 
In terms of \textit{\#Catagory}, their differences are similar to those in \textit{\#AdvLabel}. Readers may refer to Table \ref{tab: rq6_fuzz_adv_trained_model_test_suites_variants} for the details.
On average, in terms of \textit{\#CC}, \textsc{Clover} is $1.61\times$, $1.40\times$, $1.14\times$, and $1.36\times$ of \textsc{Clover+FOL}.
At the axes for $n_3=$ \textit{All} in the eight charts in Figures \ref{fig: rq6_statistics_advlabel_category_variants} and \ref{fig: rq6_statistics_cc_variants}, \textsc{Clover+FOL} produces 339, 335, 510, and 564 fewer unique adversarial labels in terms of \textit{\#AdvLabel} and 477, 323, 378, and 398 fewer unique categories in terms of \textit{\#Category} for cases \circled{5}--\circled{8}, respectively; the mean \textit{\#CC} values of \textsc{Clover} and \textsc{Clover+FOL} are similar.
\textsc{Clover+Gini} produces an order of magnitude fewer unique adversarial labels in \textit{\#AdvLabel} and fewer unique categories in \textit{\#Catagory} for the four cases \circled{5}--\circled{8}. Its CC values are also lower than that of \textsc{Clover} by a large extent.
The result indicates that both variants are less effective than \textsc{Clover} in constructing test suites to a large extent in terms of test case diversity.

The mean \textit{\#AdvLabel}, the mean \textit{\#Category}, and the mean \textit{\#CC}, each across all combinations of the four cases (\circled{5}--\circled{8}) and $n_3 {\in} \{1000, 2000, 4000, 6000\}$ are 
2993.35, 2946.15, and 0.65 for \textsc{Clover}, 
283.25, 231.19, and 0.17 for \textsc{Clover+Gini}, and
1949.38, 1868.94, and 0.50 for \textsc{Clover+FOL}, respectively.
They indicate that \textsc{Clover} produces test cases with wider ranges of unique adversarial labels and unique categories by $10.57\times$ and $12.74\times$ compared to \textsc{Clover+Gini} and by $1.54\times$ and $1.58\times$ compared to \textsc{Clover+FOL} across the benchmarks on average.
Comparing the results of \textsc{Clover}'s two variants with the results of \textsc{Adapt} and \textsc{RobOT} presented in Section \ref{sec: rq6_results_a_to_c}, \textsc{Clover+FOL} also outperforms each of \textsc{Adapt} and \textsc{RobOT} in \textit{\#AdvLabel} and \textit{\#Category}.

\begin{tcolorbox}[
enhanced, breakable,
attach boxed title to top left = {yshift = -2mm, xshift = 5mm},
boxed title style = {sharp corners},
colback = white, 
title={Answering RQ6}
]
\textsc{Clover} in Configuration \textit{B} outperforms \textsc{Adapt} and \textsc{RobOT} on fuzzing adversarially trained models in terms of the numbers of unique adversarial labels and unique categories by $1.72\times$--$4.41\times$ and $1.97\times$--$6.42\times$ for \textsc{Adapt}, and $3.98\times$--$14.69\times$ and $4.40\times$--$16.59\times$ for \textsc{RobOT}, respectively. 
\textsc{Clover} is more effective in test case generation on fuzzing adversarially trained models than its variant configured with Gini or FOL as the guided metric. The variant of \textsc{Clover} with FOL is also more effective than \textsc{Adapt} and \textsc{RobOT} in terms of \textit{\#AdvLabel} and \textit{\#Category}.
\end{tcolorbox}

\section{Threats to Validity}
\label{sec: threats_to_validity}
\textit{Threats to Internal Validity:}
We adopt the original implementations of \textsc{Adapt} and \textsc{RobOT} and port them to our framework.
The implementation of \textsc{Clover} is developed on top of the implementation of \textsc{RobOT} straightforwardly.
Our test framework, the original implementations \cite{adapt_github, robot_github} of \textsc{Adapt} and \textsc{RobOT}, and the tools to create and run these implementations may contain bugs. We have carefully tested our framework and inspected the code. 
The experiment uses representative datasets \cite{fashion_mnist, svhn, cifar}, model architectures \cite{very_deep_cov, lenet5, resnet20}, attackers \cite{fgsm, pgd}, and top-1 accuracy, which existing experiments for DL testing, maintenance, and robustness improvement \cite{robot, quote_robot_extend, deephunter, tensorfuzz, npc} widely use them to create the models under test and generate test cases from them. The hyperparameter values are typical for these models, datasets, and attack techniques.
We follow \cite{adaptfuzz} and \cite{robot} to specify the parameters used by \textsc{Adapt} and \textsc{RobOT}, respectively. If there is still any unclear part, we follow their original implementations \cite{robot_github, adapt_github}.

Table \ref{tab: retrained_model_std_acc} shows the mean test accuracy of the retrained models produced by \textsc{DeepGini}, \textsc{be-st}, \textsc{km-st} and \textsc{Clover} reported in Fig. \ref{fig: rq1_1},
and produced by \textsc{Adapt}, \textsc{RobOT} and \textsc{Clover} reported in Table \ref{tab: rq3_fuzzing_compare_by_time}. Each mean test accuracy is similar to the test accuracy of the model under test (see Table \ref{tab: clean_model}).
We do not observe an abnormality from the table.

\begin{table}[t]
\caption{Mean Test Accuracy of Retrained Models Generated by Different Techniques}
\label{tab: retrained_model_std_acc}
\resizebox{\textwidth}{!}{
\begin{tabular}{|l|c|c|c|c|c|c|c|}
\hline
\multirow{3}{*}{Benchmark Case} & \multicolumn{7}{c|}{Test Accuracy} \\ \cline{2-8}
& \multicolumn{4}{c|}{Configuration \textit{A}} & \multicolumn{3}{c|}{Configuration \textit{B}} \\ \cline{2-8}
& DeepGini & \textsc{be-st} & \textsc{km-st} &\textsc{Clover} & \textsc{Adapt} & \textsc{RobOT} &\textsc{Clover} \\ \hline
Case \circled{1}: FashionMnist+VGG16 & 93.61 & 93.62 & 93.66 & 93.66 & 93.56 & 93.59 & 93.60 \\ \hline
Case \circled{2}: SVHN+LeNet5 & 89.28 & 88.87 & 89.27 & 89.09 & 89.40 & 89.24 & 89.53 \\ \hline
Case \circled{3}: CIFAR10+ResNet20 & 87.52 & 87.48 & 87.58 & 87.60 & 87.53 & 87.77 & 87.56 \\ \hline
Case \circled{4}: CIFAR100+ResNet56 & 60.99 & 60.70 & 60.78 & 61.04 & 60.57 & 60.51 & 60.59 \\ \hline
\end{tabular}
}
\end{table}

\textit{Threats to Construct Validity:}
Following the literature on fuzzing DL models \cite{robot, quote_robot_extend, sensei, deepxplore, fuzzing_roadmap}, we focus on evaluating the robust accuracy improvement on clean models because the literature on techniques to fuzz DL models extensively and primarily use this metric to evaluate the effectiveness of fuzzing techniques \cite{robot, quote_robot_extend, sensei, deepxplore, fuzzing_roadmap} and, to our best knowledge, a vast majority of, if not all, existing work in the fuzzing literature exclusively evaluates them on clean models. 
We have also extended the data analysis to include additional metrics ($\#AdvLabel$ and $\#Category$) to evaluate the test case generation dimension without retraining the models under fuzzing \cite{adaptfuzz, deephunter, ATS2022, drfuzz}. 
As we will discuss below, we also count the number of test cases generated by each fuzzing technique for each model under fuzzing.
We have not evaluated fuzzing techniques using coverage criteria (e.g., neuron coverage) because there is still lacking evidence to show that these criteria strongly correlate to the robustness improvement in the literature, and \textsc{Clover} is not a coverage-based technique.
Having said that, the experiment also measures the effects of different fuzzers on adversarially trained models.

We use a paired test for hypothesis testing whenever applicable, and if not, we use an unpaired test. We use Cohen's $d$ to measure the effect size.
The Wilcoxon signed-rank test, Mann-Whitney U test, Spearman's correlation coefficient, and Cohen's $d$ are statistical test methods that are widely used in software engineering experiments.
Using other test methods may obtain other test results.

Due to the small number of generated test cases produced by \textsc{Adapt} and \textsc{RobOT}, we only analyze the result for Experiment 6a--6c with $n_4=18000$ because  \textsc{Adapt} and \textsc{RobOT} cannot produce test pools with sufficient test cases to select at least $n_3$ test cases for many combinations of $n_3$ and $n_4$ where $n_3$ = 1000, 2000, 4000, and 6000  and $n_4$=1800, 3600, and 7200 to make the comparison with \textsc{Clover} fair.

In our experiment, even to select 6000 test cases for the largest pool of test cases generated by \textsc{Clover}, which contains 0.47 million test cases for case \circled{2} (see Table 9), the time cost is less than 10 seconds, less than 0.056\% compared to the fuzzing time of 18000 seconds. 

\textit{Threats to External Validity:}
We have extended the evaluation to include adversarially trained models as the models under fuzzing. 
As discussed in Section~\ref{sec:discussion-adv-models}, we have attempted to include pretrained models as subjects. Due to the weaknesses of peer techniques in our experiment, which could not generate any test cases from them within the time budget affordable to us, we have excluded them in our data analysis to facilitate comparisons among fuzzing techniques on the same ground. In our pre-experiment, the original \textsc{Sensei} tool \cite{sensei} was found to be very computationally expensive and was 20x slower than \textsc{Clover} for case \circled{2} and cannot converge in our experimental setting for example. 
We also found that if keeping the resulting clean test accuracy within 1\% of the original clean test accuracy, the robust accuracy improvement achieved by \textsc{Sensei} was similar to \textsc{Random}. So, we do not include \textsc{Sensei} in our evaluation.

We only apply a typical adversarial retraining procedure in the testing-retraining pipeline and only apply one such pipeline adopted from \cite{robot} to conduct our experiments. 
We have also not evaluated the fuzzing and metric-based techniques in other adversarial training settings, such as \textsc{GradAlign} \cite{fast_ae_train}, which is a regularization strategy over the backpropagation for seeds rather than test cases to incorporate the concept of adversarial training into standard training. Applying \textsc{Clover} requires a further investigation on how to model the selected or generated test cases of higher CC values back to their regularization processes. 
We leave the generalization of the experiment in this aspect as future works.

The baseline models to apply a testing-retraining pipeline are obtained through standard training by training from scratch. An alternative way to obtain an adversarially trained model is to train such a model with adversarial training directly. 
Our experiment has not compared the robustness of the models delivered through the standard training followed by the testing-retraining pipeline with such adversarially trained models. 
To compare them, a way is to integrate \textsc{Clover} to generate test cases in each epoch of an adversarial training procedure and compare it with other adversarial training methods.
We leave the comparison in a future work.

We have repeated Experiments 3 and 6a--6c with repeated trials. We have found that the result of individual repeated trials is within 1\% (in \textit{\#AdvLabel}, \textit{\#Category}, \textit{\#CC} and robust accuracy improvement) of the result reported in Sections \ref{sec: rq3_result} and \ref{sec: rq6_results_a_to_c}.
We have also conducted smaller experiments in answering the other RQs due to our limited efforts and the scale of the experiment by repeating them with different (but not all) combinations of $n_1$, $n_2$, $n_3$, and $n_4$ in the respective experiments. 
Moreover, many existing testing works \cite{dlfuzz, robot, adaptfuzz, deepgini, mode, surprise_adequacy, deepgauge} also run their experiments once.
Having said that, conducting more repeated trials for each experiment may obtain different results.

The experimental results may be different and more generalizable if the experiment includes more datasets, more model architectures, more peer techniques, more variety of testing-retraining pipelines, more types of model retraining, wider ranges of machine learning hyperparameters and $n_1$ to $n_4$, and more variety of seed lists, selection universes, and benchmark datasets.
Moreover, line 4 of Algorithm \ref{alg: context_fuzz} and lines 5 and 7 in Algorithm \ref{alg: context_translate} can be easily configured to use another attacker technique to perturb samples.
Similar to the extension \cite{quote_robot_extend} of \textsc{RobOT}, extending \textsc{Clover} to improve the other quality attributes of DL models via testing in the testing-retraining pipeline could be interesting. We have such an extension to future work. 

We have evaluated the fuzzing techniques on DL models under test with the same architecture and trained on the same task with different degrees of test and robust accuracy with limited variety.
In the literature on fuzzing techniques, we are not aware of experiments to evaluate a fuzzing technique in this setting. Thus, our evaluation using adversarially trained models as models under fuzzing is more like a kind of exploratory study.
It seems to us that to support such an evaluation, a novel methodology to systematically produce and sample such a representative set of models could be of interest. We leave the formulation of the methodology and the corresponding comprehensive evaluation as future work. 

The three selection universes may differ in robustness-oriented quality due to different implementations and parameters.
The values of \textit{\#AdvLabel}, \textit{\#Category}, and \textit{\#CC} for $P_{train}^{\textsc{FGSM+PGD}}$ are 
57686, 51679, 0.45 for \circled{1}, and
65072, 56463, 0.89, for \circled{2}, and
47062, 45294, 0.88, for \circled{3}, and
45579, 33944, 0.71, for \circled{4}, respectively. 
The values of \textit{\#AdvLabel}, \textit{\#Category}, and \textit{\#CC} for $P_{train}^{\textsc{Adapt}}$ are 
4001, 1435, 0.83 for \circled{1}, and
11220, 1607, 0.79, for \circled{2}, and
4304, 1538, 0.87, for \circled{3}, and
3130, 675, 0.66, for \circled{4}, respectively.
The values of \textit{\#AdvLabel}, \textit{\#Category}, and \textit{\#CC} for $P_{train}^{\textsc{RobOT}}$ are 
9808, 6225, 0.69, for \circled{1}, and
10358, 8461, 0.84, for \circled{2}, and
8313, 5702, 0.92, for \circled{3}, and
3993, 2704, 0.85, for \circled{4}, respectively.
We tend to believe that they present different types of scenarios.

The result of the experiment may be different due to the implementation of different fuzzing and metric techniques.
Different fuzzing techniques generate different numbers of test cases within the same time budget. 
Selecting the same number of test cases ($n_3$) from different test pools of different sizes may affect the results.

We observe that \textsc{Clover} is more efficient than \textsc{Adpat} and \textsc{RobOT} as summarized in Table \ref{tab: throughout}, \textsc{Clover} achieves higher throughputs than \textsc{Adpat} and \textsc{RobOT} in generating test cases.
We conjecture that this relatively higher efficiency will be retained when fuzzing on other models. 
However, more experiments should be conducted to accept or reject the conjecture.
We have evaluated the key hyperparameters in answering RQ5 and found that the key hyperparameters $m$, $k$, $\delta$, and $p$-norm in the algorithm do not significantly affect the performance of \textsc{Clover}.
By factoring out the difference in metric adoption and these key hyperparameters, it appears to us that the main difference between \textsc{Clover+FOL} and \textsc{RobOT} for \textsc{Clover+FOL} to generate more test cases is the algorithmic design. 
In the experiment for answering RQ6, configuring \textsc{Clover} with CC has a higher throughput than its two variants, but the differences are not as drastic as their results on clean models.
We leave the study on the integration between \textsc{Clover} and other test case selection metrics as future work.

\begin{table}
    \centering
    \caption{Numbers of Test Cases Generated in 18000 Seconds. \textsc{Clover} and its Two Variants Achieve Higher Rates of Successfully Generating Test Cases}
    \label{tab: throughout}
    \begin{tabular}{|c|c|r|r|r|r|}
    \hline
    Model Type & Technique & Case \circled{1} & Case \circled{2} & Case \circled{3} & Case \circled{4}  \\ \hline
    \multirow{5}{*}{Clean}  & \textsc{Adapt} & 21044 & 205003 & 61417 & 24251 \\ 
     & \textsc{RobOT} & 45349 & 106040 & 45833 & 18602 \\ \cline{2-6}
     & \textsc{Clover+Gini} & 189367 & 531735 & 214958 & 95560 \\
     & \textsc{Clover+FOL} & 91187 & 218330 & 130081 & 27741 \\ 
     & \textsc{Clover} & 111894 & 471795 & 144247 & 54763 \\\hline \hline
    
    Model Type & Technique & Case \circled{5} & Case \circled{6} & Case \circled{7} & Case \circled{8}  \\ \hline
    \multirow{5}{*}{Adversarial}  & \textsc{Adapt} & 87590 & 68530 & 41832 & 30041 \\
     & \textsc{RobOT} & 18834 & 19214 & 7694 & 7075 \\ \cline{2-6}
     & \textsc{Clover+Gini} & 171350 & 154259 & 75139 & 67573 \\
     & \textsc{Clover+FOL} & 161293 & 145520 & 71255 & 68855 \\
     & \textsc{Clover} & 173842 & 160530 & 78874 & 70237 \\ \hline
    \end{tabular}
\end{table}

In Configuration $A$, test cases are generated by FGSM/PGD and fuzzing techniques over the whole training dataset.
For instance, the selection universe generated by FGSM/PGD contains 100000 test cases.
From Table \ref{tab: throughout}, there are only limited entries producing at least 100000 test cases.
At the same time, in Configuration $B$, owing to the scale of the experiment and the lower throughputs of fuzzing techniques, we cannot scale  $n_3$, $n_4$, and the sizes of the input seed lists to all fuzzing techniques to cover larger ranges due to our hardware platform and human effort constraints.
For instance, for case \circled{1}, to make \textsc{Adapt} to produce 100000 test cases for the whole training dataset, we estimate it to take \textit{test case ratio $\times$ time spent $\times$ repeated trials} = $\nicefrac{100000}{21044} \times 18000 \times 3 = 0.26$ million seconds. 
We leave the study using a larger-scale experiment as future work. 

Our experiment has not evaluated \textsc{Clover} on regression models. 
We leave the evaluation of regression models as future work after generalizing the current notion of seed equivalence (based on discrete class label pairs) to cover seed equivalences over a real number range. 

In the experiment, most of the test suites produced by the DL testing technique are combined with the retraining task without triggering AM to perform test suite reduction. The results may be different if there is a test suite reduction task in between the test suite construction task and the retraining task.

\section{Related Work} 
\subsection{Deep Learning Testing Metrics}
In the literature, many metrics have been proposed to assess DL models. 

Diverse structural coverage criteria for assessing the test adequacy of neural network models have been proposed.
For brevity, we classify them into the following categories.

The first category is for coverage criteria covering a broad range of feature-map elements (e.g., Neuron Coverage \cite{deepxplore}, Neuron Boundary Coverage \cite{deepgauge}, $k$-multisection Neuron Coverage \cite{deepgauge},  Strong Neuron Activation Coverage \cite{deepgauge},  Modified Condition/Decision Coverage \cite{deepconclic}, top-$k$ Neuron Coverage \cite{deepgauge}, top-$k$ Neuron Patterns \cite{deepgauge}).
These criteria aim to assure the forward passes of DL models under test against the possible use and non-use of neurons, which are white-box.

Our CC does not apply to this assurance scenario and is not a coverage criterion (or test adequacy criterion). 
Rather than applicable to any forward pass, it targets assessing the output effects of a set of forward passes centric around individual test cases.
It is a black-box technique because each perturbation added to the test case is obtained through uniform sampling on each dimension of the input feature vector of the test case and only measures the prediction output induced by the forward pass.
On the other hand, \textsc{Clover} has a notion of even spreading the test case selection from different seeds in Algorithm \ref{alg: context_selection}. 
However, it does not enforce any notion of test adequacy, such as terminating the test case selection procedure if all seeds have at least one test case in the constructing test suite.
It also does not judge whether a constructed test suite covering more seeds is more adequate.

The second category is for coverage criteria covering elements exhibiting values in the outliner activation ranges (e.g., likelihood-based Surprise Coverage \cite{surprise_adequacy}, and distance-based Surprise Coverage \cite{surprise_adequacy}).
They compute the statistical distributions of activation values of hidden layers on a large dataset and identify specific neurons that once exhibit activation values as outliers of the computed distributions.

Unlike them, our CC metric (see Eq. (\ref{eq: cc_short})) computes the alignment rather than outliners and involves no activation value.
It is not used as a covering criterion. Furthermore, \textsc{Clover} does not use it to formulate the requirements on the coverage items. 

Moreover, recent empirical studies \cite{robot, quote_robot_extend, is_nc_useful, structual_coverage, coverage_model_quality} show that test suites fulfilling these structural coverage criteria are not correlated with exposing the failures of DL models or improving the robustness after retraining with test suites fulfilling a high degree of these adequacy criteria.
Moreover, these criteria are either too easy or too hard to satisfy \cite{deepgini, is_nc_useful}.

The third category is for these criteria having the concept of explainable Artificial Intelligence (AI), such as \cite{npc}.
They can provide a white-box witness, such as a coverage path across the principal sets of neurons in different hidden layers in a forward pass \cite{npc}, for a sample to facilitate developers to understand the core internal pathway going through in the forward pass for a sample.
The work \cite{npc} also shows that the average ratios of such principal sets of neurons in hidden layers to the set of all neurons in the respective layers are different to some (small to moderate) extent between benign and test cases.

CC neither generates any witness nor explains the internal malfunctions of DL models.
It does not distinguish between benign samples and test cases in assessing test cases.

Another line of research in DL model metrics is assessing samples' relative quality. 
It has been widely applied to sample prioritization for labeling cost reduction or robustness improvement~\cite{deepmutation, deepmutation++, effimap, deepgini, prima}.
DeepGini \cite{deepgini} prioritize a test suite through the Gini index \cite{theoretical_comparison} and select the top section of the prioritized list for labeling and robustness improvement. 
The first-order loss \cite{robot} measures the extent of a test case correlates with the robustness of DL models. 
It deems the current state of a sample with a smaller loss value from the last state higher in quality.
They are representative and state-of-the-art (metric-based) black-box and white-box (loss-based) metrics, respectively. 
Our experiment has extensively compared these two metrics with CC.
Mutation-based metrics~\cite{prima, effimap} require a higher computational overhead to compute a mutation score or its variant for each sample.
Fuzzing easily generates tens of thousands of test cases.
It is interesting to measure the mutation scores of \textsc{Clover}'s generated test suites and prioritize the test cases therein by mutation-based metrics.
We leave the empirical study as future work.
The unique adversarial label \cite{adaptfuzz, ATS2022} and unique category \cite{adaptfuzz, deephunter} are another two metrics measuring the quality of generated test suites, and we used them to evaluate \textsc{Clover} and peer techniques in our experiments.

\subsection{Fuzzing Deep Learning Models}
Many fuzzing techniques to generate test cases for the purpose of robustness improvement of DL models via testing-retraining pipelines have been developed. 

A vast majority of them \cite{deephunter, tensorfuzz, dlfuzz, adaptfuzz} focus on the application of structural coverage criteria, such as neuron coverage in Tensorfuzz \cite{tensorfuzz}, DeepXplore \cite{deepxplore}, DLFuzz \cite{dlfuzz}, $k$-multisection neuron coverage, neuron boundary coverage, and other multi-granularity coverage criteria in DeepHunter \cite{deephunter}.
Their fuzzing processes are similar to one another and largely greedy strategies to maximize coverage while sequentially evolving the current state of a sample iteratively.
\textsc{Adapt} adopts a genetic algorithm strategy to adaptively select a small and focus group of coverage items to guide generating perturbed samples from each seed. It is a representative coverage-based and adaptive fuzzing technique.
Another type of technique is mutation-based metrics \cite{duo_differential_fuzzing, effimap} or mutation-based fuzzing techniques \cite{deepmutation, deepmutation++, prima}. They generate model mutants and perturbed samples in the fuzzing process to prioritize test cases and thus slow.
\textsc{Clover} evolves the states of each seed sequentially.
Unlike them, \textsc{Clover} guides its novel fuzzing process on each seed by the novel notion of seed equivalence and shares the knowledge of the adversarial front among different fuzzing rounds on different seeds.

Since structural coverage criteria may not correlate to robustness improvement, newer fuzzing techniques \cite{robot, quote_robot_extend, sensei} explore the loss aspect (against a loss function) among the variants of each seed.
\textsc{RobOT} \cite{robot} evolves each seed toward the direction of smaller first-order loss values.
Our experiment in answering RQ4 shows that \textsc{RobOT} does not outperform \textsc{Random}. 
\textsc{Sensei} \cite{sensei} generates samples toward a larger entropy loss. 
In each retraining epoch, the publicly available \textsc{Sensei} tool generates hundreds of variants for each seed and selects merely the single variant with the largest loss.
Like adversarial training, always finetuning a model state with new samples with the largest losses in each training epoch makes the model difficult to converge.
It is interesting to further enhance \textsc{Clover} into a loss-based technique by considering whether the CC among the test cases produced in the same round of \textit{ContextTranslate} converges. We leave it as future work.

A related pipeline for the testing-retraining pipeline is the testing-repair pipeline, in which the repair task is conducted by a DL model maintenance/deubgging technique.
Many maintenance techniques to repair DL models have been proposed, which include retraining-based techniques (e.g., MODE \cite{mode}, DeepFault \cite{deepfault}, and DeepRepair \cite{deeprepair}), direct-manipulation techniques (e.g., Apricot \cite{apricot}, Arachne \cite{arachne}, and Provable Polytope Repair \cite{provable_repair}), and techniques making structural changes to the models under test (e.g., DeepCorrect \cite{deepcorrect}, DeepPatch \cite{deeppatch}, and PatchNAS \cite{PatchNAS}).
We leave the study on the testing-repair pipeline via \textsc{Clover} as future work.

\subsection{Retraining in Testing-Retraining Pipeline}
In our experiment, we measure the robust accuracy of a retrained model output by a testing-retraining pipeline, in which a DL testing technique is configured to produce a test suite for the retraining subtask.  

In the testing-retraining pipeline, the retraining task with adversarial examples is a kind of adversarial retraining \cite{adversarial_training_1, adversarial_training_2}, which aims to retrain the given DL model with the given adversarial examples for several epochs.
In general, these adversarial examples can be generated by many types of techniques or collected elsewhere rather than generation. 
Like the general adversarial training, retraining can be configured with automated test case generation of adversarial examples (e.g., applying $PGD$, $FGSM$, or $C\&W$), but is slow.
For instance,
Shafahi et al. \cite{ae_train_free} 
argue that many adversarial training techniques are time-consuming and not applicable to large-scale datasets. 
They propose to reuse the gradients with respect to the model's parameters computed on the backward propagations in the perturbation of inputs at the same pass to produce test cases, which trade the efficiency between adversarial training (compared to PGD-based ones) and standard training.
Wong et al. \cite{fast_better_free} target to speed up adversarial training and propose a faster method for FGSM-based adversarial training attacking combined with random initialization, which is as effective as PGD-based adversarial training with lower cost.
Adversarial training methods mentioned above may result in models with the catastrophic overfitting problem, and Andriushchenko and Flammarion \cite{fast_ae_train} propose GradAlign to address it through regularization, which explicitly maximizes the alignment between the gradient of the original inputs and its perturbation set.

Since \textsc{Clover} is not an adversarial retraining method, it has not yet considered the efficiency problem when embedding it into the general adversarial (re)training and the minimization problem to reduce the overall model loss with respect to test case generation.
Wang et al. \cite{robot} further express that DL testing is a strategy to complement adversarial training techniques by generating a more diverse set of adversarial examples.
A tight integration of a fuzzing technique, such as \textsc{Clover}, with such an adversarial retraining technique in a testing-retraining pipeline could be interesting.

\section{Conclusion}
To enhance the robustness property of DL models, a testing-retraining pipeline can be employed. 
In such a pipeline, a DL testing technique generates a test suite for the retraining task to retrain the given model under test. 
The validation of a DL model produced by the pipeline is left to a robust validation dataset independent of the test suites used for retraining.
This paper has proposed a novel context-aware fuzzing technique for such a pipeline.
\textsc{Clover} generates test cases by a novel seed-equivalent sequence-to-sequence fuzzing algorithm guided by a set of adversarial front objects corresponding to those representative historic test cases measured through our novel metric \textit{Contextual Confidence}.
It layers and prioritizes the pool of the generated test cases using the abovementioned metric to construct a test suite. 
The evaluation results have shown that \textsc{Clover} outperforms the state-of-the-art coverage-based fuzzing technique \textsc{Adapt} and the state-of-the-art loss-based technique \textsc{RobOT} in the same testing-retraining pipeline, in terms of the numbers of generated test cases, unique adversarial labels, and unique categories in the generated test suites with $2.5\times$, $2.0\times$, and $3.5\times$ and $3.6\times$, $1.6\times$, and $1.7\times$ of these of \textsc{Adapt} and \textsc{RobOT} on fuzzing clean models, respectively.
\textsc{Clover} also outperforms \textsc{Adapt} by $2.1\times$, $3.4\times$, and $4.5\times$ and outperforms \textsc{RobOT} by $9.2\times$, $9.8\times$, and $11.0\times$ on fuzzing adversarially trained models in these three measurement metrics, respectively.
\textsc{Clover} is also more effective than \textsc{Adapt} and \textsc{RobOT} in robustness improvements on clean models by 72\%--154\% and 58\%--127\%, respectively, in ratio.
It has achieved a significantly higher robust accuracy improvement by selecting test cases with higher values scored by its metric Contextual Confidence through the pipeline.
Moreover, configuring \textsc{Clover} with the current metric is more effective than configuring it with the metrics in each of \textsc{DeepGini} and \textsc{RobOT}.
Configuring \textsc{Clover} with the metrics adopted by \textsc{RobOT} is also more effective than \textsc{RobOT}.
The major future work includes designing an improved notion of seed equivalence with both successful and unsuccessful test case generation attempts and designing Contextual Confidence to be aware of the seed labels of test cases.

\section{Acknowledgments}
This research is partly supported by the CityU Grant with project number 9678180.

\bibliographystyle{ACM-Reference-Format}
\bibliography{main}

\newpage
\appendix
\section{Reviews on Peer Techniques Used in Our Experiment}
\label{appendix: A}
In this section, we revisit the peer techniques we have used in the experiment presented in the paper.

\subsection{Attack techniques}
FGSM \cite{fgsm} and PGD \cite{pgd} are well-known adversarial example generation techniques. 
FGSM finds an adversarial example $x'$ from a seed sample $x_0$ by increasing the value of the loss function through a single-step scheme: 
$a = x_0+ \epsilon {\cdot} \text{sgn}(\nabla_{x_0} L(f,x_0,c_g))$ where $\epsilon$ is the parameter to fit $a$ within the perturbation bound, $\nabla_{x_0}$ is the gradient of $x_0$, \textit{sgn} is the sign function (sgn($d) = 1, 0$, and $-1$ if $d > 0, d = 0$, and $d < 0$, respectively), and $L$ is the loss function of $f$ to compute the loss of $x_0$ against the label $c_g$.
\textbf{PGD} is an iterative version of FGSM.
It has more and finer attacking steps than FGSM starting from $x' \in x_0+\epsilon$ where $x_0+\epsilon$ is the $L_{\infty}$-ball around $x_0$. Then, it iteratively adds a small step size $\alpha$ of the change direction of a sample to that sample. 
Specifically, if $x_0+\epsilon$ is a closed convex set, the projection on $x_0+\epsilon$ (in the sense of projection operator $\prod_{x_0+\epsilon}(.)$ in convex optimization) is defined as $x^{t+1} = \prod_{x_0+\epsilon}(x^t + \alpha \cdot \text{sgn}(\nabla_{x^t} L(f,x^t,c_g)))$ and 
$\prod_{x_0+\epsilon}(x^{t+1}) = \mathop{\arg\min}_{x^{t+1} \in x_0+\epsilon} \frac{1}{2} {\parallel}x^{t+1} - x^{t} {\parallel}_{2}^{2}$, where $x^t$ is initialized as $x'$ at the starting point.

\subsection{Fuzzing techniques}
\textbf{\textsc{RobOT}} \cite{robot} is the state-of-the-art robustness-oriented testing framework.
It generates test cases from the same seed until their first-order losses (FOLs) converge. 
It then retrains the model under test with a subset $A$ of the generated test cases.

{\textsc{RobOT}} includes two test case selection techniques (\textsc{be-st} and \textsc{km-st}), a metric called first-order loss (FOL), a fuzzing algorithm (FOL-Fuzz), and a model retraining step. 
Suppose $x_0$ is a seed. 
The FOL value for a test case $x' \in x_0+\epsilon$ (which is the $\epsilon$-ball of $x_0$) is defined as FOL($x'$) = $\epsilon \cdot {\parallel}\nabla_{x_0} L(f,x',c_g){\parallel}_{2}$, which is generated by FOL-Fuzz.

FOL-Fuzz generates test cases iteratively, which can be expressed recursively as follows. Let $S^t$ and $x^t$ be the working list and working sample in the $t^{th}$ iteration, and $S^0 = \langle x_0 \rangle$.
Let C1($x^t$) denote the condition FOL($x^t$) larger than the maximal FOL value of the samples in $S^{t-1}$ and C2($x^t$) denote the condition FOL($x^t$) < $\xi$ (where $\xi$ = $10^{-18}$ in \cite{robot}).
$S^t$ is computed as $S^{t-1} + \langle x^t\rangle$ if either C1($x^{t}$) or C2($x^t$) is satisfied, otherwise $S^{t-1}$. $x^t$ is perturbed from the sample $S^{t-1}[0]$ along the gradient with respect to its proposed loss function \cite{robot}, which includes the FOL value FOL($S^{t-1}[0]$) as a term.
If either C1($x^t$) or C2($x^{t}$) is satisfied and the prediction label of $x^t \neq$ the prediction label of $x_0$, then $x^t$ is marked as a test case. FOL-Fuzz repeats the above process several times for each seed. \textsc{RobOT} then puts all the marked test cases of all seeds into a list $P$ followed by selecting a subset $A$ from $P$.

\textbf{\textsc{Adapt}} \cite{adaptfuzz} is the state-of-the-art coverage-based technique that uses a genetic algorithm to explore the coverage space.
It designs a set of neuron-level and activation-level features to measure the coverage of each test case. 
It proposes a genetic-algorithm-based fuzzing technique to adaptively select neurons for those designed features to evolve test cases towards higher coverage.

The details of \textsc{Adapt} are as follows.
A chromosome in \textsc{Adapt} is a vector of 29 real numbers within [-1, 1].
\textsc{Adapt} first generates a set $P$ of random chromosomes.
Let $x^t$ be a working sample in the $t^{th}$ iteration to generate a test case for a given seed $x_0$ where $x^0 = x_0$.
For each chromosome $p \in P$, \textsc{Adapt} computes a score as the dot product of $p$ and a feature vector of each neuron when the model $f$ predicts an output for $x^t$. 
(For this purpose, \textsc{Adapt} designs 29 neuron-level Boolean features, such as whether the neuron is located in the first 25\% layers under the measure, to produce the feature vector for the neuron.)
It then selects the top-$m$ (where $m=10$ in \cite{adaptfuzz}) neurons with the highest score and perturbs $x^t$ into $x^{t+1}$ against the loss of these $m$ neurons.
\textsc{Adapt} measures the neuron coverage achieved by $x^{t+1}$.
It next reduces $P$ to a minimal subset $S$ that retains the same coverage as $P$. If $|S|$ is smaller than a required threshold, chromosomes in $P$ covering most coverage items are added to $S$ until the threshold is met.
\textsc{Adapt} then iteratively (1) crossovers two randomly-picked chromosomes in $S$ followed by adding Gaussian noise to construct a new chromosome $A$ and (2) places $A$ into $S$ until $|S| = |P|$. 
The resultant $S$ is assigned to $P$, and $x^{t+1}$ becomes the working sample.
The iteration to process $P$ (the $(t+1)^{th}$ iteration for $x_0$) repeats until the fuzzing budget is exhausted. 
The original experiment \cite{adaptfuzz} shows that \textsc{Adapt} can achieve higher coverage than fuzzing without using the adaptive strategy.

\subsection{Metric-based techniques}
\textbf{\textsc{DeepGini}} is a metric-based test case prioritization technique.
Given a test case $x$ and the probability output predicted by a model ${\langle}p_{x,1}, p_{x,2}, \dots, p_{x,N} {\rangle}$, where $N$ is the number of classes and $\sum_{i=1}^{N}p_i = 1$, the Gini metric in \textsc{DeepGini} is denoted by $\xi(.)$, where $\xi(x)$ = $1 - \sum_{i=1}^{N}{p_{x,i}^2}$.
\textsc{DeepGini}  \cite{deepgini} interprets $\xi(.)$ as follows: 
(1) Given two test cases $x_1$ and $x_2$, ``$\xi(x_1) > \xi(x_2)$ implies that $x_1$ is more likely to be misclassified'', and
(2) ``The tests prioritized by \textsc{DeepGini} at the front are more effective to improve DNN quality than the tests prioritized at the back.''

\textbf{\textsc{be-st}} \cite{robot} generates $A$ with $n$ samples by reordering test cases in $P$ with their FOL values followed by selecting the top-$\frac{n}{2}$ and bottom-$\frac{n}{2}$ test cases from the reordered $P$. 
On the other hand, \textbf{\textsc{km-st}} \cite{robot} generates $A$ by equally dividing the reordered $P$ into $k$ sections and randomly picking $\frac{n}{k}$ samples from each section.
Finally, \textsc{RobOT} retrains the model with the original training dataset of $f$ and $A$.
In the experiment \cite{robot}, \textsc{RobOT} sets a short fuzzing time budget and a small iteration bound of 3 for $x$ to limit the number of generated test cases (to alleviate the loss in standard accuracy for less than 1\% as it retrains the original model with all generated test cases after selection). 

\subsection{Defender technique}
\textbf{\textsc{nic}} \cite{nic} proposes a technique to detect adversarial examples by formulating statistical invariants among hidden layers over benign samples and monitoring their violations. 
It extracts an internal state of each hidden layer and each pair of consecutive hidden layers in the forward passes of $f$ for benign samples.
It then develops machine-learning models for each such state and sets up an ensemble of these models to maximize the prediction probability of a sample being benign. 
It effectively detects a wide range of adversarial examples with low false positive rates in its experiment.

\textsc{nic} extracts the activation values $av_l(x)$ of each hidden layer $l$ in the model $f$ for each sample $x$ in a training dataset $D$. 
It then constructs a regression model VI($l$) for the layer $l$ by learning the weight $w$ of the VI($l$) model through solving $min_{x \in D} L(av_l(x) \cdot w^T -1)$.
It then constructs a submodel of $f$ for layer $l$ by trimming off all the layers after $l$ and appending the trimmed submodel with a softmax layer.
It only trains the softmax layer of this extended submodel on $D$ to produce a trained model $P_l$ for layer $l$. 
Similar to the construction of VI($l$), for each consecutive pair of hidden layers ${\langle}l, l+1{\rangle}$ in $f$, \textsc{nic} constructs a regression model PI($l$) by learning the weight $w$ of the PI($l$) model through solving $min_{x \in D} L([P_l(x) \ P_{l+1}(x)]\cdot w^T -1)$.
Finally, \textsc{nic} trains a one-class support vector machine \cite{osvm} (OSVM) model with the radial basis function kernel. 
The OSVM model takes the outputs of VI($l$) and PI($l$) for all hidden layers $l \in f$ on each sample in $D$ as an input sample. 
It learns to predict the similarity score of each sample in $D$ to 1 (benign).

In the detection time, if the output OSVM($x$) for a test case $x$ is smaller than a threshold, $x$ is deemed as an adversarial example.

\subsection{Neural Network Invariant for Abnormal State Detection}
\label{sec: nic}
\textbf{\textsc{nic}} \cite{nic} is an approach to producing a defender to detect adversarial examples by formulating a set of statistical invariants among the hidden layers over a set of samples and monitoring their violations through a one-class classifier. 
For each hidden layer $l$ in a given model $f$, \textsc{nic} extracts the activation values $av_l(x)$ for each sample $x$ in the training dataset $D$ of $f$. 
It trains a regression model $\textit{Inv}_l$ for every hidden layer $l$ of $f$ with $\cup_{x \in D}\, av_l(x)$ as the training dataset and a regression model $\textit{Pro}_{l,l+1}$ for every consecutive pair of hidden layers ${\langle}l, l+1{\rangle}$ of $f$ with $\cup_{x \in D} \langle \textit{Inv}_l(x), \textit{Inv}_{l+1}(x)\rangle $ as training dataset.
The outputs of all these regression models are used to train a one-class support vector machine (OVSM) \cite{osvm} with the radial basis function kernel to learn to predict the similarity score of each sample in $D$ to 1.
In the detection time, if the output of the OSVM with a test case $t$ as input is smaller than a threshold, $t$ is reported as an adversarial example. 
We refer to each regression model ($\textit{Inv}_l$ or $\textit{Pro}_{l,l+1}$) as a \textbf{neural network invariant}.
A sample violating more neural network invariants tends to deviate more from being benign. 
We refer to a violation of a neural network invariant as an exposure of an \textbf{abnormal neural network state}. 

If a model $f$ contains $R$ hidden layers, then \textsc{nic} must train and generate $2R - 1$ regression models (invariants for short) and one OSVM model. 
Thus, the training process and detecting whether a test case is an adversarial example are both slow.
Their experiment \cite{nic} shows that \textsc{nic} can correctly detect 92\% to 100\% adversarial examples on many models.
 
Intuitively, a stronger defender may be obtained by designing more precise kernel functions to cover desirable inputs with low regression values output by the invariants or strengthening the invariants.
As such, test suites that expose more abnormal neural network states provide more data points for more accurate estimations on the low regression value patterns for developers to design more expressive kernel functions or source desirable benign samples to improve the invariants.

Let $X$ be a set of seeds and $P$ be a test suite generated from $X$ to test a model $f$.
Each such test case in $P$ is within the $\epsilon$-ball of its corresponding seed measured in a $p$-norm distance.
Let $t$ and $t'$ in $P$ be two test cases perturbed from the same seed $x$ (denoted by $t \approx t'$). 
Suppose further $t'$ is in $\delta$-ball of $t$ (i.e., $\parallel t - t'\parallel_p < \delta$) where $\delta \ll \epsilon$.
Suppose we find a subset $A \subseteq P$ such that $A$ simulates $P$ in the sense that each sample $t'$ in the set $P - A$ is in the $\delta$-ball of the same sample $t$ in $A$, and the prediction vectors $\vec{f}(t)$ and $\vec{f}(t')$ are similar. 
If a defender (e.g., \cite{nic}) of $f$ can reject $t$, it has a good chance to reject $t'$; and a patched architecture of $f$ produced by a model maintainer (e.g., \cite{PatchNAS, deeppatch}) that can infer $t$ correctly may infer $t'$ correctly as well.

\newpage
\section{Detailed Results on Test Case Generation}
\label{appendix: rq3_fuzz_clean_model_detail}
The values depicted in Figures~\ref{fig: rq3_statistics_advlabel_category} and \ref{fig: rq3_statistics_cc} for answering RQ3 can be found in Tables \ref{tab: rq3_fuzz_clean_model_test_suites} and \ref{tab: rq3_fuzz_clean_model_universe}.
The values depicted in Figures \ref{fig: rq4_statistics_advlabel_category} and \ref{fig: rq4_statistics_cc} for answering RQ4 are shown in Table \ref{tab: rq4_fuzz_clean_model_variants_test_suites} and \ref{tab: rq4_fuzz_clean_model_variants_universe}.
The values depicted in Figures \ref{fig: rq6_statistics_advlabel_category_variants} and \ref{fig: rq6_statistics_cc_variants} for answering RQ6 are shown in Tables \ref{tab: rq6_fuzz_adv_trained_model_test_suites_variants} and \ref{tab: rq6_fuzz_adv_trained_model_universe_variants}.

\begin{table}[tbh]
\caption{Mean Results of Test Suites Generated by 3 Techniques on Fuzzing Clean Models in Configuration \textit{B} for 3 Runs with $N_3$ and $n_4=18000$} 
\label{tab: rq3_fuzz_clean_model_test_suites}
\resizebox{\textwidth}{!}{
\begin{tabu}{|l|c|c|c|c|c|c|c|}
\hline
\multirow{2}{*}{Benchmark Case} & \multirow{2}{*}{Technique} & \multicolumn{3}{c|}{$n_3$ = 1000} & \multicolumn{3}{c|}{$n_3$ = 2000} \\ \cline{3-8}
 &  & \textit{\#AdvLabel} & \textit{\#Category} & \textit{\#CC} & \textit{\#AdvLabel} & \textit{\#Category} & \textit{\#CC} \\ \hline 
\multirow{3}{*}{\begin{tabu}[l]{@{}c@{}}\circled{1}: FashionMnist+VGG16\end{tabu}}
 & \textsc{Adapt} & 780 & 669 & 0.81 & 1297 & 959 & 0.83 \\ 
 & \textsc{RobOT} & 917 & 896 & 0.68 & 1713 & 1633 & 0.68 \\ 
 &\textsc{Clover} & 1000 & 1000 & 0.99 & 2000 & 2000 & 0.99 \\ \tabucline[1.2pt]{1-8} 
\multirow{3}{*}{\circled{2}: SVHN+LeNet5}
 & \textsc{Adapt} & 770 & 604 & 0.79 & 1384 & 930 & 0.78 \\ 
 & \textsc{RobOT} & 797 & 789 & 0.84 & 1561 & 1542 & 0.84 \\ 
 &\textsc{Clover} & 1000 & 1000 & 0.99 & 2000 & 2000 & 0.99 \\ \tabucline[1.2pt]{1-8} 
\multirow{3}{*}{\circled{3}: CIFAR10+ResNet20} 
 & \textsc{Adapt} & 678 & 615 & 0.87 & 1207 & 999 & 0.87 \\ 
 & \textsc{RobOT} & 916 & 800 & 0.92 & 1474 & 1419 & 0.92 \\ 
 &\textsc{Clover} & 1000 & 1000 & 0.99 & 2000 & 2000 & 0.99 \\ \tabucline[1.2pt]{1-8} 
\multirow{3}{*}{\circled{4}: CIFAR100+ResNet56} 
 & \textsc{Adapt} & 658 & 467 & 0.65 & 1078 & 632 & 0.67 \\ 
 & \textsc{RobOT} & 772 & 729 & 0.85 & 1316 & 1182 & 0.86 \\ 
 &\textsc{Clover} & 1000 & 1000 & 0.99 & 2000 & 2000 & 0.99 \\ \tabucline[1.2pt]{1-8} 
\multirow{5}{*}{\textbf{\begin{tabu}[l]{@{}l@{}} Mean Results \end{tabu}}} 
 & \textsc{Adapt} & 721.50 & 588.75 & 0.78 & 1241.50 & 880.00 & 0.79 \\ 
 & \textsc{RobOT} & 850.50 & 803.5 & 0.82 & 1516.00 & 1444.00 & 0.82 \\ 
 & \textbf{\textsc{Clover}} & 1000.00 & 1000.00 & 0.99 & 2000.00 & 2000.00 & 0.99 \\  \cline{2-8}
 & \textbf{\textsc{Clover}$\div$\textsc{Adapt}} & 1.39 & 1.70 & 1.27 & 1.61 & 2.27 & 1.26 \\ 
 & \textbf{\textsc{Clover}$\div$\textsc{RobOT}} & 1.18 & 1.24 & 1.20 & 1.32 & 1.39 & 1.20 \\ \hline \hline

\multirow{2}{*}{Benchmark Case} & \multirow{2}{*}{Technique} & \multicolumn{3}{c|}{$n_3$ = 4000} & \multicolumn{3}{c|}{$n_3$ = 6000} \\ \cline{3-8}
 &  & \textit{\#AdvLabel} & \textit{\#Category} & \textit{\#CC} & \textit{\#AdvLabel} & \textit{\#Category} & \textit{\#CC} \\ \hline 
\multirow{3}{*}{\begin{tabu}[l]{@{}c@{}}\circled{1}: FashionMnist+VGG16\end{tabu}}
 & \textsc{Adapt} & 1890 & 1172 & 0.83 & 2359 & 1281 & 0.83 \\ 
 & \textsc{RobOT} & 2995 & 2740 & 0.69 & 4003 & 3517 & 0.69 \\ 
 &\textsc{Clover} & 4000 & 4000 & 0.99 & 6000 & 6000 & 0.99 \\ \tabucline[1.2pt]{1-8} 
\multirow{3}{*}{\circled{2}: SVHN+LeNet5}
 & \textsc{Adapt} & 2526 & 1261 & 0.79 & 3384 & 1415 & 0.78 \\ 
 & \textsc{RobOT} & 2810 & 2750 & 0.84 & 3747 & 3615 & 0.84 \\ 
 &\textsc{Clover} & 4000 & 4000 & 0.99 & 6000 & 6000 & 0.99 \\ \tabucline[1.2pt]{1-8} 
\multirow{3}{*}{\circled{3}: CIFAR10+ResNet20} 
 & \textsc{Adapt} & 1911 & 1314 & 0.87 & 2282 & 1410 & 0.87 \\ 
 & \textsc{RobOT} & 2513 & 2317 & 0.92 & 3432 & 3079 & 0.91 \\ 
 &\textsc{Clover} & 4000 & 4000 & 0.99 & 6000 & 6000 & 0.99 \\ \tabucline[1.2pt]{1-8} 
\multirow{3}{*}{\circled{4}: CIFAR100+ResNet56} 
 & \textsc{Adapt} & 1590 & 662 & 0.67 & 1861 & 672 & 0.66 \\ 
 & \textsc{RobOT} & 2093 & 1787 & 0.85 & 2551 & 2104 & 0.85 \\ 
 &\textsc{Clover} & 4000 & 4000 & 0.99 & 6000 & 6000 & 0.99 \\ \tabucline[1.2pt]{1-8} 
\multirow{5}{*}{\textbf{\begin{tabu}[l]{@{}l@{}} Mean Results \end{tabu}}} 
 & \textsc{Adapt} & 1979.25 & 1102.25 & 0.79 & 2471.50 & 1194.50 & 0.79 \\ 
 & \textsc{RobOT} & 2602.75 & 2398.50 & 0.82 & 3433.25 & 3078.75 & 0.82 \\ 
 & \textbf{\textsc{Clover}} & 4000.00 & 4000.00 & 0.99 & 6000.00 & 6000.00 & 0.99 \\  \cline{2-8}
 & \textbf{\textsc{Clover}$\div$\textsc{Adapt}} & 2.02 & 3.63 & 1.25 & 2.43 & 5.02 & 1.26 \\ 
 & \textbf{\textsc{Clover}$\div$\textsc{RobOT}} & 1.54 & 1.67 & 1.20 & 1.75 & 1.95 & 1.20 \\ \hline
\end{tabu}
}
\end{table}

\begin{table}[t]
\caption{Mean Results of \textit{All} Test Cases Generated by Three Techniques on Fuzzing Clean Models in Configuration \textit{B} for 3 Runs with $n_4=18000$} 
\label{tab: rq3_fuzz_clean_model_universe}
\resizebox{0.7\textwidth}{!}{
\begin{tabu}{|l|c|c|c|c|}
\hline
\multirow{2}{*}{Benchmark Case} & \multirow{2}{*}{Technique} & \multicolumn{3}{c|}{\textit{All}} \\ \cline{3-5}
 &  & \textit{\#AdvLabel} & \textit{\#Category} & \textit{\#CC} \\ \hline 
\multirow{3}{*}{\begin{tabu}[l]{@{}c@{}}\circled{1}: FashionMnist+VGG16\end{tabu}}
 & \textsc{Adapt} & 4001 & 1435 & 0.83 \\ 
 & \textsc{RobOT} & 9808 & 6225 & 0.69 \\ 
 &\textsc{Clover} & 23638 & 17457 & 0.96 \\ \tabucline[1.2pt]{1-5} 
\multirow{3}{*}{\circled{2}: SVHN+LeNet5}
 & \textsc{Adapt} & 11220 & 1607 & 0.79 \\ 
 & \textsc{RobOT} & 10358 & 8461 & 0.84 \\
 &\textsc{Clover} & 21777 & 15975 & 0.94 \\ \tabucline[1.2pt]{1-5} 
\multirow{3}{*}{\circled{3}: CIFAR10+ResNet20} 
 & \textsc{Adapt} & 4304 & 1538 & 0.87 \\ 
 & \textsc{RobOT} & 8313 & 5702 & 0.92 \\ 
 &\textsc{Clover} & 27407 & 17169 & 0.98 \\ \tabucline[1.2pt]{1-5} 
\multirow{3}{*}{\circled{4}: CIFAR100+ResNet56} 
 & \textsc{Adapt} & 3130 & 675 & 0.66 \\ 
 & \textsc{RobOT} & 3993 & 2704 & 0.85 \\ 
 &\textsc{Clover} & 20237 & 13596 & 0.87 \\ \tabucline[1.2pt]{1-5} 
\multirow{5}{*}{\textbf{\begin{tabu}[l]{@{}l@{}} Mean Results \end{tabu}}} 
 & \textsc{Adapt} & 5663.75 & 1313.75 & 0.79 \\ 
 & \textsc{RobOT} & 8118.00 & 5773.00 & 0.83 \\ 
 & \textbf{\textsc{Clover}} & 23264.75 & 16049.25 & 0.94 \\ \cline{2-5}
 & \textbf{\textsc{Clover}$\div$\textsc{Adapt}} & 4.11 & 12.22 & 1.19 \\ 
 & \textbf{\textsc{Clover}$\div$\textsc{RobOT}} & 2.87 & 2.78 & 1.13 \\ \hline
\end{tabu}
}
\end{table}

\begin{table}[thb]
\caption{Results of Test Suites Generated by \textsc{Clover} and its Variants on Fuzzing Clean Models in Configuration \textit{B} with $n_3 \in N_3$ and $n_4=18000$} 
\label{tab: rq4_fuzz_clean_model_variants_test_suites}
\resizebox{\textwidth}{!}{
\begin{tabu}{|l|c|c|c|c|c|c|c|}
\hline
\multirow{2}{*}{Benchmark Case} & \multirow{2}{*}{Technique} & \multicolumn{3}{c|}{$n_3$ = 1000} & \multicolumn{3}{c|}{$n_3$ = 2000} \\ \cline{3-8}
 &  & \textit{\#AdvLabel} & \textit{\#Category} & \textit{\#CC} & \textit{\#AdvLabel} & \textit{\#Category} & \textit{\#CC} \\ \hline 
\multirow{3}{*}{\begin{tabu}[l]{@{}c@{}}\circled{1}: FashionMnist+VGG16\end{tabu}}
 & \textsc{Clover} & 1000 & 1000 & 0.99 & 2000 & 2000 & 0.99 \\ 
 & \textsc{Clover+Gini} & 928 & 859 & 0.30 & 1778 & 1621 & 0.34 \\ 
 &\textsc{Clover+FOL} & 868 & 853 & 0.90 & 1580 & 1542 & 0.89 \\ \tabucline[1.2pt]{1-8} 
\multirow{3}{*}{\circled{2}: SVHN+LeNet5}
 & \textsc{Clover} & 1000 & 1000 & 0.99 & 2000 & 2000 & 0.99 \\ 
 & \textsc{Clover+Gini} & 448 & 357 & 0.27 & 887 & 692 & 0.30 \\ 
 &\textsc{Clover+FOL} & 861 & 856 & 0.91 & 1703 & 1692 & 0.89 \\ \tabucline[1.2pt]{1-8} 
\multirow{3}{*}{\circled{3}: CIFAR10+ResNet20} 
 & \textsc{Clover} & 1000 & 1000 & 0.99 & 2000 & 2000 & 0.99 \\ 
 & \textsc{Clover+Gini} & 977 & 975 & 0.37 & 1897 & 1882 & 0.42 \\ 
 &\textsc{Clover+FOL} & 812 & 795 & 0.95 & 1472 & 1426 & 0.96 \\ \tabucline[1.2pt]{1-8} 
\multirow{3}{*}{\circled{4}: CIFAR100+ResNet56} 
 & \textsc{Clover} & 1000 & 2000 & 0.99 & 2000 & 2000 & 0.99 \\ 
 & \textsc{Clover+Gini} & 976 & 997 & 0.23 & 1896 & 1980 & 0.27 \\ 
 &\textsc{Clover+FOL} & 694 & 604 & 0.78 & 1231 & 1065 & 0.78 \\ \tabucline[1.2pt]{1-8} 
\multirow{5}{*}{\textbf{\begin{tabu}[l]{@{}l@{}} Mean Results \end{tabu}}} 
 & \textsc{Clover} & 1000.00 & 1000.00 & 0.99 & 2000.00 & 2000.00 & 0.99 \\ 
 & \textsc{Clover+Gini} & 832.25 & 797.00 & 0.29 & 1614.50 & 1543.75 & 0.33 \\ 
 & \textsc{Clover+FOL} & 808.75 & 777.00 & 0.89 & 1496.50 & 1431.25 & 0.88 \\ \cline{2-8} 
 & \textsc{Clover}$\div$(\textsc{Clover+Gini}) & 1.20 & 1.25 & 3.39 & 1.24 & 1.30 & 2.97 \\ 
 & \textsc{Clover}$\div$(\textsc{Clover+FOL}) & 1.24 & 1.29 & 1.12 & 1.34 & 1.40 & 1.12 \\ \hline \hline

\multirow{2}{*}{Benchmark Case} & \multirow{2}{*}{Technique} & \multicolumn{3}{c|}{$n_3$ = 4000} & \multicolumn{3}{c|}{$n_3$ = 6000} \\ \cline{3-8}
 &  & \textit{\#AdvLabel} & \textit{\#Category} & \textit{\#CC} & \textit{\#AdvLabel} & \textit{\#Category} & \textit{\#CC} \\ \hline 
\multirow{3}{*}{\begin{tabu}[l]{@{}c@{}}\circled{1}: FashionMnist+VGG16\end{tabu}}
 & \textsc{Clover} & 4000 & 4000 & 0.99 & 6000 & 6000 & 0.99 \\ 
 & \textsc{Clover+Gini} & 3411 & 3073 & 0.40 & 4698 & 4102 & 0.45 \\ 
 &\textsc{Clover+FOL} & 2550 & 2442 & 0.89 & 3229 & 3043 & 0.89 \\ \tabucline[1.2pt]{1-8} 
\multirow{3}{*}{\circled{2}: SVHN+LeNet5}
 & \textsc{Clover} & 4000 & 4000 & 0.99 & 6000 & 6000 & 0.99 \\ 
 & \textsc{Clover+Gini} & 1710 & 1301 & 0.34 & 2364 & 1762 & 0.36 \\ 
 &\textsc{Clover+FOL} & 3143 & 3083 & 0.89 & 4330 & 4243 & 0.89 \\ \tabucline[1.2pt]{1-8} 
\multirow{3}{*}{\circled{3}: CIFAR10+ResNet20} 
 & \textsc{Clover} & 4000 & 4000 & 0.99 & 6000 & 6000 & 0.99 \\ 
 & \textsc{Clover+Gini} & 3629 & 3569 & 0.49 & 5234 & 5087 & 0.55 \\ 
 &\textsc{Clover+FOL} & 2466 & 2334 & 0.96 & 3054 & 2849 & 0.96 \\ \tabucline[1.2pt]{1-8} 
\multirow{3}{*}{\circled{4}: CIFAR100+ResNet56} 
 & \textsc{Clover} & 4000 & 4000 & 0.99 & 6000 & 6000 & 0.99 \\ 
 & \textsc{Clover+Gini} & 3573 & 3889 & 0.33 & 5070 & 5595 & 0.38 \\ 
 &\textsc{Clover+FOL} & 2196 & 1889 & 0.76 & 2922 & 2562 & 0.76 \\ \tabucline[1.2pt]{1-8} 
\multirow{5}{*}{\textbf{\begin{tabu}[l]{@{}l@{}} Mean Results \end{tabu}}} 
 & \textsc{Clover} & 4000.00 & 4000.00 & 0.99 & 6000.00 & 6000.00 & 0.99 \\ 
 & \textsc{Clover+Gini} & 3080.75 & 2958.00 & 0.39 & 4341.50 & 4136.50 & 0.44 \\ 
 & \textsc{Clover+FOL} & 2588.75 & 2437.00 & 0.87 & 3383.75 & 2174.25 & 0.88 \\ \cline{2-8} 
 & \textsc{Clover}$\div$(\textsc{Clover+Gini}) & 1.30 & 1.35 & 2.54 & 1.38 & 1.45 & 2.27 \\ 
 & \textsc{Clover}$\div$(\textsc{Clover+FOL}) & 1.55 & 1.64 & 1.13 & 1.77 & 1.89 & 1.13 \\ \hline
\end{tabu}
}
\end{table}

\begin{table}[t]
\caption{Results of \textit{All} Test Cases Generated by \textsc{Clover} Variants on Fuzzing Clean Models in Configuration \textit{B}}
\label{tab: rq4_fuzz_clean_model_variants_universe}
\begin{tabu}{|l|c|c|c|c|}
\hline
\multirow{2}{*}{Benchmark Case} & \multirow{2}{*}{Technique} & \multicolumn{3}{c|}{\textit{All}} \\ \cline{3-5}
 &  & \textit{\#AdvLabel} & \textit{\#Category} & \textit{\#CC} \\ \hline 
\multirow{3}{*}{\begin{tabular}[c]{@{}c@{}}\circled{5}: CIFAR10+ResNet20\end{tabular}} 
 &\textsc{Clover} & 23638 & 17457 & 0.96 \\
 &\textsc{Clover+Gini} & 24891 & 17471 & 0.96 \\
 &\textsc{Clover+FOL} & 14565 & 13092 & 0.89 \\ \tabucline[1.2pt]{1-5} 
\multirow{3}{*}{\begin{tabular}[c]{@{}c@{}}\circled{6}: CIFAR10+ResNet20 \end{tabular}} 
 &\textsc{Clover} & 21777 & 15975 & 0.94 \\
 &\textsc{Clover+Gini} & 21902 & 15988 & 0.93 \\
 &\textsc{Clover+FOL} & 19597 & 15704 & 0.86 \\ \tabucline[1.2pt]{1-5} 
\multirow{3}{*}{\begin{tabular}[c]{@{}c@{}}\circled{7}: CIFAR100+ResNet56\end{tabular}}
 &\textsc{Clover} & 27407 & 17169 & 0.98 \\
 &\textsc{Clover+Gini} & 28117 & 17169 & 0.97 \\
 &\textsc{Clover+FOL} & 18966 & 16517 & 0.94 \\ \tabucline[1.2pt]{1-5} 
\multirow{3}{*}{\begin{tabular}[c]{@{}c@{}}\circled{8}: CIFAR100+ResNet56\end{tabular}} 
 &\textsc{Clover} & 20237 & 13596 & 0.87 \\
 &\textsc{Clover+Gini} & 26211 & 13707 & 0.91 \\
 &\textsc{Clover+FOL} & 14202 & 13438 & 0.72 \\ \tabucline[1.2pt]{1-5} 
\multirow{5}{*}{\textbf{Mean Results}} &
\textsc{Clover} & 23264.75 & 16049.25  & 0.99 \\ &
\textsc{Clover+Gini} & 25280.25 & 16083.75 & 0.94 \\ &
\textsc{Clover+FOL} & 16832.50 & 14687.75 & 0.85 \\ \tabucline[1.2pt]{2-5} &
\textsc{Clover} $\div$ (\textsc{Clover+Gini}) & 0.92 & 1.00 & 0.99 \\ & 
\textsc{Clover} $\div$ (\textsc{Clover+FOL}) & 1.38 & 1.09 & 1.10 \\ \hline
\end{tabu}
\end{table}

\begin{table}[thb]
\caption{Results of Test Suites Generated by \textsc{Clover} and its Variants on Fuzzing Adversarially Trained Models in Configuration \textit{B} with different $n_3 \in N_3$ and $n_4=18000$} 
\label{tab: rq6_fuzz_adv_trained_model_test_suites_variants}
\resizebox{\textwidth}{!}{
\begin{tabu}{|l|c|c|c|c|c|c|c|}
\hline
\multirow{2}{*}{Benchmark Case} & \multirow{2}{*}{Technique} & \multicolumn{3}{c|}{$n_3$ = 1000} & \multicolumn{3}{c|}{$n_3$ = 2000} \\ \cline{3-8}
 &  & \textit{\#AdvLabel} & \textit{\#Category} & \textit{\#CC} & \textit{\#AdvLabel} & \textit{\#Category} & \textit{\#CC} \\ \hline 
\multirow{3}{*}{\begin{tabu}[l]{@{}c@{}}\circled{5}: CIFAR10+ResNet20\end{tabu}}
 & \textsc{Clover} & 1000 & 1000 & 0.91 & 2000 & 2000 & 0.85 \\ 
 & \textsc{Clover+Gini} & 57 & 46 & 0.23 & 96 & 79 & 0.24 \\ 
 &\textsc{Clover+FOL} & 898 & 898 & 0.60 & 1632 & 1631 & 0.61 \\ \tabucline[1.2pt]{1-8} 
\multirow{3}{*}{\circled{6}: CIFAR10+ResNet20}
 & \textsc{Clover} & 1000 & 1000 & 0.79 & 2000 & 2000 & 0.69 \\ 
 & \textsc{Clover+Gini} & 49 & 37 & 0.20 & 91 & 70 & 0.21 \\ 
 &\textsc{Clover+FOL} & 874 & 873 & 0.50 & 1523 & 1519 & 0.51 \\ \tabucline[1.2pt]{1-8} 
\multirow{3}{*}{\circled{7}: CIFAR100+ResNet56} 
 & \textsc{Clover} & 1000 & 1000 & 0.82 & 2000 & 2000 & 0.71 \\ 
 & \textsc{Clover+Gini} & 154 & 127 & 0.11 & 295 & 235 & 0.12 \\ 
 &\textsc{Clover+FOL} & 873 & 873 & 0.49 & 1510 & 1502 & 0.49 \\ \tabucline[1.2pt]{1-8} 
\multirow{3}{*}{\circled{8}: CIFAR100+ResNet56} 
 & \textsc{Clover} & 1000 & 1000 & 0.68 & 2000 & 2000 & 0.56 \\ 
 & \textsc{Clover+Gini} & 128 & 101 & 0.09 & 253 & 197 & 0.10 \\ 
 &\textsc{Clover+FOL} & 850 & 844 & 0.40 & 1447 & 1436 & 0.40 \\ \tabucline[1.2pt]{1-8} 
\multirow{5}{*}{\textbf{\begin{tabu}[l]{@{}l@{}} Mean Results \end{tabu}}} 
 & \textsc{Clover} & 1000.00 & 1000.00 & 0.80 & 2000.00 & 2000.00 & 0.70 \\ 
 & \textsc{Clover+Gini} & 97.00 & 77.75 & 0.16 & 183.75 & 145.25 & 0.17 \\ 
 &\textsc{Clover+FOL} & 873.75 & 872.00 & 0.50 & 1528.00 & 1522.00 & 0.50 \\ \cline{2-8} 
 & \textsc{Clover}$\div$(\textsc{Clover+Gini}) & 10.31 & 12.86 & 5.08 & 10.88 & 13.77 & 4.20 \\ 
 & \textsc{Clover}$\div$(\textsc{Clover+FOL}) & 1.14 & 1.15 & 1.61 & 1.31 & 1.31 & 1.40 \\  \hline \hline

\multirow{2}{*}{Benchmark Case} & \multirow{2}{*}{Technique} & \multicolumn{3}{c|}{$n_3$ = 4000} & \multicolumn{3}{c|}{$n_3$ = 6000} \\ \cline{3-8}
 &  & \textit{\#AdvLabel} & \textit{\#Category} & \textit{\#CC} & \textit{\#AdvLabel} & \textit{\#Category} & \textit{\#CC} \\ \hline 
\multirow{3}{*}{\begin{tabu}[l]{@{}c@{}}\circled{5}: CIFAR10+ResNet20\end{tabu}}
 & \textsc{Clover} & 4000 & 4000 & 0.72 & 5595 & 5550 & 0.66 \\ 
 & \textsc{Clover+Gini} & 202 & 167 & 0.25 & 283 & 240 & 0.26 \\ 
 &\textsc{Clover+FOL} & 2630 & 2614 & 0.61 & 3333 & 3307 & 0.60 \\ \tabucline[1.2pt]{1-8} 
\multirow{3}{*}{\circled{6}: CIFAR10+ResNet20}
 & \textsc{Clover} & 4000 & 4000 & 0.56 & 4560 & 4431 & 0.56 \\ 
 & \textsc{Clover+Gini} & 170 & 135 & 0.22 & 239 & 194 & 0.23 \\ 
 &\textsc{Clover+FOL} & 2424 & 2401 & 0.50 & 2900 & 2874 & 0.50 \\ \tabucline[1.2pt]{1-8} 
\multirow{3}{*}{\circled{7}: CIFAR100+ResNet56} 
 & \textsc{Clover} & 4000 & 4000 & 0.57 & 5213 & 4911 & 0.53 \\ 
 & \textsc{Clover+Gini} & 545 & 442 & 0.14 & 817 & 678 & 0.16 \\ 
 &\textsc{Clover+FOL} & 2431 & 1402 & 0.49 & 2973 & 2925 & 0.49 \\ \tabucline[1.2pt]{1-8} 
\multirow{3}{*}{\circled{8}: CIFAR100+ResNet56} 
 & \textsc{Clover} & 4000 & 4000 & 0.44 & 4525 & 4245 & 0.42 \\ 
 & \textsc{Clover+Gini} & 466 & 380 & 0.11 & 687 & 571 & 0.12 \\ 
 &\textsc{Clover+FOL} & 2202 & 2174 & 0.41 & 2690 & 2630 & 0.40 \\ \tabucline[1.2pt]{1-8} 
\multirow{5}{*}{\textbf{\begin{tabu}[l]{@{}l@{}} Mean Results \end{tabu}}} 
 & \textsc{Clover} & 4000.00 & 4000.00 & 0.57 & 4973.42 & 4784.58 & 0.54 \\ 
 & \textsc{Clover+Gini} & 345.75 & 281.00 & 0.18 & 506.50 & 420.75 & 0.19 \\ 
 &\textsc{Clover+FOL} & 2421.75 & 2147.75 & 0.50 & 2974.00 & 2934.00 & 0.50 \\ \cline{2-8} 
 & \textsc{Clover}$\div$(\textsc{Clover+Gini}) & 11.57 & 14.23 & 3.19 & 9.82 & 11.37 & 2.81 \\ 
 & \textsc{Clover}$\div$(\textsc{Clover+FOL}) & 1.65 & 1.86 & 1.14 & 1.67 & 1.63 & 1.09 \\ \hline
\end{tabu}
}
\end{table}

\begin{table}[t]
\caption{Results of \textit{All} Test Cases Generated by \textsc{Clover} and its Variants on Fuzzing Adversarially Trained Models in Configuration \textit{B}}
\label{tab: rq6_fuzz_adv_trained_model_universe_variants}
\begin{tabu}{|l|c|c|c|c|}
\hline
Benchmark Case & Technique & \textit{\#AdvLabel} & \textit{\#Category} & \textit{\#CC} \\ \hline
\multirow{3}{*}{\begin{tabular}[c]{@{}c@{}}\circled{5}: CIFAR10+ResNet20\end{tabular}} 
 &\textsc{Clover} & 5627 & 5550 & 0.59 \\
 &\textsc{Clover+Gini} & 5293 & 5068 & 0.57 \\
 &\textsc{Clover+FOL} & 5288 & 5073 & 0.60 \\ \tabucline[1.2pt]{1-5} 
\multirow{3}{*}{\begin{tabular}[c]{@{}c@{}}\circled{6}: CIFAR10+ResNet20 \end{tabular}} 
 &\textsc{Clover} & 4624 & 4431 & 0.49 \\
 &\textsc{Clover+Gini} & 4305 & 4105 & 0.48 \\
 &\textsc{Clover+FOL} & 4289 & 4108 & 0.49 \\ \tabucline[1.2pt]{1-5} 
\multirow{3}{*}{\begin{tabular}[c]{@{}c@{}}\circled{7}: CIFAR100+ResNet56\end{tabular}}
 &\textsc{Clover} & 5375 & 4911 & 0.49 \\
 &\textsc{Clover+Gini} & 4898 & 4530 & 0.49 \\
 &\textsc{Clover+FOL}  & 4865 & 4533 & 0.49 \\ \tabucline[1.2pt]{1-5} 
\multirow{3}{*}{\begin{tabular}[c]{@{}c@{}}\circled{8}: CIFAR100+ResNet56\end{tabular}} 
 &\textsc{Clover} & 4646 & 4245 & 0.40 \\
 &\textsc{Clover+Gini} & 4130 & 3844 & 0.40 \\
 &\textsc{Clover+FOL}  & 4082 & 3847 & 0.40 \\ \tabucline[1.2pt]{1-5} 
\multirow{5}{*}{\textbf{Mean Results}} &
\textsc{Clover} & 5067.92 & 4784.58 & 0.49 \\ &
\textsc{Clover+Gini} & 4656.50 & 4386.75 & 0.49 \\ &
\textsc{Clover+FOL} & 4631.00 & 4390.25 & 0.50 \\ \tabucline[1.2pt]{2-5} &
\textsc{Clover} $\div$ (\textsc{Clover+Gini}) & 1.09 & 1.09 & 1.01 \\ & 
\textsc{Clover} $\div$ (\textsc{Clover+FOL}) & 1.09 & 1.09 & 0.99 \\ \hline
\end{tabu}
\end{table}

\newpage
\section{Additional Exploratory Study}
In this section, we describe an exploratory study to explore the ability of \textsc{Clover} to generate test cases that violate the neural network invariants extracted from defenders.  
We recall that given a coverage criterion (e.g., neuron coverage), covering more coverage items in a DL model by a test suite does not necessarily indicate that the test suite will be more useful in robustness improvement. Moreover, the grounded theory between a coverage criterion and robustness improvement remains to be developed.
Thus, through this exploratory study, we do not aim to give a conclusion on the correlation between abnormal neuron network state and how these violations relate to the quality of the test cases generated from the model under test in general. 
Rather, 
we would like to lay the groundwork for future research by exploring whether neural network invariant violation may be another evaluation dimension for fuzzing techniques.

Recall that a neural network invariant extracted from a model under fuzzing is constructed from the dataset for training the model under fuzzing.
Intuitively, it captures the essence of the model under fuzzing against the training task. 
We assume that fuzzing techniques are not able to obtain the outputs from the defenders and know their algorithms and defense strategies. Thus, the fuzzing techniques only generate test cases based on the models under fuzzing.
Violating an invariant might be seen as an indirect indicator of whether a test case has the potential to be further developed to escape from the monitor of the defender in question.

Readers may want to know the implications of invariant violations.
A test suite with more test cases passing through the invariants can seem to be more demanding for the defender to defend with (note that all test cases are already adversarial examples of the model under fuzzing).
So, from the defender's viewpoint, such test suites will be more interesting. 
Nonetheless, the focus of fuzzing is on the model under fuzzing.
If a test case violates fewer invariants, intuitively, its feature vector is closer to a typical value in the latent space of the defender over the training dataset to be within the normal range.
A defender has a series of neural network invariants, each invariant having its own perspective to determine whether a feature vector is within a normal range. 
Thus, a test case violating a higher number of invariants indicates that the test case contains more feature combinations that violate these normal ranges simultaneously. 
Since all these invariants are constructed on the training dataset of the model under fuzzing and the latent space of the model under fuzzing,
intuitively, the test case, as an adversarial example, contains more features that the model under fuzzing should be learned against with so that the improved version of the model (probably after retraining) can avoid classifying the test case wrongly. In this regard, such a test case is a harder adversarial example than the one violating a lower number of invariants.  At the same time, to improve a DL model, a test suite rather than a single test case is required. 
Nonetheless, how the features of a test case in one invariant are connected to the feature space of the model under test and its retrained version is yet to be discovered. Thus, in the exploratory study, we do not claim that a higher number of violations achieved by a test suite indicates that the test suite is better for the model under fuzzing.

We aim to study the following two research questions:
\begin{itemize}
\item[RQ7:]
To what extent is \textsc{Clover} in Configuration \textit{A} effective in exposing abnormal neural network states of defenders \cite{nic}  for the models under test?

\item[RQ8:]
To what extent is \textsc{Clover} in Configuration \textit{B} effective in exposing abnormal neural network states of defenders \cite{nic} for the models under test?
\end{itemize}

\textbf{Experiment 7 (for Answering RQ7):} 
For \textsc{nic}, we follow the description and parameters in \cite{nic} and adopt the OSVM \cite{osvm} model implemented in the \emph{sklearn} library with radial basis function kernel, $nu = 1e^{-5}$, and the default values for all the other parameters.

To measure the effectiveness of test suites either generated by fuzzing techniques or test case selection techniques on the neural network states of a model under test, for each such test case, we count the number of abnormal neural network states (generated by \textsc{nic} from the model under test) in the forward pass for the test case. We refer to the counted value for each test case as the \textit{\textbf{number of abnormal neural network states}}.

For each model under test, we implement \textsc{nic} (see Section~\ref{sec: nic}) and configure the parameters following the description in \cite{nic} because their source code is not publicly available in its project repository \cite{nic_github}.

We first apply \textsc{nic} on each model and the training dataset of the model to output the set of neural network invariant models for the model.
There are 25, 9, 39, and 111 invariant models for the four cases (\circled{1} to \circled{4}), respectively. 
We then apply each test case selection technique (\textsc{Random}, \textsc{DeepGini}, \textsc{be-st}, \textsc{km-st} and \textsc{Clover}) to select 2000 test cases from the selection universe $P_{train}^{\textsc{FGSM+PGD}}$ for the model under test.
We finally run the model under test to infer each selected test case and measure the number of abnormal neural network states incurred by its forward pass. 
Owing to the large number of extracted feature maps required to produce each invariant model and the many invariant models needed to be created, the application of \textsc{nic} on each model under test is time-consuming in both code development and execution.

\textbf{Experiment 8 (for Answering RQ8)}:
We repeat Experiment 3 with $n_4$=18000, except that we use the test suites with 6000 test cases generated by each fuzzing technique.

\textbf{Answering RQ7:} 
Fig. \ref{fig: rq7} summarizes the result of Experiment 3. 
There are four plots, one for each case (\circled{1} to \circled{4}).
Each boxplot in each plot is the number of abnormal neural network states (indicated by the $y$-axis) of the technique indicated by the $x$-value.
The $y$-axis in subplots (a)--(b) and (c)--(d) are the number of abnormal neural network states and the logarithmic value (with base 2) of the number of abnormal neural network states, respectively.
The numbers of abnormal neural network states exposed by \textsc{Random}, \textsc{DeepGini}, \textsc{be-st}, \textsc{km-st}, and \textsc{Clover} are summarized (from left to right for each case) as follows. Across the board, in general, \textsc{Clover} is more effective than \textsc{be-st}, followed by the group for \textsc{km-st} and \textsc{Random} and then \textsc{DeepGini}.
\begin{enumerate}[leftmargin=2cm]
    \item[case \circled{1}:] 5754, 2400, 8942, 7645, \textbf{9656}.
    \item[case \circled{2}:] 607, 45, 1303, 374, \textbf{1160}.
    \item[case \circled{3}:] 3841, 716, 7661, 3009, \textbf{15623}.
    \item[case \circled{4}:] 7831, 6415, 9577, 6551, \textbf{24695}.
\end{enumerate}

In all four cases, observed from Fig. \ref{fig: rq7}, \textsc{Clover} also consistently achieves a higher median in the number of abnormal neural network states than the other techniques.

\begin{figure*}[t]
    \subfigure[FashionMnist+VGG16]{\includegraphics[width=0.3\linewidth]{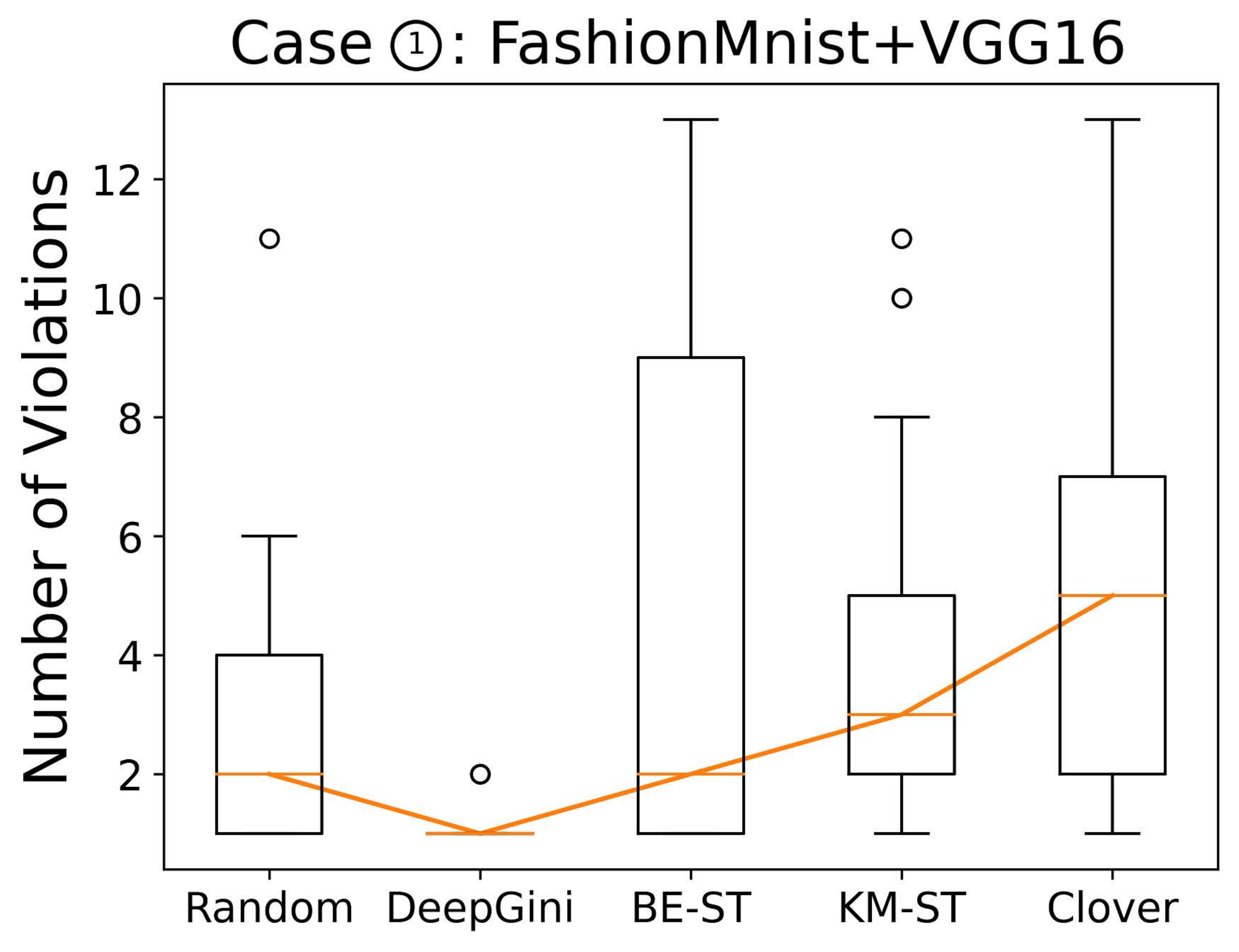}} \hspace*{1cm}
    \subfigure[SVHN+LeNet5]{\includegraphics[width=0.3\linewidth]{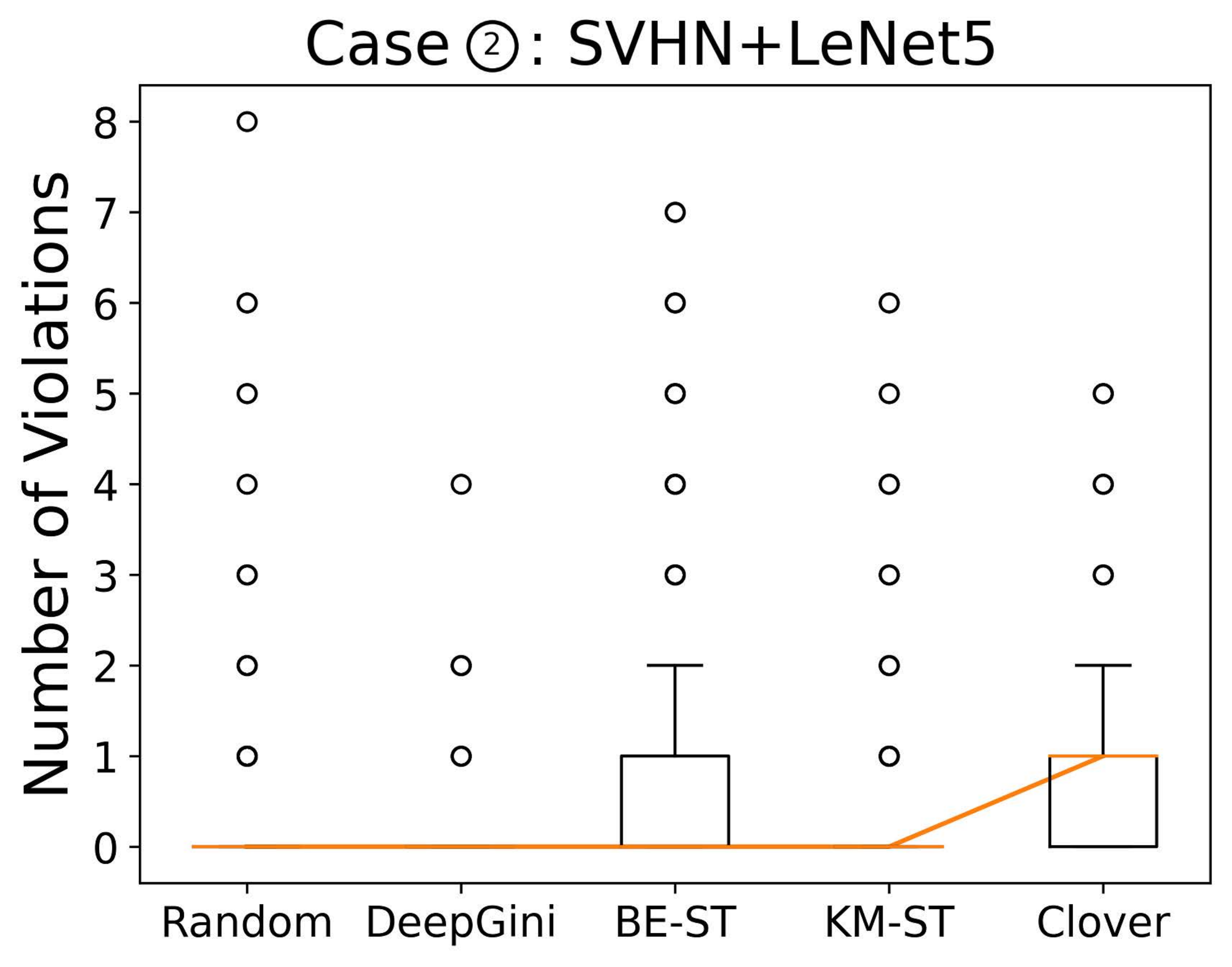}} \vspace*{2ex} \\
    \subfigure[CIFAR10+ResNet20]{\includegraphics[width=0.3\linewidth]{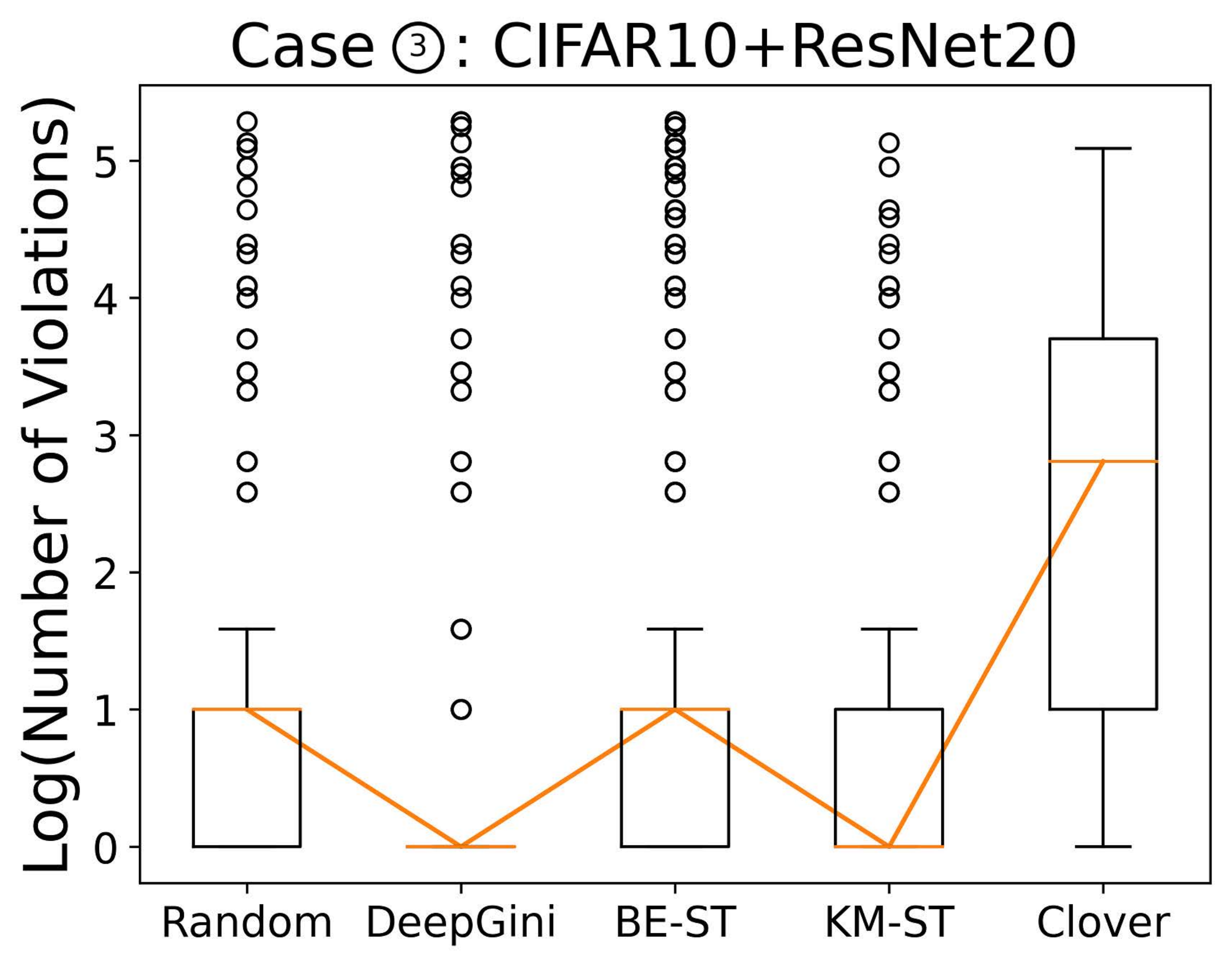}} \hspace*{1cm}
    \subfigure[CIFAR100+ResNet56]{\includegraphics[width=0.3\linewidth]{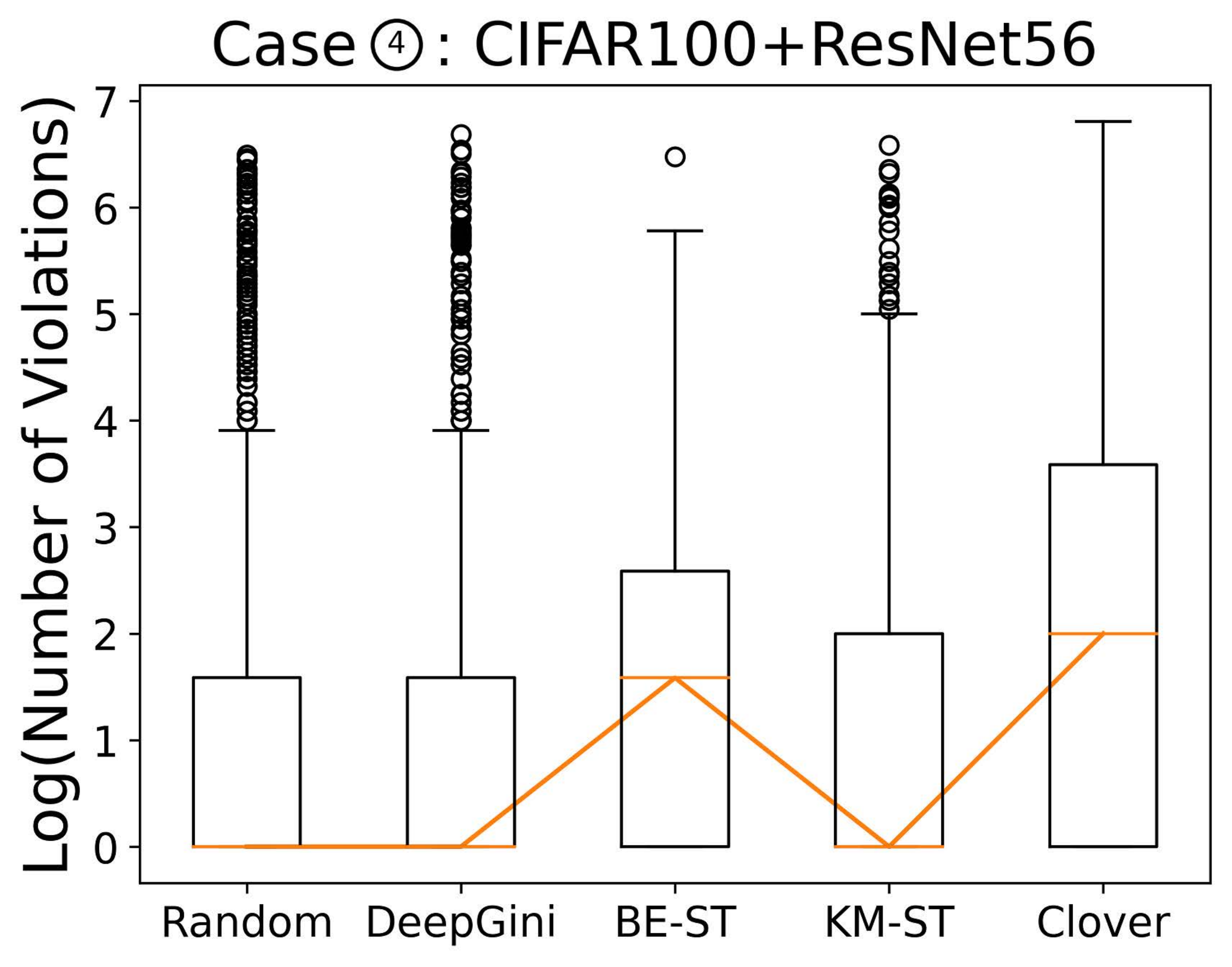}}
    \caption{Numbers of Abnormal Neural Network States Exposed by Different Techniques in Configuration \textit{A}}
    \label{fig: rq6}
\end{figure*}

We also conduct the Mann-Whitney U test with Bonferroni correction at the 5\% significance level and calculate the effect sizes by Cohen's $d$ to check whether the differences in the number of exposed abnormal neural network states between \textsc{Random} and the other four techniques are statistically meaningful. 

We summarize the main findings from the statistical comparison below.

In all four cases, \textsc{Clover} is significantly different from \textsc{Random}, where the $p$-values are all smaller than $1e^{-5}$, and its means are larger than the means for \textsc{Random}.
\textsc{be-st} has the widest range of outlier points above the box.
In all other comparisons between \textsc{Clover} and each peer technique, the means for \textsc{Clover} are always larger.
The effect size of comparing \textsc{Clover} to \textsc{Random} for cases \circled{1} to \circled{4} are 0.51, 0.43, 1.09, and 0.53. 
The observed differences in effect size between \textsc{Clover} and \textsc{Random} is medium or higher in three out of four cases, and the last is close to medium (case \circled{2}). Their observed differences are not negligible.
We also find that the pair of \textsc{Clover} and \textsc{Random} produces a larger effect size than each pair of \textsc{Random} and a peer technique.
    
\textsc{DeepGini} is significantly different from \textsc{Random} ($p$-values all smaller than $1e^{-5}$) except in case \circled{3}.
In case \circled{3}, the means for \textsc{DeepGini} are always smaller than those for \textsc{Random} in all four cases. The effect sizes for the \textsc{Random} and \textsc{DeepGini} pair are 0.17, 0.59, 0.19, and 0.08 for cases \circled{1} to \circled{4}, meaning that the differences are small in all but case \circled{2} and could be negligible. In case \circled{2}, \textsc{DeepGini} is smaller than \textsc{Random} in both mean and median,  \textsc{Random} is a better choice.

\textsc{be-st} is significantly different from \textsc{Random} with $p$-values equal to 0.01, $\leq 1e^{-5}$, $2.41e^{-3}$, and $\leq 1e^{-5}$ for the four cases, respectively.
However, their effect sizes are only 0.11, 0.35, 0.33, and 0.11, respectively, meaning that the differences are either very small or small, which are negligible.
    
\textsc{km-st} is significantly different from \textsc{Random} in two cases (\circled{2} and \circled{3}) out of the four.
The $p$-values are smaller than $1e^{-5}$ in these two cases.
The means for \textsc{km-st} are smaller than those for \textsc{Random} in all cases except case \circled{1}.
In cases \circled{1} and \circled{4}, we cannot find there are any significant differences between \textsc{km-st} and \textsc{Random} with the $p$-values equal to 0.61 and 0.37, respectively.
In the remaining two cases (\circled{2} and \circled{3}), the effect sizes are 0.20 and 0.05 only, respectively, meaning that the differences are small and could be negligible.
 
\textsc{Clover} is significantly different from \textsc{DeepGini}, \textsc{be-st}, and \textsc{km-st} ($p$-values all smaller than $1e^{-5}$) in all cases.
The Cohen's $d$ effect sizes between \textsc{Clover} and \textsc{DeepGini}, \textsc{be-st}, and \textsc{km-st} are:
0.58, 0.40, 0.48 for Case \circled{1};
1.17, 0.07, 0.67 for Case \circled{2};
1.46, 0.51, 1.27 for Case \circled{3}; and 
0.59, 0.50, 0.60 for Case \circled{4}, respectively.
Based on the effect size, the difference between \textsc{Clover} and \textsc{be-st} in Case \circled{2} could be negligible because the effect size is only 0.07.
The effect sizes between \textsc{Clover} and \textsc{be-st}, \textsc{km-st} for Case \circled{1} are 0.40 and 0.48, respectively, which are close to medium and could not be ignored.
The effect sizes in the remaining cases are all greater than the medium level (and even greater than the very large level in some cases).

The overall result indicates that \textsc{Clover} produces test cases that expose a higher number of abnormal neural network states than the peer techniques.
The effect of \textsc{Clover} relative to other techniques is not due to randomness in a statistical meaning way ($p$-value < 0.0001, and medium or higher effect size in three cases and close to the medium in the remaining case).

\begin{figure*}[t]
    \subfigure[FashionMnist+VGG16]{\includegraphics[width=0.3\linewidth]{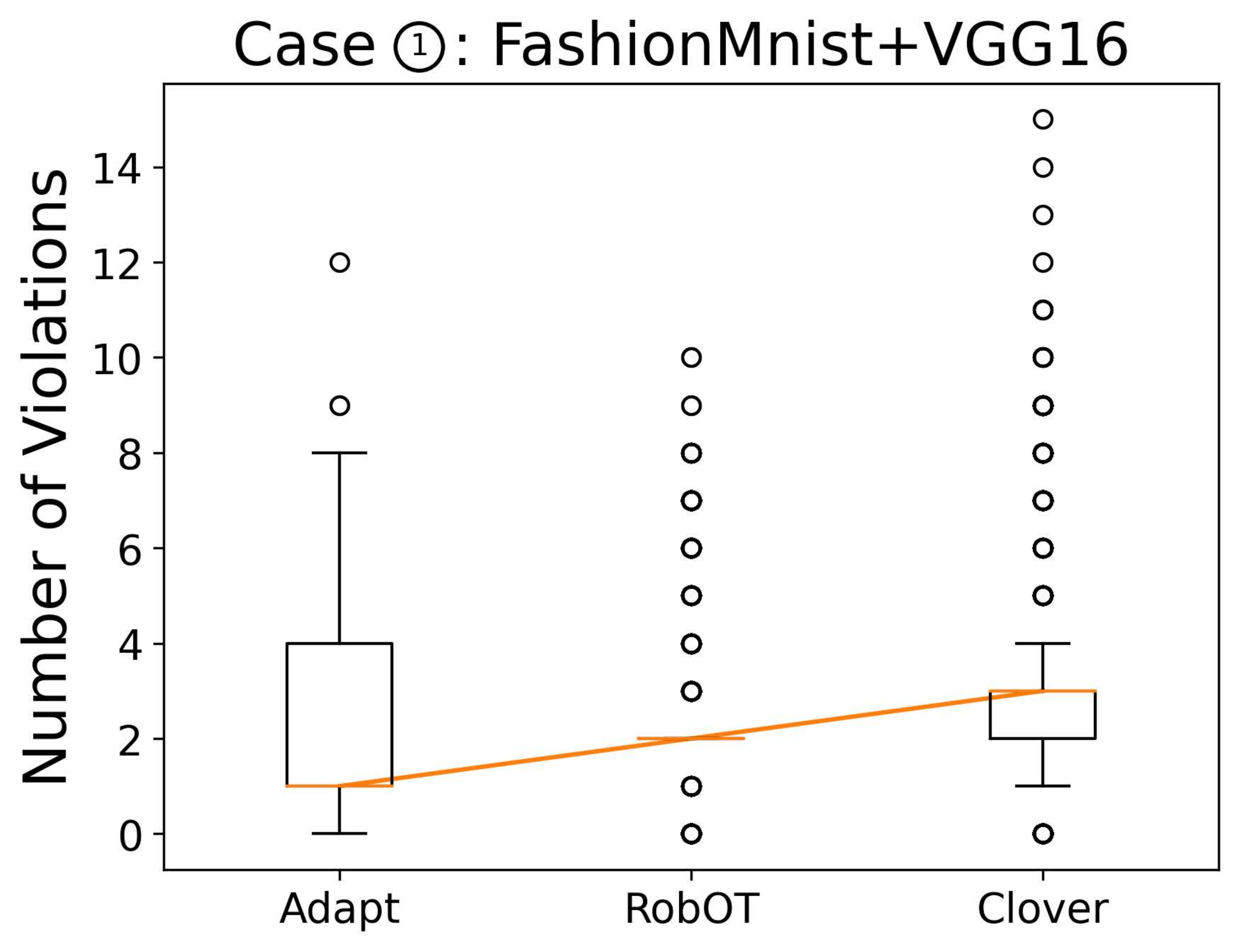}} \hspace*{1cm}
    \subfigure[SVHN+LeNet5]{\includegraphics[width=0.3\linewidth]{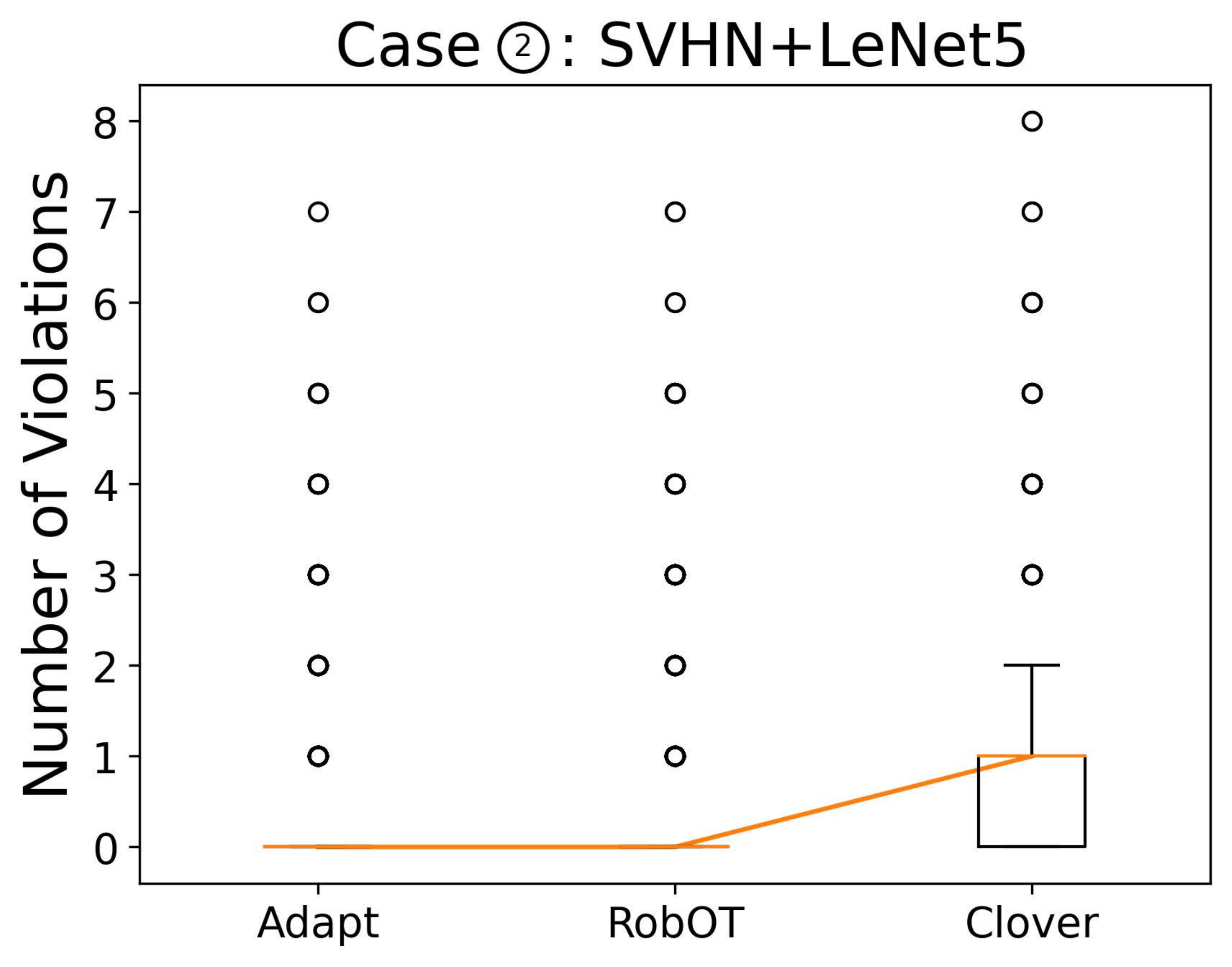}} \vspace*{2ex} \\
    \subfigure[CIFAR10+ResNet20]{\includegraphics[width=0.3\linewidth]{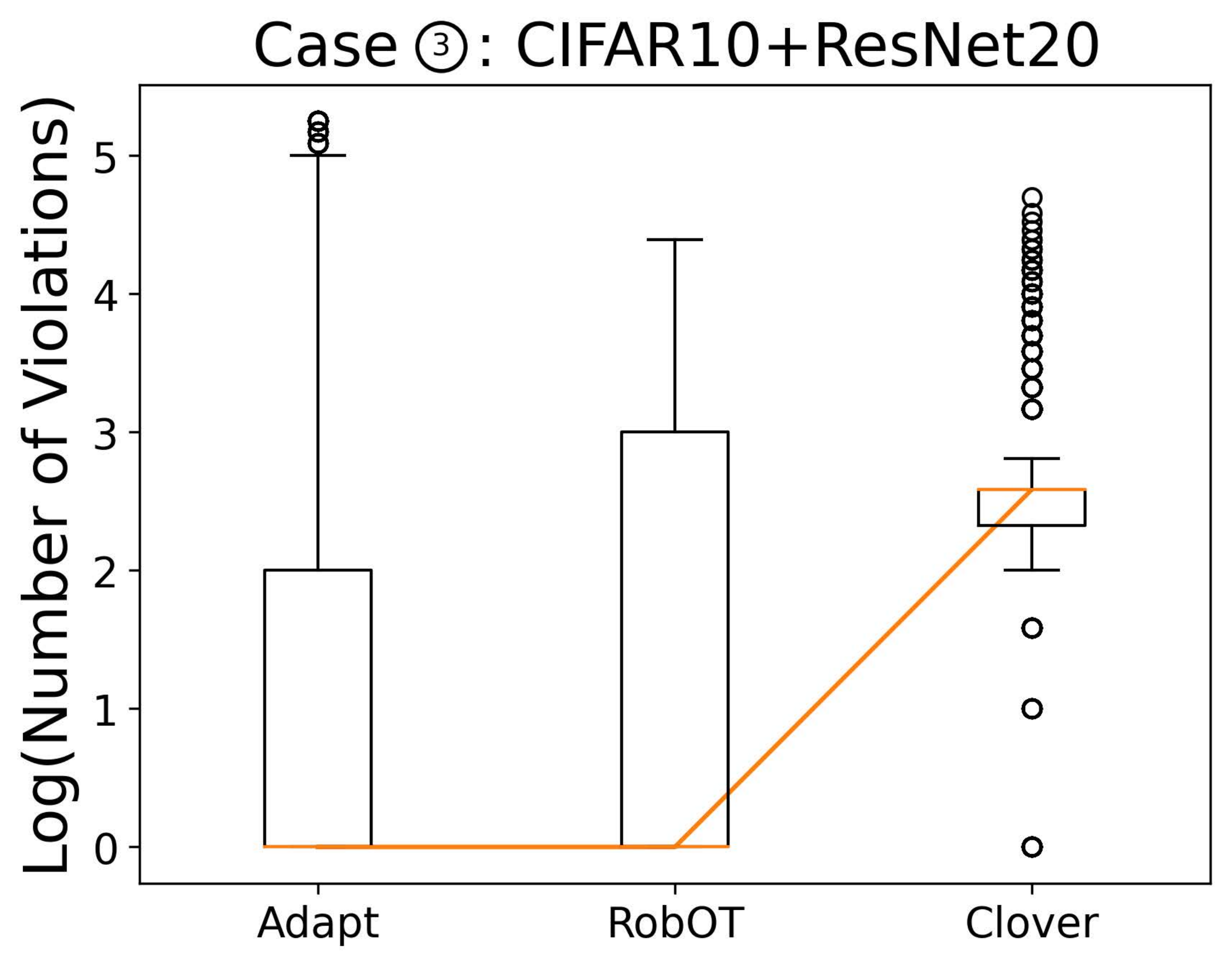}} \hspace*{1cm}
    \subfigure[CIFAR100+ResNet56]{\includegraphics[width=0.3\linewidth]{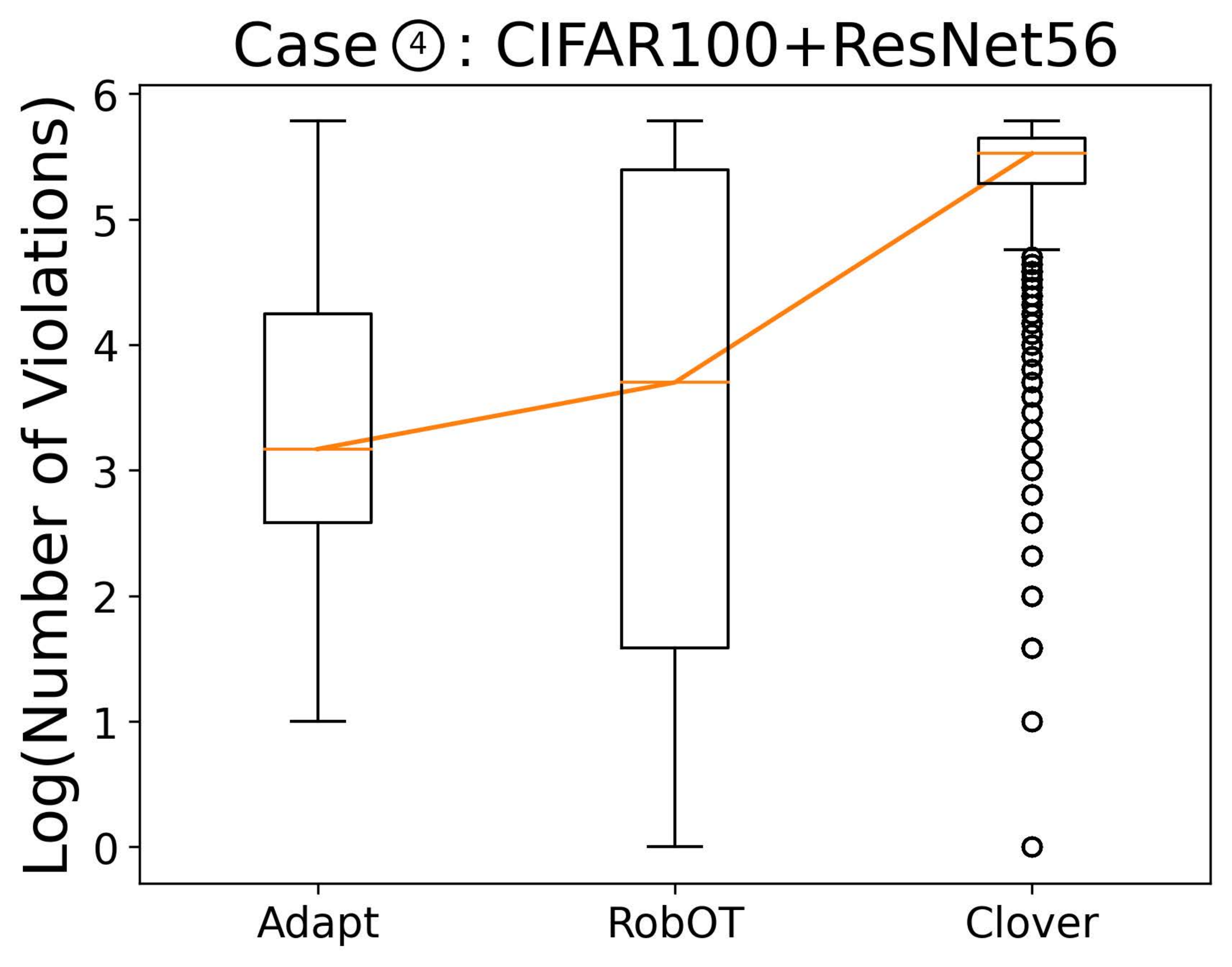}}
\caption{Numbers of Abnormal Neural Network States Exposed by Different Techniques in Configuration \textit{B}}
\label{fig: rq7}
\end{figure*}

\textbf{Answering RQ8}:
Fig. \ref{fig: rq7} summarizes the result of Experiment 6.
Its $x$-axis and $y$-axis can be interpreted like Fig. \ref{fig: rq7}.
The three techniques along the $x$-axis from left to right are \textsc{Adapt}, \textsc{RobOT}, and \textsc{Clover}. 
The numbers of abnormal neural network states exposed by \textsc{Adapt}, \textsc{RobOT}, and \textsc{Clover} in each of the four cases are summarized below from left to right.

\begin{enumerate}[leftmargin=2cm]
    \item[case \circled{1}:] 12539, 13896, \textbf{15227}.
    \item[case \circled{2}:] 1043, 1460, \textbf{3542}.
    \item[case \circled{3}:] 22274, 22182, \textbf{38122}.
    \item[case \circled{4}:] 99695, 127521, \textbf{247737}.
\end{enumerate}

In all four cases, \textsc{Clover} achieves higher medians in the number of abnormal neural network states than the other techniques.
On average, \textsc{Clover} exposed 125\% and 90\% more abnormal neural network states compared to \textsc{Adapt} and \textsc{RobOT}, respectively.

We conduct the Mann-Whitney U test with Bonferroni correction and calculate the effect size by Cohen's $d$ to check whether the difference in the number of abnormal neural network states between \textsc{Clover} and each of the other two techniques (\textsc{Adapt} and \textsc{RobOT}) is statistically meaningful.

The $p$-values for cases \circled{1} to \circled{4} are all smaller than $1e^{-5}$, meaning there are significant differences between \textsc{Clover} and the other two techniques (\textsc{Adapt} and \textsc{RobOT}) at the 5\% significance level.
The effect sizes for cases \circled{1} to \circled{4} are: 0.27, 0.64, 0.54, and 1.70 between \textsc{Clover} and \textsc{Adapt}, and 0.15, 0.52, 0.59, and 1.24 between \textsc{Clover} and \textsc{RobOT}. 
Only the effect sizes (case \circled{1} for \textsc{Clover} and \textsc{RobOT}) are at the small level. 
All the other effect sizes are at the medium or higher level, which means the observed differences are not negligible.

We also recall that FGSM/PGD and fuzzing techniques are used to generate the selection universes in Experiment 3 of Configuration $A$. In contrast, the test suites are generated by \textsc{Clover} in Experiment 6 of Configuration $B$.
On average, each test case of \textsc{Random} and \textsc{Clover} expose 2.25 and 6.39 abnormal neural network states in Experiment 3 for Configuration $A$, respectively.
On the other hand, each test case of \textsc{Clover} exposes 12.69 abnormal neural network states in Experiment 6 for Configuration $B$ on average. The difference is large.
We also conduct the Mann-Whitney U test with Bonferroni correction and calculate the effect size by Cohen's $d$ among all four cases to further compare the performance between \textsc{Clover} in Experiments 3 and 6 and between \textsc{Random} in Experiment 3 and \textsc{Clover} in Experiment 6.
The $p$-values are both $1e^{-5}$. The effect sizes are 0.62 (medium) and 0.67 (medium) in the two comparisons, respectively, indicating that the higher effectiveness of \textsc{Clover} in Experiment 6 than both \textsc{Clover} and \textsc{Random} in Experiment 3 is statistically meaningful.

\begin{tcolorbox}[
enhanced, breakable,
attach boxed title to top left = {yshift = -2mm, xshift = 5mm},
boxed title style = {sharp corners},
colback = white, 
title={Answering RQ7}
]
\textsc{Clover} in Configuration \textit{A} is more discriminative than current state-of-the-art test case selection techniques in terms of the number of abnormal neural network states.
\end{tcolorbox}

\begin{tcolorbox}[
enhanced, breakable,
attach boxed title to top left = {yshift = -2mm, xshift = 5mm},
boxed title style = {sharp corners},
colback = white, 
title={Answering RQ8}
]
\textsc{Clover} in Configuration \textit{B} is more discriminating than other robustness-oriented fuzzing techniques in exposing the abnormal neural network states of DL models.
\end{tcolorbox}

\end{document}